\documentstyle[12pt]{article}
\newcommand{\nc}{\newcommand}

\nc{\dbar}{\bar{\partial}}
\catcode`\@=11

\@addtoreset{equation}{section}

\def\@normalsize{\@setsize\normalsize{15pt}\xiipt\@xiipt
\abovedisplayskip 14pt plus3pt minus3pt%
\belowdisplayskip \abovedisplayskip
\abovedisplayshortskip  \z@ plus3pt%
\belowdisplayshortskip  7pt plus3.5pt minus0pt}
\def\small{\@setsize\small{13.6pt}\xipt\@xipt
\abovedisplayskip 13pt plus3pt minus3pt%
\belowdisplayskip \abovedisplayskip
\abovedisplayshortskip  \z@ plus3pt%
\belowdisplayshortskip  7pt plus3.5pt minus0pt
\def\@listi{\parsep 4.5pt plus 2pt minus 1pt
            \itemsep \parsep
            \topsep 9pt plus 3pt minus 3pt}}

\def\underline#1{\relax\ifmmode\@@underline#1\else
        $\@@underline{\hbox{#1}}$\relax\fi}
\@twosidetrue
\relax

\catcode`@=12

\catcode`\@=11

\def\section{\@startsection{section}{1}{\z@}{3.5ex plus 1ex minus
   .2ex}{2.3ex plus .2ex}{\large\bf}}

\def\ps@headings{\def\@oddfoot{}\def\@evenfoot{}
\def\@oddhead{\hbox{}\hfill
        \makebox[.5\textwidth]{\raggedright\ignorespaces --\thepage{}--
        \hfill }}
\def\@evenhead{\@oddhead}
\def\subsectionmark##1{\markboth{##1}{}}
}

\ps@headings

\catcode`\@=12

\relax

\def\figcap{\section*{Figure Captions\markboth
        {FIGURECAPTIONS}{FIGURECAPTIONS}}\list
        {Fig. \arabic{enumi}:\hfill}{\settowidth\labelwidth{Fig. 999:}
        \leftmargin\labelwidth
        \advance\leftmargin\labelsep\usecounter{enumi}}}
 \relax
\def\tablecap{\section*{Table Captions\markboth
        {TABLECAPTIONS}{TABLECAPTIONS}}\list
        {Table \arabic{enumi}:\hfill}{\settowidth\labelwidth{Table 999:}
        \leftmargin\labelwidth
        \advance\leftmargin\labelsep\usecounter{enumi}}}
 \relax
\def\reflist{\section*{References\markboth
        {REFLIST}{REFLIST}}\list
        {[\arabic{enumi}]\hfill}{\settowidth\labelwidth{[999]}
        \leftmargin\labelwidth
        \advance\leftmargin\labelsep\usecounter{enumi}}}
 \relax

\catcode`\@=11

\def\ps@headings{\def\@oddfoot{}\def\@evenfoot{}
\def\@oddhead{\hbox{}\hfill
        \makebox[.5\textwidth]{\raggedright\ignorespaces --\thepage{}--
        \hfill }}
\def\@evenhead{\@oddhead}
\def\subsectionmark##1{\markboth{##1}{}}
}

\ps@headings

\relax

\begin{document}

%%%\begin{titlepage}
\nopagebreak
%%%\title
{\begin{flushright}
%        \vspace*{-1.8in}
      \vspace*{-1.3in}
        {\normalsize  CK-TH-98-002/December 98 }\\[-5mm]
\end{flushright}
%\vfill
\vspace{10mm}
%%%{\large \bf #3}}
\vspace{1cm}
%%%\author{\large #4 \\ #5}
%%%\maketitle
\vskip -1mm
\nopagebreak
%%%\begin{abstract}
%%%{\noindent #6}
%%%\end{abstract}
%\vfill
%%%\vspace{1cm}
%\begin{flushleft}
%\rule{16.1cm}{0.2mm}\\[-3mm]
%$^{1}${\small E--mail: c.kokorelis@sussex.ac.uk}
%\end{flushleft}
%%%\thispagestyle{empty}
%%%\end{titlepage}}
\newcommand{\dal}{\raisebox{0.085cm}
{\fbox{\rule{0cm}{0.07cm}\,}}}
\newcommand{\bb}{\begin{eqnarray}}
\newcommand{\ee}{\end{eqnarray}}
\newcommand{\p}{\partial}
\newcommand{\bp}{{\bar \p}}
\newcommand{\bR}{{\bf R}}
\newcommand{\bC}{{\bf C}}
\newcommand{\bZ}{{\bf Z}}
\newcommand{\bS}{{\bar S}}
\newcommand{\bT}{{\bar T}}
\newcommand{\bU}{{\bar U}}
\newcommand{\bA}{{\bar A}}
\newcommand{\bh}{{\bar h}}
\newcommand{\bu}{{\bf{u}}}
\newcommand{\bv}{{\bf{v}}}
\newcommand{\D}{{\cal D}}
\newcommand{\s}{\sigma}
\newcommand{\Sg}{\Sigma}
\newcommand{\ket}[1]{|#1 \rangle}
\newcommand{\bra}[1]{\langle #1|}
\newcommand{\non}{\nonumber}
\newcommand{\ph}{\varphi}
\newcommand{\la}{\lambda}
\newcommand{\ga}{\gamma}
\newcommand{\ka}{\kappa}
\newcommand{\m}{\mu}
\newcommand{\n}{\nu}
\newcommand{\th}{\vartheta}
\newcommand{\Lie}[1]{{\cal L}_{#1}}
\newcommand{\eps}{\epsilon}
\newcommand{\bz}{\bar{z}}
\newcommand{\bX}{\bar{X}}
\newcommand{\om}{\omega}
\newcommand{\Om}{\Omega}
\newcommand{\we}{\wedge}
\newcommand{\La}{\Lambda}
\newcommand{\bOm}{{\bar \Omega}}
\newcommand{\CA}{{\cal A}}
\newcommand{\CF}{{\cal F}}
\newcommand{\CbF}{\bar{\CF}}
\newcommand{\CAM}{\CA^{(M)}}
\newcommand{\CAS}{\CA^{(\Sg)}}
\newcommand{\CFS}{\CF^{(\Sg)}}
\newcommand{\I}{{\cal I}}
\newcommand{\al}{\alpha}
\newcommand{\be}{\beta}
\newcommand{\cm}{Commun.\ Math.\ Phys.~}
\newcommand{\pr}{Phys.\ Rev.\ D~}
\newcommand{\pl}{Phys.\ Lett.\ B~}
\newcommand{\ibar}{\bar{\imath}}
\newcommand{\jbar}{\bar{\jmath}}
\newcommand{\np}{Nucl.\ Phys.\ B~}
\newcommand{\e}{{\rm e}}
\newcommand{\gsi}{\,\raisebox{-0.13cm}{$\stackrel{\textstyle
>}{\textstyle\sim}$}\,}
\newcommand{\lsi}{\,\raisebox{-0.13cm}{$\stackrel{\textstyle
<}{\textstyle\sim}$}\,}
\date{}
\vspace{2cm}
\begin{center}
{\Large Theoretical and Phenomenological Aspects of Superstring}
\end{center}
\begin{center}
{\Large Theories} 
\end{center}
\vspace{1.5cm}
\begin{center}
{ Christos Epameinonda Kokorelis$^{1}$}
\end{center}
\begin{center}
{\normalsize
School of Mathematical and Physical Sciences, Sussex University\\
\normalsize Brighton BN1 9QH, U. K}
\end{center}
\begin{center}
{\normalsize
D. Phil Thesis,\\
December 1997}
\end{center}
\vspace{9cm}
\begin{flushleft}
\rule{16.1cm}{0.2mm}\\[-3mm]
$^{1}${\small E--mail: c.kokorelis@sussex.ac.uk}
\end{flushleft}

\newpage
\begin{center}
{\bf Abstract}
\end{center}
{ We discuss aspects of heterotic string effective field theories in
orbifold constructions of the heterotic string. We calculate the moduli 
dependence of threshold corrections to gauge couplings in $(2,2)$ symmetric 
orbifold compactifications . We perform the calculation of the threshold 
corrections for a particular class of abelian $(2,2)$ symmetric 
non-decomposable orbifold models, where the internal twist is realized as 
a generalized Coxeter automorphism. We define the limits for the existence 
of states causing singularities in the moduli space in the perturbative 
regime for a generic vacuum of the heterotic string. The 'proof' provides 
evidence for the explanation of the stringy 'Higgs effect'. Furthermore, we 
calculate the moduli dependence of threshold corrections as target space 
invariant free energies for non-decomposable orbifolds, identifying the 
'Hauptmodul' functions for the relevant congruence subgroups. The required 
solutions provide for the $\mu$ mass term generation in the effective low 
energy theory and affect the induced supersymmetry breaking by gaugino 
condensation. In addition, we discuss one loop gauge and gravitational 
couplings in $(0,2)$ non-decomposable orbifold compactifications. In the 
second part of the thesis, the one loop correction to the  K\"ahler metric 
for a generic $N=2$ orbifold compactification of the heterotic string is 
calculated as solution of a partial differential equation. In this way, 
with the use of the one loop string amplitudes, the prepotential of the vector
multiplets of the $N=2$ effective low-energy heterotic strings is calculated 
in decomposable toroidal compactifications of the heterotic string in 
six-dimensional $N=1$ string vacua. This method provides the solution for the 
one loop correction to the prepotential of the vector multiplets for the 
heterotic string compactified on the $K_3 \times T^2$ manifold. Moreover, 
using modular properties of the one loop prepotential, we calculate it for 
$N{=}2$ heterotic strings coming from $N{=}1$ orbifold compactifications of 
the heterotic string based on non-decomposable torus lattices. These sectors 
appear in $N=1$ orbifold compactifications of the heterotic string on 
non-decomposable torus lattices. In the third part of the thesis we discuss 
supersymmetry breaking through gaugino condensates in the presence of the 
subgroups of the modular group $SL(2, \bf Z)$. We examine the way we can 
modify a known semirealistic model to incorporate S- and T- dualities in the 
superpotential of its effective action. We show how the discrete isometries of 
the effective theory restrict the K\"{a}hler potential and the superpotential 
of the effective theory together with implications for the globally 
supersymmetric case. Finally, we discuss the effect of our one-loop 
computation of the first part of the thesis on the $\mu$ term in orbifold 
compactifications.
}

\newpage
\vspace{2cm}
\begin{center}
CONTENTS
\end{center}
\vspace{1in}
Where the * sign appears, is is a sign of indication that at this part
of the thesis, new results are abtained.}
\newline
CONTENTS{\hfill{3}\rightmargin 1cm}
\newline
ACKNOWLEDGEMENTS{\hfill{5}\rightmargin 1cm}
\newline
CHAPTER 1{\hfill{6}\rightmargin 1cm}
\newline
1 {\bf Introduction} {\hfill{7}\rightmargin 1cm}
\newline
CHAPTER 2{\hfill{20}\rightmargin 1cm}
\newline
2  {\bf Compactification on Orbifolds}{\hfill{21}\rightmargin 1cm}
\newline
{\hfill{}\leftmargin 1in}2.1$\;\;$  Toroidal Compactifications
{\hfill{21}\rightmargin 1cm}
\newline
2.2 General Theory of Orbifold Compactifications
{\hfill{24}\rightmargin 1cm}
\newline
2.3 Duality Symmetries{\hfill{34}\rightmargin 1cm}
\newline
CHAPTER 3{\hfill{38}\rightmargin 1cm}
\newline
3 {\bf Aspects of Threshold Corrections to Low Energy
Effective string Theories compactified on Orbifolds}
{\hfill{39}\rightmargin 1cm}
\newline
3.1 Introduction{\hfill{39}\rightmargin 1cm}
\newline 
3.2 Threshold corrections to gauge couplings
{\hfill{45}\rightmargin 1cm}
\newline
3.2.1 Introduction{\hfill{45}\rightmargin 1cm}
\newline
3.3 * Low Energy Threshold Effects and Physical
Singularities
{\hfill{49}\rightmargin 1cm}
\newline
3.4 * Target Space Automorphic functions from string
compactifications{\hfill{52}\rightmargin 1cm}
\newline
3.5 * Threshold corrections to gauge and gravitational
couplings
{\hfill{75}\rightmargin 1cm}
\newline
3.5.1 * Threshod corrections to gauge couplings{\hfill{75
}\rightmargin 1cm}
\newline
3.5.2 * Threshold corrections to gravitational couplings
{\hfill{83}\rightmargin 1cm}
\newline
3.6 * Threshold corrections for the $Z_8$ orbifold{\hfill{85
}\rightmargin 1cm}
\newline
CHAPTER 4{\hfill{91}\rightmargin 1cm}
\newline
4 {\bf Introduction}{\hfill{92}\rightmargin 1cm}
\newline
4.1 Properties of $N=2$ heterotic string and Calabi-Yau vacua
{\hfill{99}\rightmargin 1cm}
\newline
4.2 Special Geometry and Effective Actions{\hfill{107}\rightmargin 1cm}
\newline 
4.3 Low Energy Effective theory of N=2 heterotic Supestrings
and related issues{\hfill{112}\rightmargin 1cm}
\newline 
4.4.4 * One loop correction to the prepotential from string
amplitudes
{\hfill{117}\rightmargin 1cm}
\newline
4.4.1 One loop contribution to the K\"ahler
metric-Preliminaries
{\hfill{117}\rightmargin 1cm}
\newline
4.4.2 * Prepotential of vector multiplets/K"ahler metric
{\hfill{119}\rightmargin 1cm}
\newline
4.5 * One loop prepotential-perturbative aspects
{\hfill{125}\rightmargin 1cm} 
\newline
CHAPTER 5{\hfill{132}\rightmargin 1cm}
\newline
5 {\bf Superpotentials with T and S-duality and effective
$\mu$ terms}{\hfill{133}\rightmargin 1cm}
\newline
5.1 Introduction{\hfill{133}\rightmargin 1cm}
\newline
5.2 * Constraints from duality invariance on the
superpotential
and the K\"ahler potential for the globally and the locally 
supersymetric theory{\hfill{137}\rightmargin 1cm}
\newline
5.3 *  S- and T-dual supersymmetry breaking
{\hfill{140}\rightmargin 1cm}
\newline
5.4 * Effective $\mu$ term in orbifold compactifications
{\hfill{146}\rightmargin 1cm}
\newline
5.4.1 Generalities
{\hfill{146}\rightmargin 1cm}
\newline
CHAPTER 6{\hfill{156}\rightmargin 1cm}
\newline
6 {\bf Conclusions and Future Directions}
{\hfill{157}\rightmargin 1cm}
\newline
CHAPTER 7{\hfill{161}\rightmargin 1cm}
\newline
7.1 {\bf Appendix A}{\hfill{162}\rightmargin 1cm}
\newline
7.2 {\bf References}{\hfill{164}\rightmargin 1cm}

\newpage
\vspace{2cm}
\begin{center}
ACHNOWLEDGEMENTS
\end{center}
\vspace{2cm}
I would like to thank my supervisor Prof. D. Bailin, for giving me
the opportunity to be engaged in research in superstring theory.
For his interest in my work and his support, patience and hospitality.
I also enjoyed his reaction each time I gave him a new outline
of my thesis.

I would like to thank th Department of Physics for its excellent
facilities.

I would also to thank A. Clark, A. Lahiri, M. Hindmarsh from the Physics
department, J. Bennings from the Library, P. Croyden and G. Mills from
the computed center and also I acknowledge discussions during the D.
Phil degree from the following people : D. Bailin, W. Sabra, I.
Antoniadis, S. Thomas, R. Lewes, D. Lust, A. J Scholl, C. Kounnas, J.
Casas,
D. Zagier and B. De Carlos.
 
I would like to thank my friends at Brighton, who made my stay in
England,
exciting and enjoyable.

Finally I would like to express my gratitude to my family, my father
Epaminondas, my mother Maria,my brother Konstantinos for their Moral,
Psychological and Financial support, without whom this work would have
never
been accomplished.

\newpage
\vspace{5cm}
\begin{center}
{\bf CHAPTER 1}
\end{center}

\newpage
\section{\bf Introduction}

String theory still remains, after a period of over ten years of
development, the only candidate for a quantum theory of gravity, which 
we believe to finally describe consistently and accurate the particle 
world. 
A great obstacle, in the interpretation of properties of string theory, 
that could exhibit the novel futures
of the formalism and allow us to extract the physical information,
remains the mathematical structure and our ability
to improve the calculability of perturbative and non-perturbative
aspects of the theory.
However, {\em string theory is unique}. It gives partial or complete 
answers
to the biggest problems of particle physics. Namely, {\em spacetime
supersymmetry breaking}, {\em cosmological constant}, {\em strong and 
weak CP
violation}, {\em flavour changing neutral currents}, {\em determination 
of the unbroken gauge group} of the theory. 
In this Thesis, we will be discussing in detail some of the above
problems, while we will mention the rest.

In addition, in string theory there is another 
another problem, the determination of
the underlying principle able to choose the correct vacuum, of the 
final theory, among the huge degeneracy of string 
vacua\cite{fer1,fer2,fer3,fer4,dhvw}.  
Determination of the correct vacuum will give the realistic
three generation
model which may be extension of the standard model and may include the new
physics beyond the electroweak scale.
This problem is largely unanswered, as string vacua 
provide us with a huge number of possible semirealistic models, which 
all but failed to satisfy all the phenomelogical requirements.
In the most popular scenaria of recent years\cite{iblu,bim,ILR,AEHN} 
 the minimal supersymmetric standard model(MSSM) 
was used a priori as an effective low energy theory 
of string vacua. Its presence gives quantative results for string vacua
since it has not been 
proved to be coming from a particular string vacuum.
We should note that in all the model constructions up to now, where the
gauge group contains part of the standard model gauge group, the particle 
content of string models is not in any case of the MSSM.
When it does appear it is corrected 
by additional particles.

The problem of proliferation of the string vacuum
is left untouched from string perturbation theory.
 In particular, supersymmetric models
coming from superstring vacua appear in great numbers
and there is no undelying principle to
distinguish between different categories apart from phenomenological
criteria\cite{iblu,bim}.

The simplest string model, the classical bosonic string represents
a generalization of the one-dimensional point particle action\footnote{
where the invariant integral is given by
$ds^2 =-g_{\mu \nu}(x) dx^{\mu} dx^{\nu}$.}
 $\textstyle{S = -m \int ds}$, to an n dimensional 
object, which sweeps an $(n+1)$ worldvolume as it moves into space,
described by the metric $h_{\alpha \beta}(\sigma)$ and the
action $S=-(T/2) \int d^{n+1}\sigma \sqrt{h} h^{\alpha \beta}(\sigma)
g_{\mu \nu} \partial_{\alpha} X^{\mu} \partial_{\beta} X^{\nu}$. Here, 
 $h^{\alpha \beta}$ is the inverse of $h_{\alpha \beta}$ and 
$\sigma^0 (\tau)=\tau$ and $\sigma^i, i=1,\dots,n$ is an
n-dimensional object\footnote{
For $n=0$ we have the point particle, for  $n=1$ we have the string,
while for $n=2$ we have the membrane and so on $\dots$.}
, while T is the string tension with coefficient of
(mass)$^{n+1}$. 
Especially, for strings the tension T is given in terms of the Regge
slope $T \propto \frac{1}{\alpha^{\prime}}$. 
In this Thesis, we will discuss properties of oriented closed strings(OCS).
For oriented closed string the physical spectrum is invariant under
the level matching\footnote{See the footnote following \ref{lmc}.}
constraint \ref{lmc}.  
For non-orientable clodes strings physical states must also be invariant
under the operator $\cal F$, which exchanges e.g, for the bosonic 
string, the $\alpha$ and ${\tilde \alpha}$ oscillators of the 
left and right moving number operators 
$N = \sum_{i=1}^{\infty}n \alpha_{n  \dagger}^k 
\alpha_n^k$, ${\tilde N}=\sum_{i=1}^{\infty} n {\tilde \alpha}^{k
\dagger}_n {\tilde \alpha}^k_n$,  respectively.

In the early days of its development string theory was used as a theory 
of hadrons.
The important step in realizing string theory as a theory of
fundamental interactions was taken in \cite{hone,sce}
where it was shown that the effective action of a massless spin two state
is in the zero slope limit given by the Eistein-Hilbert lagrangian.
However, it was not until 1984 when Green and Schwarz\cite{gs1}
proved that the type I theories\footnote{Unoriented 
open and closed strings with $N=1$ supersymmetry.}
are free of anomalies which made the 
physics community realized the importance of superstring theories.  
They proved that, by adding non-gauge
invariant local counterterms for the two form B in the effective low energy
theory\footnote{
The ten dimensional $N=1$ supergravity coupled to matter has anomalies 
coming from hegagon diagrams.} as
\begin{equation}
\delta B= tr(A d\Lambda) - tr (\omega d\theta),
\label{amonal1}
\end{equation}
where $\Lambda$, $\theta$ are infinitesimal gauge and Lorentz trasformations 
of the ten dimensional background gauge field and spin connection A and 
$\omega$ respectively, the theory remains free of gauge, 
gravitational and mixed gauge and 
 gravitational anomalies when the Yang-Mills gauge group is $SO(32)$ 
or $E_8 \times
E_8$. While type I theories with group  $SO(32)$ where known at the time,
the theory corresponding to $E_8 \times E_8$ was not known. It was 
soon after the heterotic string was build, where gauge
symmetries coming from the group $E_8 \times E_8$ were observed, even not
chiral,
that a Kaluza-Klein origin of gauge symmetries as isometries 
of the internal manifold was suggested\cite{na}.
However, compactification of the heterotic string on toroidal backgrounds   
produces unwanted non-chiral models with extended $N=4$ supersymmetry.
Soon after\cite{Calabi1}, by examining the low energy lagrangian 
of $N=1$ supergravity, the low energy limit of the heterotic string when
the Regge slope $\alpha^{\prime} \rightarrow $0, supersymmetric 
solutions in four dimensions were found. They correspond to 
classical solutions of the string equations\footnote{They examined 
the field theory limit of the heterotic string compactified
on the space $M^4 \times K$, where K a compact manifold.}
where the internal manifold is a smooth manifold, the so called
Calabi-Yau manifold. The phenomenologically interesting $N=1$
supersymmetry comes by demanding\footnote{see chapter three}
that the manifold has $SU(3)$ holonomy\cite{yau}.
The theory is subject to gauge symmetry breaking by twisting 
the boundary conditions in a way that does not preserve charges
corresponding to the broken symmetries of the world sheet.
The problem of determination of the gauge group of the theory,
consistently attached to the grand unified theories\cite{La} and their
supergravity successors\cite{nille,lah} in string models is solved.
However, it remains the problem of determing the derivation of 
the standard model from it.  
A number of different methods of producing consistent compactifications
of the heterotic string has been
constructed\cite{dhvw,imnq,fer3,fer4,ge1,kasu,senda1,fer1,fer2} which
all but one will 
not be reviewed in this Thesis. The research carried out in this Thesis,
will be based in the theory of orbifolds\cite{dhvw}.
In this Thesis, we will discuss the results of our research of
 phenomenological and theoretical aspects of 
the orbifold constructions of the heterotic string.
In chapter two, we discuss
elements of the basic theory which is extensively well known\cite{dhvw}.

{\em Orbifolds} are constructed, by twisting boundary conditions to break 
Lorentz symmetries, so that spacetime coordinates of the $x^{\mu}(\sigma)$
are not periodic functions of $\sigma$ but periodic up to 
Lorentz transformations. 
Mathematically, this means that compactifying the heterotic string
on a six dimensional torus T and dividing by a non freely acting
symmetry group G. There are sectors in the theory, called twisted
sectors,  where the 
coordinates of the string obey not $X^i(\sigma+ 2 \pi)= X^i(\sigma)$ but 
$ X^i(\sigma+ 2 \pi)=g X^i(\sigma)$ for $g \in G$.

Superstring theory offers a large number of theoretical arguments 
which single out its uniqueness.
It combines quantum field theory with general relativity.   
It possesses a minimal number of parameters, namely the string 
scale\footnote{
at one loop level the string scale is corrected by moduli dependent 
effects}, and in 
addition a number of fields called moduli, whose vacuum expectation values 
enter the calculation of the basic functions which determine the
fermion and boson terms of the low energy energy $N=1$ lagrangian.
The low energy limit of orbifold models\cite{dhvw} is that of $N=1$ 
supergravity coupled to super Yang-Mills modified by moduli dependent 
effects characterizing the orbifold vacuum\cite{Calabi1,HV,ANTO,ovr1}.

In this way, the running gauge couplings\cite{kj} exhibit a specific
moduli dependence characterized from invariances of the 
moduli variables.  
In chapter three, we discuss the calculation of threshold corrections
for a particular class of non-decomposable orbifold models.
We discuss as well, some aspect of the gauge couplings of the theory 
related, as the string analog of Seiberg-Witten theory \cite{seiwi1}, to 
special points of gauge symmetry enhancement.
It has to do with the appearance of gauge symmetry enhancement at special
points in the moduli space and its contribution to the gauge coupling
constants for different regions in the moduli space.
In general, when in a region of moduli space there is a point where a
previously massive states become massless, then the effective gauge 
couplings exhibit a behavior like in eqn.(\ref{koko1}).

In addition, in  chapter three we discuss the
calculation of {\em target} {\em space free 
energies}\cite{fklz,lust,oova} as 
moduli dependent threshold corrections, coming from
integration of massive compactification modes.
The calculation uses the free energy, the effective action coming from
the integration of the massive compactification modes.
We use this result to calculate the physical effects 
of the integration of the massive  compactification modes to the 
calculation
of the threshold corrections for the gauge and gravitational coupling 
constants in orbifold models.  This method is alternative
to the calculation of the effective action  coming from a direct string 
loop calculation,
and it must give the same result if the associated sums are
performed exactly. Unfortunately at the moment, complete
regularization of the sums is not possible.

The general method of determining the exact string effective action 
of the massless modes comes from the calculation of string diagrams
which have been performed e.g for the heterotic 
string\cite{Gr,Gr1} or for orbifolds\cite{HV,dkl2,anto3,ANTO}.
The low energy theory of $N=1$ orbifolds is fixed in terms of the  
K\"ahler potential K, the superpotential W and the function f which 
determines the kinetic terms for the gauge fields.
In $N=1$ locally supersymmetric theories the superpotential W
calculates the Yukawa couplings of the chiral matter superfields to the
Higgs scalars.
The superpotential of the effective theory of $(2,2)$ $N=1$ orbifolds 
depends on the moduli fields which have a flat potential
to all orders of string perturbation theory. 
As a result their values remain undetermined and the superpotential
is not renormalized to all 
orders of string perturbation theory due to non-renormalization
theorems\cite{marti} but is corrected from
 world sheet instantons\cite{dsww,HV}. 
 As a result moduli dependent contributions to the superpotential
 may be come from non-perturbative effects, infinite genus effects.
 While non-perturbative techniques are not yet available, it is 
 possible to calculate non-perturbative contributions to the 
 superpotential of the effective supersting action below the string 
 unification scale. By integrating out the massive modes coming from 
 compactification 
 modes\cite{fklz,oova}, and taking into account the singularities
 in the moduli space of vacua, we will obtain non-perturbative 
 contributions to the superpotential of $(2,2)$ $N=1$ orbifold 
 compactifications of the heterotic string in chapter three.
 These contributions are coming at the level of the effective
 superstring theory from locally supersymmetric F-terms
 involving more than two spacetime derivatives\cite{ANTO}.

Toroidal compactifications of the heterotic string have
in four dimensions $N=4$ supersymmetry and contain among
other fields, the dilaton $\Phi$, the antisymmetric tensor $B_{\mu \nu}$
which transforms after a duality trasformation to the axion $\tilde
\alpha$ and moduli scalars described by a matrix $M_{ab}$, which
parametrize the coset $SO(6,22)/SO(6) \times SO(22)$. The spectrum and
interactions of the perturbative phase of the effective theory are
invariant under the "duality" symmetry $R \leftrightarrow
\alpha^{\prime}/R$,
which survives in all orders in string perturbation theory\cite{alva}.

In orbifold models the physical quantities, coming from the
compactifications of the heterotic string, depend on
the moduli parameters.
For any orbifold constructions of the heterotic string
associated with toroidal compactifications, the 
specturm in four dimensions contains, among other fields, 
the U modulus is associated with             
the complex structure of the torus and
complex modulus field T given by
 $T= R^2/{\alpha^{\prime}}+ iB$, where 
the vacuum expectation value of the T modulus is associated with the
size R of the compactification 
and the antisymmetric background field B.

The low energy world observed at energies of the electroweak scale
is in perfect agreement with the standard model predictions.
The standard model does not have any supersymmetry. 
However, in the following, we insist on supersymmetric models as 
supersymmetry solves technically the {\em hierarchy problem}\cite{dere2}.
Take for example the standard model. Its lightest scale, corresponding 
to the 
spontaneusly breaking of the $SU(2) \times U(1)$ into $U(1)_{em}$
giving mass to the $W^{\pm}$ and $Z^{o}$ gauge boson carriers
of the weak force and to quarks and leptons, is about 100 GeV.
In the case that the standard model is embedded in a grand unified 
theory (GUT) with a bigger gauge group, there is another scale in the 
theory the grand unified scale which can be of order of $10^{16}$ 
GeV\cite{La}. If the standard model is coming from a superstring
vacuum then the additional scale is the superstring scale which 
can be two orders of magnitude higher than GUT scale.
The question then arises of hierarchy of scales, why the 
electroweak scale is so small compared with the other scales.
In addition, non-supersymmetric theories involving Higgs scalars, 
like the standard model, receive quadratic quantum corrections to the 
Higgs sectors of their theory which can drive the low energy scale 
as high as e.g the GUT scale\cite{dere2}.  In this form,  
renormalized low energy scale $M_{r}$ receives quadratic
quantum corrections in the form
$M^2_{r}= C\alpha^2_{r} M^2_{GUT} + M^2$, where C a number of order 
$10^{0\pm 1}$, $\alpha_{r}$ a coupling constant and $M_{GUT}$ the high 
energy scale of the theory. The quadratic quantum corrections for a typical 
GUT scale are of order $10^{30}$ GeV, something which is unphysical.
It requires in
order to keep the corrections to $M_{r}$ under control
large fine tuning of parameters C, $\alpha$.
Sypersymmetry gives a solution to the problem of quadratic divergencies,
since supersymmetric models are free of quadratic divergencies
and the contributions to $M_{r}$ are $M_{r}^2 = M^2(1+ C \alpha^2_{r} \ln
\frac{M_{GUT}^2}{M^2})$.
However, in rigid supersymmetric lagrangians we can add 
the so called soft terms which  
break supersymmetry without introducing new quadratic divergencies.
These terms arise naturally, to the models coming from 
locally supersymmetric lagrangians as it is the 
heterotic string or the type II superstring in their field theory 
limits.

Moreover, in the simplest effective rigid theory, the minimal 
supersymmetric standard model, the Higgs potential of the theory
has a $\mu$ term which mixes the two Higgs doublets.
This term when the low energy theory comes from a $(2,2)$
orbifold compactifications of the heterotic string receives\cite{ANTO}
contributions, beyond the expected ones coming from the general 
presence of mixing terms between the $27$ and $\bar 27$'s in the 
K\"ahler potential of the theory. They are coming from the so called
higher weight interactions\cite{ANTO} and represent higher derivative 
couplings of auxiliary fields and scalars.  
The phenomenologically interesting case of candidate {\em non-perturbative 
superpotentials} which contribute to the 
value of $\mu$ term and to soft terms in $(2,2)$ orbifold 
compactifications 
is exhibited appropriately in chapter five, by taking into account 
contributions coming from the calculation of the threshold corrections
as target space free energies of chapter three. The proposed solutions 
are solutions to the $\mu$ problem along the lines of \cite{ANTO,lust} and 
can easily form alternative scenaria for solutions of the $\mu$ problem
along the lines of \cite{camuno}. 
The proposed superpotentials may come from gaugino 
condensation\cite{camuno}
and 
 respect transformation properties originating from the invariance
of the one-loop superstring  effective
action in the linear representation of the dilaton.
The proposed candidate superpotentials   
provide us with the necessary $\mu$ terms which solve the strong CP
problem and are
required from the low-energy superpotential of the theory to give masses 
to d-type quarks and e-type leptons and to avoid the massless axion.

We know that phenomenologically interesting models must have 
$N=1$ or $N=0$ supersymmetry\footnote{Tachyon free non-supersymmetric 
models 
have been constructed from the heterotic string in \cite{dihar,ggmv}.}. 
However, we will only be interested to 
models coming from $(2,2)$ symmetric orbifolds with $N=1$ space-time 
supersymmetry, since 
non-supersymmetric models suffer from stability problems\cite{giva}
related to tunneling of the cosmological constant to negative values.
The cosmological constant $\Lambda$ amounts to the 
introduction of  a general constant  $\Lambda$ into the 
effective action in D-dimensions in the form 
$-\frac{1}{k_{grav}^2}\int d^D
s  \sqrt{-g} \Lambda $, where g is the determinant of the metric
and $k_{grav}$ the gravitational coupling constant in D dimensions.
Here, $k_{grav}^2 =8 \pi G$, where G is the
 Newton's constant in D dimensions.
The one loop 
contribution to the cosmological constant $\Lambda$ in string theory is 
given by $\Lambda=\frac{1}{2}k_{grav}^2 \sum_i (-1)^F \sum_i \int d^D p 
\log(p^2 + M^2_i)$,
 where F is the fermion number operator and the sum is over all
particles.
Because the value of  $\Lambda$ from dimensional reasons has upper
limit of order
$M_{Plank}^4 \approx 10^{76}$ $GeV $, while the astronomical upper limit
is $10^{-47}$ $GeV^4$, there is a huge descrepancy of order of magnitude 
$10^{123}$ between the theoretical and cosmological considerations.
This creates the {\em cosmological constant} problem\cite{weinbe}.
In supersymmetric vacua we expect $\Lambda$ to vanish.
 Because in supersymmetric vacua we have equal number of fermions and
bosons at each mass level this contribution vanishes.
However, because at low energies we should recover the standard model
 which is not supersymmetric, supersymmetry must be broken in these
models,
and the cosmological constant is different from zero.
The breaking of spacetime supersymmetry due to the
 presence of the gravitino in the effective
 low energy theory of $(2,2)$ $N=1 $ symmetric orbifolds
must be spontaneous. 
In general, we expect the cosmological
constant to be different from zero after supersymmetry breaking.
In models coming from heterotic strings the cosmological constant
may be different form zero when 
supersymmetry is broken spontaneously\cite{rohm} 
or even when the model is non-supersymmetric\cite{ggmv,giva}.
If the cosmological constant $\Lambda$ for a particular string
vacuum is different from zero this means that, there is a
non-vanishing dilaton one point function, the background is not
a solution to the string equations of motion.                              
Recently, a different mechanism was proposed to set the
cosmological constant to zero\cite{witco}. By starting with the three
dimensional local theory in three dimensions which has zero
cosmological constant\cite{witco}, we can claim that the four dimensional 
theory maintains the same property.
It was proved that it happens in three dimensions\cite{bebest} and it
was claimed in \cite{witco} that by sending
the coupling constant of the three dimensional theory to $\infty$, we
get a four dimensional theory with zero cosmological constant.

In chapter five, we will discuss supersymmetry breaking.  
Supersymmetry breaking, with or without vanishing
cosmological constant, is one of the main unsolved problems of string 
 theory but contrary to grand unified theories or supergravity theories,
 string theory can suggest that its origin may come
from stringy non-perturbative effects.
In conventional field theoretical approach to supersymmetry breaking,
supersymmetry breaking appears as a field theoretical phemonenon,
coming from {\em gaugino condensation}\cite{drsw,Nil,CLMR,Kras} in
a pure Yang-Mills sector of the theory.
In this case, the nonperturbative superpotential W appears as
$W= (1/4)U (f + 2/3 \beta \log U)$, where U is the gaugino
bilinear superfield and $\beta$ is the $\beta$-function of the theory.
In general, gaugino condensation can occur in the pure Yang-Mills
theory.
The theory is asymptotically free, and the gauge coupling becomes 
strong at some scale which is the gaugino condensation scale.
Take, for example, the auxiliary fields of $N=1$ supergravity
\begin{equation}
F_i= e^{-G/2}(G^{-1})_i^j G_j + \frac{1}{4} f_{\gamma \delta \xi}
(G^{-1})_i^{\xi}{\lambda}^{\gamma} {\lambda}^{\delta},
\label{auxilia1}
\end{equation}
where G is the function of eqn.(\ref{opi}) and $\lambda$ is the gaugino
field. When gauginos condense, then $< \lambda \lambda > \neq 0$ and 
the auxiliary field gets a non-vanishing expectation value, and thus
breaks supersymmetry. The scale of supersymmetry breakdown is
then $M_S^2 \propto < \lambda \lambda >/M^2$, where M is the Planck 
mass $M= 1/k_{grav}$.

Toroidal compactifications of the heterotic string have 
in four dimensions $N=4$ supersymmetry and contain among
other fields, the dilaton $\Phi$, the antisymmetric tensor $B_{\mu \nu}$
which transforms after a duality trasformation to the axion $\tilde
\alpha$ and moduli scalars described by a matrix $M_{ab}$, which
parametrize the coset $SO(6,22)/SO(6) \times SO(22)$.

The spectrum and
interactions of the perturbative phase of the effective theory are
invariant under the "duality" symmetry $R \leftrightarrow
\alpha^{\prime}/R$,
which survives in all orders in string perturbation theory\cite{alva}.
In $N=1$ orbifolds\cite{dhvw}, the perturbative duality symmetry is 
known under the 
name T-duality, and the spectrum and interactions are parametrized 
in terms of the T moduli, mentioned earlier, of the unrotated $N=2$ 
complex planes. The spectrum and the interactions of the theory are
invariant under the symmetry $R \leftrightarrow
\alpha^{\prime}/R$ and the form of the effective action is strongly
constrained by the T-duality symmetry\cite{flst}.
If we freeze all moduli except the T-modulus then under the general 
$PSL(2,Z)_T$ trasformations (\ref{kolosaman}), we get e.g that the
superpotential has to transform with modular weight -1.
We describe the constraints imposed by the physical symmetries of
string theory,
on the basic quantities of the low energy effective $N=1$
supergravity theory extending earlier results \cite{flst}.

Let us combine the axion and the dilaton into a complex scalar 
$\lambda ={\tilde \alpha} + iexp(-\phi)$ with the
expectation value $<\lambda>=\theta/2 \pi +i/g^2=\lambda_1 + i\lambda_2$, 
where $\theta$ is the vacuum angle and g the coupling constant.
Then the action coming from compactification of the
ten dimensional heterotic string on a 6-dimensional torus
can be written in the form 
\begin{eqnarray}
S=\frac{1}{32 \pi^2} \int d^4 x \sqrt{-g} \{R-\frac{1}{2 (\lambda_2^2)}
G^{\mu \nu}\partial_{\mu} \lambda \partial_{\nu} \lambda -
\lambda_2 F_{\mu \nu}^T \cdot L M  L \cdot F^{\mu \nu} + 
\lambda_1 F_{\mu \nu}^T \cdot L \cdot {\tilde F}^{\mu \nu} \nonumber\\
+\frac{1}{8} G^{\mu \nu} Tr(\partial_{\mu}ML\partial_{\nu} ML)\},
\label{koiujip}
\end{eqnarray}
where M a $28 \times 28$ matrix satisfying\cite{magha}
\begin{eqnarray}
M^T=M,\; M^T LM=L,\;with\;L=\left(\begin{array}{ccc} 0&I_{6} &0\\
I_{6}&0&0\\
0&0&-I_{16}
\end{array}\right).
\label{koiujip1}
\end{eqnarray}
In addition, $A^{(a)}_{\mu}, \; a=1,\dots,28$ is a set of 28 
abelian vector fields and $F_{\mu \nu}$ a 28 dimensional vector 
${\tilde F}^{\mu \nu}$ the dual of $F^{\mu \nu}$.
The action (\ref{koiujip1}) is invariant under the group 
$O(22,6)$. The moduli fields obeying (\ref{koiujip1}) parametrize the 
Narain lattice $O(22,6)/O(22) \times O(6)$. The equations of motion 
derived from the previous action are invariant under
\begin{equation}
\lambda \rightarrow \lambda + 1,\;F_{\mu \nu} \rightarrow M,
G_{\mu \nu} \rightarrow G_{\mu \nu},\;and  
\label{sfdtr1}
\end{equation}
\begin{equation}
\lambda \rightarrow 
\lambda^{\prime}=-\frac{1}{\lambda},\;F_{\mu \nu} \rightarrow 
F_{\mu \nu}^{\prime} = -\lambda_2 M L {\tilde F}_{\mu \nu}- \lambda_1
F_{\mu \nu},\;M\rightarrow M,\;G_{\mu \nu} \rightarrow  G_{\mu \nu}.
\label{sfdtr2}
\end{equation}
The transformations involving $\lambda_1$ generates\footnote{
In reality the action (\ref{koiujip}) the equations of motion
are invariant under shifts in the vacuum parameter which give
as $\lambda \rightarrow \lambda + c$, where c a real number, but it 
is believed that it is broken\cite{sha} to $SL(2,Z)$ by world sheet 
instantons.}
 the $SL(2,Z)$.
So the equations of motion, and not the action,  coming from the complex
dilaton field
are invariant under the $SL(2,Z)_S$ transformations 
\begin{equation}
\lambda \rightarrow \frac{a \lambda \ -i b}{i c \lambda  + d}.
\label{kolosamin}
\end{equation}
It was claimed\cite{SDUAL1} that
the $SL(2,Z)$ S-duality invariance group of the 
equations of motion in (\ref{kolosamin}) has to be 
promoted\footnote{In exact analogy as electric-magnetic duality in 
the $N=4$ super-Yang Mills the spectrum and 
interactions of the elementary string states 
are identical to those of the monopole sector of the theory, 
claimed to occur\cite{mool} with the spectrum and interactions of 
$N=4$ super-Yang Mills\cite{osbo}. }
to an exact symmetry group of string theory and be active in the $N=1$ 
superstring vacua coming from 
compactifications of the heterotic string.
We should note, that $SL(2,Z)_S$ is largely conjectural, as  
this group appears only as an invariance of the equations of motion 
of toroidal compactification of the heterotic string and not as an 
invariance of the action\cite{SDUAL2}. 
Since this symmetry inverts the coupling constant
is non-perturbative in nature.
Moreover, since S-duality represents strong weak coupling duality,
it can be used to 
constrain the form of the effective action coming from 
non-perturbative effects.
S-duality in string theory is associated with supporting 
evidence in $N=1$ non-abelian supersymmetric 
Yang-Mills\cite{Seiber} (SYM)or $N=4$ non-abelian SYM\cite{cva1}.
We should say here, that at the limit $\alpha^{\prime} \rightarrow 0$,
the $N=4$ SYM appears as the low energy limit of the toroidaly 
compactified heterotic string.

In chapter five we use S-duality, to allow for
S-dual superpotentials which use a single composite bilinear
gaugino condensate chiral superfield. 
They are used to break supersymmetry
in order to allow the dynamical determination of the
dilaton vacuum expectation value, which represents the string tree
level coupling constant\footnote{In perturbative string theory, the
vacuum expectation value of the dilaton is the true expansion 
parameter\cite{dimeni} around a specific vacuum.}.
This extends an earlier result \cite{nil1,nil2}, related to 
S-duality invariant gaugino condensates in the effective 
Lagrangian approach. 
The effective low-energy $N=1$ supergravity theory associated with 
the proposed generalized S-T dual superpotentials exhibit the 
same target space duality modular symmetry groups as the one's 
appearing 
in the non-decomposable $(2,2)$ symmetric orbifold models.
The appropriate use of the the bilinear condensates in an action 
invariant under the
same S-duality as target space duality group, tacitly conjectures 
the existence of $\Gamma^{0}(n)$  or  $\Gamma_{o}(n)$ S-duality 
invariance of the low-energy effective 
action. 
In the spirit of \cite{F}, we accept that this class of  
theories in its final form, 
including non-perturbative contributions, must respect $\Gamma^{0}(n)$
or $\Gamma^{0}(n)$ respectively,
as an exact symmetry of string theory at the quantum level.
Nevertheless, such a limit will exist,  
 and unless someone proves
that $SL(2,Z)$ is singled out as an exact symmetry of string theory
such a conjecture may be expected to hold.

The ten dimensional heterotic, fundamental, string admits {\em five brane} 
solutions
which correspond to a {\em dual } formulation\cite{Duffos} of the 
heterotic
string. After compactification to six dimensions the dual heterotic 
string has a dual dilaton $\tilde S$, which
obeys the string equations of motion with the opposite sign 
and is related to S of the fundamental string,  as ${\tilde S}= -S$.
Especially, for compactifications to four dimensions the roles
of $SL(2,Z)_T$ and $SL(2,Z)_S$ dualities
are interchanged $T \leftrightarrow S$ and the 
$SL(2,Z)_S$ invariance group appears  as a subroup of the target space
duality group of the dual string. In addition, the $SL(2,Z)_S$ group
which for the fundamental theory was the invariance group of the
equations of motion, now becomes the target space duality group
$SL(2,Z)_T$.
This is equivalent to the fact that $SL(2,Z)_T$ target space duality
for the dual string is equivalent to $SL(2,Z)_S$  duality 
for the fundamental string.
We should note that we must expect the S-duality of $N=4$ heterotic
string constructions to hold, i.e in heterotic strings\cite{huto} 
compactified on a four torus and its dual IIA on $K_3$, since a one loop
partition function 
test of $N=4$ super Yang-Mills(SYM) has been performed\cite{cva4}.

It appears that S-duality holds in a twisted $N=4$ SYM in the 
following\cite{cva2} sense :
the modular transformations exchange the partition function for the 
gauge groups SU(2) with SO(3)
and the partition function Z(S) of the theory transforms with modular 
weight k,
$Z(-1/S)= (\frac{S}{2})^{w} Z(S)$, where 
$w \propto {\cal \chi}$ the Euler characteristic. 
S-duality holds only when we introduce the modified partition function
${\tilde Z(S)}\stackrel{def}{\equiv} e^{-w{\log \eta} } Z(S)$.

 We refer to this version 
of SYM because it provides evidence for the appearance of the groups
$\Gamma_o(n)$ in SYM and indirectly to a possible string theory limit.
For example for $N=4$ SYM with $SU(2)$ gauge group the 
modular symmetry group $\Gamma_o(2)$ appears\cite{cva2}.
The cusp at $\tau = + \infty$ corresponds to the $SU(2)$ instanton expansion
while the cusp at $\tau = 0$ to its dual theory $SO(3)$.

Phenomena related to duality invariance do appear at the
string theory level
in different heterotic compactifications and in different
spacetime dimensions than four or ten. These phemonena involve
equivalences between different string theories\footnote{Unfortunately,
all string theories have $N=4$ supersymmetry} ,
under which different regions of the moduli space of the two theories
are matched to each other for particular limits of their coupling
constants.
 Such phenomena are widely known as {\em dualities }\cite{var}
and for a particular version of dual of theories
are the subject of chapter four.

However, it may be that the complete solution of all the 
problems, mentioned up to now,
may come from determing the way that we break spacetime supersymmetry.
However, this involves determination of the complete effective action
of the string theory vacuum.
Recently, there is accumulating evidence that type I, type II and 
Heterotic string theories
may be complementary descriptions of a more fundamental theory 
originating from higher dimensions, which in dimensions greater than ten 
is coming from M\cite{var} of F-theory\cite{ftheo}.
In chapter four, we examine this {\em duality} phenomena. 
We discuss perturbative aspects of string theory, based 
on orbifold\cite{dhvw} and manifold\cite{kava} compactifications of the 
heterotic string.
In particular, we reexamine a recently proposed equivalence of moduli
spaces between the $N=2$ heterotic string compactified on the 
$K_3 \times T_2$, we will call it A theory, 
and the $N=2$ type II superstring compactified on a Calabi-Yau three 
fold\cite{kava}, which we will call it B theory.
It has been conjectured\cite{kava} that moduli spaces of A, B theories 
coincide when all 
corrections including perturbative and non-perturbative one's are taken 
into account. In this way, $N=2$ physical quantities like, expressed 
in $N=1$ language, {\em Yukawa couplings}, {\em gauge
couplings} which are impossible to be calculated in the
weakly coupled phase of the heterotic string are mapped to
superstring vacua of type II, where their expressions are well 
known\cite{ber,hoso1,hoso2,hoson3}.

The effective low energy theory of the $N=2$ heterotic compactifications 
is described by the language of special geometry\cite{Ceresole,FVP}. 
For the part of the moduli space described by vector superfields 
the effective theory is determined completely by the knowledge of one 
particular function, the prepotential. 
The prepotential $\cal F$ of $N=2$ compactifications of the heterotic 
string receives perturbative corrections up to one loop\cite{wkll}.
The third derivative of $\cal F$ with respect of the 
T moduli appearing in two dimensional torus compactifications,
which in $N=1$ 
language represents Yukawa couplings, was calculated with the use
of modular symmetries in \cite{wkll} while directly via 
the one loop corrections of the K\"ahler metric in \cite{afgnt}.
The calculation on both cases was based on non-decomposable 
orbifolds. The same result was shown \cite{mouha} to 
be coming from the indirect, ansatz, calculation of $\cal F$
when the heterotic string is compactified on the $K_3 \times T_2$.
In chapter four, we find through a string one loop 
calculation the general differential equation obeyed by $\cal F$
and which is valid for any compactification of the heterotic string
on $K_3 \times  T_2$.
In addition, we derive the differential equation for the 
Yakawa couplings with respect of the U moduli, corresponding to
the complex structure of the $T_2$.

Our final hope is that when all perturbative and non-perturbative
corrections will be taken into account that the  
vacuum potential will be such that, it does not
only calculate the values of the moduli fields but it fixes
simultaneously
the value's of the matter fields allowed in the theory and break
the gauge group into that of the standard model.  

\newpage
\vspace{10cm}
\begin{center}
{\bf CHAPTER 2}
\end{center}
\newpage

\section{\bf Compactification On Orbifolds}

In this chapter of the Thesis, we will mainly review the
orbifold compactifications of the heterotic string\cite{dhvw}.
It is illustrative to consider before we describe the orbifold 
construction,
the compactification on a D dimensional torus $T^D$. The purpose of 
doing so is not meaningless since the orbifolds 
that we are interested in the bulk of this work, are coming from
the toroidal compactification of the heterotic string modded out 
by the appropriate action of a point group. Our presentation is 
organized as follows:
In section 2.1 we will discuss toroidal compactifications of the 
closed bosonic string\cite{gsw1}.
In section 2.2 we will discuss toroidal compactifications of the 
heterotic string\cite{Gr} together with orbifold 
constructions\cite{dhvw} of the heterotic string. 
In section 2.3 we describe the redundancy under 
global parametrizatiions in the moduli space of string vacua, in
particular the duality invariance in toroidal and orbifold
compactifications.

\subsection{Toroidal Compactifications}

We are considering bosonic strings, living in twenty six dimensions,
 propagating in a toroidal background.
If we compactify D bosonic coordinates on a D dimensional torus 
we end up with a 26-D dimensional flat space.
The D-dimensional torus $T^D$ is parametrized so that points 
on its $R^D$ space are identified\footnote{ 
In the following we follow the convention that repeated indices are
summed, e.g $e^{I}_i e^{*I}_j \equiv \sum_{i=1}^D e^{I}_i 
e^{*J}_j$. Compact coordinates will be denoted with capital 
superscript letters.} as 
\begin{equation}
X^I =X^I + \sqrt{2} \pi \sum_{i=1}^D n_i R_i e^I_i,\;i=1,\dots,D,
\;n_i\;\in\;Z. 
\label{ee1}
\end{equation}
Here, $e^I_i$ are D basis vectors with the property $e_i^2 =2$.
In addition, the quantities\cite{lustheis}
$L^I = (1/2)^{1/2} n_i R_i e^{I}_i$ can be considered as 
lattice vectors 
on a D-dimensional lattice $\Lambda^D$ with basis vectors 
$(1/2)^{1/2}R_i e^{I}_i$.
For compactifications on a circle $S^1$, (\ref{ee1}) 
can simplified as  $X^1 = X^1  + 2 \pi R L $, 
that we identify points which differ by an integer L  times
$2 \pi R$, where R the circle radius.  

The torus is identified as $T^D = R^D / 2 \pi \Lambda^D$.
The momentum vectors are defined as 
$\textstyle{p^I= \sqrt{2}  \frac{m^i}{R_i} e^{*I}_i}$,
where $e^{*I}_i$ are the dual vectors on the lattice with the 
properties $e^{I}_i e^{*I}_j = \delta_{ij}$, $e^{I}_i
e^{*J}_i = \delta^{I J}$ and $e_i^{*2} =1/2$.
The basis vectors on the dual lattice $\Lambda^{*}$ are defined
as $\textstyle{\frac{\sqrt{2}}{R_i}e_i^{*}}$.
For closed bosonic strings the compact dimensions, on the torus, 
satisfy
\begin{equation}
x^{I}(\sigma + 2 \pi , \tau) = x^{I}(\sigma, \tau) + 
2 \pi L^{I}.
\label{ksdd1}
\end{equation}
Splitting (\ref{ksdd1}) into left and right moving 
coordinates\footnote{ 
, where $p_I$ and $L^I$ the momentum and the winding numbers 
respectively,}(respectively) as
\begin{eqnarray}
X^{I}_L=\frac{1}{2} x^{I}+ (p^{I} + 
\frac{1}{2} L^{I})(\tau + \sigma)+ i \sum_{n \neq 0}{\bar a}^{I}_n 
\frac{1}{n} e^{-in(\tau + \sigma)},\nonumber\\
X^{I}_R =\frac{1}{2} x^{I} + (p^{I} -  
\frac{1}{2}L^{I})(\tau - \sigma)+ i \sum_{n \neq 0}a^{I}_n
\frac{1}{n} e^{-in(\tau - \sigma)},
\label{ksdd2}
\end{eqnarray}
we obtain for the mass formula, 
of the left and right movers in the uncompactified coordinates
\begin{eqnarray}
m_L^2 = \frac{1}{2}(p^I + \frac{1}{2} L^I)^2 + N_L -1 =\frac{1}{2}
{p}_L ^2 + N_L -1,\;and\nonumber\\
m_R^2 = \frac{1}{2}(p^I - \frac{1}{2} L^I)^2 + N_R -1 =\frac{1}{2}
{p}_R ^2 +N_R -1,
\label{ksdd3}
\end{eqnarray}
where $N_L$, $N_R$ the number of the left and right moving oscillators.
The total mass operator in the uncompactified $26-D$ dimensions is 
defined as
\begin{eqnarray}
m^2 = M_L^2 + M_R^2 = N_R + N_L -2 + \sum_{I=1}^D (p^I p^I 
+ \frac{1}{4} L^I L^I) \nonumber\\
= N_L + N_R -2 + \sum_{i,j=1}^D (m_i g_{ij}^{*} m_j + \frac{1}{4} n_i 
g_{ij} n_j),
\label{ksdd3}
\end{eqnarray}
where $g_{ij}$ and $g_{ij}^{*}=g^{-1}_{ij}$ 
\begin{equation}
g_{ij} = \frac{1}{2}\sum_{i=1}^D R_i e_i^I R_j e^I_i,\;\;
g_{ij}^{*} =2 \sum_{i=1}^{D}\frac{1}{R_i} e^{*I}_i 
\frac{1}{R_j} e^{*I}_J
\label{ksdd4}
\end{equation}
are the metrics on the lattices $\Lambda^D$ and $\Lambda^{*D}$. 
Defining the signature of the metric on the lattice in the form
$((+1)^D, (-1)^D)$, meaning $P \cdot P^{\prime} = (p_L^I 
{(p^{\prime})}^I_L - p_R^I {(p^{\prime})}_R^I)$, 
the vectors $P=(p_L, p_R)$ build an even,
$P \cdot P^{\prime} \in 2Z$, self-dual lattice $\Gamma_{D,D}$.
In addition, the theory satisfy the reparametrization invariance 
constraint\footnote{see the comment in chapter three, after relation 
(\ref{lmc}).} $ N_R - N_L = \sum_{i=1}^D m_i n_i$.
Lets us now consider the spectrum in some detail. Set the momentum and 
the winding numbers equal to zero. 
Then the tachyon comes from $N_R=N_L=0 $, $m^2=-2$, which we identify 
as $|0>$. The graviton, dilaton, antisymmetric tensor and the 
dilaton come from states at the next, massless, level, 
$a^{\mu}_{-1} {\bar a}^{\nu}_{-1} |0>$.
Let us we decompose $a^{\mu}_{-1} {\bar a}^{\nu}_{-1}$ into irreducible 
representations of the transverse rotation\footnote{Lorentz invariance 
requires that physical states are physical states are arranged into
representations of the little group of the Lorentz group $SO(d-1,1)$.
This is $SO(d-2)$ for massless particles.} group $SO(24-D)$ in the 
light cone gauge, as        
\begin{eqnarray}
 a^{\mu}_{-1} {\bar a}^{\nu}_{-1}|0>
= \{ a^{[\mu}_{-1} {\bar a}^{\nu]}_{-1}\}|)0>
+ \{  ( a^{(\mu}_{-1} {\bar a}^{\nu)}_{-1} - 
\frac{1}{(24-D)}\delta^{\mu \nu}  a^{\rho}_{-1} 
{\bar a}^{\rho}_{-1} \}|0>\nonumber\\
+ \{\frac{1}{(24-D)}\delta^{\mu \nu}  a^{\rho}_{-1}
 {\bar a}^{\rho}_{-1}\}|0>.
\label{ksdd5}   
\end{eqnarray}
The brackets in the first term denote symmetrization of
indices, while the parenthesis in the second term donote
antisymmetrization of indices. Each of the three terms corresponds
to a different kinds of particles, namely the first term to the 
graviton, the second term to the antisymmetric tensor and the 
last one to the dilaton.
 
In addition, we have 2D vectors in the form $|A_1>=
\alpha_{-1}^{\mu} {\bar \alpha}_{-1}^{I} |0>$ and $|A_2>= 
{\bar \alpha}_{-1}^{\mu} \alpha_{-1}^{I}|0>$
associated with the gauge bosons of the $U(1)_L \times U(1)_R $ gauge
symmetry of the torus. Moreover, there are $D^2$ massless scalars
$\phi^{IJ}= \alpha_{-1}^{I} {\bar \alpha}_{-1}^{J}|0>$, which 
correspond
to the moduli of the toroidal compactification.
The number of $D^2$ parameters can be interpreted as corresponding 
to the presence of the following background fields, whose vacuum
expectation values on the torus are given by
$\frac{D}{2}(D+1)$  and $\frac{D}{2}(D-1)$ remaining parameters, 
coresponding to the metric $g_{IJ}$ and the
antisymmetric tensor $B_{IJ}$ respectively.
The action corresponding to the toroidal compactification via the
background field interpretation is 
\begin{equation}
\int d^2 \sigma (\sqrt{g} \eta^{\mu \nu} \partial_{\mu} X^{I}
\partial_{\nu} X^{J} g_{IJ} + \epsilon^{\mu \nu}\partial_{\mu} X^{I}
\partial_{\nu} X^{J} B_{IJ}),
\label{sddd6}
\end{equation}
where $\epsilon$ the antisymmetric invariant in two dimensions. 
Because of the interpretation of the moduli parameters associated
with $B_{IJ}$, the mass operator of eqn.(\ref{ksdd3}) now must 
reanalysed 
such that \cite{na} it receives contributions from the non zero
$B_{IJ}$ in the form, $p_{L,R}^2 \propto \frac{1}{2} B^{IJ}
L_J$.
The Lorentzian even self dual lattices build from different values of 
the
$D^2$ parameters are obtained from each other under $SO(D,D)$ rotations
of some reference lattice, which for convenience can correspond to the
lattice with $B_{IJ} =0$ and $g_{IJ} =\delta_{IJ}$.
The exact  moduli space of torus compactification 
is build, in its exact form, after we consider the invariance of the
spectrum under the rotations $SO(D)_L$, $SO(D)_R$ of the vectors
$ p_L$, $p_R$.
All even self dual lattices in the form discussed in this
section are invariant under $SO(D,D;R)$ rotations. However, the 
Hamiltonian of the toroidal compactifications 
$\frac{1}{2}({p}_L^2 + {p}_R^2)$
is invariant under $SO(D)_L \times SO(D)_R$. This means that the scalar
fields parametrizing the moduli space take values on the coset space 
$SO(D,D;R)/SO(D) \times SO(D)$. 
In addition, for special values of the momentum and the winding numbers
$m_i = n_i= \pm 1$, $ g_{ij} = 2 \delta_{ij}$ we can get the additional 
massless vectors, $p_L^2 =2$,$N_L =0$ , $p_R=0$, $N_R=1$ and the
gauge symmetry is enhanced to $SU(2)^D_L \times U(1)_R^D$. 
This corresponds to $R_i =\sqrt{2}$, and $e_i =\sqrt{2} \delta_i^I$, 
the torus is compactified on D circles with radii $R=\sqrt{2}$.

\subsection{General theory of orbifold compactifications}  

In the main bulk of this section, we will be concentrating in the study 
of strings propagating in $Z_N$ orbifold 
backgrounds \cite{dhvw,imnq,INQ,fer4,HV}.
The orbifold compactifications that we are interested in this Thesis,
will come from toroidal compactifications of the heterotic 
string\cite{Gr}. 
The heterotic string is a construction of a left moving twenty six
dimensional bosonic string together with a right moving ten dimensional
superstring. In the bosonic construction, as left moving coordinates we 
have ten uncompactified bosonic fields $X_L^{\mu}(\tau + \sigma)$, 
$\mu=0,\dots,9$ and sixteen internal bosons which live on a sixteen
dimensional torus. The right moving degrees of freedom consist of ten
uncompactified bosons $X_R^{\mu}(\tau -\sigma)$,
$\mu=0,\dots,9$ and their fermionic superpartners the
Ramond-Neveau-Schwarz(RNS) right moving fermions $\Psi_R^{\mu}
(\tau -\sigma)$. The sixteen internal fields are compactified on
a torus of fixed radii and the momenta $P^I$ span an even self-dual 
lattice $\Gamma^{16}$. In sixteen dimensions there are two such 
lattices, the root lattice of $E _8 \times E_8$ and the weight lattice
of Spin$(32)/Z_2$ . 
Both the appearance of the two groups as well as spacetime SUSY
are enforced by modular invariance of the left and right moving sectors.
The Hilbert space of the theory is build by direct products of the
left and right moving states.
The physical massless modes correspond to ten dimensional $N=1$
supergravity coupled to $E _8 \times E_8$ (or Spin$(32)/Z_2$).
The theory contains no tachyon as in the case of the bosonic string.

The most general compactification of the heterotic string\cite{na} 
involves compactification on the torus
$R^{10-d,10-d +16}/{\Gamma_{10-d, 10-d+16}}$, where d 
dimensions are uncompactified.
Here, $\Gamma_{10-d,10-d+16}$ is the lattice coming from the 
$SO(10-d, 10-d+16)$ rotation of the lattice 
$\Gamma_{16} \otimes (P_2)^{10-d}$, where $P_2$ is the 
two dimensional Lorentz lattice with signature $((+),(-))$ and 
$\Gamma_{16}$ is the $E _8 \times E_8$ (or Spin$(32)/Z_2$).
However, toroidal compactifications of the heterotic string
in four dimensions give $N=4$ space time supersymmetry   
which is non-chiral and thus disaster for phenomenology. 
To obtain supersymmetry breaking down to $N=1$ or $N=0$ we have to
consider different compactification 
schemes\cite{dhvw,fer1,fer2,fer3,Calabi1,fer4}. Here, we will consider 
orbifolds\cite{dhvw}.
 
In general, if the action of a discrete group on a manifold acts freely, 
without fixed points, we obtain another manifold, if not then the 
resulting space is an orbifold ${\cal O}$.
Because the discrete group should preserve the metric of the space, 
then if the space is Euclidean the discrete group must be a subgroup of 
the Euclidean group consisting of translations and rotations. In this
case the discrete group is the space group S. 
The general element of S
is represented by $g=(\theta, \upsilon)$, where $\theta $ represents 
rotations and $\upsilon$ translations. 
In this way, each point is identified with its orbit under the action 
of the space group, and thus the name orbifold.
The action of the space group element $g=(\theta, \upsilon)$ on a  
vector $x \in R^d$ is $ gx = \theta x + \upsilon$, while that of the
inverse element is defined as $g^{-1}= (\theta, \upsilon)^{-1} = 
(\theta^{-1}, - \theta^{-1} \upsilon)$ and $(\theta, \upsilon) (k, j)=
(\theta k, \upsilon + \theta j)$.
The subgroup of S, consisting of elements of the form $(1, \upsilon)$,
forms the d-dimensional lattice $\Lambda$ of S.
The point group P is defined 
as the subgroup of S consisting of pure rotations $\theta $, such that 
$(\theta, \upsilon) \in S$ for some S.   
Take for example two different elements $(\theta, u)$ and
 $(\theta, \upsilon) \in S$. Then $(\theta, \upsilon) (\theta, u)^{-1}=
(1, \upsilon -u)$. So the point group has a well defined action on the
torus and its action on the torus is unique up to lattice translations.
We can conclude that we can identify its action on the torus as 
${\bar P}= S/{\Lambda} $.
So there are two equivalent ways of definning orbifolds.
This means that either the orbifold can be viewed as the D dimensional 
Euclidean space $R^D$ divided by the action of the space group, e.g
${\cal O}= R^D/S$ or as the torus divided by ${\bar P}$. Symbolically, 
\begin{equation}
{\cal O}=R^d/S = T^d/{\cal P}.
\label{karampita1} 
\end{equation}

In the bosonic formulation\footnote{In a fermionic formulation, 
the compactified coordinates can be described by worldsheet 
fermions\cite{fer1,fer2}.}, a four dimensional string is obtained, by
compactifying the heterotic string with six right-moving and twenty two 
left-moving coordinates on a torus ${T_{R}}^{6} \otimes T_{L}^{22}$, 
where -(in obvious notation) the right handed coordinates are compactified 
on the torus $T^6$, while the left coordinates are compactified 
on the torus $T^{22}$.
Modding out this torus by a discrete group, an isometry of the torus,
we obtain an orbifold. 
In general, there are different ways that the modding out operation can 
be realized.
A asymmetric orbifold can be constructed by modding the left and right 
coordinates as
\begin{equation}
{\cal O}=\frac{T_R^6}{P_R} \otimes \frac{T_L^6}{P_L} \otimes 
\frac{T_L^{16}}{G}.
\label{karampi1}
\end{equation}
Here, we modded the extra six dimensions with different point groups P.
The extra sixteen coordinates were modded by G, which we will call it
the gauge twisting group.
A symmetric orbifold can be obtained by modding the extra six dimensions 
with the same point group, $P_L =P_R$.
Orbifolds with non-abelian point group are called non abelian
orbifolds.
However, in the case where the point group $P$ is abelian the space group
in general is not.
This means that if we embed the space group in a non-abelian way
in the gauge degrees of freedom then we obtain a non-abelian action
in the gauge degrees of freedom, even if the point group is abelian. 

Here, we will be interesting in the case where $T_L^6 = T_R^6$ and 
$T_L^{16}= E_8 \times E_8.$     
In this case we will represent our orbifold as 
\begin{equation}
{\cal O}=\frac{R^6_{L+R}}{S} \otimes \frac{T^L_{E_8 \times E_8}}{G},
\label{karampi2}
\end{equation}
where P is associated with the point group action on the corresponding 
torus and S the space group in the gauge degrees of freedom.
The meaning of (\ref{karampi2}) is that the bosonic closed strings 
boundary conditions are modded out by the element $w=(\theta, v^i; 
\Theta, V^I)$
\begin{eqnarray}
X^i(\sigma=2 \pi) = \theta X^i(\sigma= 0) + v^i,\;i=1,\dots,6,
\nonumber\\
X^I(\sigma=2 \pi)= \Theta X^I(\sigma= 0) + V^I,\;I=1,\dots,16.
\label{faal1}
\end{eqnarray}
Here, $\theta$, $\Theta $ are automorphisms of the 
corresponding torus and $E_8 \times E_8$ lattices respectively.
The corresponding lattice shifts are given by $\upsilon$ and $V$.
 
Orbifolds have two types of closed strings, untwisted and twisted.
An untwisted string is the one which is closed in the torus, even before 
twisting starts. Their boundary conditions are
\begin{equation}
X^i(\sigma=2 \pi) = X^i(\sigma= 0) + w^i,\;i=1,\dots,6,
\label{faal2}
\end{equation}
where $w^i$ is a vector on the lattice. For $w^i=0$ the string is
obviously closed. The general coordinate expansion of the closed 
string for  $w^i \neq 0$ has the form
\begin{equation}
X^i(\sigma, \tau)= X^i_u
+ p^i \tau + \frac{w^i \sigma}{2 \pi} + \frac{i}{2 \pi} \sum_{n \neq 0}
\left( \frac{\alpha_n^i}{n} e^{-in(\tau - \sigma)} +
\frac{{\tilde \alpha}_n^i}{n} e^{-in(\tau  + \sigma)}\right),
\label{faal3}
\end{equation}
where $X^i_u$ is the centre of mass coordinate, and $p^i$, $w^i$ are 
momentum and winding numbers. Strings which
are not closed on the torus but are closed on the orbifold are called
twisted strings. From the general form (\ref{faal1}) of the twisted 
strings we can conclude that they don't have any momentum or winding 
numbers in the twisted directions. However, when the strings are 
twisted\cite{ibann} by the element $\theta=1$ and in addition 
there is a lattice shift the boundary conditions have the form   
\begin{equation}
X^i(\sigma, \tau)= X^i_t + \frac{1}{2}i\sum_{m=0} \left( 
\frac{\alpha^i_{m + \frac{1}{n}}}
{m + \frac{1}{n}} e^{-i(m + \frac{1}{n})(\tau - \sigma)}
+ \frac{{\tilde \alpha}^i_{m - \frac{1}{n}}}{m - \frac{1}{n}}
e^{-i(m- \frac{1}{n})(\tau  + \sigma)}\right).
\label{faal4}
\end{equation}
Here, n is the order of the twist and the centre of mass $X^i_t$ is 
quantized.
However here we confine our study of orbifolds to the effect of abelian 
$Z_N$ twists on the compactified heterotic strings \footnote{The case 
of $Z_N \times 
Z_N $ orbifold twists have been discussed in\cite{fonter}.}.
In this case, the action of the general element of the space group
S, $(\theta^k,\upsilon)$ on the closed strings takes the form
\begin{equation}
X^i(\sigma  + 2 \pi, \tau) = S X^i(\sigma, \tau)= \theta^k X^i(\sigma, 
\tau) + \upsilon^i,
\label{faal5}
\end{equation}
with mode expansions
\begin{equation}
X_k^i(\sigma , \tau) = x^{(k, f)i} + \frac{1}{2}i
\sum_{m \neq 0} \left( \frac{\alpha_{m+ \frac{k}{m}}}{m+ \frac{k}{m}}
 e^{-i(m + \frac{k}{n})(\tau - \sigma)}
+\frac{{\tilde \alpha}^i_{m - \frac{k}{n}}}{m - \frac{k}{n}}
e^{-i(m- \frac{k}{n})(\tau  + \sigma)}\right).
\label{faal6}
\end{equation}
The position of the centre of mass $x^{(k, f)i}$,
if $\theta^k \neq 1$, is not quantized and corresponds to the fixed 
points of the orbifold.
In general, the action of the space group on the twisted strings
forces the centre of masses to be the fixed points of the orbifold.
From (\ref{faal1})
we deduce that the "fixed points" of the orbifold obey
$X^{(k, f)i} = (1 - \theta^k)^{-1} v^i$.

We are interested in the effect of twistings $\theta$ which leave 
unbroken
one space-time supersymmetry. To specify the orbifold exactly we have 
to choose a lattice in which the automorphism is acting.
Different lattices may be chosen with the same point group.
We will be mainly concerned with six dimensional Lie algebra lattices,
where the point group is generated by automorphisms of these lattices.

We will now start our description of $ Z_N $ Coxeter orbifolds.
The construction of Coxeter orbifolds involves the action of the point
group generated from a Coxeter element on the six dimensional
\cite{marku} torus. 
Propagation on the six-dimensional torus is associated with boundary
conditions in the form
\begin{equation}
X^{i}(\sigma = 2 \pi) = X^{i}(\sigma = 0) + 2 \pi m^{\alpha}
e^{i}_{\alpha},\;i=1,\dots,6,
\label{closs}
\end{equation}
 where the $e^{}_{\alpha}$ are a set of simple roots of the root lattice
$\Gamma$ of
the six dimensional torus $T_{6}$ defined by $T_{6} = 
\frac{R^{6}}{\Lambda}$
\footnote{ ${\Lambda}$ is the the subgroup of  the space group,
consisting of pure translations.}.
The $ m_{\alpha}$ are integers.
The general form of the $X^{i}$ fields is \cite{kats,imnq} 
\begin{equation}
X^{i}(\sigma , \tau) = q^{i} + {\frac{1}{2}} {p^{i}}{\tau} +
 m^{\alpha} e^{i}_{\alpha}{\sigma} + {oscillators},\;i=1,\dots,6.
\label{clos}
\end{equation}
The canonical momenta $p^{i}$ take their values on the dual lattice 
${\Gamma}^{*}$ with ${p_{i}} = n_{a} {e_{i}^{* a}}$.
The ${e_{i}^{* a}}$ represent basis vectors on the lattice
${\Gamma}^{*}$ and the $n_{a}$ are integer valued.

The point group automorphisms of six-dimensional lattices
for the Coxeter orbifolds is generated from Weyl reflections in 
the form  
\begin{equation}
S_{i}(x) = x - \frac{2(e_{a},x)}{|e_{a}|^{2}}.
\label{faal7}
\end{equation}
For the phenomenologically interesting $N=1$ supersymmetric models
it is useful to represent the action of the point group
in the complex basis 
\begin{eqnarray}
Z^{i}= {\frac{1}{2}}({X^{2 i + 1}}+ i X^{2i + 2})&\;,
{{\bar Z}^{i}} = {\frac{1}{2}}({X^{2 i + 1}} - i X^{2i +
2})&:i=1,\dots,3,
\label{faal8}
\end{eqnarray}
by the Coxeter element
\begin{equation}
\theta = diag(exp [2 \pi i({\eta^{1}},{\eta^{2}}, {\eta^{3}})]),
\label{faal9}
\end{equation}
where ${\eta^{i}}$ are integers with $0<{\eta^{i}}<1$, e.g for the $Z_3$
orbifold a possible Coxeter element is $\theta_{Z_3} =(1/3)(1,1,-2)$.
Because, $\theta$ is a automorphism of the six dimensional lattices,
i.e it may act crystallografically. It must transform 
e.g $\theta \rightarrow A \theta A^{-1}$ with $A \in GL(2,R)$.
The last condition, determines the crystallographic condition, 
namely that the order of $\theta$ is 1,2,3,4 or 6.

The action of the point group produces on the orbifold two types 
of closed strings, the untwisted and the twisted strings.
The untwisted strings are given by the expressions of
eqn's (\ref{clos}) and (\ref{closs}) while for the twisted strings
the following expressions hold
\begin{equation}
Z^{i}(\sigma = 2 \pi) = {{\theta}^{k}}Z^{i}(\sigma = 0)\; mod\;2 \pi 
\Gamma,
\label{faal10}
\end{equation}
which look like
\begin{eqnarray}
Z^{i}(\sigma,\tau)=z_{i}^{(k,f)i}+{\frac{1}{2}}i \sum_{n\in Z+k 
\eta_{i}} \frac{1}{\eta_{i}} b^{i}_{n} e^{-i n (\tau- \sigma)} + 
\frac{1}{2}i\sum_{n\in Z-k\eta_{i}}  
 \frac{1}{\eta_{i}}{ c}^{i}_{n} e^{-i n (\tau + \sigma)}\nonumber\\
{\bar Z}^{i}(\sigma,\tau)={\bar z}_{i}^{(k,f)i}+{\frac{1}{2}}i +
 \sum_{n \in Z - k\eta_{i}}\frac{1}{\eta_{i}}{\bar b}^{i}_{n}
e^{-i n (\tau-\sigma)}+\frac{1}{2}i
\sum_{n \in Z + k \eta_{i}}
\frac{1}{\eta_{i}} {\bar c}^{i}_{n} e^{-i n (\tau +\sigma)}&.
\label{faal11}
\end{eqnarray}
The position of the center of mass is quantized and the center of mass 
is found from 
\begin{equation}
q^{(k,f)i} = (C^{k} q^{(k,f)})^{i} + 2 \pi m^{a} e^{i}_{a}.
\label{faal12}
\end{equation}
The value of k represents the twisted sector and $k=0,\dots,N-1$. 
For the untwisted sector $k=0$.
The twisted strings in this way are fixed by the automorphisms of the
lattice of the orbifold.
We can consider now the action of the space group in the NSR fields.
By representing the action of the point group on the lattice with the
shift $(1, u^{t}) $ we get that the NSR fields are given by,
\begin{equation}
{\phi} = q^{t} + {\frac{1}{2}}(p^{t} + k u^{t})(\tau - \sigma) +
\frac{1}{2} \sum_{(n \neq 0)} \frac{1}{n} a_{n}^{i} e^{-i n (\tau - 
\sigma)}, t=1,\dots,4.
\label{faal13}
\end{equation}
Here, we used the bosonized form of the NSR fermions, with $p^t$
taking values on the weight lattice\footnote{The weight lattice 
of $SO(8)$ can be constructed from its vector and spinor 
representations.} of $SO(8)$. 
As a result, the mass formulas for the
spectrum of the physical particles from the k-twisted sectors, are
found to be
\begin{equation}
{\frac{1}{8}}{(m_{R}^{(k)})}^{2} = {\frac{1}{2}} 
{\sum_{i=3}^{8}} {(p_{R}^{i})}^{2} {{\delta}_{k,0}} +
{\frac{1}{2}} {\sum}_{i=1}^{4} {(p^{t} + k u^{t})}^{2} +N_{R}^{(k)}
- \frac{1}{2} + c_{k},\;i=1,\dots,6
\label{faal14}
\end{equation}
and 
\begin{equation}
{\frac{1}{8}}{(m_{L}^{(k)})}^{2} = \frac{1}{2}
\sum_{i=3}^{8} {(p_{L}^{i})}^{2} {\delta}_{k,0}+
\frac{1}{2} {\sum}_{i=1}^{16} {(P^{I} + k V^I)}^{2} + N_{L}^{(k)}
- 1 + c_{k},\;i=1,\dots,6
\label{faal15}
\end{equation}
where $p_{L}^{i}$ and $p_{R}^{i}$ represent the left and right moving
momenta respectively of the $X^{i}$, while $ N_{R}^{(k)}$
and $ N_{L}^{(k)}$ are the number operators in the k-twisted sectors.
The shift embedding $V^I$ in the gauge degrees of freedom is a automorphism
of the $E_8 \times E_8$ and represents the embedding of the space group in 
the gauge degrees of freedom. In general S is embedded 
by an automorphism and/or shift in the gauge degrees of freedom.
If the automorphism is in the Weyl group of $E_8 \times E_8$
then the twist is realized through a shift\cite{ibaniqe1}.
In general, it is possible to have additional background field parameters
corresponding to topologically non-trivial directions in the gauge and
torus directions defined by 
\begin{eqnarray}
\int A^I_i dx_i= A^I_i  e^i_a \equiv {\alpha}^I_a,\;\;I=1,\dots,16,\;\;
i,a=1,\dots,6,
\label{kikiri6}
\end{eqnarray}
where $e^i_a$ are the basis vectors on the six torus.
In the case that the Wilson lines commute with the general gauge group 
element\footnote{i.e 
representing the gauge group element in the form  $G \times P$, where G, P 
denote
the gauge and point group elements, we may have 
$g \rightarrow (\Theta,0;1,V^I),\; A \rightarrow (1, e^i_a;1,
{\alpha}^I_a)$. Then $[g, A] =0$.}, we can realize the gauge element and 
the Wilson lines through 
shifts in the $E_8 \times E_8$ lattice. 
In this case, the shift in the momentum in the twisted sectors
will be modified to $ p^I + V^I + \sum_{a=1}^6 m_a^f {\alpha}^i_a$, where
$m_a^f$ integers is associated with the fixed points.
If the Wilson lines do not commute with shifts on the lattice, they
can lower the rank of the gauge group\cite{ibaniqe1}. We will not describe,
additional properties of the Wilson lines since we will not need them in 
this Thesis.

The numbers $c_{k}$ represent the contribution of the oscillators to 
the zero point energy in the form
\begin{equation}
c_{k} = {\frac{1}{2}} \sum_{i=1}^{3}{\left( |k \eta_{i}| - Int(|k 
\eta_{i})\right)}{ (1 - |k \eta_{i}| + Int(|k\eta_{i}|)},
\label{faal16}
\end{equation}
where $Int(|k\eta_{i}|)$ denotes the integer part of the expression 
in parenthesis.
In addition,  physical states have to satisfy criteria that are 
coming from the requirement that the physical Hilbert space of the
theory must keep states which are only $Z_{N}$ invariant.
In the untwisted sector, we will have to project to $S \times P$ invariant 
states in order to implement the orbifold projection of the heterotic string
spectrum.

The right moving excitations of the heterotic string are described 
by the ten dimensional superstring (SP).  The tachyon is projected 
out by the GSO projection, and the vacuum states are those of the SP.
As a result, the chirality of the physical states in the Hilbert space comes 
form the right moving sector.   
The latter are the ten dimensional\footnote{
The mass operator for the heterotic string in the right moving sector
is that of the NS-R superstring\cite{gsw1}.   namely $N=\sum_{n=1}^{\propto}
(a^i_{-n} {\tilde a}^i_n + n S_{-n}^i S^i_n)$, where  
vector $b_{\frac{1}{2}} |0>$ and 
spinor $|S^a>$. }
The supersymmetry charges surviving the orbifold projection are given 
by the condition
\begin{equation}
[V(\upsilon), P_{\phi}] =0,\;\;\; V(u)= e^{2i \upsilon^t \phi^t},\;\;\; 
P_{\phi} = e^{-2 \pi i u^t p^t}, 
\label{proje1}
\end{equation}
where $ V(\upsilon)$ is the supercharge and $\upsilon^t \in
8_c$, and  $P_{\phi}$ is a translational operator acting on the
bosonic fields and represents a $Z_N$ rotation on the
bosonic coordinates, $X^i(\sigma + 2\pi, \tau) =e^{ 2 \pi i u^i}
X^i(\sigma, \tau)$.
As a result the number of supersymmetries is equal to the number
of supercharges which satisfy the previous condition.
The number of unbroken supersymmetries
is given by the number of vectors which satisfy 
(\ref{proje1}), i.e  $ \sum_{i=1}^{4} u^{i}
{\upsilon}^i = integer $ with ${\upsilon^i} \in 8_{c}$. The $8_{c}$
corresponds to the conjugate spinor representation of $SO(8)$.
For vacua with $N=1$ supersymmetry unbroken the previous condition
becomes $ \pm  u^1 \pm u^{2} \pm u^{3} \pm u^{4} = 0$, where
we must set $u^1 =0$ since space time degrees of freedom are not 
rotated. 
%This condition gives . 
Now the point group is embedded in the standard way in a $SO(6)$ subgroup 
of $E_8$. Because its eight eigenvalues acting on the spinor of $SO(8)$
are in the form $e^{i \pi(\pm u^1 \pm u^2 \pm u^3)}$, there are at least two 
zero 
modes for the right handed spinor fermions in the Green-Schwarz formalism.
So at least one unbroken supersymmetry in four dimensions remains.

We will now describe constraints from modular invariance of the one loop
vacuum amplitude.
The space of physical states in the Hilbert space of the orbifolds is given
by the
direct product of the Hilbert spaces of the left and right sectors.
The constraints of the spectrum coming from modular invariance are best 
described from the 
examination of the properties of the vacuum amplitude $Z(g,h)$.
At one loop the string world sheet describing the one loop amplitude
with no external legs is a torus. The torus is described by the modular 
parameter
 $\tau$\footnote{ If we want to describe the group of global diffeomorphisms
 on the torus, the space of inequivalent tori, then the parameter space for
  
 $\tau$ is the group of modular tranformations under which $\tau \rightarrow 
 \frac{a \tau + b}{c \tau + d}$, with $ad-bc=1$, a, b, c, d $\in Z$,
 the modular group of the torus $PSL(2,Z)$. See appendix A.}.
If we parametrize the torus by $(\sigma_1, \sigma_2)$, where a point on the torus
corresponds to the complex number $  \sigma_1 + \tau  \sigma_2$, then we can 
consider the bosonic string variables on the orbifold in the form
\begin{equation}
X( \sigma_1 + 2 \pi, \sigma_2) =h  X(\sigma_1,\sigma_2),\;
X( \sigma_1, \sigma_2 + 2 \pi)= g  X(\sigma_1,\sigma_2),\; 0 \leq \sigma_1,
\sigma_2 \leq 2 \pi.
\label{proje2}
\end{equation}
Here, g and h represent twisted boundary conditions, which can be periodic 
or antiperiodic. For consistency of boundary conditions h and g must commute.

We are introducing the one-loop $Z_{N}$ invariant vacuum amplitude as
\begin{equation}
Z(g,h) = \int \frac{{d}^{2} \tau}{{(Im \tau)}^{2}} {\cal H}(\tau).
\label{faal17}
\end{equation}
Here, ${\cal H} = \int {\cal D}X  {\cal D}\Psi e^{-S} $ is the path integral of 
the action $S$ on the world 
sheet torus 
over the fermionic $\Psi$ and bosonic X degrees of freedom.
In orbifolds, ${\cal H}$ receives contributions from the untwisted and the
twisted sectors while its general form is 
 \begin{equation}
 {\cal H} = \frac{1}{|{\cal P}|} \sum_{[h,g]=0} \epsilon(h,g) I(h,g)(\tau).
 \label{faal171}
 \end{equation}
Here, $\epsilon(h,g)$ are the discrete torsion phases\cite{vafas}, while
the interpretation of the sum in (\ref{faal171}) is as follows.
For abelian group G, the path integral is calculated with boundary 
conditions h (g) in the $\sigma_1$ $(\sigma_2)$ directions. 
The various components of $I(h,g)$ are related one another by modular 
transformations of the world sheet parameter $\tau$.  Under 
$\tau$ modular transformations, i.e $\tau \rightarrow \frac{a \tau + b}{
c \tau + d}$, $(h, g) \rightarrow (h^d h^a, h^b g^a)$
and ${\cal H}(h, g) \rightarrow {\cal H}(h^d g^c, h^b g^a)$.
For
$Z_N$ orbifolds with $\cal P$ elements $\theta^m, m=0,\dots,N-1$, we have
$\epsilon(\theta^m, \theta^n) =1$ for every m, n.
The various contributions to $I(h,g)$ are given by
\begin{equation}
Tr \int \frac{{d}^{2} \tau}{{(Im \tau)}^{2}} 
 (G^{(h)} e ^{2 \pi i \tau L_{o}^{h}} e^{- 2 \pi
i{\bar \tau}
{\bar  L}_{o}^{h}}),
\label{faal1712}
\end{equation}
where the Tr is over all modes while the $L_{o}^{h}$ and ${\bar L}_{o}^{h}$
are the 
Virasoro operators for the h-twisted sectors, right and left
respectively. Eqn.(\ref{faal1712}) 
is associated with the sum over g twisted sectors, twisted by h. 
Moreover,
\begin{equation} 
{\hat G}^{(h)} = \frac{1}{N} \sum_{g=0}^{N-1} {\tilde \psi}(h,g)
{(\triangle_{(g)})}^h ,
\end{equation}
where, N is the order of the point group,  ${\hat G}^{(h)}$ implements
the projection operator ${(\triangle_{(g)})}^h$ in the g twisted sector.
For $Z_N$ orbifolds, 
\begin{eqnarray}
{\hat G}{(\theta^m)}= \frac{1}{N} \sum_{n=0}^{N-1}
{\tilde \psi}(\theta^m, \theta^n)\triangle_{g}(\theta^m)^n,
\label{kikiri1}
\end{eqnarray}
\begin{eqnarray}
\triangle_{g}(\theta^m) = e^{(2 \pi i
[ \frac{1}{2} n(\sum {- {(u^{t})}^{2} 
 + \sum {(V^{t})}^{2}} )+ \sum(P^{I} + mV^{I})V^{I}
 - \sum(p^{t} + m u^{t})u^{t}])}.
 \label{kikiri2}
\end{eqnarray}
Here, $V$ is the embedding of the point group in the gauge 
degrees of freedom. 
In the case, that there are massless states with left oscillators
(\ref{kikiri2}) must be modified with the addition of the term 
$exp^{2 \pi i \xi} $, which is the eigenvalue of the oscillator 
under $\theta$.  
The number of fixed points under $\theta$  gives us the 
degeneracy of the corresponding twisted vacuum and
appears in the one loop partition function as an overall factor.
The factor ${\tilde \psi}(\theta^m, \theta^n)$ is the degenerary factor.
The explicit formula for the ${\tilde \psi}$ reads
\begin{eqnarray}
{\tilde \psi}(\theta^m, \theta^n)= {\psi}(\theta^m,\theta^n),\;for\;
{\psi(\theta^m)} \neq 0,&\nonumber\\
{\tilde \psi}(\theta^m, \theta^n)= \frac{\psi(\theta^m, \theta^n)}
{\displaystyle\mathop{\Pi_j }
\psi_j (\theta^n)},\;for\;\psi(\theta^m)=0,\;n \neq 0&\nonumber\\
{\tilde \psi}(\theta^m, \theta^n)= \Pi_{k} \psi_k (\theta^m),\; 
for\;\psi(\theta^m)=0\;and\;n=0,
\label{kikiri3}
\end{eqnarray}
where ${\psi}(\theta^m,\theta^n)$ is the number of points simultaneously 
fixed by the point group elements $\theta^m, \theta^n$ and the subscript j
runs over untwisted components and k over twisted components.
If the rotation $\theta^m$ leaves fixed tori, $\psi(\theta^m)=0$.
 We should notice that the number of points fixed under the 
automorphism $\theta^{\rho}$ depends only 
on the automorphism, not on the specific lattice and can
be calculated using the Lefschetz fixed point theorem \cite{dhvw}
\begin{equation}
n= \psi(\theta^{\rho}) = det(1 - \theta^{\rho})= {\Pi_j} 4 sin^{2} m \eta_{j},
\end{equation}
where the determinant is in the vector $\bf 6$ representation of $SO(6)$ 
for compactifications that preserve one four-dimensional
supersymmetry.

For $m=0$, $ {\tilde \psi}(1,\theta^m )=1$, since we are considering 
untwisted strings. For example, the total partition function
can be written as
\begin{equation}
P =P_{Untwisted} + P_{Twisted},\;P_U = \sum_{n=0}^{n-1} I(1,\theta^n),\;
P_T=\sum_{n=0}^{n-1} \sum_{
[h,g]=0,m \neq 1} I(\theta^m, \theta^n).
\label{kikiri4}
\end{equation}

The invariance of the vacuum amplitude  $(\theta, 1)$ under modular 
transformations $\tau \rightarrow \tau + N$ gives\cite{vafas,senda1}
\begin{equation}
N \left( \sum {(u^{t})}^{2} - \sum {(V^{t})}^{2} \right) = mod \;n,
\label{faal20}
\end{equation}
\begin{equation}
N \sum_{i=1}^{4} u^{t} = N \sum_{J=1}^{8} V^{J} = N {\sum_{J=9}^{16}}
V^{j'} = 0 \bmod 2.
\label{koularisa}
\end{equation}
The level matching constraint (\ref{faal20})\cite{vafos,dhvw} holds for n 
even, while for n odd becomes mod 2n.

Up to know we have discussed properties of the
 space group of the orbifold . In general S will be embedded
 in the gauge degrees of freedom by acting with an 
automorphism $\theta$ and or shift V on the gauge
degrees of freedom. In the absence of Wilson line
backgrounds a
Weyl automorphism is equivalent to a lattice shift\footnote{The reverse 
argument is not valid.}.

If one acts on the gauge degrees of freedom with
exactly in the same way as in the $SU(3) \subset SO(6)$ subgroup of
extra six dimensions this is called the standard embedding.
This scheme is reminiscent of the identification of the spin connection 
with 
the gauge connection\cite{Calabi1} in Calabi-Yau manifolds. There, 
by setting the H field equal to zero\footnote{See some relevant discusion 
in chapter four.} the Bianchi identity, coming from compactification
of the heterotic string in a six dimensional Calabi-Yau manifold, 
 becomes  
$F^a_{[MN}F^a_{KL]}=R^{cd}_{[MN} R^{cd}_{KL]}$, where the capital letters 
take
values in the tangent space of the ten-dimensional space time. 
Picking up as a solution, for manifolds of $SU(3)$ holonomy
the identification of the $SU(3)$ spin connection 
with the $SU(3)$ gauge field $A_M$, breaks the $E_8$ down to $E_6$.
In the same way, in orbifolds we can embed the $SO(6)$, in the fermionic 
representation of $E_8$ where the $E_8$ degrees of freedom are 
described by sixteen world-sheet fermions transformimg as $\bf 16$ under
the $SO(16) \subset E_8$, action of P in S such that the
embedding of the $SO(6)$ in its Cartran subalgebra, i,e 
$exp^{2 \pi i(a J_{12} + bJ_{34} + c J_{56})}$  is identified 
with a shift $(a, b, c, 0^5)$ on the group torus\cite{dhvw}.
For the $Z_3$ orbifold, coming by the standard embedding, choosing the
the point group embedding $r_1= \frac{1}{3}(1,1,-2)$ and the gauge group 
embedding $V=\frac{1}{3}(1,1,,0^5)$ the gauge group breaks to 
$E_6 \times SU(3) \times E_8$.
The gauge group of the
theory will come from the untwisted sector while the matter supermultiplets
come from the twisted sectors. The gauge and matetr content is a consequence 
of the massleness condition (\ref{faal14},\ref{faal15}) and the orbifold 
projection (\ref{kikiri2}).

Our main interestcis to the $Z_N$ twists of six-dimensional
Lie-algebra
lattices - with the point group to be generated by automorphisms of
these lattices - which leave unbroken space-time    
supersymmetry. These automorphisms
are realized as inner or outer. The inner
automorphisms are given by the Weyl group of the algebra. 
A special class of inner automorphisms
are the
Coxeter elements, which can be written as products of Weyl reflections
with respect of all the simple roots, that will be discussed in more detail
 in chapter three. 

\subsection{ Duality symmetries}

 The simplest way to get a four-dimensional (or more generally
a D-dimensional, $D<10$) theory from the ten-dimensional
heterotic string is the compactification of six (or $d$)
space dimensions on a torus \cite{na}:
${R^{10}} \longrightarrow {M^{D}} \times T^{d}$,
where
\begin{equation}
T^{d} = \frac{R^{d}}{\Lambda},\;\;\;d = 10 - D,
\end{equation}
and $\Lambda$ is a $d$-dimensional lattice with basis $\{{\bf e}_{i}\}$.
A state on the compactified sector is labeled by its
continuous space-time momentum $p^{\mu}$, $\mu = 1,\ldots,D$,
its oscillation state,its gauge quantum numbers $p^{I}, I=1,\ldots,16$
and by its winding vector
\begin{equation}
{\bf w} = w^{i}{\bf e}_{i} \in \Lambda,\;\;i=,\dots,d
\end{equation}
describing how the string
wraps around the internal dimensions. In addition, it is labeled 
by its discrete internal momentum
\begin{equation}
{\bf p} =  p_{i}{\bf e}^{*i} \in \Lambda^{*},;i=1,\dots,d,
\end{equation}
where
$\{ {\bf e}^{*i} \}$ is the standard basis of the dual lattice
 $\Lambda^{*}$ of $\Lambda$. The metrics of the two lattices are
\begin{equation}
G^{ij} =  {\bf e}^{*i}\cdot {\bf e}^{*j} \mbox{ and }
G_{ij} =  {\bf e}_{i}\cdot {\bf e}_{j}, \mbox{   where  }
G^{ij}G_{jk} = {\bf e}^{*i}\cdot {\bf e}_{k}  = \delta^{i}_{k}.
\end{equation}
All the internal quantum numbers  of the D--dimensional theory
can be combined into the left-- and right--moving momenta, which  
form the momentum lattice $\Gamma_{16+d;d}$:
\begin{equation}
\left( {\bf p}_{L} ; {\bf p}_{R} \right) =   
\left( p_{L}^{I} {\bf e}_{I} + p_{L}^{i} {\bf e}_{i} ;
p_{R}^{i} {\bf e}_{i}  \right) \in \Gamma_{16+d;d},
\end{equation}
\begin{equation}
p_{L}^{I} = p^{I} + {A^{I}}_{i} w^{i},\;{\bf A} ={A^{I}}_{i} e_I e^i,
\end{equation}
\begin{equation}
p_{L}^{i} =  \frac{1}{2}p^{i} - B_{\;k}^{i}w^{k}
- \frac{1}{2} p^{J}A^{Ji} - \frac{1}{4}
A^{Ki}A^{K}_{j}w^{j} + w^{i}
\end{equation}  
with $w^{i} \in Z$ 
\begin{equation}
p_{R}^{i} =  \frac{1}{2}p^{i} - B_{\;k}^{i}w^{k}   
- \frac{1}{2} p^{J}A^{Ji} - \frac{1}{4}.
       A^{Ki}A^{K}_{j}w^{j} - w^{i}
\end{equation}
Here, ${\bf A}$ are the Wilson line background fields\cite{naaa1} and the 
$e^I$ are basis vectors for the self-dual lattice $\Gamma_{16+6;6}$.
Toroidal compactification of the heterotic string with
p left moving and (16 + p) right moving coordinates compactified, 
gives a moduli space parametrized from 
p(16 + p) components. The antisymmetric tensor 
corresponds to $(1/2)p(p-1)$ components, the metric tensor $g_{ij}$
to $(1/2)p(p+1)$ components. The remaining 16p parameters are 
associated with the Wilson lines $A^I_i$. 
The presence of the Wilson lines are neccesary since
even if the condition for finding a vacuum solution for the Yang-Mills 
strength is $F_{ij}=0$, the Yang-Mills field on the torus can still have 
a non-trivial holonomy associated with the Wilson lines. Of course, this is
in exact analogy with the instanton solitions of Yang-Mills equations
in gauge theories.

In the case of vanishing Wilson lines $A^{I}_{i}$, I can write for the
momenta
\begin{equation}
\begin{array}{ccccr}
{p_{L}} = \frac{1}{2}m + (G - B) n&,&{p_{R}} = \frac{1}{2}m - (G + B) n,
\end{array}     
\end{equation}
where $G$, $B$ represent the background metric and the
antisymmetric tensor field.
The moduli space of heterotic string compactified
on a D-dimensional torus\cite{na}
is locally isomorphic to the coset manifold  $\frac{SO(D,D)}
{SO(D) \times SO(D)}$.
However in order to reveal the global geometry of the moduli space we
have
to find how the discrete  modular symmetries modify its coset
structure.
In this part of the thesis we will describe the way, that the six-
dimensional part of the compactification lattice fixes the background
deformation parameters, namely we  will find  the conditions
imposed to the background fields from the requirement of duality
invariance on the physical
spectrum of orbifold compactifications.

The target space duality symmetries now are defined as those discrete
transformations acting 
on the the quantum numbers which leave the spectrum and the
interactions invariant.

Duality in its simplest version, is the compactification of a
closed bosonic string on a circle,
represents the invariance of the spectrum under the inversion of the 
radius $ R \rightarrow \frac{1}{2R}$ 
and a simultaneous interchange of the momentum and winding numbers
$m \leftrightarrow n$. This result can be seen directly
from the discussion in section 2.1, but 
for simplicity let us review this result\cite{losathoma}.
Compact bosonic strings with one of the compact dimensions, e.g $X^{25}$
 compactified
on $S^1$, a circle of radius R, means that $X^{25} \equiv X^{25} + 2 \pi R n$.
In this case, the wave equation for the compact coordinate splits into
left and right movers  
\begin{equation}
 X_R = x_R - \frac{1}{2} p_R (-\tau +\sigma)
 +\frac{1}{2} \sum_{k \neq 0} a_k e^{-ik(\tau -\sigma)},
\label{koipoio1}
\end{equation}
\begin{equation}
 X_L = x_L - \frac{1}{2} p_L(\sigma + \tau) +\frac{1}{2}\sum_{k \neq 0},
 {\tilde a}_k e^{-ik(\sigma + \tau)},
 \label{koipoio12}
\end{equation}
where $a_k$ , ${\tilde a}$ are oscillators. The left and right moving momenta
 the Hamiltonian and the spin are given by
\begin{equation}
p_L=\frac{m}{2R} + nR,\;p_R=\frac{m}{2R}-nR,\;H=\frac{m^2}{4 R^2} + n^2 R^2,\;
S= \frac{p_L^2}{2}  -\frac{p_R^2}{2}=  mn.
\label{koipoio112}
\end{equation}
Obviously, the spectrum is invariant under the transformations 
$R \leftrightarrow \frac{1}{2R}$ and $m \leftrightarrow n$.

For the heterotic string,
the Hamiltonian and the spin of the vertex
operator are defined by\cite{na}
\begin{equation}
H = \frac{1}{2}\left({p_{L}}^{\tau} G^{-1} p_{L} + {p_{R}^{\tau}}
G^{-1} p_R \right) = \frac{1}{2} \left( u^{t} \Xi u \right),
\label{invar1}
\end{equation}
\begin{equation}
S = \frac{1}{2} \left({p_{L}}^{\tau} G^{-1} p_{L} - {p_{R}^{\tau}}
G^{-1} p_R \right) = \frac{1}{2} \left( u^{t} \eta u \right),
\label{invar2}
\end{equation} 
where
\begin{eqnarray}
u=\left(\begin{array}{ccr}
n\\
m\\
\end{array} \right),\;
\eta=\left(\begin{array}{ccrccr}
0&{{\bf 1}_{d}}\\
{{\bf 1}_{d}}&0\\
\end{array} \right),\; 
\Xi= \left(\begin{array}{ccrccr}
2\left(G - B\right) G^{-1} \left(G + B \right)&B G^{-1}\\
-B G^{-1}&\frac{1}{2} G^{-1}\\
\end{array} \right),
\end{eqnarray}
and ${\bf 1}_{d}$ the identity matrix in d-dimensions.
The duality transformations $\Omega$ act as
\begin{equation}
\Omega : u \rightarrow\ {S_{\Omega}} \left(u \right)={{\Omega}^{-1}}u.
\end{equation}   
Invariance of  eqn's.(\ref{invar1},\ref{invar2}) under target space duality
transformations gives the conditions for the background
$H$, $S$ to remain invariant namely :
\begin{equation}
{\Omega}^{\tau} \eta \Omega = \eta
\label{invar3}
\end{equation}
\begin{equation}
\Xi \rightarrow {{\Omega}^{\tau}} \Xi  \Omega,
\label{invar4}
\end{equation}
so that $\Omega$ is an element of $O(d,d,Z)$  and (\ref{invar4}) defines
the action of the duality group on the moduli fields.
In the case of toroidal orbifolds\cite{imnq} the quantum numbers
transform as 
\begin{equation}
u  \rightarrow {u'} = R u,\;R^N=1.
\label{invar5}
\end{equation}
Because the point group acts on the compactified coordinates $x^{\mu}$ as
$x^{\mu}( 2 \pi, \tau) = \theta^{\mu}_{\nu}x^{\nu}(\sigma, \tau) + 
2 \pi w^{\nu}$,
where $\mu=1,\dots,d$ and the $\theta^{\mu}_{\nu}$ the twist
matrix in the space time basis, we have that
\begin{equation}
R\stackrel{def}{=}\left(\begin{array}{ccccr}
Q&0\\
0&Q^{*}\\
\end{array} \right),
\label{invar6}
\end{equation}
where $w^{\nu}$ is the winding number and $Q$ by definition is
\begin{equation}
{\theta^{\mu}_{\nu}} e^{\nu}_{i} \rightarrow e^{\mu}_{j} Q_{ij}
\end{equation}  
Because the twist is an automorphism of the lattice, Q must have integer
entries.

Finally the condition that the point group to be an a lattice
automorphism
gives\cite{spa2}
\begin{equation}
{Q^{t}} G Q = G,
\end{equation}
\begin{equation}
{Q^{t}} B Q = G.
\end{equation}  
We will end the discussion of the toroidal duality symmetries  
in orbifold models by discussing one more condition that have to be
satisfied by the momentum and winding numbers in an orbifold
background.
For the $Z_N$ orbifolds the target space duality symmetries  
of the untwisted sector that are
surviving the orbifold projection, i.e the torus has to commute with
the twist, have to satisfy\cite{ersp} ,
$\Omega R - R^k \Omega = 0 , k =0,1,..N $,
while if the point group action $\theta^{k}$  leaves  a
complex plane invariant $ Q^k n = n , {Q^*}^{k} m = m $, where * denotes
inverse and transpose.
For twisted sectors of $Z_N$ Coxeter orbifolds modular symmetries
were examined in \cite{blst1,blst2}. In reality there are three ways
to find the modular symmetry. They are listed in \cite{lust}.

\newpage
\vspace{5cm}
\begin{center}
{\bf  CHAPTER 3}
\end{center}
\newpage

\section{\bf Aspects of Threshold corrections to  Low 
energy Effective string theories compactified on Orbifolds}

\subsection{ Introduction} 
%********************************************  A1  *******
Four-dimensional \cite{fer1,fer2,fer3,dhvw,ge1,kasu} 
superstrings \footnote{derived from  
Heterotic string 
constructions \cite{Gr} .}
represent 
at present our best ever candidates for a theory which can 
consistently
unify all interactions including gravity, even if at present, 
there is no
second quantized formalism. In order for the string theory at the 
Planck
scale to make contact with the observed world at the weak scale, one
needs 
to find the effective low energy theory of the superstring  
theory(ELET).

Corrections to the general theory of relativity(GR) coming from the
 massless 
modes of the superstring theory appear in two different forms. One is 
associated with the use of background field 
contributions of the infinite tower of massive modes leading to 
corrections in $\alpha'$, while the other corresponds to quantum loop
effects.
In the first form,
the effective lagrangian of the massless modes can be studied in
the $\sigma$ model approach\cite{SIG}. An example action involving 
bosonic backgrounds only is the following:
\begin{eqnarray}
S=\frac{1}{2\pi \alpha^{\prime}} \int \frac{d^2}{dz}[\sqrt{g} g^{ab} 
\partial_a X^{\mu} \partial_b X^{\nu} G_{\mu \nu}(x) +\epsilon^{ab} 
\partial_a X^{\mu} \partial_b X^{\nu} B_{\mu \nu}(x)\nonumber\\
+ a^{\prime} \sqrt{g}R^{(2)} \Phi(x) + \sqrt{g} f(x)],
\label{back1}
\end{eqnarray}
where $G_{\mu \nu}(x)$, $B_{\mu \nu}(x)$,  $\Phi$ and the vacuum 
expectation values of the usual background fields, namely metric, 
antisymmetric tensor, dilaton and tachyon field 
respectively\cite{SIG}. In the conformal gauge $\sqrt{g} g^{ab}=
\delta^{ab}$, imposing 
conformal invariance, namely traceless stress energy tensor,
quarantees decoupling of negative norm states. In this case,  
conformal invariance, i.e vanishing  $\beta$-functions for the 
background\footnote{By expanding the stree enrgy tensor to leading 
order in the loop coupling constant $\alpha^{\prime}$.}fields, 
leads into the correct\cite{gsw1} equations of 
motion for the massless fields. 
\begin{eqnarray}
\beta^{\Phi} = \frac{1}{\alpha^{\prime}} \frac{d-26}{48\pi^2} +
\frac{1}{16\pi^2}[4({\bigtriangledown}^2 \Phi)^2 - 
4(\bigtriangledown^2 \Phi)-R+\frac{1}{12}H^2]+
{\cal O}(\alpha^{\prime})&\nonumber\\
\beta^{G}_{\mu \nu} = R_{\mu \nu} - \frac{1}{4} H_{\mu}^{\xi \rho}
H_{\nu \xi \rho}+2\bigtriangledown_{\mu} \bigtriangledown_{\nu}\Phi+
{\cal O}(\alpha^{\prime} &\nonumber\\
\beta^{B}_{\mu \nu}= \bigtriangledown_{\xi} H^{\xi}_{\mu \nu} - 2(
 \bigtriangledown_{\xi} \Phi)H^{\xi}_{\mu \nu}+{\cal O}
(\alpha^{\prime}),\;\;H_{\xi\mu\nu}=3\bigtriangledown_{[\xi}B_{\mu\nu]}
\label{back2}
\end{eqnarray}
  
In the other form, corrections to GR come via the S-matrix 
approach\cite{Gr1}. In the S-matrix approach,
the
calculation of the effective lagrangian of the massless modes proceeds
through the calculation of string scattering amplitudes. It has been 
shown\cite{Gr1}, that the tree level action of the heterotic string 
corresponds to the bosonic part of the Chapline-Manton lagrangian with
the
field strength of the antisymmetric tensor field appropriately
generalized to include Chern-Simons three forms. The latter account for 
the cancellation of the gauge and gravitational anomalies in the 
10D heterotic string action.

For the $N=1$ heterotic string the massless sector consists of the 
Yang-Mills\footnote{Here $A_M^{\alpha},\;\chi^{\alpha}$ represent
the gauge 
field
and its gaugino in 10 dimensions respectively.}
supermultiplet $(A_{\mu}^{\alpha}, \chi^{\alpha})$ in the adjoint
representation of the gauge group and the $N=1$ supergravity 
multiplet(SM) consisting partly of the graviton, the dilaton 
scalar ${\phi}$, and the antisymmetric tensor $B_{\mu \nu}$.
It's effective lagrangian in 10D describes 
$N=1$ 
supergravity
coupled to supersymmetric Yang-Mills.
The low energy lagrangian coming from the calculation of superstring 
scattering
amplitudes describes the massless mode excitation dynamics of the
heterotic string\cite{dkl1,kap} at the string unification scale.

The effects of string theory in our low energy world at the weak scale,
are becoming apparent via the running of physical couplings through the
evolution of the Renormalization group equations(RGE).
Particular role in this respect is played by the gauge coupling
constants whose properties we will examine later at this chapter.  In
string theory physical couplings and masses are field dependent.  They
depend explicitly on the vevs of some massless scalar fields, the so
called moduli   
\footnote{The continuous deformations, 
of a superstring solution constitute its moduli space.
At the level of conformal field theory they correspond\cite{Seiberg} to 
integrably marginal operators $\Phi$, i.e. BRST invariant operators,
that we can add to the world sheet lagrangian without affecting the
equations of conformal invariance i.e. the value of the $\beta$-function
remains zero.
The general form of the deformation appears as
$\sum_i g_i\int d^2z \Phi_i(z,{\bar z})$, where the constants $g_i$
correspond to the coordinates of the moduli space, i.e they are 
moduli.}
From the point of view of effective low energy theory the moduli
are massless neutral scalar fields with a flat potential to all orders
\cite{Di1,Disei} of perturbation theory
with classically undetermined vevs that can be as
large as $M_{plank}$. Neglecting non-perturbative effects, the moduli 
fields give an infinite degeneracy of string vacua. In addition, 
the global structure of the the moduli space $\cal M$ is affected by 
the invariance
under some discrete reparametrizations of the moduli fields $\Phi$
\begin{equation} 
\Phi_{i} \rightarrow {\bar \Phi}_{i} \left({\Phi}_{i}\right) \in 
{\cal M},
\label{epi1}
\end{equation} 
the so called
 target space duality\cite{dis,Kik,ler,cmo,alva,nair,GPR} 
transformations.
They change the geometry of
the internal space and leave invariant
\cite{GPR} the massive spectrum 
and interactions. 
They are of great importance, since by 
lifting the degeneracy
 by perturbative string theory
\cite{kp,fkpz,fkp,kr} or non-perturbative
effects\footnote{See chapter 5.} we will be able to see clearly the
effects of string theory on the physical observables. On the other hand,
the moduli dependence of the effective action is important since
non-perturbative effects like gaugino condensation can provide a
potential for moduli fields which can lift the vacuum degeneracy and
provide a mechanism for supersymmetry breaking\footnote{The most
satisfactory solutions to supersymmetry breaking up to  know are coming
from non perturbative gaugino condensation mechanisms even if they fail
to 
fix correctly the dilaton-see chapter 5.}. Unfortunately,
a non-perturbative formulation of string theory is still 
lacking\footnote{For a review see reference (\cite{vafos}).}.

In this chapter we will explicitly discuss the one-loop moduli 
dependence
of effective gauge couplings\footnote{At the level of perturbative 
string theory.}. String theory demands that the massless spectrum of 
physical
particles originate from superstring excitations that are massless
at the string unification scale.
Below this scale the effective gauge couplings(EGC)\footnote{
Effective quantum field theories involve\cite{kalou1,sv,sv1} two 
kinds of couplings.
The Wilsonian gauge couplings, which are gauge couplings of an effective
lagrangian
from which the massive modes have been integrated out and
depend on the cut-off scale. In  addition, there are the
EGC which depend on the momentum scale. Since  don't depend
on the cut-off scale, they 
do not correspond to any local effective lagrangian.}
evolve according to the usual RGEs which at the one loop level 
receive a threshold correction from the ultrahigh energy 
theory, string theory.
The values of the physical couplings at the string unification scale
$M_{string}$ represent the boundary conditions of our RGEs.  At tree
level the gauge interactions $g_{a}$ are all connected to a single scale,
the string mass $M_{string}\simeq 0.527{g_{string}}10^{18}GeV$ \cite{kap,dfkz} 
as follows\cite{gi}
\begin{equation} 
g_a^2 k_a = {4\pi}{\alpha'}^{-1}{G_N} = g_{string} =
{G_N}{M^2_{string}} =\frac{\kappa^2}{2 \alpha'}, 
\label{2}
\end{equation} , 
where $\kappa$ represents the gravitational coupling, $k_a$ the 
kac-Moody level of a gauge group factor\footnote{We assume that the 
gauge group at the string unification scale is a product of
group factors $G = \Pi_{a} G_a$.}, 
${\alpha'}^{-1/2}$ the string tension and $G_{N}$ the Newton constant.
The gauge coupling constant at a mass scale $\mu$, 
receives up to one-loop level\cite{kap}
corrections according to 
\begin{equation} 
\frac{1}{g_a^2(\mu)}=\frac{k_a}{g_{string}^2} +
\frac{b_a}{16\pi^2} \ln \frac {M_{string}^2}{\mu^2} -
\frac{1}{16 \pi^2} \triangle_a , 
\label{3} 
\end{equation} 
with $\triangle_a$ given by 
\begin{equation}
\triangle_a =\int_{\cal F} \frac {d^2 \tau}{\tau_2} \left({\cal B}_{a}
(\tau,\bar \tau) - b_{a}\right) .
\label{trigo}
\end{equation} 
Here $b_{a}$ is the
 field theoretical $\beta_a$ function coefficient of the
gauge  group factor $G_a$ of the effective theory of massless modes
contributing to the threshold corrections. The quantity $\cal B$ will be
defined in detail later in eqn.(\ref{51}).
$\triangle_a$, denotes the string
theoretical threshold correction\footnote{More details will be
given in section $(3.2.1)$.},
 of the factor $G_a$ of the 
gauge group $G = \Pi_a G_a$, to the gauge coupling constants.
In addition, $\tau=\tau_1 + i \tau_2$ is the modulus of the
world-sheet torus and the integration is over the 
fundamental domain ${\cal F}=\{\tau_2  > 0,\;
|\tau_1|<1/2,|\tau|>1 \}$.

The inclusion of threshold corrections is necessary to test the old 
ideas of unification of gauge interactions in the grand unified 
models, particularly since LEP measurements\cite{La} support the 
possible existence of a grand unified theory at an energy 
of $10^{16}$ GeV. Matching the correct values of the
electroweak data at $M_Z$, by ruuning the RGEs down to energies of the
electroweak scale, can support  
the existence of Higgs scalars in the adjoint, necessary to break the
grand unified gauge group to the standard model. Unfortunately the 
gauge
interactions in string theory depend on the Kac-Moody level "k" and in the
most popular searches at $k=1$ \cite{al2} there is no way that adjoint
scalars can
appear\cite{al3}, except the construction of flipped SU(5)\cite{anto}
where the breaking of the GUT group happens without adjoint scalars. 
 The appearance of adjoint Higgs at higher Kac-Moody levels \cite{al1,al2} 
becomes possible, via the existence  
 of string models with a grand unified group such as i.e, $SU(5)$ or 
$SO(10)$ \cite{al1}. This
of course makes more appealing the testing of GUT's but complicates the
proliferation of a specific string vacuum if any\footnote{If after 
compactification  we get a gauge group as 
$G_{1} \times G_{2} \times \dots $ then from the total contribution of the 
central charge to the left moving sector, we get the constraint
$c_{G} = \sum_{i} \leq 22$ . From this relation
$c_{G} = \sum_{i} = \sum_{i} \frac{k_{i} dim G_{i}}{k_{i} + 
\rho_{i}}\leq 22$, where $dim G_{i}$ and $\rho_{i}$ are respectively 
the dimension and the dual 
Coxeter number of $G_{i}$ $(\rho=N $ for $SU(N)$; $\rho=12$ for $E_{6})$, 
we can easily derive that $SO(10)$, $E_{6}$ can be 
at most realized at levels 7 and 4 respectively.}, since the
proligeration
of the vacuum is reduced to the old GUT problem of gauge 
symmetry breaking. 

One of the big problems of string theory at the moment, is that the 
value of 
string unification scale 
which is calculated up to one loop at string level\cite{kap} in
the $\bar DR$ scheme to \footnote{${\gamma}_{\epsilon} \simeq0.57722$ is
the Euler-Mascheroni constant. Normalization of the string coupling is as    
$g_{string} \stackrel{tree level}{=}g_{GUT}$. In this case\cite{kap},
$\alpha^{\prime} M_{plank}^2 g^2_{string} =32 \pi$. } be 
\begin{eqnarray} 
M_{string}\stackrel{def}{=}& \left(
\frac{2 e^{(1 - \gamma_{\epsilon})/2} 3^{-3/4}}{(2\pi 
\alpha^{\prime})^{1/2}} \right)&=
\frac{e^{(1-\gamma)/2} 3^{-3/4}}{4\pi}g_{string} M_{plank}\\
 &&\equiv 0.527\;g_{GUT} \times 10^{18}\;GeV, 
\label{dfkz}
\end{eqnarray} 
is in apparent disagreement with the success of the gauge coupling 
unification of minimal supersymmetric 
standard model (MSSM)\cite{ep2} at an energy of about $10^{16}$ GeV. 
Since MSSM gives us full agreement with the LEP
measurements the reason for disagreement caused a lot of excitement 
and several reasons were invoked to desolve the discrepancy.
In the case of additional massless chiral fields on top 
of the spectrum of MSSM\cite{ep1,ep2,bai1,difa}, one needs an additional
intermediate scale M at $\sim 10^{12-14}$ GeV\footnote{ which is
by itself contradictory since string theory is 
a theory of only one scale.}, to lower the string unification scale down  
to $10^{16}$ GeV.
An alternative way of lowering the string unification scale, 
is to consider variations of the hypercharge 
normalization\cite{iba93},  which for the case of the $U(1)$ gauge group
is identical to $k_1$. In this case we find 
agreement as long as the $k_1 \sim 1.4$.

 A different option uses the target space modular invariant
constraints necessary for cancellation of target space $\sigma$model
duality anomalies 
of the effective lagrangian to find the necessary range of
modular weights of matter fields which account for anomaly cancellation
associated with completely rotated planes\footnote{Planes for which the 
eigenvalue of the point group embedding is equal to -1} and minimal string 
unification. In \cite{iblu} it was
found that for (0,2) orbifolds only the $Z_{8}^{\prime}$ and $Z_{N} \times
Z_{M}$ orbifold survive this test.
 Unfortunately the values of the moduli which satisfy the constraints 
of duality cancellation and minimal string unification have values 
near 16, very far from the values obtained from gaugino 
condensation at
their self-dual points.  Of course it remains to be seen if further
superstring corrections to physical quantities of interest will improve
this analysis.  Finally, we would like to mention the recent
attempts \cite{bim,kloui} which use the soft terms\footnote{ Being 
left
over after 
\cite{sowe}the spontaneous supersymmetry breaking of the effective 
supergravity 
theory of our heterotic vacuum.}and minimal 
unification of gauge coupling constants to calculate at the weak scale 
the
masses.  In such an approach, one uses as effective low-energy 
particle content of the theory
the MSSM and the minimal unification approach,
to make predictions for the low-energy
$\alpha_{em}$ and the particle masses. This will test 
string theory in the near future.

 Various calculations of string threshold corrections to the
gauge coupling constants have been performed in the literature for 
different classes of heterotic strings. 
Initially calculations were performed \cite{kap} for $Z_3$ models with 
(2,2) supersymmetry and
with the presence of no Wilson line background fields. Further
investigation of the one-loop moduli dependence of the gauge coupling 
constants
was performed in \cite{dkl2}.
Application to fermionic constructions\cite{laca} was performed for the  
$Z_2 \times Z_2$ orifolds and especially for the flipped $SU(5)$ model,
where the corrections 
were incompatible with minimal unification.
An  analogous calculation was applied in the case of type-II superstrings 
in \cite{do} where the moduli sit at the 
enhanced symmetry point \footnote{This is just an example of a CFT 
since type-II superstring cannot incorporate the standard 
model \cite{kdva}.} .
The same investigation was applied for various Calabi-Yau manifolds
in \cite{ber,kj}. In addition, for symmetric (2,2)decomposable orbifold 
compactifications in \cite{dkl2} and 
for the non-decomposable orbifolds in\cite{deko}.

    The value of the $g_{string}$ in eqn. (\ref{trigo}) includes a 
universal-gauge group independent and moduli dependent contribution 
$\Delta^{universal}$ in the form\cite{kj}
\begin{equation}
g_{string}= Re S +\frac{1}{16\pi^2}\Delta^{universal}(\phi,{\bar \phi}),
\label{stringy1}
\end{equation}
which was discarded in all the 
previous
calculations.
By the way, the practical use of eqn.(\ref{3}) was in the calculation
of the one-loop threshold corrections coming  
by taking differences between different gauge groups. In this way the
value of the universal term didn't really matter.
The infrared regulator part and the contribution due to gravity of 
were recently calculated in \cite{kikou}. 
In\cite{mns,che} a numerical calculation of the value of the
gauge group independent universal term Y term\footnote{Writing the 
threshold corrections $\bigtriangleup_a = b_a \bigtriangleup + k_a Y$, 
where $\bigtriangleup$ the moduli 
dependent contribution, $k_a$ the Kac-Moody level and $b_a$ the $N=2$ 
$\beta$-function.} was reported for a variety of backgrounds 
with Wilson lines.
The value of the universal term for $Z_2 \times Z_2$ was recently 
calculated \cite{ri} using the work of reference \cite{kikou,mouha}.
It's value represents the exact contribution to the 
threshold corrections since in this case the underlying fields
$F_a^{\mu \nu}$ are exactly marginal and therefore their deformation
behaviour exactly calculable. 
The value of the universal term reflects the contribution\cite{kap}
of the gravitational back-reaction to the Einstein equations of motion
when the non-zero $F_{\mu \nu}$ background field is turned on.

Furthermore the full moduli dependence for orbifolds where the 
underlying lattice is assumed to decompose into a direct sum of a
two dimensional and a four dimensional sublattices namely
${\Lambda _{2}} \bigoplus \Lambda _{4}$ with the unrotated plane
lying in $\Lambda _{2}$ together with the 
inclusion of  Wilson line background fields  have been derived 
in\cite{ms2}. 

the inclusion of the 
Wilson line dependence on the one-loop string threshold corrections,
 can 
provide us with the the correct values of the Weinberg's angle and
strong coupling constant $\alpha_s$ at the scale $M_Z$\cite{nisti}.
Nevertheless, it is
introducing gauge group dependence. It is
unlikely \footnote{In my opinion.} that a proliferation of
the string vacua will involve any Wilson-line dependence since 
in this case there must be a principle to proliferate over the 
particular choice of Wilson lines which breaks the $(0,2)$ 
compactification to the $SU(3) \times SU(2) \times U(1)$ gauge 
group at $M_Z$.

An interesting development is the 
direct calculation of the threshold corrections in \cite{mouha} of
N=2 theories exhibiting
exactly the behaviour of the threshold corrections at the enhanced
symmetry points, as was predicted on the basis of symmetry
 arguments in \cite{lust}\footnote{via the calculation of 
the topological free energy.}.
It is interesting threfore to perform the same kind of calculation
for the case of non-decomposable orbifolds.

  In this part of the thesis we will calculate the full moduli
dependence of the threshold corrections for the case of the 
$Z_{8}^{\prime}$ non-decomposable orbifold \cite{erkl}, for 
the case where no Wilson line background fields are involved . This 
particular class of orbifold models was singled out from the list
of $ Z_N $ orbifolds, in the analysis 
of \cite{iblu} on the basis of satisfying the constraints from modular
anomaly cancellation and unification of coupling constants, if the 
low-energy particle content was that of the MSSM. Of course there is
always the  counter-argument: why is there no string theory with
just this particle content?
\newline
\newline
\subsection{\bf Threshold corrections to gauge couplings }

\subsubsection{Introduction}

The calculation of the threshold corrections to the gauge coupling
constants in N=1 orbifold compactifications was performed 
explicitly in \cite{dkl2}.
For the calculation of the threshold corrections to gauge couplings for 
the $Z_{8}^{\prime}$
non-decomposable orbifold\cite{erkl}, we need the full expression
of the $\triangle$ quantity in eq.(\ref{3}).

This quantity appears in the calculation of the one-loop amplitude
involving two gauge bosons. 
In analytic form,
\begin{equation}
{\cal B}_{a}(\tau,\bar {\tau}) = \frac{2}{|\eta(\tau)|^{4}}
\sum_{{even}\;s} \frac{(-1)^{s_1 + s_2}}{2 \pi i} 
\frac{d Z_{\psi}({\bf s},\bar \tau)}{d \tau} 
\times tr_{s_1} \left[ Q_a^{2} (-1)^{s_{2} F} q^{H} \bar q^{\bar H} 
\right]_{int} .
\label{51}
\end{equation}

The factor $1/{|\eta(\tau)|^{4}}$ is associated with the contribution
of the light cone partition functions of the space-time bosonic
coordinates $X_{\mu}$, while $Z_{\psi}$ is the corresponding quantity
for the space-time bosons $\Psi_{\mu}$. 

 The spin
structure of the fermions is denoted by ${\bf s}=(s_1,s_2)$, where
$s_1,s_2 \in \{0,1\}$.  A zero (one) refers to 
anti-periodic(periodic) boundary condition on the torus.
The partition function for one complex
fermion is given\footnote{$Z_{\psi}({\bf s},\tau)\stackrel{def}{=}
\{\frac{1}{\eta({\bar \tau})} \theta_i(\bar \tau)$, where
$i=3$ for ${\bar s}=(0,0)$, $i=4$ for ${\bar s}=(0,0)$, i=2 for
${\bar s}=(0,0)$ and $i=0$ for $s=(1,1)$.\} }
by $Z_{\psi}({\bf s},\tau)= \frac{1}{\eta(\tau)}
\theta \left[^{s_{2}}_{s_{1}}\right](\tau)$.
The trace over the internal sector receives contributions only for
massless particles since the operator $(-1)^{s_2 F} $ determines
the chirality of the massless fermions. Note, that massive fermions
are not chiral as they form complete supermultiplets. 
In addition, H and ${\bar H}$ are the right and left moving sector 
Hamiltonian operators in the corresponding internal sectors.    
Furthermore, $q=e^{2 \pi i \tau}$ and $Q_a$ is the charge of a state
with respect to a generator of the gauge group labeled by $a$,
and $\tau =\tau_1 + i \tau_2$ is the modulus of the
world-sheet torus over the fundamental region defined 
by ${\cal F}=\left\{ \tau_{2}>0,-{1\over 2}\le {1\over 2},\tau_{1}  
\le 0 ,|\tau | > 1 \right\}$.

In addition, in the infrared limit $ \tau_2 \rightarrow \infty$, the
${\cal B}$ becomes
\begin{equation} 
b_a = \lim_{\tau_{2} \rightarrow\infty} 
{\cal B}_{a}(\tau, \bar \tau) = \frac{-11}{3} tr_{V} Q_a^2 +
{\frac{2}{3}} tr_{F} Q_a^2 + {\frac{1}{3}} tr_{S} Q_a^2  .
\label{6}
\end{equation}
The individual traces receive contributions from massless vectors
($tr_{V}$),
 where
$(H,{\bar H})_{int}=(1/12,-3/8)$; from massless fermions ($tr_{F}$) having
$(H,{\bar H})_{int}=(1/12,0)$ and massless scalars ($tr_{S}$) having 
$(H,{\bar H})_{int}=(1/12,1/8)$. 

This formula is
valid for any tachyon free vacuum of the heterotic string. The trace on
the above formula is over the internal sector and the partition 
function
involved in the sum must be separated into sums depending on the
number of supersymmetries which are preserved by the world-sheet
boundary conditions. 
The trace is model dependent. For orbifold constructions 
of the heterotic string, this trace decomposes into sectors with
boundary conditions $(g,h)$ along the cycles of the
world-sheet torus as:
\begin{equation}
Tr_{s_1}(Q{_{a}}^2 {(-)^{s_2 F}}q^{H -11/12} 
{\bar q}^{{\bar H} -3/8})_{int}=
\frac{1}{|G|} {\displaystyle\mathop{{\sum_{\begin{array}{ccc}
{g,h\; \in \;G}\\
gh\; = \;hg
\end{array}}}}}
Tr_{(g,s_1)} (Q_{a}^2(-)^{s_2 F} q^{L_o -11/12} 
{\bar q}^{{\bar L}_o -3/8}).
\label{koroil}
\end{equation}
Here, $L_o$ and ${\bar L}_o$ represent the generators of dilatations in the 
complex plane for the left and right moving sectors respectively.
Only sectors which are not completely rotated from
orbifold twists are nonvanishing in the sum.  For the $N=4$ sectors 
where 
all the fermions have to be considered as untwisted, the sum over 
spin 
structures decomposes as 
\begin{equation} 
\sum_{{s_1,s_2}={0,1}}(-1)^{s_1 + s_2} \theta(^{s_1}_{s_2})' 
\theta^3(^{s_1}_{s_2})\frac{d}{d \tau} [\sum_{s_1,s_2}(-1)^{s_1+s_2} 
\theta^4(^{s_1}_{s_2})] = 0,
\label{9}
\end{equation}
where $[\sum_{s_1,s_2}(-1)^{s_1+s_2}\theta^4(^{s_1}_{s_2})]=
\theta_2^4(0|\tau) - \theta_3^4(0|\tau)+\theta_4^4(0|\tau)=0$,
because of the zero identity of the $\theta$ functions.
The only terms that give non-vanishing contributions to the
moduli dependent sum in eqn.(\ref{koroil}) are the sectors, where the
point group G which divides the six dimensional torus is
a subgroup of $SU(2)$, i.e sectors with $N=2$ supersymmetry. 
The union of all twists associated with $N=2$ planes of the 
abelian $Z_N$ orbifolds form    
the little groups $G_i \subset G$ of the unrotated planes. In this case,
the moduli dependent threshold corrections become
\begin{equation}
\Delta_a = \frac{b^{\prime}_a |G'|}{|G|} {\Delta^{\prime}}_a,
\label{thre1}
\end{equation}
where $\Delta^{\prime}_a$ is the moduli dependent 
contribution corresponding to the $N=2$ 
$T_6/G^{\prime}$ orbifold.

In addition, for orbifold compactifications where the internal torus
decomposes in the form $T_6 = T_2 \oplus T_4$, the internal sector sum 
on ${\cal B}_a$ splits into different factors coming from the internal 
superconformal field theories (SCFT)
with central charges $(c,{\bar c})=(20,6)$ and $(2,3)$.  Remember, that 
any heterotic vacuum is obtained by tensoring in the light-cone gauge
two left moving free bosons $X_{\mu}$
together with their right moving fermionic superpartners,
corresponding to the space-time
coordinates
and the internal SCFT with central charge $(22,9)$.

In this case, ${\cal B}_a$ becomes\cite{dkl2}

\begin{equation} 
{\cal B}_a = Z^{torus} {\cal K}_a ,\;\;
Z^{torus} = \sum_{(P_L,P_R) \in \Gamma_{(2,2)}} q^{P_L^2} 
{\bar q}^{P_R^2},
\label{10}
\end{equation}
with
\begin{equation} 
{\cal K}_a \equiv \eta(\tau)^{-4} Tr(\frac{1}{2}(-)^F Q_a^2 q^{L_o-5/6}
{\bar q}^{{\bar L}_o -1/4})_{(c,{\bar c})=(20,6)}.
\label{11}
\end{equation}

It is obvious from eqn.(\ref{10}) that the moduli 
dependence\footnote{This dependence will be elaborated in 
section $(3.6)$. }of the threshold corrections is included in 
the term $Z^{torus}$. 
To be precise, the exact form of the threshold corrections to gauge
couplings is given\cite{anto2} by the equation
\begin{equation}
{\cal B}_{a}(\tau,\bar {\tau}) = \frac{2}{|\eta(\tau)|^{4}}
\sum_{{even}\;s} \frac{(-1)^{s_1 + s_2}}{2 \pi i} 
\frac{d Z_{\psi}(\bf s,\bar \tau)}{d \tau} 
 tr_{s_1} \left[ (Q_a^{2} - \frac{k_a}{8 \pi\tau_2}) 
(-1)^{s_{2} F} q^{H} \bar q^{\bar H} \right]_{int} .
\label{522}
\end{equation}

Comparing the eqns.(\ref{51}) and (\ref{522}) we notice the 
presence of the additional term $-k_{\alpha}/8\pi \tau_2$.
This term was included in the calculation of the threshold corrections
in \cite{anto2}.
We will comment on this term in chapter four, where we will
explain its connection to the universal term.

The general form of the moduli dependent threshold corrections is 
\begin{equation}
\Delta_{\alpha}=  \int_{\Gamma} \frac{d^2 \tau}{\tau_2}
\sum_{(g,h)}b_{\alpha}^{(g,h)}{Z}^{torus}_{(g,h)}(\tau,{\bar \tau})
-b_{\alpha}^{N=2} \int_{{cal F}} \frac{d^2 \tau}{\tau_2},
\label{klol10}
\end{equation}
where the ${Z}^{torus}_{(g,h)}$ refer to the moduli dependent part
of the $N=2$ sector of the $(g,h)$ orbifold invariant under
the group $SL(2,Z)$ as it happen for the 
decomposable orbifolds. The integration is over the fundamental domain
$\cal F$ of the inhomogenous modular group\footnote{
 For a definition see appendix A. } $PSL(2,Z)$.

The sum is over the $N=2$ orbit of the
orbifold sectors created by the $N=2$ sectors.

For non-decomposable orbifolds eqn.(\ref{klol10}) can be rewritten as
\begin{equation}
\Delta_{\alpha}= \sum_{(g_o,h_o) \in {\cal O}} b_a^{(h_o,g_o)}
\int_{\tilde {\cal F}} \frac{d^2 \tau}{\tau_2}
Z_{(h_o,g_o)}(\tau,{\bar \tau}) -b_a^{n=2}\int_{\cal F} 
\frac{d^2 \tau}{\tau_2}.
\label{klol11}
\end{equation}
Here, ${(g_o,h_o)}$ denotes the set of twisted sectors which 
be created from the representative fundamental element $Z_{(h_o,g_o)}$,
by exactly those modular transformations which create the 
modular group of $Z_{(h_o,g_o)}$ from the fundamental region of
$PSL(2,Z)$.

 For non-decomposable orbifolds the moduli dependent sum
in eqn.(\ref{klol11}) is invariant under the modular group
but under some congruence of $\Gamma$, namely $\Gamma_0(n)$ or 
$\Gamma^0(n)$.

Here, ${\tilde {\cal F}}$ is the enlarged region defined as a left coset
decomposition of the fundamental region $\tilde F$, namely
${\tilde {\cal F}}= \cup a_i {\cal F}$. 
For the group $\Gamma_0(p)$ the union $\cup$ of transformations 
$a_i{\cal F}$ is represented from the set of transformations\cite{apostol}
\begin{equation}
a_i =\{1, S,ST,\dots,ST^{p-1} \}.
\label{sfo1}  
\end{equation}

\subsection{ * Low Energy Threshold Effects and Physical 
Singularities}

In general, if one wants to describe globally the moduli space
and not just the small field deformations
of an effective theory around a specific vacuum solution, one
has to take into account the number of massive states 
that become massless at a generic point in moduli space.
This is a necessary, since the full duality group
$SO(22,6;Z)_T$ mixes massless with massive modes\cite{GPR}.
It happens because there are transformations of O(6,22,Z) acting as
automorphisms of the Lorentzian lattice metric of
$\Gamma^{(6,22)}=\Gamma^{(6,6)} \oplus \Gamma^{(0,16)}$
that transform massless states into massive 
states\footnote{
Construction of effective actions invariant under the $O(6,22,Z)$ 
duality group reproducing $N=4$ low energy effective actions of the 
heterotic string were constructed in \cite{giio}.}.

Let us consider the $T_2$ torus, coming from the decomposition
of the $T_6$ orbifold into the form $T_2 \oplus T_4$.
The $T_2$ torus can be defined on a two 
dimensional lattice $\Gamma^{(2,2)}$ which is generated from the basis 
vectors ${\vec e}_1$ and ${\vec e}_2$ .
The metric $G_{ij} \stackrel{def}{=}{\vec e}_1 \cdot 
{\vec e}_2$ has three independent components, while the 
antisymmetric tensor $B \stackrel{def}{=}b \epsilon_{ij}$
one. 
In total we have four independent real components which define the 
moduli of the string compactification on $T^2$.
The moduli can be further combined in the form of two complex moduli 
as $U =\frac{|{\vec e}_1|}{|{\vec e}_2|} e^{if}$ and
$T= 
2(b+iA)$, with $0 \leq f < \pi$ the angle
between the basis vectors and $A=\sqrt{|det G|}$ is the area of the
unit cell of the lattice $\Gamma$.
At the large radius limit 
It was noticed in \cite{cfilq} that in the presence of states that
become massless at a point in moduli space 
e.g, when the  $T \rightarrow U$, the threshold corrections 
to the gauge coupling constants receive 
a dominant logarithmic contribution\cite{cfilq} in the form, 
\begin{equation}
{\triangle}_a(T,{\bar T})\;\; {\approx}\;\; b_{a}^{'} \;\;
\int_{\Gamma} {\frac{d^2 \tau}{\tau_{2}}} e^{-M^2(T){\tau_{2}}}
\;\;{\approx} - b_a {\log{M^2\left( T \right)}},
\label{koko1}
\end{equation}
where $b_ {a}$ is the contribution to the $\beta$-function from the 
states that become nassless at the point $T=U$.

Strictly speaking the situation is sightly different. We will argue 
that if
we want to include in the string effective field theory large field
deformations and to describe the string Higgs effect 
\cite{abak,illt,gipo} and not only small field fluctuations, 
eqn.(\ref{koko1}) must be modified.

We will see that massive states which become massless at specific
points in the moduli space do so, only if the values of the untwisted 
moduli dependent masses are 
between certain limits.
In \cite{cfilq} this point was not emphasized and it was presented 
in a way
that the appearance of the singularity had a general validity for 
generic 
values of the mass parameter.
We will complete the picture by giving more details on the exact
behaviour of the contribution to the threshold corrections to the gauge
coupling constants.
We introduce the function Exponential Integral $E_1(z)$
\begin{equation}
E_1 (z) = \int_z^\infty \frac{e^{-t}}{t} dt\;\;\;\;\;(|arg\;z|)<\pi),
\label{tion11}
\end{equation} 
with the expansion
\begin{equation}
E_1 (z)= - \gamma - lnz - \sum_{n=1}^\infty \frac{(-)^n z^n}{nn!}.
\label{tion12}
\end{equation}
It can be checked that for values of the parameter $|z|>1$, the lnz term is
not the most
dominant, while for $0<|z|<1$ it is. 
In the latter case\cite{hand} the $E_1(z)$ term is 
approximated\footnote{where \begin{eqnarray}
{\alpha}_{00} =-.577&{\alpha}_{10} =0.999&{\alpha}_{20}=-0.249\nonumber\\
{\alpha}_{33} =-0.551&{\alpha}_{44} =-0.009&{\alpha}_{55}=0.001
\end{eqnarray}
and $\epsilon(z) < 2 \times 10^{-7}$.}
as
\begin{equation}
E_1 (z) = -ln(z) +a_{00} +a_{11} z +a_{22} z^2 +a_{33} z^3 +a_{44} z^4 + 
a_{55} z^5 + \epsilon(z).
\label{tion14}
\end{equation}
Take now the form of eqn.(\ref{koko1}) explicitly
\begin{equation}
\triangle(z,{\bar z}) = b^{\prime}_a \int_{|\tau_1|<{1/2}} 
d\tau_1 \int_{\tau_2 > 1} \frac{d\tau_2}{\tau_2}
e^{-M^2(T)\tau_2}.
\label{lop12}
\end{equation}
Then by using eqn.(\ref{tion11}) in
eqn.(\ref{lop12}),
we can see that the $- b^{\prime}_a\;ln M^2(T)$ indeed arise.
 
Notice now, that the limits of the integration variable $\tau_1$ 
in the world-sheet integral in eqn.(\ref{koko1}) are between $-1/2$
 and $1/2$.
Then especially for the value $|1/2|$ the lower limit in the 
integration 
variable $\tau_2$ takes its lowest value e.g 
$(1-\tau_1^2)^{1/2} =(1 -(1/2)^2)^{1/2} = \sqrt(3)/2$.
Use now eqn.(\ref{tion12}). 
Rescaling the $\tau_2$ variable in the integral, 
and using the condition $0<z<1$
which is necessary for the logarithmic behaviour to be
dominant, we get\footnote{Restoring units in the Regge slope 
parameter $a^{\prime}$.}
\begin{equation}
0<M^2(T)<\frac{4}{{\sqrt3}a^{\prime}}.
\label{wqe12}
\end{equation}
This means that the
dominant behaviour of the threshold corrections appears in the form 
of a logarithmic singularity, only when the moduli scalars 
satisfy the limit
$ M^{2} < 4/(\sqrt{3}a^{\prime})$. 
 
We know that for particular values of the moduli 
scalars, the low energy effective theory appears to have 
singularities, which are due to the appearance of charged massless 
states in the physical spectrum.
At this stage, the contribution of the mass to the low
energy gauge coupling parameters is given by\cite{lust}
\begin{equation}
M^2 \rightarrow  - n_H|T-p|^{2},
\end{equation}
where the $n_{H}$ represents the number of states $\phi_{H}$ which 
become
massless at the point $p$. This behaviour is consistent with large
field
deformations of the untwisted moduli.
It is obvious at this point that the behaviour of the threshold effects 
over the whole area of the moduli space can not consistently described 
by the behaviour of the latter equation. The singular limit of this 
expression $T=p$ "can not be reached"\footnote{
When it is reached perturbation theory is not valid. This is a signal 
that new states become massless at this point.}

Of course,  and as a consequence 
neither can 
the enhanced symmetry point. 
The parameter $(a'\sqrt{3}/{4}) M^2$ must always be between the 
limits 
zero and one in order that the contribution of the physical 
singularity to $\triangle$ become dominant.
Therefore, the complete picture of the threshold effects reads
\begin{equation} 
\frac{1}{g_a^2(\mu)}=\frac{k_a}{g_{string}^2} + \\
\frac{b_a}{16\pi^2} \ln \frac {M_{string}^2}{\mu^2} -\\
\frac{1}{16 \pi^2} |\triangle_a(T_{i})|^2 - \\
\Theta(- M^2 + \frac{4}{{\sqrt 3}a'}) b'_{a} \log M^2(T)
\label{gatos1}
\end{equation} 
where $\Theta$ in eqn.(\ref{gatos1}) is the step function. 
The logarithmic contribution at this stage is actually the threshold
effect of the contribution of the states which become light.
Their direct effect on the low energy effective theory is
the appearance of the automorphic functions of the 
moduli dependent masses, after the integration of the 
the massive modes.

The same threshold effect dependence on the $\Theta$ 
function, takes place in Yang - Mills theories, via the
decoupling theorem \cite{algaa}. The contribution of the 
various thresholds decouples from the full theory, and the 
net effect
is the appearance of mass suppressed corrections to the 
physical quantities. 

So far, we have seen that the theory can always approach the 
enhanced symmetry point behaviour from a general massive point on the 
moduli space under specific conditions.
For "large" values of the moduli masses the enhanced symmetry point
can only be reached if its mass is inside the limit (\ref{wqe12}).
 Remember that at the point $T=p$ 
eqn.(\ref{gatos1}) breaks down, since at this point perturbation theory
is not valid any more.
The upper limit (\ref{wqe12}) is in the 
strongly coupled regime of the perturbative $\sigma$-model
coupling expansion parameter.

The previous result is particularly important in view of the 
fast development of the subject of dualities in superstring theory,
It should be noted that in the case of Calabi-Yau manifolds, the 
appearance of singularities in the target space 
 can provide for the web of connectivity \cite{stro1,stro2} through
the entire moduli space.

Interestingly enough the presence of the logarithmic term was confirmed 
in the 
calculation of
the target space \cite{lust} free energies for the massive modes and 
recently
with an exact \cite{mouha} calculation with the calculation of 
threshold effects from BPS states. 
An interesting example of the appearance of the singularity in
heterotic strings will be described in the next section 
when a non-decomposable lattice is involved in the $(2,2)$ symmetric 
orbifold compactifications.

%************************************************************
%******************************************************************
%*********************************************************************
%*********************************************************************
\subsection{ * Target space automorphic functions from string 
compactifications} 

In this part of the thesis, we will discuss the contribution of massive 
moduli dependent masses of
the heterotic compactification to the threshold corrections of 
the gauge coupling constants. 
  
In addition, we discuss the appearance of the
extended non-abelian gauge group in
particular Narain orbifolds, namely those that the internal 
lattice involved is a $Z_N$ non-decomposable orbifold. Initially,
we will be describe how the moduli dependence becomes visible in the 
mass operators, in untwisted subspaces of orbifold compactifications 
of the heterotic string. Later on, we will focus our attention to the
calculation of the moduli dependent threshold corrections, coming from
direct integration of the massive untwisted states of the 
compactification.

For orbifold compactifications, where the underlying internal torus 
does 
not decompose into a $ T_6 = T_2 \oplus T_4 $ , the $ Z_2 $ twist 
associated with the reflection $ - I_2 $ does not put any
additional constraints on the moduli $U$ and $T$. As a consequence
the moduli space of the untwisted subspace is the same as in 
toroidal compactifications. Orbifold sectors which have the lattice 
twist acting as a $Z_2$, give non-zero threshold one-loop corrections  
to the gauge coupling constants in $N=1$ supersymmetric orbifold 
compactifications.                      
 
When the heterotic string is compactified on a six dimensional torus,
the physical states have their mass given by
\begin{equation} 
\frac{\alpha'}{2} M^{2} = N_{L} + N_{R} + \frac{1}{2}({\cal P}_{L}^2 + 
{\cal P}_{R}^{2}) - 1 ,
\label{we1}
\end{equation}
where ${\cal P}_{L(R)}$, $N_{L(R)}$ are the left and right moving 
momentum and number operators respectively.
In addition, invariance\footnote{
This is equivalent
to the argument to the following argument:
in closed string theory, physical states must be
invariant under global shifts in the space-like coordinate $\sigma$
of the world-sheet.
The operator $e^{i \lambda (L_o - {\bar L}_o)}$ 
satisfies, $U_{\lambda}^{\dagger} X^{\mu}(\sigma,\tau)U_{\lambda}
=x^{\mu}(\sigma+\lambda,\tau)$ which means that it generates 
translations in the world-sheet variable $\sigma$. However, in closed 
string theory there is no distinguished point in the world-sheet.
This condition forces us to define the condition 
$(L_o - {\bar L}_o)|phys> = 0$ or $L_o - {\bar L}_o$.}
of the one-loop vacuum amplitude under 
the modular transformations $T \rightarrow T + 1$ gives the level 
matching constraint
\begin{equation}                                            
\frac{\alpha'}{2} M_{L}^{2} = N_{L} + \frac{1}{2}{\cal P}_{L}^{2} -1 =
N_{R} + \frac{1}{2}{\cal P}_{R}^{2} =  \frac{\alpha'}{2}M_{R}^{2}.
\label{lmc}
\end{equation}
From the above equations we deduce
\begin{equation}
\frac{\alpha'}{2} M^{2} ={\cal P}_{R}^{2} + 2 N_{R}.
\label{we22}
\end{equation}                                            

For the calculations which we will describe in this chapter,
we will need the moduli dependence of the 
mass operator.  
In order to display the moduli 
dependence\footnote{In the following we will follow closely 
the article of (\cite{lust}).}
of (\ref{we22}), we need
the general form of the Narain lattice vector ${\cal P}_R$ in the presence 
of the Wilson lines as given in chapter two. 

This is manifestly  
exhibited by expressing ${\cal P}_R$ in terms of the quantum numbers 
of the Lorentzian Narain lattice ${\Gamma}_{22;6}$ 
and then projecting into an orthonormal basis.  
In the orthonormal basis, the untwisted moduli space factorizes into
factors corresponding to the different twist eigenvalues.

The Narain vector ${\cal P}_R$ of the untwisted sector 
of the $N=1$ orbifold
is then parametrized in the usual way\cite{gigi1}, by expressing it 
in terms of the 
$28$ integer quantum numbers, namely the winding numbers 
$n^{i}$,
the momentum numbers $m_{i}$ of the compactification and the charges 
$q_{I}$ of the sixteen 
dimensional Euclidean even-self dual lattice of the leftmoving current 
algebra.

Initially, we project the momentum vectors
\begin{equation}
P = q^{I} l_{I}+n^{i}{\bar k}_{i}+m_{j} k^{j}\;\;i=1,\dots,6\;\;
\;I=1,\dots,16
\label{vect1}
\end{equation}
into the vectors ${\bf e}_{\mu}^{R} = (0_{16},\;0_{6},\;{\bf e}_{\mu})$
of the orthonormal basis of $R^{6}$ with the result
\begin{equation}
{\cal P}_{R} = (q^{I},\;n^{i},\;m_{j})\left( \begin{array}{c}
l_{I} \cdot {\bf e}_{\mu}^{(R)}\\{\bar k}_{i} \cdot {\bf e}_{\mu}^{(R)}\\
k^{j} \cdot {\bf e}_{\mu}^{(R)} \end{array}\right) {\bf e}_{\mu}^{R}.
\label{proj}
\end{equation}

In this way, the norm of the right moving momentum factorises as
\begin{equation}
{\cal P}_{R}^{2} = v^{T} \Phi {\Phi}^{T} v,
\end{equation}
where\footnote{Here M represents a 
$1 \times 28$ matrix with integer coefficients.}
\begin{equation}
v^{T}=(q^{I},n^{i}, m_{j})\; \in M(1,28,{\bf Z}) \simeq {\bf Z}^{28}
\label{proj0}
\end{equation}
and 
\begin{equation} 
{\Phi}= \left( \begin{array}{c}
l_{I} \cdot {\bf e}_{\mu}^{(R)}\\{\bar k}_{i} \cdot {\bf e}_{\mu}^{(R)}\\
k^{j} \cdot {\bf e}_{\mu}^{(R)} \end{array} \right)
= {\frac{1}{2}} \left( \begin{array}{c}
- A E^{*}\\DE^{*}\\E^{*}
\end{array}\right)\; \in M(28,\;6,\;{\bf R})
\label{proj2}
\end{equation}
which contains all the moduli dependence.
Here A represents the Wilson lines namely
\begin{equation}
A = (\;A_{Ii}\;)=(\;{\bf e}_{I} \cdot {\bf A}_{i}\;)\; \in M(16,6,{\bf R}),
\label{proj3}
\end{equation}
while D is the moduli matrix in the lattice basis
\begin{equation}
D = (D_{ij})=2(\; B_{ij} - G_{ij}- \frac{1}{4} {\bf A}_i {\bf A}_j ) \in
M(6, 6, {\bf R})
\label{proj4}
\end{equation}
and the 6-bein of the dual lattice ${\Lambda}^{*}$ is  
\begin{equation} 
E^{*}\; = \;(E^{i}_{\nu})\; = \;({\bf e}^{i} \cdot {\bf e}_{\nu}).
\label{proj5}
\end{equation}
With the heterotic string further compactified on an orbifold,
the action of the twist on the moduli is subject to 
compatibility conditions\cite{imnq,erja,moh1}. 
These equations which are satisfied from the continuous parts 
of the
metric, antisymmetric and Wilson line background fields
are as follows:
\begin{equation}
D_{ij} {\theta}^{j}_{k} = {\theta}_{i}^{j} D_{jk}\;,\;A_{Ij} = 
{\theta}_{I}^{J} A_{Jk}.
\label{klut}
\end{equation}
The quantities $ {\theta}_{i}^{j}$ and $ {\theta}^{j}_{k}$ represent 
the twist action on the $\Gamma_{6;6}$ and its dual,
while ${\theta}_{I}$ represents the 
action of the gauge twist in the gauge degrees of freedom of the 
$E_{8} \times E_{8}$ current algebra.
The moduli variable $\Phi$ satisfies\footnote{
This means that the untwisted moduli space of the 
$D=4$, $N=4$ toroidal compactifications of the heterotic string is the 
coset space
$\frac{SO(22,6,)}{SO(22) \times SO(6)}$ .}
the well known equation for the 
$SO(22,6)$ coset space
\begin{equation}
{\Phi}^{T} H_{22,6} {\Phi} = -I_{6},
\label{qwa1}
\end{equation}
with $H_{22,6}$
the pseudo-euclidean lattice metric\footnote{Here
$C_{(16)}$ is the Cartran matrix of $E_{8} \times E_{8}$.}
 of the Narain
lattice $\Gamma_{22,6}$
\begin{equation}
H_{22,6} =  \left( \begin{array}{ccc}
C^{-1}_{(16)}&0&0\\
0&0&I_{6}\\
0&I_{6}&0\\
\end{array}
\right).
\label{uiop}
\end{equation}

The solution of the compatibility conditions (\ref{klut}) for the 
moduli 
in the orthogonal basis, results in 
the decomposition of the untwisted moduli space of the orbifold 
into
factors corresponding to different twist eigenvalues.

Following this procedure, we now factorize the moduli space 
into subspaces corresponding to
the different internal and gauge twist eigenvalues.
We perform the change of variables from the lattice
to an orthonormal basis by
\begin{equation}
{\Phi}=\frac{1}{2}
\left(
\begin{array}{c}
-A_{Ij}E^{j}_{\nu}\\
D_{ij}E^{j}_{\nu}\\
E^{i}_{\nu}\\
\end{array}
\right)
={\frac{1}{2}}
\left(
\begin{array}{ccc}
{\cal E}_{I}^{M}&0&0\\
0&T_{i}^{\mu}&0\\
0&0&T_{i}^{\mu}\\
\end{array}
\right)
{\hat \Phi}
\end{equation}
creating the variable ${\hat \Phi}$ which exhibits no moduli 
dependence\footnote{The quantities $T_{i}^{\mu}$ and 
$T_{i}^{\mu}$ are moduli independent and obey the relation :
$E_{i}^{\nu} = S_{i}^{j} T_{j}^{\nu} = T_{i}^{\mu} S_{\mu}^{\nu}$, with
S a deformation matrix parameter connecting lattice to 
lattice$({S_{i}^{j}})$ or orthonormal$({S_{\mu}^{\nu}})$ basis.}.
In this form, the variable ${\hat \Phi}$ satisfies the equation
\begin{equation}
{\hat \Phi} {\hat H}_{22,6} {\hat \Phi} = - I_{6}
\label{ty1}
\end{equation}
with
\begin{equation}
{\hat H}_{22;6} = \left( \begin{array}{ccc}
I_{16}&0&0\\
0&0&I_6\\
0&I_6&0\\
\end{array}
\right).
\label{rt1}
\end{equation}
In this way, the component form of the $\hat \Phi$ variable becomes block 
diagonal 
with factors corresponding to the different twist eigenvalues, namely
\begin{equation}
{\hat \Phi} \rightarrow  {\hat \phi}^{(j)} \oplus  {\hat \phi}^{(+1)}
\oplus {\hat \phi}^{(-1)},
\label{rt2}
\end{equation}
while the mass operator factorises  as 
${\cal P}_{R}^2 = {\hat v}^{T}\;{\hat \phi}\;{\hat \phi}^{T}\;{\hat v}$
with
\begin{equation}
{\hat v}^{T} = \left( q^{I},n^{i},m_{j} \right)
\left(
\begin{array}{ccc}
{\cal E}_{I}^{M}&0&0\\
0&T_{i}^{\mu}&0\\
0&0&T_{i}^{\mu}\\
\end{array}
\right).
\label{rt3}
\end{equation}

The dimensions of the matrix variable  $\hat \Phi$, 
depend on the multiplicities of the eigenvalues 
of the internal and gauge twists in their block diagonal form.
Especially, for the subspace of the twists corresponding\footnote{For 
the gauge twist, we assume an orthogonal decomposition
into subspaces corresponding to the 
the complex eigenvalues $e^{\pm k_{j}}$
and the real eigenvalues $-1$ and $+1$, with multiplicities 
$R_{j}\;,q\;,p$ correspondingly.
For the internal twist, we assume an orthogonal decomposition
into subspaces corresponding to the complex eigenvalues
$e^{\pm {\rho}_{j}}$ and the real ones $-1,\;+1$. 
The corresponding multiplicities are assumed to be $Q_{j}$,s,$\xi$} to 
the eigenvalue
$-1$, the dimensions of the variable $\hat \Phi$ are
$(q\;+\;s,\;s)$.

However, the coset space structure of the moduli space becomes 
obvious in the standard metric ${\eta}_{22;6}$, i.e $(+)^{22}(-)^{6}$.
The transition to this metric can be made obvious 
by an appropriate transformation\cite{lust} on the 
${\hat \phi} \rightarrow {\tilde \phi}$ variable and 
${\hat v} \rightarrow {\tilde v}$, in such a way that
${{\hat v}^{T}}{\hat \phi} = {{\tilde v}^{T}}{\tilde \phi}$.
The ${\tilde \phi}$ variable satisfies the equation
\begin{equation}
{\tilde \phi}^{T} {\eta}_{q+2,2} {\tilde \phi} = -I_{2},
\label{arxidi5}
\end{equation}
while the introduction of complex variables as
\begin{equation}
{\phi}_{c} = \left( \begin{array}{c}
{\phi}^{1}_{(1)} + i {\phi}^{1}_{(2)}\\
\dots\\{{\phi}_{(1)}^{q=4}} + i {\phi}_{(2)}^{q=4}
\end{array} \right)
\label{arxidi6}
\end{equation}
restructures the (\ref{arxidi5}) equations into the constraint 
equations for the $SO(q+2,2)$ coset, namely 
\begin{eqnarray}
{\phi}_{c}^{\dagger}\;{\eta}_{q+2,2}\;{\phi}_{c} = -2 \;,&
{\phi}_{c}^{T}\;{\eta}_{q+2,2}\;{\phi}_{c} = 0.
\label{arxidi7}
\end{eqnarray}

The direct result is that the mass takes the form
\begin{equation}
{\cal P}_{R}^{2}={\tilde v}^{T} {\tilde \phi}{\tilde \phi}^{T}{\tilde v}= 
{\tilde v}^{T} {\phi}_{c} {\tilde \phi}_{c}^{\dagger}
{\tilde v}=|{\tilde v}^{T} {\phi}_{c}|^{2}.
\end{equation}

The solution of the coset equations (\ref{arxidi7}) 
eliminates the redundant degrees of freedom .
By defining $y \in  {\bf C}^{q+4}$ the eqn's (\ref{arxidi7})
become
\begin{equation}
{\sum_{i=1}^{q+2}}\;|y_{i}|^{2} -|y_{q+3}|^{2} -|y_{q+4}|^{2} = -2Y
\label{arxidi8}
\end{equation}
\begin{equation}
{\sum}_{i=1}^{q+2}\;y_{i}^{2}-y_{q+3}^{2}-y_{q+4}^{2}=0.
\label{arxidi9}
\end{equation}

Then e.g for the 
coset $SO(4,2)$\footnote{It is associated with the 
two dimensional torus lattice $T_{2}$ of the untwisted subspace 
of a $Z_{2}$ orbifold, for which a two component Wilson line 
is turned on. The $T_{2}$ is a sublattice decomposition
of the Narain lattice as $\Gamma_{2} \oplus \Gamma_{4} + \dots$,
with the internal twist acting  as $-\;I_2$ on ${\Gamma}_2$.},
the derivation of the mass operator for the untwisted subspace
results from the solution of the 
the constraint equations with the ansatz
\begin{eqnarray}
y_{1}=(B_{1}+C_{1}),&y_{2}=(B_{1}-C_{1})\nonumber\\
y_{3}=\frac{1}{2}(T-2U),&y_{4}=-i(1-\frac{1}{4}(2TU-2BC))\nonumber\\
y_{5}=\frac{1}{2}(T+2U),&y_{6}=i(1+\frac{1}{4}(2TU-2BC))
\end{eqnarray}
and

\begin{equation}
Y=\frac{1}{2}(T+{\bar T})(U + {\bar U})-\frac{1}{2}(B+{\bar C})(C+{\bar B})
\label{arxidi10}
\end{equation}
giving the mass formula
\begin{equation}
\frac{\alpha'}{2}M^{2}=\frac{|{\tilde v}^{T}y|^{2}}{Y}.
\label{arxidi11}
\end{equation}
The general contibution\cite{lust,stiebe1} to the mass formula for 
the $Z_2$ orbifold 
plane\footnote{for a general twist embedding in the gauge degrees of
freedom 
with $k_{1},\dots, k_{d}$ gauge lattice quantum numbers in the
invariant directions.},
\begin{equation}
p_{R}^{2}=|m_{2} -iU m_{1} + iT n^{1} - (T U - B C)n^{2} + 
r_{1}f_{1} (k_{i})(B+C) + r_{2} f_{2} (k_{i})(B-C)|^{2},
\label{arxidi12}
\end{equation}
where $r_{1},r_{2} \in R$ and $f_{1} (k_{i}),\;f_{2} (k_{i})$ 
functions of the gauge quantum numbers .
The above formula involves perturbative BPS states which 
preserve $1/2$ of the supersymmetries, which belong to short
multiplet representations of the supersymmetry algebra.

In the study of the untwisted moduli space, we will assume 
initially that
under the action of the internal twist there is a
sublattice
of the Narain lattice ${\Gamma}_{22;6}$ in the form
$ \Gamma_{2} \oplus \Gamma_{4} \supset {\Gamma}_{22;6}$ with the 
twist acting  as $-\;I_2$ on $\Gamma_{2}$.    
In the general case, we assume that there is always a 
sublattice\footnote{this 
does not 
correspond to a decomposition of the Narain lattice as
${\Gamma}_{22;6} = {\Gamma}_{q+2;2} \oplus \dots$ since
the gauge 
lattice $\Gamma_{16}$ is an Euclidean even self-dual lattice. 
So the only way for it to factorize as $\Gamma_{16} = {\Gamma_{q}}
\oplus {\Gamma_r}$, with $\;q+r=16$, is when $q=r=8$.}
${\Gamma}_{q+2;2} \oplus {\Gamma}_{r+4;4} \subset {\Gamma}_{16;6}$
where the twist acts as $-\;I_{q+4}$\footnote{on $ {\Gamma}_{q+2;2}$}
 and with eigenvalues different than -I on ${\Gamma}_{r+4;4}$.

In this case, the mass formula for the untwisted subspace 
${\Gamma}_{q+2;2}$ depends on 
the factorised form $P_{R}^{2} = v^{T} {\phi}{\phi^{T}}v$,
with $v^{T}$ taking values as the row vector
\begin{equation}
v^{T}=(a^{1},\dots,a^{q};n^{1},n^{2};m_{1},m_{2}).
\label{kln1}
\end{equation}
The quantities in the parenthesis represent the lattice
coordinates of the untwisted sublattice ${\Gamma}_{q+2;2}$,
with $a^{1} \dots a^{q}$ the Wilson line quantum numbers and
$n^{1},n^{2},m_{1},m_{2}$ the winding and momentum quantum numbers of
the
two dimensional subspaces.

When Wilson lines are present, the variable 
\footnote{in the case that the
the untwisted subspace where the twist acts as $-\;I_{q+4}$.}
${\phi} = {\phi}^{-1}$ satisfies the coset equation
\begin{equation}
{\phi}^{T} H_{q+2;2} {\phi} = -I_{2}.
\label{kln2}
\end{equation}
The meaning of the previous equation  is that the untwisted moduli space
is that of an $SO(q+2,2)/\left( SO(q+2) \times SO(2)\right)$
coset.
For decomposable orbifolds with 
continous Wilson lines turned on, the untwisted moduli space
is $\frac{SO(q+2,2)}{SO(q+2)\times SO(2)}$ when the 
twist has two eigenvalues $-1$.

 Shortly, we will discuss the case of $Z_{6}-IIb$ non-decomposable  
orbifold. In this example\footnote{with continous Wilson lines turned on}
the untwisted moduli space is as before, i.e  
in the form $SO(4,2)/\left( SO(4) \times SO(2)\right)$. The internal 
twist 
acting on the 6-torus    
has two eigenvalues  $-1$. The action of the internal twist can
be made to act as $-\;I_2$ on a $T_2$ by appropriate parametrization 
of the momentum quantum numbers.

The moduli metric $H_{q+2;2} $ in (\ref{kln2}) is given by 
the matrix

\begin{equation}
 H_{q+2;2}=\left( \begin{array}{ccc}
C_{q}^{-1}&0&0\\
0&0&I_{2}\\
0&I_{2}&0
\end{array} \right)
\label{kln3}
\end{equation}
where the matrix variable $\phi$ in (\ref{kln1})is a 
$(q+2,2)$ matrix with integer values and $C_{q}^{-1}$ the lattice metric
for the 
invariant directions in the gauge lattice\footnote{Compatibility of the 
untwisted moduli with the twist action on the gauge coordinates
comes from the non-trivial action of
the twist in the gauge lattice. This means that that the untwisted
moduli of the orbifold have to  
 the equation M A = A Q, where
Q ,A and M represent the internal, Wilson lines and gauge twist 
respectively.}.

Let us consider first the generic case of an orbifold
where the internal 
torus\footnote{at the end of this
discussion we will comment on the difference of the 
mass operator for the non-factorizable case}
factorizes into the orthogonal sum $T_6 = T_2 \oplus T_4$ with the 
$Z_2$ twist acting on the 2-dimensional torus lattice.
We will be interested  in the mass formula of the untwisted subspace
associated with the $T_2$ torus lattice. 
We consider as before that there is a sublattice of the Euclidean 
self-dual lattice 
${\Gamma}_{22,6}$ as 
${\Gamma}_{q+2,2} \oplus {\Gamma}_{20-q,4} \subset {\Gamma}_{22,6}$.
In this case, the momentum operator factorises into the
orthogonal components of the sublattices with 
$(p_{L};p_{R}) \subset  {\Gamma}_{q+2;2}$ and             
 $(P_{L};P_{R}) \subset  {\Gamma}_{20-q;4}$ . And as a result the 
mass operator (\ref{we22}) factorises into the form
\begin{equation}
\frac{\alpha'}{2} M^{2} = p_{R}^{2} + P_{R}^{2} + 2N_{R}.
\label{arxidi1}
\end{equation}
On the other hand, the spin operator S for the ${\Gamma}_{q+2;2}$ 
sublattice becomes
\begin{equation}
p_{L}^{2} - p_{R}^{2}= 2(N_{R} + 1 - N_{L}) + \frac{1}{2} P_{R}^{2} -
\frac{1}{2} P_{L}^{2}=2 n^{T} m + b^{T} C b,
\label{arxidi2}
\end{equation}
where C is the Cartran metric operator for the invariant
directions of the sublattice ${\Gamma}_{q}$ of the $\Gamma_{16}$
even self-dual lattice.
The spin S can be expressed more elegantly in matrix form as
\begin{equation}
p_{L}^{2} - p_{R}^{2}= {\frac{1}{2}}u^T {\eta}u,
\label{arxidi3}
\end{equation}
where 
\begin{equation}
u=\left(\begin{array}{c} b\\n\\m
\end{array}\right),\;\;\;and\;  
{\eta} = \left(\begin{array}{ccc}
0&0&C_{q}\\0&0&I_{2}\\0&I_{2}&0
\end{array}\right) 
\label{arxidi4}
\end{equation}

Let us now consider the $Z_6-IIb$ orbifold. For this particular 
orbifold we will discuss a number of issues. In particular, gauge 
symmetry 
enhancement and calculation of the modular orbits resulting in the 
generation of the non-perturbative superpotential.
This orbifold is defined on the torus lattice 
$SU(3) \times SO(8)$ and 
the twist in the complex basis is defined as 
$\Theta = exp(2,-3, 1) \frac{2 \pi i}{6}$. 
The twist in the lattice basis is defined as 
\begin{equation}        
Q=\left(\begin{array}{ccccccr}
0&-1&0&0&0&0 \\
1&-1&0&0&0&0 \\
0&0&0&1&-1&-1 \\
0&0&1&1&-1&-1 \\
0&0&0&1&-1&0 \\
0&0&0&1&0&-1
\end{array}\right) .
\label{uua}
\end{equation}

 This orbifold is non-decomposable 
in the sense that the action of the lattice twist does not decompose in
the orthogonal sum  
$ T_6 = T_2 \oplus T_4$ with the fixed plane lying in $T_2$. 
The orbifold twists $\Theta^2$ and $\Theta^4$, leave the second  
complex plane unrotated . The lattice in which the twists $\Theta^2$
and $\Theta^4$ act as an lattice automorphism is the $SO(8)$. In addition 
there
is a fixed plane which lies in the $SU(3)$ lattice and is associated with
the $\Theta^3$ twist.

Consider now a k-twisted sector of a six-dimensional orbifold of the
the heterotic string associated with the twist $\theta^{k}$.
If this sector has an invariant complex plane then its
twisted sector quantum numbers have to satisfy 
\begin{eqnarray}
Q^{k}n=n,\;\;Q^{*k}m=m,\;\;M^{k}l=l,
\label{axid1}
\end{eqnarray}
where Q defines the action of the twist on the internal lattice and
M defines the action of the gauge twist on the $E_{8} \times E_{8}$
lattice.
In general, if $E_{\alpha},\;{\alpha}=1,\;2$ is a set of basis vectors
for the fixed directions of the orbifold\footnote{assuming that the
internal lattice has basis vectors $e_{i},\;i=1,\dots,6$ 
and dual $e_{i}^{*}$ with $e_{j}^{*} \cdot e_{i} = \delta_{ij}$}
and 
${\cal E}_{\mu},\;{\mu}=1,\dots,d$ is a set of basis vectors for the 
fixed directions in the gauge lattice then the momentum and 
winding numbers for the invariant directions of the twisted states,
are found to have the general form 
\begin{equation}
P={\breve m}_1 {\breve E}_1 + {\breve m}_2 {\breve E}_2,\;
L={\breve n}_1 {E}_1 + {\breve n}_2 {E}_2,\;({\breve m}_{1},{\breve
m}_{2},{\breve n}_{1},{\breve n}_{2}\;\in\; Z),
\label{axid2}
\end{equation}
with ${\breve E}_1,{\breve E}_2$ particular linear combinations of the 
dual basis $e_{i}^{*}$ and 
${\breve E}_{a} \cdot E_b \stackrel{def}{=}{\rho}_{ab}$. 
This 
means that with
\begin{eqnarray}
n=\left(\begin{array}{c}{\breve n}_1 \\{\breve n}_2
\end{array}\right),\;m= \left(\begin{array}{c}{\breve m}_1\\
{\breve m}_2\end{array}\right),\;l= \left(\begin{array}{c}{\breve l}_1\\
{\breve l}_2
\end{array}\right)
\label{bre1}
\end{eqnarray}
the momenta take the form
\begin{equation}
P'_{L} = ({\rho}\frac{\breve m}{2}+(G_{\bot}-B_{\bot}-
\frac{1}{4}A_{\bot}^{t}C_{\bot}A_{\bot}){\breve n}-\frac{1}{2}
A_{\bot}^{t}C_{\bot}{\breve l},\;{\breve l}+A_{\bot} 
{\breve n})=(p'_{L},\;p'_{R})
\label{kij2}
\end{equation}
\begin{equation}
P'_{R} =(\;\;\;{\rho}\frac{\breve m}{2}\;\;\; +
(\;\;G_{\bot}\;\;-\;\;B_{\bot} + \frac{1}{4}A_{\bot}^{t} C_{\bot} 
A_{\bot}){\breve n} - \;\;\frac{1}{2} A_{\bot}^{t} C_{\bot} 
{\breve l}, 0 )  =  (p'_{R},\;0 ).
\label{kij1}
\end{equation}
Here $\rho$ is the $\rho_{ab}$ matrix, $C_{\bot}$ is the Cartran matrix
for the fixed directions and  $A_{\bot}^{Ii}$ is the matrix for the
continous Wilson lines in 
the invariant directions $i=1,2,\;I=1,\dots,d$. $G_{\bot}$ and $B_{\bot}$  
are
$2 \times 2$ matrices and ${\breve n},\;{\breve m}\;{\breve l}$ are the 
quantum
numbers in the invariant directions.

For the $Z_{6}-IIb$ orbifold,  
${\rho} = \left( \begin{array}{cc}
1&0\\0&3\end{array}\right)$
In addition, the values of the
momentum and winding numbers parametrizing
the $\theta^{2}$ subspace
are
\begin{equation}
w=\left( \begin{array}{c}
0\\0\\n^{1}\\0\\n^{1}-n^{2}\\n^{2}
\end{array}\right)\;with\;n^{1},n^{2}\;\in Z\;,
p=\left( \begin{array}{c}
0\\0\\m_{1}\\-m_{1}\\m_{2}\\m_{1}-m_{2}
\end{array}\right) \;with\;m_{1},m_{2}\;\in\;Z.
\label{oxi11}
\end{equation}

The mass formula \cite{lust,fklz} for the $\Theta^2$ subspace reads 
\begin{equation}
m^2 = \sum_{m_1, m_2 \atop n^1,n^2} \frac{1}{Y} 
|-T U'n^2+iTn^1-iU'm_1+3m_2|^2_{U'= U-2i} = {|{\cal M}|^2}/{(Y/2)},
\label{kavou} 
\end{equation}
with 
\begin{equation}
Y = (T +{\bar T})(U+{\bar U}).
\label{kavo1} 
\end{equation}

The quantity $Y$ is connected to 
the K\"{a}hler potential, $ K = - \log Y $ .
The target space duality group is $\Gamma^0(3)_T \times 
\Gamma^0(3)_{U'}$ where $U' = U -2i$.

In eqns(\ref{arxidi1},\ref{arxidi2},\ref{arxidi3}), we  
discussed the level matching condition in the case of
a $T_6$ orbifold admitting an orthogonal decomposition.
Mixing of these equations gives us the following equation
\begin{equation}
p_{L}^{2} - \frac{\alpha'}{2}M^{2} =  2(1-N_{L}-\frac{1}{2}
P_{L}^{2})  =  2 n^{T} m + q^{T} C q.
\label{oki1}
\end{equation}

The previous equation gives us a number of different orbits
invariant under $SO(q+2,2;Z)$ transformations :

$i)$ the untwisted orbit with 
\begin{equation}
2 n^{T} m + q^{t} C q = 2.
\label{siga1}
\end{equation}
In this orbit, 
$ N_{L}=0,\;P_{L}^{2}=0 $. When $M^{2}=0$, this orbit is 
associated with the "stringy Higgs effect". The "stringy Higgs 
effect appears as 
a special solution of the equation (\ref{oki1}) 
at the point where $p_{L}^{2}=2$, where additional massless
particles may appear.  

$ii)$ the untwisted orbit where 
\begin{equation}
2 n^{T} m + q^{t} C q = 0.
\label{siga2}
\end{equation}
Here $ 2 N_{L}+P_{L}^{2} = 2 $. This is the orbit relevant 
to the calculation of threshold corrections to the gauge couplings,
without the need of enhanced gauge symmetry points, as may happen
in the orbit $i)$.

iii) The massive untwisted orbit with 
\begin{equation}
2N_{L} +P_{L}^{2} \geq 4
\label{siga3}
\end{equation}
Here always $M^{2} \geq 0 $.

 Let us now consider\footnote{for the orbifold $Z_{6}-IIb$}
the modular orbit \cite{lust} 
associated with the 'stringy Higgs effect'. It corresponds to  
certain points in the moduli space where singularities 
associated with the additional 
massless 
particles appear and have as a result
gauge group enhancement.
This point
correspond to $T=U$ with $ m^2 = n^2 = 0 $  and
 $ m^1 = n^1 = \pm 1$.
At this point the 
gauge symmetry is enhanced to $SU(2) \times U(1)$.
In particular, the left moving momentum 
for the two dimensional untwisted subspace gives

\begin{equation}
p_{L}^{2}= \frac{1}{2T_{2}U_{2}}
|{\bar T}Un_{2} - i{\bar T} n_{1} -i U' m_{1} + 3 m_{2}|^{2}=2,
\label{oki3}
\end{equation}
while 
\begin{equation}
p_{R}^{2} = \frac{1}{2T_{2}U'_{2}}|-{T}U'n^{2} + i{T} n^{1}
-iU'm_{1} + 3 m_{2}|^{2} = 0.
\label{oki2}
\end{equation}

At the fixed point of the modular group $\Gamma^{o}(3)$,
$\frac{\sqrt{3}}{2}(1 + i \sqrt{3})$, there are no additional massless
states,
so there is no further enhancement of the gauge symmetry.

We will now describe the calculation of threshold corrections
from the target space free energies of the massive untwisted states.
In quantum field theory, when we are interested in the calculation of the
effective lagrangian, then we have to deal with the generator of
the 1PI Feynman diagrams\footnote{Feynman diagrams that cannot become
disconnected by cutting off one of their internal lines. See for
example \cite{ramo}.}, the generating function $\Gamma$.
In general, if our theory
contains a number of fields, including light fields $\phi$ and
heavier fields $\Phi$, then the effective action ${\hat \Gamma}$ for
the light fields $\phi$, is  
the sum 
of the 1PI Feynman diagrams with respect of the light fields $\phi$.
In other words, the effective action in this case, is given by

\begin{equation}
e^{-\; {\hat \Gamma}} = \int [d{\Phi}] e^{-I(\phi, \Phi)}
\label{qua2}
\end{equation}
where the "superheavy" fields $\Phi$ are "integrated out".  
The previous quantity $e^{-\; {\hat \Gamma}}$, is known to be related to 
the
topological free energy\footnote{There is a distinction between
bosonic or fermionic free energy, depending on whether the functional 
integration is over bosonic or fermionic states $\Phi$ respectively.
}
through the definition
\begin{equation}
e^{-\; {\hat \Gamma}} = e^{F}
\label{qua3}
\end{equation}

For example from the definition of the bosonic free energy, by 
expansion, we get
\begin{equation}
e^{F_{(bosonic)}}= \int [{\cal D}\phi]e^{-{\phi} M_{\phi}^{2} {\phi} +
\dots}
\label{qua4}
\end{equation}
Here the ellipsis represent the usual derivative expansion terms
of the effective action i.e $i \frac{\partial}{\partial{M_{\phi}}}$
and higher order $\phi$ terms. 
The action $\hat \Gamma$ contains an infinite number of
non-renormalizable interactions\cite{wein1}, suppressed at 
energies $E<<M_{\phi}$ by 
powers of $\frac{E}{M_{\phi}}$.

The fermionic free energy - for supersymmetric backgrounds coming from 
the integration of massive fermions - is defined as the
negative of the bosonic free energy 
\begin{equation}
F_{fermionic}=\log det {M}^{\dagger}{M}.
\label{qua5}
\end{equation}

Here ${M}$ represents the fermionic mass matrix.
Working in this way, we define the free energy as the one
coming from the 
integration of the massive compactification modes,i.e. 
Kaluza-Klein and
winding modes. We exclude non-compactification modes like
massive oscillator modes.
In this sense, the free energy is topological\cite{fklz,oova}.

 Of particular importance to us,
will be the calculation of the non-perturbative superpotential.
 We will 
calculate it, through
its identification with the target space free energy coming 
from the 
massive compactification modes.
In particular it was argued\cite{fklz}, that the
target space partition function Z, defined as
\begin{equation}
Z=e^{-F_{fermionic}}= - det(M^{\dagger}M) = - \frac{|W|^2}{Y},
\label{qua71}
\end{equation}
when appropriately regularized\footnote{We comment on the 
regularization procedure after relation (\ref{polop}).},
provides
us with 
the non-perturbative
superpotential coming from
the integration of the massive chiral compactification modes.

Here M is the mass matrix of all chiral masses 
of the particles involved in the compactification process and Y 
is connected to the K\"{a}hler potential K, via the relation 
$K=-\log Y$(see eqn.(\ref{kavo1})).
For the two-dimensional toroidal compactification  
with moduli space $SO(2,2)/SO(2) \times SO(2)$ 
and modular group $SL(2,Z)$, the topological free energy is 
equal to
$F= \sum_{mom.\;and\;wind.\;numbers} \log{M^{\dagger}M}$, where
$M^{\dagger}M = ({\cal M}^2 /Y) $ and 
${\cal M}=|-T Un^2+iTn^1-iUm_1+m_2|$.

In this way, the non-perturbative superpotential is identified as
\begin{equation}
W = det {\cal M}
\label{qua72}
\end{equation}  
Here W is the mass matrix $\cal M$ of the chiral masses 
of the compactification modes.

 Especially, for the case where the 
calculation of the free energy is that of the moduli space of the 
manifold $\frac{SO(2,2)}{SO(2) \times SO(2)}$, in a factorizable 
2-torus $T_{2}$, the topological\cite{fklz} bosonic free energy
is exactly the same as the one,
coming from the string one loop calculation in \cite{dkl2}.

We will describe now the low-energy behaviour of the $N=1$
orbifold string effective field theory.

Recall the general form of the bosonic non-local effective 
lagrangian in
four-dimensions, up to two space-time derivatives\cite{cre,dkl2}
\begin{equation}
{\cal L}_{eff} = \frac{R}{2k^{2}} +
\left(\frac{1}{4g^{2}(\phi)} \right)_{ab} F_{\mu \nu}^{\alpha}
F^{\mu \nu\;\;b} +
i\frac{{\Theta(\phi)}}{32 {\pi}^{2}}F_{\mu \nu}^{\alpha}
{\tilde F}^{\mu \nu\;\;b} +
\frac{1}{2} G_{ij}(\phi)D^{\mu} {\phi}^{i}D_{\mu}{\phi}^{j} +
V(\phi).
\label{qua6}
\end{equation}

The matrices $g_{\alpha b}^{-2}(\phi)$ and
${\Theta}_{\alpha b}(\phi)$ are field dependent inverse gauge 
couplings
and vacuum angles respectively, 
$G_{ij}=\frac{\partial^2 K(\phi,{\bar \phi})}{{\partial \phi^i}{\bar 
\partial} {\phi}^{\bar i}} $,with $K(\phi,{\bar
\phi})$ the K\"{a}hler potential,  is the metric on the
Riemannian manifold of the scalar fields, V is the scalar potential,
and R is the scalar curvature of the space-time metric $G_{\mu \nu}$.

If the low energy theory is that of a $D=4$, $ N=1$ 
supergravity\cite{cre}, then it is completely 
determined
from the knowledge of three functions, namely the K\"{a}hler potential,
the superpotential\footnote{Their effect on 
the physical parameters will become obvious in chapter five.}
and the gauge kinetic function f. The latter is defined as

\begin{equation}
f_{ab}({\tilde \phi})= \left( \frac{1}{g^{2}({\tilde \phi})} \right)_{ab}- 
i \frac{{\Theta}_{ab}({\tilde \phi})}{8{\pi}^{2}}
\label{qua7}
\end{equation}
and it has to be a holomorphic function
of the complex coordinates $\tilde \phi$ of the  K\"{a}hler scalar 
manifold.
For $N=1$ supersymmetric orbifolds the dependence on the complex scalar 
coordinates
arises from the moduli fields, defined later. 
The study of the $f_{ab}$ function is particularly important, in 
view of the fact that its derivatives are involved in
various non-renormalizable interactions\footnote{A number of 
$f_{ab}(\phi)$ derivatives,  
contributes to the effective superpotential which may cause
supersymmetry breaking, through formation of gaugino 
condensates\cite{fegini}.}.

A general problem of a quantum field theory that involves 
massless particles\cite{dkl2} appears when we expand the 
Wilson's effective action 1PI diagrams, 
in terms involving a power series in particles momenta.
Because the radius 
of convergence of the series is that of the lightest
particle, if there are massless particles 
the radius is equal to zero. In this case, 
there is no local 
effective
lagrangian, and the effective renormalized gauge 
couplings $1/g_{a}^2(p^2 = 0)$, cannot be defined.

However, by studying the running gauge couplings 
$ \left(\frac {\partial{g_{\alpha}^{-2}}}
{\partial {\Phi}_{i}} \right)$ at some off-shell
 momentum ${p^2} \neq 0$,
this problem is avoided
and the $p^2 =0$ limit can be reached.
For this reason, the supersymmetric
one loop graph involving two gauge fields and one scalar 
field, with charged fermions contributing in the loops,
gives the following relations\cite{dkl2} between the gauge 
coupling constants
and the effective axionic couplings at one-loop
\begin{equation}
8 {\pi}^{2} \left(\frac {\partial{g_{\alpha}^{-2}}}
{\partial {\tilde \Phi}_{i}} \right)^{(1-loop)}= -i \left(
\frac{\partial \Theta_{\alpha} }{\partial {\tilde \Phi}_{i}}
\right)^{(1-loop)} = \frac{-1}{2}Tr\left( Q_{\alpha}^{2}
\frac{\partial M}{\partial {\Phi}_{i}} M^{\dagger}
\frac{1}{MM^{\dagger} + O(p^{2})} \right).
\label{qua8}
\end{equation}
Here $Q_{\alpha}$ is the generator of the gauge group $\alpha$.
Furthermore, as matter as it concerns the integrability 
conditions, the following relations hold\cite{dkl2}:

$i)$ the integrability condition for the running axionic
couplings\footnote{The subscripts denote derivatives 
with respect to moduli fields, namely 
$\Theta_{\alpha\;,i} \equiv \frac{\partial}{\partial \phi^{i}}
\Theta_{\alpha}$.} 
 \begin{equation}
{\Theta}_{{\alpha}\;,ij}(p^2\;;<{\tilde \phi}>) \neq 
{\Theta}_{{\alpha}\;,ji}(p^2\;;<{\tilde \phi}>)
\label{qua9}
\end{equation}
is satisfied only in the $p^{2}=0$ limit.
This means, that there is no well defined running
axionic $\Theta$ coupling off-shell. In general,
integrability is retained for $p^2 << mass^2$ of the 
lightest charged fermion, bearing in mind that the 
gauge symmetry must not be chiral(there must not be massless
fermions) in order for the $M^{\dagger} M$ matrix to be 
invertible.

In the $p^{2}=0$ 
limit
\begin{equation}
({\Theta}_{\alpha}^{2})^{1-loop}_{(p^{2}=0)}=-Tr(Q_{\alpha}^{2}
Im \log M) +\; constant = 8 {\pi}^2 Im f_{\alpha}.
\label{qua11}
\end{equation}
For QCD the above relation becomes the $\Theta$ angle,
$\Theta = ArgDet(M_{quark})$.

$ii)$ As a result of the non-integrability of the 
axionic $\Theta$ couplings, the one-loop corrections to 
the gauge couplings are non-holomorphic.

By further integration of (\ref{qua8})-at the 
infrared limit $p^{2}=0$ limit-
we obtain
\begin{equation}
16 {\pi}^{2} \left({g_{\alpha}^{-2}} \right)=
- tr(\log (Q_{\alpha}^{2} M^{\dagger} M))+\;constant  .
\label{qua10}
\end{equation}
Here M is the field dependent mass matrix for the charged
fermions.

The equation (\ref{qua10}) is the supersymmetric version of Weinberg's
formula\cite{wein1} for the one-loop gauge coupling constants.

The stringy version of the Weinberg's formula for the one-loop
correction to the gauge 
couplings constant may come\footnote{The same relation was used in the 
calculation of the threshold corrections as target space free energies
in \cite{lust}.} from the relation
\begin{equation}
{\Delta}_0={\sum}_{n\;,m }{\log {\cal M}^{2}}\stackrel{def}{=} 
{\sum}_{n\;,m }
{\log {\cal M}} + {\sum}_{n\;,m }{\log {\cal M}^{\dagger}}.
\label{uqaa1}
\end{equation}
We will now use eqn. (\ref{uqaa1}) to calculate the stringy 
one-loop 
threshold 
corrections to the gauge coupling constants coming from the
integration of the massive compactification modes with
$(m,m',n,n') \neq (0,0,0,0)$.
The total contribution to the threshold corrections, coming 
from modular orbits associated with the presence of massless 
particles, is connected with the existence
of the following\footnote{We calculate only the ${\sum}
{\log {\cal M}}$ since the  $\sum \log {\cal M}^{\dagger}$ quantity 
will give only the complex conjugate.}
orbits\cite{fklz,lust}
\begin{eqnarray}
{\Delta}_0&=&{\sum}_{2 n^t m + {q^T {\cal C} q} =2 }
{\log {\cal M}}|_{reg} \nonumber\\ 
{\Delta}_1 &=&{\sum}_{2 n^t m+{q^T{\cal C}q}=0}{\log \cal M}|_{reg}. 
\label{polop}
\end{eqnarray} 
 
In the previous expressions, a regularization procedure is assumed that
takes place, which renders the final expressions finite, as infinite
sums are included in their definitions. The regularization is
responsible for the subtraction\footnote{More details 
of this precedure can be found in \cite{fklz}. } 
of a moduli independent quantity
from the infinite sum e.g ${\sum}_{n\;,m \in orbit}
{\log {\cal M}}$. 
We demand that the regularization procedure for $exp[\Delta]$
has to respect both modular invariance  and 
holomorphicity.

In eqn.(\ref{polop}), $\Delta_0$ is the   
orbit relevant for the stringy Higgs effect . This orbit is associated 
with
the quantity  $2 n^T m + q^T {\cal C} q = 2$ where 
$n^T m = m_1 n^1 + 3 m_2 n^2$ for the $Z_6 - IIb$ orbifold.
This specific orbit will be used as well, in the second part of the
thesis, to 
discuss the threshold correction contribution to the gravitational 
couplings
from the point of view of extended gauge group enhancement.

The total contribution\footnote{We use a general embedding 
of the gauge twist in the gauge degrees of freedom.}
from the previously mentioned orbit is
\begin{eqnarray}
{\Delta}_{0}{\propto}{\sum}_{n^T m + q^2 =1} \log {\cal M} 
= {\sum}_{n^T m =1 , q^{T} C q=0} \log {\cal M} +
\sum_{n^T m =0 , q^T C q = 2} \log {\cal M} +&\nonumber\\
\sum_{n^T m = -1 , q^T C q = 4} \log {\cal M} + \dots
\label{wzerooo}
\end{eqnarray}

We must notice here that we have  written the sum \cite{lust} over 
the states associated with the $ SO(4,2) $ invariant orbit 
$ 2 n^T m + q^T {\cal C} q = 2 $ in terms of a sum over
$\Gamma^0(3)$ invariant orbits $n^T m = constant$ .  
We will be first consider the contribution from the orbit
$2 n^T m + q^T {\cal C} q  = 0 $. We will be working in analogy with
calculations associated  with topological free energy 
considerations \cite{oova}.
From the second equation in eqn.(\ref{polop}), considering in general 
the $S0(4, 2)$ coset, we get for example that
\begin{eqnarray}
\Delta_{1} \propto  \sum_{n^T m + q^2 = 0} \log {\cal M} =
\sum_{n^T m =0 , q= 0 }\log {\cal M} + 
\sum_{n^T m =-1,q^2 = 1}\log {\cal M} + \dots
\label{wzerooooa}
\end{eqnarray}

Consider in the beginning the term
$\sum_{n^T m =0 , q = 0 } {\log {\cal M}} $. 
We are summing up initially the 
orbit with  $n^T m = 0; n, m \neq 0$, i.e $\Delta_1$ 
\begin{equation}
{\cal M} = 3 m_2 - im_1 U' + in^1 T + n^2 (-U' T + B C)+
{q\;dependent\;terms}. 
\label{poli}
\end{equation}
We calculate the sum over the modular orbit $n^T m + q^2 = 0$.
As in \cite{lust} we calculate initially the sum over massive
compactification states with $q_1\;= q_2\;= 0$ and $(n, m) \neq 0$. 
Namely, the orbit 
\begin{eqnarray}
\sum_{n^T m =0,\; q = 0} \log {\cal M}=
\sum_{(n,m)\neq (0,0)}\log(3m_2 -im_1 U'+in_1 T+n_2 (-U'T))
& \nonumber\\
+\;BC \sum_{(n,m)\neq (0,0)}{\frac{n_2}{(3m_2 -im_1 U'+in_1 T-n_2 U'T)}}
+{\cal O}((BC)^2).
\label{polia}
\end{eqnarray}
The sum in relation (\ref{polia}) is topological(it excludes 
oscillator excitations) and is subject to the 
constraint $3m_2 n^2 + m_1 n^1 = 0$. 
Its solution receives contributions 
from the following sets of integers:

\begin{equation}
m_2 = r_1 r_2 \;,\;n_2 = s_1 s_2\;,\;m_1 = - 3 r_2 s_1\;,\;n_1 = r_1 s_2
\label{polib}
\end{equation} 
and
\begin{equation}
m_2 = r_1 r_2 \;,\;n_2 = s_1 s_2\;,\;m_1 = - r_2 s_1\;,\;n_1 = 3 r_1 s_2.
\label{polibb}
\end{equation}
So the sum becomes,
\begin{eqnarray}
\sum_{n^T m =0} \log {(3m_2 -im_1 U'+in_1 T-n_2 U'T)} =\sum_{(r_1,s_1) 
\neq (0,0)} \log \left(3(r_1 +is_1 U')\right)\times&\nonumber\\
\sum_{(r_2,s_2) \neq (0,0)}\log(r_2 +is_2 {\frac{T}{3}})
+\sum_{(r_1,s_1)\neq (0,0)} \log 3(r_1 +is_1{\frac{U'}{3}})
{\sum_{(r_2,s_2)\neq (0,0)}\log(r_2 +is_2 T)}&
\label{koukiol}
\end{eqnarray}

Substituting explicitly in eqn.(\ref{polia}), the values for the 
orbits in equations (\ref{polib}) 
and (\ref{polibb}) together with eqn.(\ref{koukiol}), we obtain

\begin{eqnarray}
\sum_{n^T m =0;q=0} \log {\cal M}= 
\log\left({\frac{1}{3}\eta^{-2}(U') \eta^{-2}({\frac{T}{3}})}
\right)+\log{\left(\frac{1}{3}\eta^{-2}({\frac{U'}{3}})\eta^{-2}(T)
\right)}\;+&\nonumber\\
\left(BC \left( \sum_{(r_1 ,s_1 )\neq (0,0)}\frac{s_1}{r_1 +is_1 U'}
\right) \left( \sum_{(r_2 ,s_2 )\neq (0,0)} \frac{s_2}{3(r_2 +is_2 
\frac{T}{3})}\right) \right)\;+&\nonumber\\
\left(BC \left(\sum_{(r_1 ,s_1 )\neq (0,0)}\frac{s_1}{3(r_1 +i
s_1\frac{U'}{3})}\right)\left(\sum_{(r_2 ,s_2 )\neq (0,0)}\frac{s_2}
{r_2 +is_2 T}\right)\right)+ {\cal O}((BC)^2)&
\label{polid}
\end{eqnarray}

Notice that we used the relation
\begin{equation}
\sum_{(r_1,s_1) \neq (0,0)} \log{3}=\log\frac{1}{3}
\label{kro}
\end{equation}
with $\sum_{(r_1,s_1) \neq (0,0)} = -1$.
We substitute  
$\sum_{(r_1,s_1) \neq (0,0)} \stackrel{def}{=} \sum^{'}$
and $\sum_{(r_2,s_2) \neq (0,0)} \stackrel{def}{=} \sum^{"}$.
Remember that $\sum^{'} \log(t_1 + i t_2 T) = 
log\;\eta^{-2}(T)$,
with $\eta(T)= exp^{\frac{-\pi T}{12}} \Pi_{n >0}(1- exp^{-2 \pi n T})$
This means that eqn.(\ref{polid}) can be rewritten as
\begin{eqnarray}
\log_{n^T m =0;q=0} \log{\cal M}=
\log\left(\eta^{-2}(U') \eta^{-2}(\frac{T}{3})( \frac{1}{3})\right)+
\log{\left(\frac{1}{3}\eta^{-2}({\frac{U^{\prime}}{3}})\eta^{-2}(T)
\right)}\;+&\nonumber\\
-BC\left(\left(\partial_{U'}\sum^{'}\log(r_1 + is_1 U')\right)\left(
\partial_{T}\sum^{"}\log(r_2 +is_2 \frac{T}{3})\right)\right)\nonumber\\
-BC\left(\left(\partial_{U'}\sum^{'} \log(r_1 + is_1 \frac{U'}{3})\right)
\left(\partial_{T}\sum^{"}\log(r_2 + is_2 T)\right)\right) + 
{\cal O}((BC)^{2})
\label{eikos1}
\end{eqnarray}

Finally,
\begin{eqnarray}
\sum_{n^T m=0;\; q = 0} \log {\cal M} = \log \left({\eta^{-2}(T)}
{\eta^{-2}(\frac{U'}{3})})({\frac{1}{3}})\right)+\log\left({\eta^{-2}(U')}
{\eta^{-2}(\frac{T}{3})}{\frac{1}{3}}\right)\;&-\;&\nonumber\\
- B C\;  
\left((\partial_T \log {\eta^{-2}(T)})(\partial_U' \log \eta^{-2}(
\frac{U'}{3})) + (\partial_T \log  {\eta^{-2}(\frac{T}{3})})
(\partial_U' \log \eta^{-2}(U')) \right)&+&\nonumber\\
+\; {\cal O}((BC)^2)&&\nonumber\\
\label{kolloii}
\end{eqnarray}
So
\begin{eqnarray}
\sum_{n^T m=0;\; q = 0} \log {\cal M} = \log[\left({\eta^{-2}(T)}
{\frac{1}{3}}{\eta^{-2}{(\frac{U'}{3})}}\right)(1-\;BC\;
(\partial_T \log {\eta^{2}(T)})&\times&\nonumber\\
(\partial_U' \log {\eta^{2}(\frac{U'}{3}}))]\;\; +
\log [\;(({\eta^{-2}(U')}{\frac{1}{3}}){\eta^{-2}(\frac{T}{3})})
(1-BC(\partial_T \log &\times&\nonumber\\
{\eta^{2}(\frac{T}{3})})(\partial_U' \log\eta^{2}(U')))\;]\; +
{\cal O}((BC)^2)&&\nonumber\\
\label{plio}
\end{eqnarray}
or
\begin{eqnarray}
\sum_{n^T m=0;\; q = 0} \log {\cal M} = \log[\left({\eta^{-2}(T)}
{\frac{1}{3}}{\eta^{-2}{(\frac{U'}{3})}}\right)(1- 4\;BC\;
(\partial_T \log {\eta(T)})&\times&\nonumber\\
(\partial_U' \log {\eta (\frac{U'}{3}}))]\;\; +
\log [\;{\frac{1}{3}}(({\eta^{-2}(U')}){\eta^{-2}(\frac{T}{3})})
(1- 4BC(\partial_T \log &\times&\nonumber\\
{\eta(\frac{T}{3})})(\partial_U' \log\eta(U')))\;]\; +
{\cal O}((BC)^2)&&\nonumber\\
\label{plio1}
\end{eqnarray}
The last expression   
provides us 
with the 
non-perturbative \cite{fklz,lust} generated superpotential $\cal W$, by
direct integration of the string massive modes.
In fact \cite{lust} the corresponding expression for the
decomposable orbifolds, was found to be the same as the 
expression argued to exist in \cite{ANTO}, for the non-perturbative 
superpotential. The latter was
obtained from the requirement that the one loop effective action 
in the linear formulation for the dilaton be invariant
under the full $SL(2,Z)$ symmetry up to quadratic order
in the matter fields. In exact analogy, we expect our
expression in eqn.(\ref{plio}), 
to represent the non-perturbative
superpotential of the $Z_{6}-IIb$ orbifold\footnote{ Further
discussion of our results and related matters will be presented
in chapter 5, which is related to supersymmetry breaking mechanisms
in string theory.}.
The contribution of this term could
give rise to a direct Higgs mass in the effective action and represents
a particular solution to the $\mu$ term problem.
In $(2,2)$ compactifications of the heterotic string,  
 a superpotential mass term in the form
 ${\mu}_{\alpha \xi}{D^{\alpha}}{E}^{\xi}$ is generated\cite{ANTO} 
in the observable 
sector below the supersymmetry breaking scale. Here, ${D^{\alpha}},\;
{E}^{\xi}$
correspond to singlet superfields(moduli), which are in one to one
correspondence with the ${\bar{\bf  27}}$, $\bf 27$ supermultiplets of 
matter fields of the $E_6 \times E_8$ gauge group. The dependence of 
the $\mu$ term on the
non-perturbative superpotential appears through the relation
\begin{equation}
{\mu} \propto e^{G/2}{\cal W}_{DE},
\label{tion1}
\end{equation}
where ${\cal W}_{DE}$ e.g represents the quantity
\begin{equation}
{\cal W}_{DE} = \frac{-4}{\eta^{2}(T) \eta^{2}(\frac{U'}{3}) 3}
(\partial_{T} \log \eta(T))(\partial_{U'} \log \eta(\frac{U'}{3}))
\end{equation}
and transforms correctly under the
required modular transformations. Here, G is the 
gauge kinetic function. More details on the $\mu$ term generation
can be found in chapter 5.

The exact form of the non-perturbative superpotential
for the $Z_{6}-IIb$ orbifold is given by(see chapter 5)
\begin{eqnarray}
{\cal W}e^{-3S/2b} = [\;( \eta^{-2}(T)(\frac{1}{3}){\eta^{-2}}
(\frac{U'}{3}))(1 -\; B C \;(\partial_T \log {\eta^{2}
(T))}(\partial_U \log \frac{1}{3}&\times&\nonumber\\ 
\eta^{2}(\frac{U'}{3})))\;]{\tilde W}\;+\; [\;({\eta^{-2}(U^{\prime})}
({\eta^{-2}(\frac{T}{3})}){\frac{1}{3}})( 1 -\;B C 
\;((\partial_T \log {\eta^{2}(T))}&\times&\nonumber\\
(\partial_U \log \frac{1}{3} \eta^{2}(\frac{U'}{3})))\;]{\tilde W}
\;+\;{\cal O}((BC)^2),&&\nonumber\\
\label{first}
\end{eqnarray}
where S is the dilaton and b the $\beta$ function of the
condensing gauge group, and ${\tilde W}$ depends on the moduli of 
the other planes,e.g the third invariant complex plane, when 
there is no cancellation of anomalies by the Green-Schwarz 
mechanism.

We will see later in chapter five that
the exact form of the induced, $\mu$-term, 
depends explicitly on the details of the non-perturbative
generated 
superpotential we propose. 

We have calculated the topological free energy,
as a sum of the effective theory of the massive compactification modes.
Alternatively, the previous calculation could be performed 
directly at string theory level. The general result for a 
vacuum associated with D compactified coordinates ,
is that the free energy is the ratio of\cite{fklz}
the world-sheet determinants of the 
$\partial_{k}{\partial_{\bar k}}$ operator for the D-dimensional space
$R^{D}$ and the D-dimensional internal space $M^{D}$.
Explicitly             
\begin{equation}
F = {\int}_{\cal F} \frac{d^{2} \tau}{{(Im \tau)}^{2}} 
\left( \frac{{(det {\partial}_{k} {\partial}_{\bar k})}_{R^{D}}}{V(R^{D})
{(det {\partial}_{k} {\partial}_{\bar k})}_{M^{D}}} -\;1 \right).
\label{caroto}
\end{equation}

We turn now our discussion to the contribution from the first equation in
(\ref{polop}) which is relevant to the stringy Higgs effect.  
Take for example the expansion (\ref{wzerooo}. Let's examine the first orbit
corresponding to the sum ${\Delta}_{0,0} =\sum_{n^T m =1, q=0} 
\log {\cal M}$ . This orbit is the orbit for which some of the 
previously 
massive states, now become massless. At these points
the ${\Delta}_{0,0}$
has to exhibit the logarithmic singularity. 
In principle we could predict, in the simplest case when the Wilson 
lines have been switched off the form of ${\Delta}_{0,0}$.
The exact form, when it will be calculated has to respect that that
the 
quantity $e^{\Delta_{0,0}} $ has modular\footnote{This point
was not explained in \cite{lust} but it is obvious that it 
corresponds to the superpotential and thus transforming with
modular weight $-1$.}
weight $-1$ and reflects
exactly the 
presence of the physical singularities of the theory.
At this point it is appropriate to introduce 
the quantity ${\omega(T)}$ where ${\omega(T)}$ is given explicitly by
\begin{equation}
\omega(T) = (\frac{\eta(T/3)}{\eta(T)})^{12}
\end{equation}
and represents the Hauptmodul
for $\Gamma^0(3)$,
the analogue of $j$ invariant for $SL(2,\bf Z)$.
It is obviously automorphic under $\Gamma^0(3)$ and possess 
a double pole at infinity and a double zero at zero.
It is holomorphic\cite{shoe} in the upper complex plane and at 
the points
zero and infinity has the expansions
\begin{eqnarray}
{\omega(T)}= t_{\infty}^{-1} {\sum}^{\infty}_{{\lambda}=0}
a_{\lambda} t^{\lambda}_{\infty}&\;, {a_{o} \neq 0} \nonumber \\
{\omega(T)}= t_{o}^{-1} {\sum}^{\infty}_{{\lambda}=0}
b_{\lambda} t^{\lambda}_{o}&\;, {b_{o}\neq 0}
\end{eqnarray}
at ${\infty}$ and $0$ respectively with $ t = e^{- 2\pi T}$.

For ${\Delta}_{0,0}$ we predict
\begin{equation}
{\Delta}_{0,0} \propto (\omega(T) - \omega(U'))^{\gamma}  \times \dots
\end{equation}

In full generality, the Hauptmodul
functions for the $\Gamma^0(p)$ are the functions\cite{apostol}
\begin{equation}
\Phi(\tau) = \left( \frac{\eta(\frac{\tau}{p})}{\eta(\tau)}\right)^r
\label{haupt1}
\end{equation}
 Here, p=2,3,5,7 or 13 and $r=24/(p-1)$. For these values of p
the function in eqn.(\ref{haupt1}) remains modular invariant, i.e it 
is a modular function.

The corresponding functions for 
the group $\Gamma_0(p)$ are represented by the
$(\frac{\eta({\tau})}{\eta(p\tau)})^r$.
\begin{eqnarray}
\Delta_{0,0} = \sum_{n^T m =1}\log(TU'n^2 +Tn^1-U'm_1 +3m_2)&
=&\log\{(\omega(T)-\omega(U'))^{\xi} \eta(T)^{-2}\nonumber\\ 
\times \eta(U'/3)^{-2}\}+\log \{(\omega(T)-\omega(U'))^{\chi} 
\eta(T/3)^{-2}\eta(U' )^{-2}&+&\dots
\label{repjj}
\end{eqnarray}
The behaviour of the $\Delta_0$ term reflects the\footnote{in the
following we will be using the variable $U$ instead of $U'$.}
fact that at the points with $T=U$, generally previously massive states 
becoming massless, while the $\eta$
terms are needed for consistency under modular transformations. 
Finally, the integers $\xi, \chi$ have to be calculated from a string 
loop calculation or by directly performing the sum.

After this parenthesis, we continue our discussion by turning on, Wilson
lines.
When we turn the Wilson lines on, for the $SO(4,2)$ orbit of the 
relevant untwisted two dimensional subspace $\Delta_{0,0}$ becomes
\begin{equation}
\Delta_{0,0} =
\sum_{n^T m =1} \log \{ 3m_2 - i m_1 U +  i n_1 T - n_2 ( U T
-  B C )\}
\label{delzero}
\end{equation}

The sum after using an ansatz similar to \cite{lust}
and keeping only lowest order terms has the form 
\begin{eqnarray}
\Delta_{0,0} & = &
\log \left(\omega(T) - \omega (U) +B C\;X (T,U) 
\right)^{\xi} +\log\left(\eta(T)^{-2} \;\eta(U/3)^{-2} + 
B C \;Y (T,U)\right) \nonumber\\
&+ &\log\left(\eta(T/3)^{-2} \;\eta(U)^{-2} + B C 
\;W(T,U)\right)\;+\; \dots
\label{w00bcc}
\end{eqnarray}
The functions $X(T,U)$,$Y(T,U)$, $W(T,U)$ will be 
calculated by the demand of duality invariance. 

Demanding invariance \cite{ANTO} of the first term in (\ref{w00bcc})
under the target space duality transformations which leave the tree 
level K\"ahler potential
invariant
\begin{eqnarray}
U &\rightarrow& \frac{\alpha U -i \beta}{i \gamma U + \delta}
\;,\;\;\; T \rightarrow T - {i \gamma} \frac{BC}{i \gamma U +
 \delta} \;,\;\;\; \alpha \delta - \beta \gamma = 1 \nonumber\\
B &\rightarrow& \frac{B}{i \gamma U + \delta} \;\;,\;\;
C \rightarrow \frac{C}{i \gamma U + \delta},\;\;\;\beta=0\;mod\;3,
\label{utbctransf}
\end{eqnarray}
we get that $X (T,U)$ has to obey - to lowest order in B C -
the transformation  
\begin{equation}
\omega(T) - \omega(U) \stackrel{\Gamma^o(3)_{U}}{\rightarrow} \;\;
\omega(T) - \omega(U) \; - \; {i \gamma} \; \frac{BC}{i 
\gamma U + \delta} \;\; \omega(T)'.
\end{equation}
As a consequence
\begin{equation}
X(T,U) \stackrel{\Gamma^o(3)_{U}}{\rightarrow}(i\gamma U +\delta)^2 
\;X(T,U)+ i \gamma (i\gamma U+\delta) \; \omega'(T).
\label{kourada1}
\end{equation}

In the same way, demanding invariance under $\Gamma^o(3)_T$
transformations we find that $X(T,U)$ has to transform as
\begin{equation}
X(T,U)\stackrel{\Gamma^o(3)_{T}}{\rightarrow} 
(i\gamma T +\delta )^2 \; X(T,U)- i\gamma(i\gamma T +\delta)\; 
\omega'(U).
\label{kourada2}
\end{equation}

Because, the first term in (\ref{w00bcc})
has to exhibit the logarithmic singularity at the point $T=U$,
$X(T,U)$ in turn has to vanish at the same point.
All the previous mentioned properties,
are properly exhibited from the function
\begin{eqnarray}
X(T,U) = \partial_U \{\log \eta^6(\frac{U}{3})\} 
\; \omega'(T) -  \partial_T \{\log \eta^6(\frac{T}{3})
\}  \; \omega'(U)& +\nonumber\\
\beta \{\omega (T) - \omega(U) \} \eta^4(\frac{T}{3})
\eta^4(\frac{U}{3}) + {\cal O}((BC)^2)
\label{kourada3}
\end{eqnarray}
The  $\beta$ is a constant which may be decided 
from a loop calculation. The exact calculation of the 
threshold corrections   
involving the presence of the logarithmic term may come
from a calculation similar to the one performed 
in \cite{mouha}.
Let us discuss now the term $Y(T,U)$.
Demanding the term $\eta(T)^{-2} \;\eta(U/3)^{-2} + 
B C \;Y (T,U)$ to transform under the $\Gamma^o(3)_U$ transformations
of eqn.(\ref{utbctransf}) with modular weight $-1$, gives that
$Y(T,U)$ has to transform
\begin{eqnarray}
Y (T,U) \stackrel{{\Gamma}^{o}(3)_{U}}{\rightarrow}(i \gamma U +\delta)
 Y (T,U) + i\gamma (i \gamma U + \delta)\eta(\frac{U}{3})^{-2}
(\partial_{T} \eta^{-2}(T)). 
\label{kourada4}
\end{eqnarray}
Under $\Gamma^o(3)_T$, $Y(T,U)$ has to transform as
\begin{eqnarray}
Y (T,U) \stackrel{{\Gamma^o(3)}_{T}}{\rightarrow}(i \gamma U +\delta)
 Y (T,U) +i\gamma (i \gamma U + \delta)(\partial_{U} 
\eta(\frac{U}{3})^{-2}) \eta(T)^{-2}. 
\label{kourada5}
\end{eqnarray}
The following function satisfies all requirements up 
to order $(BC)^2$,
\begin{eqnarray}
Y (T,U)=-\eta^{-2}(T)\eta^{-2}(\frac{U}{3})(\partial_{T} \log
\eta^{2}(T))(\partial_{U}\log\eta^{2}(\frac{U}{3}))+\upsilon_{1} 
\eta^{4}(T)\eta^{4}(\frac{U}{3})\nonumber\\
+\upsilon_{2} \eta^{4}(T)\eta^{4}(U).
\label{kourada51}
\end{eqnarray}
The transformation
behaviour under the proper modular transformnations
is not enough to determine the constants $\upsilon_{1}$ and
$\upsilon_{2}$.
They may be decided from a string loop calculation\cite{mouha}.
In a similar way, demanding the term
$\eta(T/3)^{-2} \eta(U)^{-2} + BCW(T,U)$ to transform with modular
weight
$-1$, we find that $W(T,U)$ has to transform as
\begin{equation}
W(T,U)\stackrel{\Gamma^o(3)_{U}}{\rightarrow} W(T,U)(i\gamma U+\delta) 
+ i \gamma (\partial_{T} \eta^{-2}(\frac{T}{3})) \eta^{-2}(U),
\label{kourada6}
\end{equation}
and
\begin{equation}
W(T,U) \stackrel{\Gamma^o(3)_{T}}{\rightarrow} W(T,U)(i \gamma T
+ \delta)+i\gamma (\partial_{U} \eta^{-2}(U)) \eta^{-2}(
\frac{T}{3}).
\label{kourada7}
\end{equation}
The following function satisfies the requirements of eqn.(\ref{kourada6})
and (\ref{kourada7}),
\begin{eqnarray}
W =-\eta^{-2}(\frac{T}{3})\eta^{-2}(U)(\partial_{T}\log \eta^{2}
(\frac{T}{3}))(\partial_{U}\log \eta^{2}(U))+ 
\lambda_{1} \eta^{4}(\frac{T}{3})\nonumber\\
\times \eta^{4}(U) + \lambda_{2} \eta^{4}(T)\eta^{4}(U),
\label{kourada8}
\end{eqnarray}
where $\lambda_{1},\;\lambda_{2}$ will be decided from the string 
loop calculation similar that in \cite{mouha}.
\newline
\newline

%************************************************************************
%************************************************************************
%
%******************************************************************
%********************************************************************
\subsection{\bf  * Threshold corrections to gauge and gravitational 
couplings}
\subsubsection{* Threshold corrections to gauge couplings}
Let us now complete our previous discussions, by
considering the contributions of gravitational threshold corrections
due to the integration of the massive modes of the heterotic string.
We will be concentrating our discussion on $(2,2)$ symmetric
non-decomposable orbifolds for which 
an explicit calculation of moduli dependence of the threshold corrections
to the gauge coupling constants exists.

We will analyze the case of gravitational threshold corrections
in the case of $N=2$ heterotic string compactifications ,
up to one loop and for the case of non decomposable orbifold 
compactifications
of the heterotic string. 
Before we examine the threshold contributions to the gravitational 
threshold corrections, we will study their effect on the 
gauge coupling constants.
It will help us to understand properly the connection between
the calculation of the free energy we performed before,
and the the effective gauge couplings.

When considering an effective locally supersymmetric field 
theory,
we have to distinguish between two kinds of renormalized physical 
couplings involved in the theory. These are the cut-off dependent 
Wilsonian gauge couplings and the moduli and momentum dependent
effective 
gauge couplings(EGC)\cite{SV}.

Let us follow a field theoretical approach for the calculation of 
contributions
of the physical modes of our theory to the effective gauge couplings.
We demand our physical theory at the high energy threshold,
to be a product of several gauge groups 
namely $G = \otimes G_{a}$.
Then, 
the one loop corrected effective gauge couplings obey
the following formula 
\begin{equation}
\frac{1}{g_a^2(p^2)} = \frac{k_a}{g_a^2(M^2_{X})}
+ \frac{b_a}{16 \pi^2} \log \frac{M^2_{X}}{p^2} +
\frac{\Delta_{a}}{16 \pi^2} +
\frac{{\tilde \Delta}_a}{16 \pi^2},
\label{grun}
\end{equation}
when\footnote{For convinience, we will set the Kac-Moody level
equal to one.}
\begin{equation}
\Delta_{a} =[- 2 \sum_{i} T_{a}(r) \log det g_{r} ) + c_{a} K].
\label{grun1}
\end{equation}

In the formula (\ref{grun}) which is valid at energies 
$p^2 \ll M_X^2$, 
we have tacitly assumed that 
the light particles of the theory are exactly massless, while
the massive charged fields decouple at the high energy threshold
$M_X^2$.
 
Here, the $N=1$ $\beta$ function coefficient is given by 
$b_{a} = -3 c(G_{a}) + {\sum_{\omega} T_{a}(r_{\omega})}$, where 
$c(G_{a})$ is the 
quadratic
Casimir of the gauge group and the sum is over the massless charged 
chiral matter superfields transforming under the representation 
$r_{\omega}$ of the gauge group $G_{a}$.
In addition,
$c_{a} = \left(- c(G_{a})  +
{\sum_{\omega}} T_{a}(r_{\omega}) \right)$ 
and $T_{a}(r)$ is given 
by $Tr(T^{\mu}_{a} T^{\nu}_{a}) = {\delta}^{\mu \nu} T_{a}(r)$, 
where $T^{\nu}_{a}$ is the generator of the gauge group $G_{a}$ 
and the sum is over massless fermions transforming under $G_{a}$. 
Finally, K is the K\"ahler potential of our low 
energy theory and $g_{r}$ is the $\sigma $-model metric of the 
massless subsector of the charged matter fields transforming in the 
representation r of the gauge group.

The contributions $\tilde \Delta$ of the threshold corrections
describe the tower of massive modes that decouple\footnote{In
N=1 orbifold compactifications the high energy threshold 
coincides\cite{dkl2,kap} with
the string unification scale.}
at the high energy threshold $M_X$. 
The non-holomorphic threshold contribution of the term in the 
brackets comes from the contributions of  
the K\"ahler and $\sigma$ model anomalies. 
Its contribution to the four dimensional effective one-loop 
string action is associated to triangle diagrams involving 
two \cite{L1,ovr1,dfkz,kalou1} gauge and moduli fields as 
external legs while massless particles running in the loops.
The $\sigma$ model anomalies are similar to the local gauge 
anomalies but now one of the external legs of the triagle diagram
is a
composite $\sigma$ model connection or a K\"ahler connection.
In the ${\sigma}$-model description\cite{cre,dere1} of $N=1$ 
supergravity, fermion kinetic terms 
\begin{equation}
\frac{i}{2} g_{ij}{\bar \psi}_{j} \gamma^{\mu} \partial_{\mu}
{\psi} + 
\frac{i}{2}  Re f_{ab} {\bar \lambda}^{a} \gamma^{\mu}
\partial_{\mu}{\lambda}^{b}
\label{fer1}
\end{equation}
are accompanied by the interaction terms
\begin{eqnarray}
\left( \frac{i}{2}  Re f_{ab} {\bar \lambda}^{a}_{L} \gamma^{\mu}
\partial_{\mu}{\lambda}^{b}_{L}
-
\frac{i}{2} g_{ij}{\bar \psi}_{Lj} \gamma^{\mu} \partial_{\mu}
{\psi}_{Li} \right) \frac{1}{2} V_{\mu}^{Kahler} \nonumber\\
+\left( \frac{i}{2} g_{ij}{\bar \psi}_{Lj}
\gamma^{\mu} \partial_{\mu}
{\psi}_{Li}(-i \Gamma_{ikl} \partial_{\mu} z_k) + h.c\right),
\label{fer2}
\end{eqnarray} 
with the $\sigma$-model connection is given by
$\Gamma{ijk} = \frac{\partial}{\partial \phi^{i}}g_{k {\bar m}} $
and the K\"ahler connection is given by
\begin{equation}
{\alpha}_{\mu}= -i\left[ \frac{\partial}{\partial \phi_{i}}
K(\phi,{\bar \phi}) {\partial}_{\mu} \phi^{i} -
\frac{\partial}{\partial{\bar
\phi}_{i}}K(\phi,{\bar \phi}) {\partial}_{\mu}{\bar \phi}^{i} 
\right].
\label{sig1}
\end{equation}

The composite K\"ahler connection is analogous to $K(\phi,{\bar
\phi})$ and\footnote{Here $K(\phi,{\bar \phi})$ represents the moduli
field dependent part of the K\"ahler potential. Of course, we concentrate
our discussion in the $N=2$ sectors of the $N=1$ $(0,2)$ orbifold 
compactification of the heterotic string.} couples to gauginos 
${\lambda}_{L}$ as well 
to chiral matter fields ${\psi}_{L} \equiv A_{\alpha}(r_{\omega})$. 
It's presence is a 
reflection of the tree level invariance of the theory under K\"ahler 
transformations.

The contributions from the K\"ahler and $\sigma$ model 
connections lead to the following one-loop modification of the tree 
level supersymmetric non-linear $\sigma$- model moduli Lagrangian:
\begin{eqnarray}
{\cal L}_{non-local} =\sum_{\alpha}\int d^2 \theta \frac{1}{4}
W^{\alpha}W_{\alpha}\{S - \frac{1}{16{\pi}^2}\frac{1}{16}{\Box}^{-1}
{\overline{\cal D}}{\overline{\cal D}}{\cal D}{\cal D}&\nonumber\\
\left([c(G_{\alpha})- \sum_{r_{\omega}}T(r_{\omega})]K(\phi,{\bar \phi}) 
+ 2\sum_{r_{\omega}}T(r_{\omega})\log det K_{\alpha \beta}(\phi,{\bar 
\phi})\right)\}+\;h.c,&
\label{sig2}
\end{eqnarray}
with $K_{\alpha \beta}(A,{\bar A})$ the K\"ahler\footnote{The
K\"ahler metric of the matter fields, appears when we expand
the K\"ahler potential for the matter fields in lowest order in
the matter fields, as $K^{matter}= K_{\alpha \beta}^{matter}A_{\alpha}
{\bar A}_{\beta}$.}metric of the matter 
fields, the chiral superfield $W^{\alpha}\stackrel{ def}{=} -(1/4)
\overline{DD}e^{-V}D_{\alpha}e^{V}$
and V is the vector superfield\footnote{See for example 
\cite{dere2,dere3}.}.
In this form the general field theoretical contribution to the
threshold corrections appears to be

\begin{equation}
\frac{1}{g_a^2(p^2)} = \frac{1}{g_a^2(M^2_{X})}
+ \frac{b_a}{16 \pi^2} \log \frac{M^2_{X}}{p^2} +
\frac{\Delta_{a}}{16 \pi^2} 
\label{grunn1}
\end{equation}

\begin{equation}
\Delta_{\alpha}= 16 \pi^2 Re f^{1-loop} -
[- 2 \sum_{i} T_{a}(r) \log det g_{r} ) + c_{a} K]
\label{sig2a}
\end{equation}
and $Ref^{1-loop}$ is induced from the integration of massive modes
that decouple at the scale $M^2_{X}$.

Notice that the general form
of the gauge coupling dependence in a $N=1$ supersymmetric gauge theory
appears in the form
\begin{equation}
\frac{1}{4} \sum_{a} \int d^2 \theta f_{a}(\phi)(W^{\alpha}
W_{\alpha})_{a} + h.c = -\frac{1}{4}\{ \sum_{a}(Ref)_{a}(F_{\mu \nu}
F^{\mu \nu})_{a} - Imf_{a}(F{\tilde F})_{a}\},
\label{sder}  
\end{equation}
where the index a labels the different group factors of the high energy
gauge group $G =\otimes  G_{a}$. Obviously
\begin{equation}
f_a^{tree} = \{\frac{1}{g_a^2} - \frac{i \theta_a}{8 \pi^2}\}^{tree} =
k_a S,
\label{sderaa}
\end{equation}
with  $k_a$ the Kac-Moody level.

By looking at eqn.(\ref{sig2}), we can see that the $\sigma$ -model
lagrangian is not invariant under the duality K\"ahler transformations
\begin{equation}
K(\phi,{\bar \phi})\rightarrow K(\phi,{\bar \phi})+g(\phi)+
g({\bar\phi}),
\label{sig3}
\end{equation}
and reparametrizations which act on the matter metric as
\begin{equation}
K_{\alpha \beta} \rightarrow h_{\alpha \gamma}(\phi)^{-1}h_{\beta
\delta}^{-1}({\bar \phi}) K_{\alpha \beta}  
\label{sig4}
\end{equation}
The non-invariance of eqn.(\ref{sig2})is reflected in the presence
of the additional term
\begin{equation}
\delta{\cal L} = \frac{-1}{16{\pi}^2}\sum_{\alpha}\int d^2 \theta
\frac{W^{\alpha}W_{\alpha}}{4} \left([c(G_{\alpha}) -
\sum_{r_{\omega}} T(r_{\omega})]g(\phi)+2
\sum_{r_{\omega}}T(r_{\omega})
\log det h_{\alpha\beta}(\phi,{\bar\phi})\right)+h.c
\label{sig5}
\end{equation}

Take for example $(0,2)$ abelian orbifolds. 
The K\"ahler  potential for the matter fields, when it is expanded
around the $<A_{\alpha}=0>$ classical vaccum becomes 
\begin{equation}
K^{matter}= \delta_{\alpha \beta}
{\displaystyle\mathop{\Pi}_{i=1}^{h_{(1,1)}}} (T+{\bar
T})_i^{n^i_{\alpha}}\;{\displaystyle\mathop{\Pi}_{m=1}^{h_{(2,1)}}}
(U+{\bar U})_m^{l^m_{\alpha}}
A_{\alpha} {\bar A}_{\beta} 
\label{sig6}
\end{equation}
which means that every matter field is characterized by 
$(h_{(1,1)} + h_{(2,1)})$ rational numbers, the modular weights, 
which are represented in vector form as $\vec{n}^i_{\alpha}=
(n^1_{\alpha},n^2_{\alpha}\dots,n^{h_{(1,1)}}_{\alpha})$ and
$\vec{l}^m_{\alpha}=(l^1_{\alpha},l^2_{\alpha}\dots, 
l^{h_{(2,1)}}_{\alpha})$.
In addition, invariance of the kinetic energy for the matter fields
under e.g $SL(2,Z)_{T}\times SL(2,Z)_{U}$ target space duality 
transformations, produces the requirement
\begin{equation}
A_{\alpha}\rightarrow
A_{\alpha}{\displaystyle\mathop{\Pi}_{i=1}^{h_{(1,1)}}}
(ic_T + d)_i^{n^i_{\alpha}}{\displaystyle\mathop{\Pi}_{m=1}^{h_{(2,1)}}}
(icT + d)_m^{l^m_{\alpha}}.
\label{sig7}
\end{equation}
For abelian orbifolds and untwisted matter fields associated with the
$j-th$
complex plane the modular weights are given\cite{iblu} by 
$n_j^i=-\delta_j^i$ 
and $l_j^i=-\delta_j^i$,
while for twisted states associated with the order N twist vector
$\vec{\theta}=(\theta^1,\theta^2,\theta^3)\;(0\leq \theta^i<1,
{\displaystyle\mathop {\sum}_{i=1}^3} \theta^i =1)$
and having a complex plane not being fixed in two or all three 
complex planes, the modular weights are $n^i_{\alpha}=-(1-\theta^i),\;
l^i_{\alpha}=-(1-\theta^i),\;\theta^i\neq 0$ and
$n^i_{\alpha}=l^i_{\alpha}=0,\;for\;\theta^i= 0$.

Substituting explicitly in eqn.(\ref{sig2}) the values of 
K\"ahler potential and the matter metric we get
\begin{eqnarray}
{\cal L}=\sum_{\alpha}\int d^2 \theta \frac{W^{\alpha}W_{\alpha}}{4}
\{S -\frac{1}{16{\pi}^2}\frac{1}{16}\frac{{\overline{\cal D}}
{\overline{\cal D}}{\cal D}{\cal D}}{\Box}[\sum_{i=1}^{h_{(1,1)}}
{b'}_{\alpha}^i \log(T+{\bar T})_i +\nonumber\\ 
\sum_{m=1}^{h_{(2,1)}}{b'}_{\alpha}^m \log(U+{\bar U})_{m}]\}+h.c,
\label{sig8}
\end{eqnarray}
with 
${b'}_{\alpha}^i = -c(G_{\alpha}) + \sum_{r_{\omega}} T(r_{\omega})
(1+2n^i_{r_{\omega}})$ and ${b'}_{\alpha}^m= -c(G_{\alpha}) +
\sum_{r_{\omega}} T(r_{\omega})(1+2l^i_{r_{\omega}})$.

The final contribution to the gauge kinetic terms including
one-loop corrections coming from the heavy modes that decouple at 
the 
string unification scale is found to be\cite{dfkz,L1}
\begin{eqnarray}
{\cal L}=\sum_{\alpha}\int d^2 \theta \frac{W^{\alpha}W_{\alpha}}{4}
\{S -\frac{1}{16{\pi}^2}\frac{1}{16}\frac{{\overline{\cal D}}
{\overline{\cal D}}{\cal D}{\cal D}}{\Box}[\sum_{i=1}^{h_{(1,1)}}
{b'}_{\alpha}^i \log(T+{\bar T})_i \eta^{4}(T)+\nonumber\\ 
\sum_{m=1}^{h_{(2,1)}} {b'}_{\alpha}^m \log(U+{\bar U})_{m}
\eta^{4}(U)]\}+h.c.
\label{sig81}
\end{eqnarray}

It appears finally\cite{dfkz,ovr1,iblu,kalou} that the 
non-invariance of the lagrangian (\ref{sig2},\ref{sig81}) under 
$SL(2,Z)_T$ modular transformations
\begin{equation}
T \rightarrow \frac{a T - i b}{i c T + d}\;,\;ad-bc=1;a,b,c,d\;\in\;Z,
\label{sig9}
\end{equation}
can only be compensated by the use of the Green-Schwarz(GS) mechanism.

In
the presence of the Green-Schwarz
mechanism\footnote{,used for the cancellation
of anomalies of the ten dimensional theory or better to
provide for a Fayet -Iliopoulos D-term in order to break
supersymmetry,}in the four dimensional one-loop effective
string action, the previous considerations have to be modified.
In that case we will have to subtract the contribution of the
Green-Schwarz term from the total anomaly coefficient.
The Green-Schwarz mechanism in four-dimensions induces the following 
modification to the lagrangian (\ref{sig81}) 
\begin{eqnarray}
{\cal L}=\sum_{\alpha}\int d^2 \theta \frac{W^{\alpha}W_{\alpha}}{4}
\frac{1}{16{\pi}^2}\frac{1}{16}\frac{{\overline{\cal D}}
{\overline{\cal D}}{\cal D}{\cal D}}{\Box} \{[(S+{\bar S})+
\frac{1}{4\pi^2}\sum_{i=1}^{(h_{(1,1)}} \delta^i_{GS}\log 
(T+{\bar T})&\nonumber\\
+\frac{1}{4\pi^2}\sum_{m=1}^{h_{(2,1)}}\delta^m_{GS}
\log (U+{\bar U})_m]+
\frac{1}{4 \pi^2} \sum_{i=1}^{h_{(1,1)}}(b'^{i}_{a}- \delta^i_{GS})
\log(T_i+{\bar T}_i) \eta(T_i)^4 + &\nonumber\\
\sum_{m=1}^{h_{(2,1)}}
\frac{1}{4 \pi^2}(b'^{m}_a-\delta^m_{GS})\log(U_m+{\bar U}_m) 
\eta(U_m)^4\}
\label{sig82}
\end{eqnarray}
Notice that the presence of duality anomalies under the transformation 
of eqn.(\ref{sig9})can be cancelled
by a transformation of the dilaton field as
\begin{equation}
S \rightarrow S - \frac{1}{8 \pi^2} \sum_{i=1}^{h_{(1,1)}}  
\delta^i_{GS} \log(icT + d)_i 
\label{sig93}
\end{equation}
 
In equation (\ref{sig82}) the term in brackets involving 
the dilaton induces a mixing in the one-loop K\"ahler potential 
between dilaton and the moduli fields.
This mixing corresponds to the following K\"ahler potential
\begin{equation}
K = - \log[ S + {\bar S} -\frac{1}{4 \pi^2} \sum_{i=1}^{h_{(1,1)}}
\delta^i_{GS}\log(T + {\bar T})].
\label{sig94}
\end{equation}
The induced one-loop correction to the tree level K\"ahler potential
corresponds to a supergravity lagrangian formulated in the 
linear representation\footnote{See chapter five.}of the dilaton.

We may now complete our discussion of the contributions
to the gauge couplings by considering  contributions
from a $(2,2)$ symmetric non-factorizable orbifold theory. 
We will apply our discussion to the $Z_{6}-IIb$ orbifold.
We examine the contributions from the subsector $(1,{\Theta^2})$ only.
Contributions coming from the  $(1,{\Theta^3})$ sector
are invariant under $SL(2,Z)_{T} \times SL(2,Z)_{U}$, they are identical
to those listed in \cite{lust}, 
and we will not include them here as their result is additive to the
gauge coupling constants.

We consider the contribution to the effective gauge coupling 
constants\footnote{A similar
methodology was followed in \cite{lust} by
turning on moduli fields B,C in the invariant subsector.}. 

The moduli 
will play the role of Higgs fields, breaking the original gauge group.
The original gauge group of our theory is supposed to be sited at the 
string 
unification scale $M_{s}$ and it is composed from gauge group
factors $G_a$, as $G = \oplus G_a$. 
We assume that space-time supersymmetry remains unbroken while the
moduli fields spontaneously break the original gauge group in
the form ${\tilde G}= \oplus {\tilde G}_a$. The running of the   
subgroups ${\tilde G}_a$ includes the presence of additional 
threshold scales coming from gauge group enhancement at special 
points and as a reverse consequence previously massless particles 
become massive and decouple. 
In describing the running of effective gauge couplings of those 
subgroups,
we substitute explicitly the tree level 
values for the gauge coupling constants 
at the string unification scale ${(S + {\bar S})}/2$, the 
value of the ${\sigma}$-model matter metric in the case of vanishing
Wilson lines $\left( (T+{\bar T})(U + {\bar U}) \right)^{-1}$.
We find that the full 
contribution to the gauge couplings \footnote{Remember that we
consider a $(2,2)$ theory for which the gauge group is always 
$E_8 \times E_6$ and the matter fields are consisting of 
$h_{(1,1)}\;27$
multiplets and $h_{(2,1)}\;{\bar 27}$ multiplets in the $E_6$ 
part of the gauge group.} 
from the modular orbit $2 n^{T} m + q^{T} {\cal C} q = 0$ 
is
\begin{eqnarray}
\frac{1}{g_{E_8}^{2}(p^2)} = \frac{S + \bar S}{2} +
 \frac{b_{E_8}}{16 \pi^2} \log \frac{M_{string}^2}{p^2}
+\frac{c_{E_8}}{16\pi^2}\log\left((T+{\bar T})(U^{\prime}+
{\bar U}^{\prime})-
(B+{\bar C})(C+{\bar B})\right)&\nonumber\\
+\frac{c_{E_8}}{16 \pi^2}\log \left(9 \eta^{4}(T)\eta^{4}
(\frac{U^{\prime}}{3})
|(1 - BC(\partial_{T} \log \eta^{2}(T))(\partial_{U^{\prime}}
 \log \eta^{2}(\frac{U^{\prime}}{3})))|^{-2} \right)&\nonumber\\
+{\frac{c_{E_8}}{16 \pi^2}} 
\log\left(9\eta^{4}(U^{\prime}) \eta^{4}(\frac{T}{3})
|(1 - BC(\partial_{T} \log \eta^{2}(\frac{T}{3}))(\partial_{U^{\prime}}
 \log 
\eta^{2}({U^{\prime}})))|^{-2}\right),&\nonumber\\
\label{pre1}&
\end{eqnarray}
with 
$b_{E_8} = - 3 c(E_8)$ and we have assumed for simplicity that
$\delta_{GS}=0$.

The expression (\ref{pre1}) represents the gauge couplings which 
are not affected from the presence of the Wilson line moduli.
These can be for example the gauge couplings which belong 
to the $E_8$ hidden sector. By making the correspondence with 
the general superfield behaviour of the charged matter in 
$(2,2)$ theories, we conclude that a priori this has to be the
case since the charged matter in $(2,2)$ theories are always
$E_8$ neutral. 
This built in property of the theory, can be used in the gaugino
condensation
approach to supersymmetry breaking by taking the hidden $E_8$ sector to
be associated with the pure Yang-Mills gauge sector.
In the previous description of the gauge couplings we have tacitly 
assumed that the original gauge group of the theory,
spontaneously breaks at a product of subgroups and as a consequence
previously massless particles decouple.

In the presence of the Green-Schwarz mechanism, (\ref{pre1})
becomes
\begin{eqnarray}
\frac{1}{g_{E_8}^{2}(p^2)}=\frac{\Pi}{2}+
\frac{b_{E_8}-{\delta_{GS}}}{16\pi^2}\log\frac{M_{string}^2}
{p^2}+\frac{c_{E_8}-{\delta_{GS}}}{16\pi^2}\log((T+{\bar T})
(U^{\prime}+{\bar U}^{\prime})-(B+{\bar C})&\nonumber\\
\times (C+{\bar B}))+\frac{c_{E_8}-{\delta_{GS}}}{16\pi^2}\log
\left(9\eta^{4}(T)\eta^{4}(\frac{U^{\prime}}{3})|1-BC(\partial_{T}
\log\eta^{2}(T))(\partial_{U^{\prime}}\log\eta^{2}(\frac{U^{\prime}}
{3}))|^{-2}\right)&\nonumber\\
+{\frac{c_{E_8}-\delta_{GS}}{16\pi^2}}\log\{9\eta^{4}(U^{\prime})
\eta^{4}(\frac{T}{3})|1-BC(\partial_{T}\log \eta^{2}(\frac{T}{3}))
(\partial_{U^{\prime}}\log\eta^{2}(U^{\prime})|^{-2}\},\;\;\;\;\;
\label{pre22}
\end{eqnarray}
with 
\begin{equation}
{\Pi} = S + {\bar S} -\frac{\delta_{GS}}{8 \pi^2} \log \left(
(T + {\bar T})(U^{\prime}+{\bar U}^{\prime})-(B+{\bar C})(C+{\bar B})
\right) +(modular function).
\label{pre3}
\end{equation}
We will give the unspecified  
modular function later, in chapter 4. We will only say at this stage 
that its value
is universal, and gauge group independent.
In the $N=2$ part of the internal superconformal
field theory, the gauge couplings depend on scalars belonging
to vector multiplets and not on the hypermultiplet moduli.

We will now discuss the effect on the gauge couplings on the vector
multiplets of the invariant subspace.
In the case of six dimensional compactifications of heterotic string
 vacua,
the moduli of the invariant subspace belong to vector multiplets.
In such a case, the gauge couplings of the vector multiplets result 
in
\begin{eqnarray}
&{\frac{1}{g_{U(1)}^2(p^2)} ={\frac{1}{g_{tree}^2}}
+{\frac{\hat{b}_{U(1)}}{16\pi^2}}\log{\frac{M^2_{string}}{p^2}}
+{\frac{({\hat{b}}_{U(1)}-b_{U(1)})}{16{\pi}^2}}}
{\log{({\omega(T)}-{\omega(U^{\prime})})}^{2}}-{\frac{a_{U(1)}}
{16{\pi}^{2}}}\nonumber \\
&\times\{\log\left((T+{\bar T})(U^{\prime}+{\bar U}^{\prime})
|9\eta(\frac{U}{3} \eta(T)|^4 \right) + \log\left(T+{\bar T})
(U^{\prime}+{\bar U}^{\prime}){|9\eta(U^{\prime)}
\eta(\frac{T}{3})|}^{4} \right)
\label{ret}
\end{eqnarray}
with ${\hat{b}_{U(1)}} =0$, since $c_{U(1)}=0$ and there are no 
hypermultiplets charged under the $U(1)$.
Here, the without hat quantities correspond to the running of the 
gauge 
couplings between the string unification scale and the 
intermediate threshold $M_I$, while the hat quantities correspond
 to the running of the 
gauge
couplings between the threshold $M_I$ and the momentum scale
$p^2$, with $p^2 < M_I^2$.
Explicitly,
\begin{eqnarray}
\frac{1}{g_{U(1)}^2(p^2)}=\frac{1}{g^2(M^2_I)}+\frac{{\hat b}_{U(1)}}
{16\pi^2}\log|\omega(T)-\omega(U^{\prime})|^2+\frac{{\hat b}_{U(1)}}
{16\pi^2}\log\frac{M_{string}}{p^2}+&\nonumber\\
+ \frac{\hat \triangle}{16\pi^2}
\label{ret1}
\end{eqnarray}
where $\hat \triangle$ the contribution from K\"ahler and
$\sigma$-model anomalies.
The sum is over the chiral multiplets of massless particles
between the thresholds.
Furthermore, for the running of the couplings between the threshold
$M_I^2$ and the string unification scale 
\begin{equation}
\frac{1}{g_{U(1)}^2(M_I^2)}=\frac{1}{g_{U(1)}^2(M_{string}^2)}+
\frac{b_{U(1)}}{16\pi^2}\log\frac{1}{|\omega(T)-\omega(U^{\prime})|^2}
+\frac{\triangle}{16\pi^2},
\label{ret2}   
\end{equation}
where $\triangle$ are the contribution coming from integration of the 
massive modes. Combining eqn.(\ref{ret1}) and eqn.(\ref{ret2}), we obtain
eqn.(\ref{ret}) .

In the same way as in \cite{lust},
$a_{U(1)}=0= -c_{U(1)} + \sum_{r_C} T_{r_C}$, since
the gauge group under the additional threshold scale $M_I$ is 
abelian. The coefficient $b_{U(1)} $ equals ${\hat b}_{U(1)}
+ 2 b_{vec}$, where $2 b_{vec}$ the contribution from 
 the two $\beta$-function coefficients of the two 
 vector multiplets which are massless above the threshold scale.
  The additional threshold scale beyond the traditional
string tree level unification scale is the one associated with the
term $\omega(T) - \omega(U)$. The threshold scale now becomes
$M_{I} = |\omega(T) - \omega(U^{\prime})| M_{string}$ and is associated 
with the enhancement of the abelian part of the gauge group to $SU(2)$.
The appearance of the threshold scale is specific at the point where the
non-abelian $SU(2)$ factor is being broken to $U(1)$.

\subsubsection{ * Threshold corrections to gravitational couplings}

In this part of the thesis we will discuss briefly contributions to
the running gravitational couplings in $(2,2)$ symmetric $Z_N$ orbifold 
constructions of the heterotic string.

For $(0,2)$$\;$ $Z_{N}$ orbifolds the effective 
low energy action of the  
heterotic string is\cite{grsl,zwie1,tsey1,thei1}
\begin{equation}
{\cal L} = -{\frac{1}{2}}{\cal R} + {\frac{1}{4}}{\frac{1}{g_{grav}}
{\cal C} + {\frac{1}{4}}{(\Re S)}}GB + {\frac{1}{4}}{(\Im
S)}{\cal R}_{abcd}{\cal R}^{abcd}
\end{equation}
and ${\Re S} \equiv (S + {\bar S})/2$, $\Im S  \equiv (S-{\bar S})/2 $.
We have used as the gravitational couplings 
$1/{g_{grav}}\; \equiv \;{\Re S}$ , while GB is the Gauss-Bonnet 
combination
\begin{eqnarray}
4GB = {{\cal C}^{2}}-2{\cal R}_{ab}^{2} +{\frac{2}{3}}{{\cal R}^{2}},
&{\cal C} = {\cal R}_{abcd}{\cal R}^{abcd} -2{\cal R}_{ab}
{\cal R}^{ab}+\frac{1}{3}{\cal R}^2,
\label{gauss1}
\end{eqnarray}
and ${\cal C}$ the conformal Weyl tensor ${\cal C}_{abcd}$.
The Weyl tensor C can occur only through the Gauss-Bonnet (GB)
combination in eqn.(\ref{gauss1}), since single $C^2$ terms coupled to
Einstein gravity can violate\cite{zwie1} unitarity. In other words,
presence of powers of GB terms may quarantee the absence of
ghost particles in the effective low energy limit of string theories.
When the above relation is written in the form 
\begin{equation}
 L_{grav} = {{\triangle}^{grav}}(T,{\bar T})({\cal R}^{2}_{abcd} -
4 {\cal R}^{2}_{ab} + {\cal R}^{2}) + \nonumber\\
{\Theta}^{grav}(T,{\bar T})
{\epsilon}^{abcd} {\cal R}_{abef} {\cal R}_{cd}^{ef}
\end{equation}
then the one-loop corrections \cite{anto2} to the gravitational 
action, in the absence of Green-Schwarz mechanism, give
${{\triangle}^{grav}} = {\frac{{\tilde \beta}^{grav}}{32{\pi}^{2}}} 
\log(T + {\bar T})|\eta(iT)|^{4}$, e.g for a $Z_4$ orbifold.
The graviattional $\beta$-function coefficient 
${\tilde \beta}^{grav}$ equals 
\begin{equation}
{\tilde \beta}^{grav}=\frac{1}{45}(N_S +\frac{7}{4}-13N_V -\frac{113}{2}
N_F^L N_V^R +304N_V^L N_V^R).
\label{gravico1}
\end{equation}
Here, $N_S,N_V,N_F$ is the number of scalars, vectors and 
Majorana fermions contributing to the gravitational
$\beta$-function. The coefficients in front of
$N_S,N_V,N_F$ represent the contributions\cite{duff1}
of the various
fields to the  
integrated conformal anomaly. This is 1 for scalars,
$7/4$ for spin $1/2$ fermions, 33 for vector bosons and $-233/4$ 
for gravitinos and 212 for the graviton.
Finally, 304 is the contribution in the trace anomaly for the 
graviton, dilaton and antisymmetric tensor $B_{\mu \nu}$, while
$N_F^L N_V^R$ is associated with the contribution of
the gravitino together with a Majorana fermion.
 
The corrections to the gravitational couplings considered up to now
in the literature, are concerned with the decomposable orbifolds.
We will complete the discussion 
of corrections to the gravitational couplings by examining
non-decomposable
orbifolds. We focus our attention to the case of ${Z_6}-IIb$ 
orbifold. We consider the case of vanishing Wilson lines.
Working in the field theoretical approach \cite{caluo},
in the presence\cite{lust} of the threshold 
${p^2} \ll {M_{I}^{2}} \ll {M_{string}^{2}}$,  we get 
\begin{eqnarray}
\frac{1}{g_{grav}^2 (p^2)}= \frac{S+{\bar S}}{2}+\frac{{\hat b}_{grav}}
{16\pi^2} \log\frac{M_{I}^2}{p^2}-\frac{b_{grav}}{16\pi^2}
\log\frac{M_{I}^2}{M_{string}^2}- \frac{{\hat a}_{grav}}{16\pi^2}
\log[(T+{\bar T})(U^{\prime}+{\bar U}^{\prime})]&\nonumber\\
- \frac{a_{grav}}{16\pi^2}\log[{\eta^{2}(T)}
{\eta^{2}(\frac{U^{\prime}}{3})}{9}] 
-\frac{a_{grav}}{16 \pi^2}\log[\eta^{2}(\frac{T}{3})
\eta^{2}(U^{\prime})9].&\nonumber\\
\label{tyui}
\end{eqnarray}
Let me explain more on eqn.(\ref{tyui}).
Here, ${\hat b}_{grav},\;{\hat a}_{grav}$ are associated with the 
running of the gravitational couplings below the additional 
threshold scale
$M_I^2$, while the bare quantities, e.g $b_{grav},\;a_{grav}$ are 
associated with the running in the area $M_I^2 \ll p^2 \ll M_{string}$.
The following equation is valid for energies $p^2 \ll M_I^2$.
\begin{equation}
\frac{1}{g_{grav}^2(p^2)} = \frac{1}{g_{grav}^2(M_I^2)}+
\frac{{\hat b}_{grav}}{16\pi^2} \log \frac{M_I^2}{p^2}
 -{\hat a}_{grav}\frac{{\tilde \triangle}_{grav}}{16\pi^2},
\label{katoura1}
\end{equation}
where ${\hat a}_{grav}$ is given below and ${\tilde \triangle}_{grav}$
is the moduli dependent contribution from the K\"ahler and $\sigma$-model 
anomalies.
For decomposable orbifolds, this contribution\cite{anto2}
is the usual one as for the gauge coupling\cite{dkl2}.
For energies $M_I^2 \ll p^2 \ll M_{string}^2$ the following
equation is valid\cite{lust,caluo}
\begin{equation}
\frac{1}{g_{grav}^2(M_I^2)} =\frac{S+{\bar S}}{2}+
\frac{b_{grav}}{16\pi^2}\log\frac{M_{string}^2}{M_I^2} -
a_{grav} \frac{\tilde 
\triangle}{16\pi^2},
\label{katoura2}
\end{equation}
where $\tilde \triangle$ is the moduli dependent contribution coming 
from the integration of massive modes.
Substituting eqn.(\ref{katoura2}) in eqn.(\ref{katoura1}) we get
\begin{eqnarray}
&{\frac{1}{g_{grav}^2(p^2)}} =\frac{S + {\bar S}}{2}
+ \frac{{\hat b}_{grav}}{16\pi^2} \log \frac{M^2_{string}}{p^2}
+\frac{[-{\hat b}_{grav} +b_{grav}]}{16\pi^2}
\log|\omega(T)-\omega(U^{\prime})|^2 &\nonumber\\
&-K_1\frac{a_{grav}}{16\pi^2}\log[|9\eta(T)\eta 
\frac{(U^{\prime})}{3}]^{4}-K_2\frac{a_{grav}}{16\pi^2} 
\log[9\eta^{2}(U^{\prime})\eta^{2}(\frac{T}{3})]^{4} &\nonumber\\
&-K_1\frac{{\hat a}_{grav}}{16\pi^2}\left((T+{\bar T})(U^{\prime}+
{\bar U}^{\prime})\right)-K_2\frac{{\hat a}_{grav}}{16\pi^2}
\left((T+{\bar T})(U^{\prime}+{\bar U}^{\prime})\right).
\label{gau1}
\end{eqnarray}
The coefficients $K_1,K_2$ are numerical coefficients that may appear in
front of the moduli dependent threshold corrections after
performing the string calculation\cite{anto2} of moduli dependence of
threshold
corrections in analogy to the calculation of the threshold corrections
to gauge couplings in \cite{deko}.

Here ${\hat a}_{grav}$ comes from non-holomorphic contributions from
K\"{a}hler and ${\sigma}${-model anomalies and is given by
$\hat{\alpha}_{grav} = \frac{1}{24}
(21 + 1 - \dim G + \gamma_M + {\sum}_{\hat C}
(1+2n_{\hat C}))$.
The ${\hat a}_{grav},g_{grav}$ coefficients have been calculated in the 
absence of\cite{anto2} Green-Schwarz 
mechanism, as coefficients of the Gauss-Bonnet
term in the gravitational action and represent the contribution
of the completely rotated plane. In that case, 
${\hat a}_{grav} \equiv a_{grav}= {\tilde \beta}_{grav}$.

The gauge couplings receive contributions from
states appearing at points in the moduli space where $T=U^{\prime}$,
namely, one additional $N=2$vector multiplet, namely at the 
additional threshold scale $M_I=\left(\omega(T)-\omega(U^{\prime})
\newline
\right)M_{string}$. Furthermore, $b_{grav} - {{\hat b}_{grav}} =
{\delta^{V}}_{grav} +{\delta^C}_{grav}$, where ${\delta^{V}}$ and 
${\delta^C}$ are the contributions
to the gravitational $\beta$-function coming from the N=1 vector and
chiral decompositions of the $N=2$ vector multiplet.
A $N=2$ vector multiplet of a supersymmetric gauge theory has in total 
$2^{2+1}=8$ states and
consists two vectors, two complex Majorana fermions and a complex scalar.
It can decomposed into a $N=1$ vector multiplet which has two vectors
plus their fermionic superpartners and a $N=1$ hypermultiplet with
two Majorana fermions plus their superpartner, a complex scalar. 
  
Rewriting eqn.(\ref{gau1}), we obtain
\begin{eqnarray}
&{\frac{1}{g_{grav}^2(p^2)}} ={\frac{S + {\bar S}}{2}}
+ {\frac{\hat{b}_{grav}}{16 \pi^2}} \log {\frac{M^2_{string}}{p^2}}
- {\frac{{\delta^{V}}_{grav} + {\delta^C}_{grav}}{16 \pi^2}}
\log |\omega(T)-\omega(U^{\prime})|^2 &\nonumber\\
&-K_1\frac{{\hat a}_{grav}}{16\pi^2}\log[|9\eta(T)\eta
\frac{(U^{\prime})}{3}]^{4}-K_2\frac{{\hat a}_{grav}}{16\pi^2}
\log[9\eta^{2}(U^{\prime})\eta^{2}(\frac{T}{3})]^{4} &\nonumber\\
&-K_1\frac{ a_{grav}}{16\pi^2}\log\left((T+{\bar T})(U^{\prime}+
{\bar U}^{\prime})\right) -K_2\frac{{ a}_{grav}}{16\pi^2}
\log \left((T+{\bar T})(U^{\prime}+{\bar U}^{\prime})\right).
\label{gau2}
\end{eqnarray}
Finally,
\begin{eqnarray}
{\frac{1}{g_{grav}^2(p^2)}} ={\frac{S + {\bar S}}{2}}
+ {\frac{\hat{b}_{grav}}{16 \pi^2}} \log {\frac{M^2_{string}}{p^2}}
- {\frac{{\delta^{V}}_{grav} + {\delta^C}_{grav}}{16 \pi^2}}
\log |\omega(T)-\omega(U^{\prime})|^2 &\nonumber\\
-K_1\frac{a_{grav}}{16\pi^2}\log\{(T+{\bar T})(U^{\prime}+
{\bar U}^{\prime})|9\eta(T)\eta\frac{(U^{\prime})}{3}]^{4}\}
&\nonumber\\
-K_2\frac{a_{grav}}{16\pi^2}\log\{(T+{\bar T})(U^{\prime}+
{\bar U}^{\prime})9\eta^{2}(U^{\prime}) \eta^{2}(\frac{T}{3})\}.
\label{gau3}
\end{eqnarray}

% ********************************************************************
%*********************************************************************
%*********************************************************************
%*********************************************************************
%*********************************************************************
\subsection{ * Threshold corrections for the $Z_{8}$ orbifold}
\label{z4}

We will present now the calculation of the threshold corrections for 
the 
class of orbifolds defined by the Coxeter twist 
$(e^{\frac{i\pi}{4}},e^{\frac{3i\pi}{4}},-1)$
on the root lattice of 
$A_3 \times A_3$ .
This orbifold is non-decomposable, in the sense that the action of
the lattice twist on the $T_6$ torus does not decompose into the
orthogonal sum $T_6 = T_4 \oplus T_4$ with the fixed plane lying 
on the $T_2$ torus. 
Similar calculations\footnote{of threshold corrections for 
non-decomposable orbifolds} have been performed in 
\cite{deko}. 
Our calculation completes the calculation of the threshold
corrections 
for the list of 
orbifolds defined\footnote{In \cite{erkl} a classification of
six orbifold compactifications with $N=1$ supersymmetry was performed.
Similar calculations for non-decomposable $Z_N \times Z_M$ orbifolds 
were examined in \cite{blsth}. }
in \cite{erkl}.

The twist can equivalently be realized 
through the generalised Coxeter automorphism 
$S_{1} S_{2} S_{3} P_{35} P_{36} P_{45}$. 
The generalized Coxeter automorphism is defined as a product 
of the Weyl reflections\footnote{The Weyl reflection $ S_{i} $ 
is defined as a reflection 
\begin{equation}  
S_{i}(x) = x - 2 {\frac{<x,e_{i}>}{<e_{i},e_{i}>}}e_{i}
\label{coxe1}
\end{equation}
with respect to the hyperlane perpendicular to
the simple root.}
$ S_{i} $ of the simple roots and the outer 
automorphisms represented by the transposition 
of the roots. A outer automorphism represented by
a transposition which exchange the  roots $i \leftrightarrow j$,
is denoted by  $P_{ij}$ and is a symmetry of the Dynkin diagram.

For the orbifold ${Z_8}$  there are four complex moduli fields.
There are three 
$(1,1)$ moduli due to the three 
untwisted generations $27$ 
and one $(2,1)$-modulus\footnote{In
Table five of \cite{erkl}, the number of the $h^{(2,1)}$ moduli
was reported for the $Z_8$ and $Z^{\prime}_8$  orbifolds to be
zero and one respectively. They were missquoted. 
Clearly, these values may be interchanged. In our calculation for
the $Z_8$ orbifold with the twist defined via 
the generalized Coxeter twist $S_{1} S_{2} S_{3} P_{35} P_{36} P_{45}$
the value of the $h^{(2,1)}$ moduli is one, confirming the results
of \cite{dkl1}.}
due to the one untwisted generation ${\bar 27}$.

The realization of the point group is generated by 
\begin{equation}
Q=\left(\begin{array}{ccccccccr}
0&0&0&0&0&-1 \\
1&0&0&0&0&0 \\
0&1&0&0&0&-1 \\
0&0&1&0&0&0 \\
0&0&0&1&0&-1 \\
0&0&0&0&1&0
\end{array}\right). 
\end{equation}
Therefore the metric $g$ 
(defined by $g_{ij}=<e_i|e_j>$) has three and the antisymmetric
tensor field $b$ an other three real deformations. The equations
$gQ=Qg$ and $bQ=Qb$
determine the background
fields in terms of the independent deformation parameters. If the
action of the  generator of the point group leaves some complex plane 
invariant then the corresponding threshold corrections have to depend 
on
the associated moduli of the unrotated complex plane.  

Solving
these equations one obtains for the metric

\begin{equation}
G=\left(\begin{array}{llccrr}
R^2&u&v&-u&-2v-R^2&-u\\
u&R^2&u&v&-u&-2v-R^2\\
v&u&R^2&u&v&-u\\
-u&v&u&R^2&u&v\\
-2v-R^2&-u&v&u&R^2&u\\
-u&-2v-R^2&-u&v&u&R^2
\end{array} \right)\;,
\end{equation}\ 
with $R,u,v \in \Re$ and the antisymmetric tensor field :
\begin{equation}
B=\left(\begin{array}{ccccccccr}
0&x&z&y&0&-y\\
-x&0&x&z&y&0\\
-z&-x&0&x&z&y\\
-y&-z&-x&0&x&z\\
0&-y&-z&-x&0&x\\
y&0&-y&-z&-x&0 
\end{array}\right)\;,
\end{equation} 
with $x,y, z \in \Re $.
The N=2 orbit is given by these sectors which contain completely
unrotated planes,
${\cal O} ={(1,\Theta^4),(\Theta^4,1),(\Theta^4,\Theta^4)}$.

The element
$(\Theta^4,1)$ can be obtained from the {\em fundamental element}
$(1,\Theta^4)$
by an $S$--transformation on $\tau$ and
similarly $(\Theta^4,\Theta^4)$ by an $ST$--transformation.
The partition functions for the zero mode parts $Z_{(g,h)}^{torus}$ 
of the fixed plane have the following form\cite{erkl}
\begin{eqnarray}
Z_{(1,\Theta^4)}^{torus}(\tau ,\bar \tau,G,B) &=& \displaystyle
\mathop{\sum_{P \in( \Lambda_{N^{\bot})}} 
q^{\frac{1}{2} {P_L}^{2}} \bar q^{{P_R}^{2}}} 
\nonumber \\
Z_{(\Theta^4,1)}^{torus}(\tau ,\bar \tau,G,B) &=& \displaystyle
\mathop{\frac{1} {V_{\Lambda_{N}^{\bot}}} 
\sum_{P \in ({\Lambda}_{N}^{\bot})^{\ast}}
q^{\frac{1}{2} {P_L}^2} \bar q^{\frac{1}{2}{P_R}^{2}}}
\nonumber \\
Z_{(\Theta^4,\Theta^4)}^{torus}(\tau ,\bar \tau,G,B)& =&\displaystyle
\mathop{\frac{1} {V_{\Lambda_{N}^{\bot}}}
\sum_{P \in ({\Lambda}_{N}^{\bot})^{\ast}}
q^{\frac{1}{2} {P_L}^2} \bar q^{\frac{1}{2}{P_R}^{2}}}
q^{i \pi(P^2_L - P^2_R)}\;,
\label{partis1}
\end{eqnarray}
with $\Lambda_{N}^{\bot}$ we denote the Narain lattice of 
$A_3 \times A_3$
which has 
momentum vectors 

\begin{equation}
P_L = \displaystyle{\frac{p}{2}+(G-B)w\ ,} \hspace{1cm}
P_R = \displaystyle{\frac{p}{2}-(G+B)w\ .}
\label{zero}
\end{equation}
$\Lambda_{N^{\bot}}$ is that part of the lattice which remains fixed
 under
$Q^4$ and $V_{\Lambda^\bot_N}$ is the volume of this sublattice.
The lattice in our case is not self dual in contrast with the case of
partition functions
$Z_{(g,h)}^{torus}(\tau,\bar \tau,g,b)$ of \cite{dkl2} . 
Stated differently the general result is - for the case of orbifolds 
similar to our's - exactly, that the 
modular symmetry group is some subgroup of $\Gamma$ and as a  
consequence the partition functions
$\tau_2Z_{(g,h)}^{torus} (\tau,\bar \tau,g,b)$ are 
invariant under the same subgroup of $\Gamma$.  

The subspace corresponding to the lattice $\Lambda^\bot_N$
can be described by the following winding and momentum vectors,
respectively:

\begin{equation}
w=\left(\begin{array}{c} n^1\\n^2\\0\\0\\n^1\\n^2
\end{array}\right)\ ,\ \ n^1,n^2 \in Z\ \ \ {\rm and}\ \ \ 
p= \left(\begin{array}{c}
m_1\\m_2\\-m_1\\-m_2\\m_1\\m_2\end{array}\right)\,\; m_1,m_2 \in Z.
\label{vecz4}
\end{equation}

They are determined by the \footnote{By * we mean inverse transpose.}
equations $Q^{4}w=w$ and $Q^{*4}p=p$.
The partition function
$\tau_2 Z_{(1,\Theta^4)}^{torus}(\tau,\bar \tau,g,b)$ is invariant 
under the group
$\Gamma_0(2)$
which belongs to the congruence subgroups of 
$\Gamma$ \footnote{see appendix A}.
The integration of the contribution of the various
sectors $(g,h)$ is over the fundamental domain for 
the group $\Gamma_{0}(2)$ which is a three fold covering of
the upper complex plane.
By taking into account the values of the momentum and winding vectors
in the fixed directions we get for 
$Z_{(1,\Theta^4)}^{torus}$ 
\begin{eqnarray}
Z_{(1,\Theta^4)}^{torus}(\tau,\bar \tau,g,b)= 
{\sum_{(P_L,P_R) \in {\Lambda_{N}^{\bot}}} }
q^{\frac{1}{2} P_L^t G^{-1} P_L}
q^{\frac{1}{2} P_R^t G^{-1} P_R} & \nonumber\\
={\sum_{p,w} e^{2 \pi i \tau p^t w}}
e^{-\pi \tau_2 (\frac{1}{2}p^t G^{-1}p - 2p^t G^{-1} B w + 2 w^t G w - 
2 w^t B G^{-1} B w - 2 p^tw)}&\;.
\label{general}
\end{eqnarray}

Consider now the the following parametrization of the torus
$T^{2}$\cite{dis}, namely define the 
the $(1,1)$  $T$ modulus  and
the $(2,1)$ $U$ modulus as:
\begin{equation}
\begin{array}{ccccl}
T&=&T_1+iT_2 &=&2(b+i \sqrt{\det G_\bot})\\ 
U&=&U_1+iU_2&=&\frac{1}{G_{\bot 11}}(G_{\bot 12}+i\sqrt{\det G_\bot}).
\end{array}
\end{equation}
Here $g_\bot$ is uniquely determined by
 ${w^t} G w =(n^1 n^2) {G_\bot} \left({n^1 \atop n^2 }\right)$.
 
Here b is the value of the $ B_{12}$ element of the two-dimensional matrix
B of the antisymmetric field.
This way one gets
\begin{eqnarray}
T & = & 4(x - y) + i 8 v\\
U & = & {i}\;.
\label{comple1}
\end{eqnarray}

The partition function
$Z_{(1,\Theta^4)}^{torus}(\tau,\bar \tau,g,b)$ takes now the form
\begin{equation}
Z_{(1,\Theta^4)}^{torus}(\tau,\bar \tau,T,U)= 
\sum_{m_1,m_2 \in 2Z \atop n^1,n^2 \in Z} e^{2 \pi i \tau (m_1
n^1 + m_2 n^2)} e^{\frac{-\pi \tau_2}{T_2U_2} |T U n^2+T
n^1-Um_1+m_2|^2}\ .
\label{partition}
\end{equation}

By Poisson resummation on $m_1$ and $m_2$,using the identity:
\begin{equation}
\displaystyle\mathop{
\sum_{p \in \Lambda \ast} e^{[- \pi(p+\delta)^{t}C(p+\delta)]
+2\pi i p^t \phi]}={V}_{\Lambda}^{-1}{\frac{1}{\sqrt det{C}} 
{\sum}_{l\in \Lambda}e^{[-\pi(l+\phi)^{t} C^{-1}(l+\phi) -
2 \pi i {\delta}^t (l+\phi)]}}\;\;\;\;\;\;\;\;\;\;\;\;}
\label{karo}
\end{equation}
we conclude
\begin{equation}
\begin{array}{lrlcr}
&\tau_2 \ Z_{(1,\Theta^4)}^{torus}(\tau,\bar \tau,T,U) &=& 
{{\frac{1}{4}} {\sum_{A \in \cal M}} e^{- 2 \pi i T \det A}\ T_2
e^{\frac{-\pi T_2}{\tau_2 U_2} \left|(1,U)A \left(\tau \atop 1\right)
\right|^2}\ ,}
\end{array}\end{equation}
where
\begin{eqnarray}
\cal M& =& \left(
\begin{array}{cc}
n_1&\frac{1}{2} l_1\\[1mm]
n_2&\frac{1}{2} \l_2 \end{array} \right) 
\end{eqnarray}
and $n_1,n_2, l_1,l_2  \in Z$.

From (3.57)  one can obtain $\tau_2Z_{(\Theta^4,1)}^{torus}
(\tau,\bar \tau)$
by
an $S$--transformation on $\tau$. After exchanging $n_i$ and $l_i$
and performing again a Poisson resummation on $l_i$ one obtains

\begin{equation}
\displaystyle\mathop{
Z_{(\Theta^4,1)}^{torus}(\tau,\bar \tau,T,U) = {\frac{1}{4}}
{\sum_{\stackrel{m_1,m_2 \in Z}{  n^1,n^2 \in Z}}} 
e^{2 \pi i \tau (m_1 \frac{n^1}{2} + m_2 \frac{n^2}{2})} 
e^{{\frac{-\pi\tau_2}{T_2U_2}}|T U{\frac{n^2}{2}}+T{\frac{n^1}{2}}-
U m_1 + m_2|^2}}
\label{pa}
\end{equation}
The factor $4$ is identified with the volume 
of the invariant sublattice in $(3.58)$.
The expression $\tau_2 Z_{(\Theta^4,1)}^{torus}(\tau,\bar \tau,T,U)$ is 
invariant under $\Gamma^0(2)$ acting on $\tau$ and is
identical to that for the $(\Theta^4,\Theta^4)$ sector.
The subgroup $\Gamma^0(2)$ of $SL(2, \bf Z)$ is defined
as with $b=0\ \bmod 2$ instead of $c=0 \bmod\ 2$.
Thus the contribution of the two
sectors $(\Theta^4,1)$ and $(\Theta^4,\Theta^4)$ to the coefficient
$b_a^{N=2}$ of the $\beta$--function is one fourth of that of the
sector $(1,\Theta^4)$.

The coefficient $b^{N=2}_a$ is the contribution to the $\beta$ functions
of the $N=2$ sectors of the $N=1$ orbifold as we already described in
the introduction of this chapter.

Including the moduli dependence of the different sectors, we conclude
that the final 
result for
the threshold correction to the inverse gauge coupling (3.3)
reads
\begin{eqnarray}
{\triangle_{a}(T,\bar T,U,\bar U)} = -b^{(1,\Theta^4)}_a &
\ln|{\frac{8\pi e^{1-\gamma_E}}{3 \sqrt{3}} }
{\frac{T_2}{4}} |\eta\left({\frac{T}{2}}\right)|^4 
U_2 |\eta(\left( U \right)|^4 |&  \nonumber \\
-  \frac{1}{2} b_a^{(1,\Theta^4)}& 
\ln\left(\frac{8 \pi e^{1-\gamma_E}}{3 \sqrt{3}} 
T_2 | \eta\left(\frac{T}{2}\right)|^4 U_2 |\eta(U)|^4\right)   
\label{rez4}
\end{eqnarray}

The value of $U_2$ is fixed and equal to one as can be easily seen from 
eqn.(\ref{comple1}). 
In general for $Z_N$ orbifolds with $ N \geq 2$ 
the value of the U modulus is fixed.
The final duality symmetry of (\ref{rez4}) is ${\Gamma^{0}_{T}(2)}
\times \Gamma_{U}$ with the value of U replaced with the constant
value i.

In the appendix A, we list details of the integration.

\newpage
\vspace{5cm}
\begin{center}
{\bf  CHAPTER 4}
\end{center}
\newpage

\section{\bf  Introduction}

Inspired from the progress\cite{seiwi1,club}
on the rigid supersymmetric Yang-Mills 
theories, recently much progress has been made towards 
understanding
non-perturbative effects in string theory\cite{kava,fhsv}.
At the level of $N=2$ supersymmetric $SU(r+1)$ Yang-Mills the
quantum moduli space was associated with a particular genus r 
Riemann surface parametrized by r complex moduli and 2r periods 
$(\alpha_{D_{i}},\alpha)$\footnote{This development was subsequently 
generalized \cite{club} for arbitrary $SU(n)$ gauge groups.}, 
while their effective theories up to two derivatives is encoded
in the following $N=2$ effective supersymmetric lagrangian of r 
abelian 
$N=2$ vector multiplets in the adjoint representation of the 
gauge group
\begin{eqnarray}
{\cal L} = \frac{1}{4\pi} Im \{ \int d^4\theta
\frac{\partial {\cal F}(\alpha)}{{\partial {\cal A}}_i}
{\bar {\cal A}}_i + 
\int d^2\theta \frac{1}{2} 
\frac{\partial^2 {\cal F({\cal A})}}{{\partial{\cal A}_i}
 {\partial{\cal A}_j}} 
W_{\alpha\;i} W_{\alpha\;j} \},&\;i,j=1,\dots,r.
\label{eq1}
\end{eqnarray}
 
When matter is not present it allows for generic values 
of the scalar field of the theory to be broken down to the 
Cartran sub-algebra and it is described from r $N=2$ abelian vector 
supermultiplets, which can be decomposed into r $N=1$ chiral
superfields $\cal A$ and r $N=1$ vector superfields $W_{a}$ . 
The theory is dominated from the behaviour of the holomorphic 
function $\cal F(\cal A)$, namely the prepotential.
The supersymmetric non-linear $\sigma$-model is described
by the K\"ahler potential 
$K(A,{\bar A})=Im\{\frac{\partial {\cal F}(A)}{\partial A} {\bar A}\}$,
while the metric in its moduli space 
$\tau(A) = Im\{{\partial^2 {\cal F}/{\partial A}^2}\}$ is
connected to the complexified variable $\theta_{eff}/{\pi} + 
{8\pi i}(g_{eff}^{-2})\equiv \tau(A)$. The metric
is connected\footnote{$\alpha$ is the 
scalar component of the superfield $\cal A$ and can play the role 
of the Higgs field.}to the interpretation of the periods 
$\cal \pi$
\begin{eqnarray}        
{\cal \pi}=\left(\begin{array}{c}{\alpha}^{i}_D\\{\alpha^i}
\end{array}
\right),\;\;\;\;\alpha^i_D =\frac{\partial{\cal F}}{\partial
{\alpha}^i},\;i=1,\dots,r
\label{period1}                                                                                                      
\end{eqnarray}
as an appropriate family of a meromorphic one-forms associated
with $\lambda$, 
\begin{equation}
{\alpha}^i_D = \oint_{\alpha_i} \lambda,\;\;\;{\alpha}^i =
\oint_{\beta_i}\lambda,\;\;\tau =\frac{{d{\alpha}^i_D}/du}{
d{\alpha_i}/du}. 
\label{curve3} 
\end{equation}
Here, ${\alpha}_i, \beta_i$ form a basis\footnote{              
The cycles $\alpha, \beta$ form a basis of the first homology group 
$H_1(E_g, Z) = Z^{2g}$, where $E_g$ a Riemann surface at 
genus $g$. The intersection of cycles in the canonical basis 
takes the form $(a_i, b_j) = - (b_j,a_i)=\delta_{ij}$.}
of the homology cycles of the hyperelliptic  curve 
which has the same moduli space as $N=2$
supersymmetric Yang-Mills theory. 

For the $SU(2)$ group, the hyperelliptic curve $E_u$ is
\begin{equation}
y^2 = (x-1)(x+1)(x-u),
\label{curve1}
\end{equation}
and 
\begin{equation}
\tau \equiv \tau_{E_u}= \frac{\oint_{\alpha_1} \lambda^1}{\oint_{\beta_1}
\lambda^1},\;\;\lambda^1 = \frac{dx}{y}
\label{period2}
\end{equation} 
The theory possess singular points with non-trivial monodromies M.
The one-loop correction to the prepotential drives the local 
monodromy M around a given singularity and transforms the 
section
\begin{eqnarray}
\left(\begin{array}{c}
{\alpha}_D\\
\alpha
\end{array}\right)\rightarrow M
\left(\begin{array}{c}
{\alpha}_D\\
\alpha
\end{array}\right),\;M\;\in\;Sp(2r,R).
\label{period4}
\end{eqnarray}
The metric in the  moduli space
\begin{equation}
(ds)^2= Im \frac{\partial^2 {\cal F}}{\partial a_i \partial a_j}
da^i d{\bar a}^j = Im\; d a_D^i d{\bar a}_i
\label{periodd1}
\end{equation}
is invariant under the monodromy transformations M.
The one loop contribution to the holomorphic prepotential is due 
to the appearance of extra massless states at special points in 
the quantum moduli space. At a generic point of the moduli space 
the semiclassical monodromies split into non-perturbative
monodromies. The pure gauge theory has singularities at the 
points $\infty$, $\pm \Lambda^2$, where $\Lambda$ is 
the dynamically generated scale where the gauge coupling becomes 
strong. The contributions to the prepotential at $\infty$ 
correspond to weak coupling and the semiclassical monodromy M
at $\infty$
splits as $M^{\infty} = M^{+ {\Lambda}^2} M^{-{\Lambda}^2}$.

The theory yet 
possess stable dyonic points, labeled by magnetic $\nu_m$ and 
electric charges $\nu_e$, the so called BPS states.
The masses of the stable particles are given by the BPS 
formula $M^2=2|Z|^2=2|{\nu}_e \alpha + {\nu}_m \alpha_D |^2$.
At the strong coupling singularity $+{\Lambda}^2$ a magnetic 
monopole become massless with quantum numbers 
$({\nu}_{e}, {\nu}_m)= \pm (0,1)$. The
corresponding cycle vanishes. At $-\Lambda^2$ a dyon 
become massless, a particle with both electric and magnetic 
charges, with $({\nu}_{e}, {\nu}_m)= \pm (-2, 1)$.
For the classical $SU(n)$ gauge theory the moduli space of 
the theory is parametrized by the parameter u$=Tr(\alpha^2)$.
\newline
The parameter $\alpha$, the complex scalar of eqn.(\ref{eq1}), 
is in the adjoint representation of the gauge group.
For non-trivial u values, the gauge group is abelian. 
In the quantum theory the last picture is modified. The theory
takes into account the singularities in the moduli space.
The gauge group around the singularities is abelian and the 
non-abelian gauge symmetry is never restored. 
Note, that the classical point $\alpha=0$, where the gauge symmetry
is non-abelian is missing in the quantum theory. 
Instead, the weak coupling
point $\infty$ is available where perturbative calculations can be 
carried out\footnote{
For various checks os Seiberg-Witten theory using traditional style 
perturbation techniques see
\cite{trad}.}. 
The general picture emerging from the study of the 
supersymmetric Yang-Mills is that the vacuum expectation 
values of the Higgs fields break the theory to its maximal abelian 
subgroup.

In general, there are only five different descriptions of
string theories which all give consistent string vacua.
The closed type IIA and type IIB superstrings,
the open type I with
gauge group $SO(32)$ and the closed heterotic 
string with gauge group 
$SO(32)$ or $E_8 \times E_8$. For these vacua we can perform 
perturbative calculations. This means that
the any amplitude representing the scattering of n-particles 
can be expanded in the form\footnote{See for
example \cite{louisne}.}
\begin{equation}
{\hat {\cal A}}=\sum_{n=0}^{\infty} g_{string}^{-{\cal \xi}} 
{\hat {\cal A}}^n,
\label{amlplit}
\end{equation}
with $\cal \xi$ the Euler number\footnote{The Euler number is 
defined
as $(1/4\pi) \int d^2\sigma \sqrt{g} R^{(2)} ={\cal \xi}=-2 h -b +2$, 
where h is the number of loops, b the 
number of external legs, $g_{string}$ the string coupling constant
of the string vacuum.}and ${\cal A}^n$ the scattering 
amplitude in genus n Riemann surface.
Eqn.(\ref{amlplit}) represents the fact that that the n-point amplitude
as an expansion in the string coupling is equivalent to the sum over all
worldsheet topologies.
 The variable $g_{string}$ is 
equal to $e^{D}$. Here, D is the dilaton field of the bosonic part of 
the 
ten dimensional N=1 heterotic string in the string frame\cite{Gr1}
\begin{eqnarray}
{\cal L}=\sqrt{g}e^{-2D}\{-\frac{1}{2{\kappa}^2}R -\frac{1}{4}
\sum_a {k_a}(F_{\mu \nu} F^{\mu \nu})_a -\frac{2}{\kappa^2}\partial_m
D\partial^m D + \frac{1}{16 \kappa^2}H_m H^m&\nonumber\\
-G_{IJ}D_m \phi^I D^m {\bar \phi}^{\bar J} -V(\phi, {\bar \phi})\}.&
\label{korikos1}
\end{eqnarray}
We have assumed that the gauge group of the low energy theory is a
product of gauge group factors in
the form $G = \otimes  G_a$.
Here, $\kappa$ is the gravitational coupling, $k_a$ the Kac-Moody 
level
and the $H_m$ field strength contains Chern-Simons terms necessary
for the anomaly cancellation of gauge and gravitational anomalies in 
ten dimensions. 
Rescaling\cite{Wit} the space time metric $g_{\mu \nu}$ and changing 
variables as $S = e^{-2D} + i{\tilde a}  $, $\tilde a$ the axion,
in the lagrangian (\ref{korikos1}) 
we bring the Einstein term in its canonical form. 
Direct comparison with eqn.(\ref{qua6}) gives that in four-dimensions
$K = -{\kappa}^{-2} \ln(S+{\bar S})+ \hat{K} {(\phi,{\bar \phi})}$ 
and $G_{IJ} = \partial_{I} \partial_{\bar J} K$ 
and $\hat{K}$ a function of the scalars fields $\phi$.

Various equivalences between the different 
theories have been proposed and the picture emerging is that
the different string theories are expansions of a
more fundamental theory around different points in 
the moduli space of string vacua.
We mention string-string duality where type IIA compactified on
$K_3$ manifold with $N=4$ supersymmetry has the same moduli space as the
heterotic string
on a $T^4$ torus\cite{huto,sen1,var,hastro,asmo,aspo1} with $N=4$ 
supersymmetry.
For $N=2$ type IIB the string generalization of Seiberg-Witten's(SW)
quantum theory is provided by the conifold 
transitions\footnote{Moduli spaces of distinct Calabi-Yau(CY) manifolds 
touch each other along certain regions. These regions are called  
conifolds and they represent smooth manifolds apart from some singular 
points. In this sence CY's can form a web of 
connected\cite{stro2,cande,cande1} 
manifolds. The neighbourhood of the singular region is described from 
the quadric $\sum_{i=1,\dots,4}(Z^{i})^2 =0$, where $Z^i$ complex 
variables in $C^4$.}
of wrapped three-branes 
on Calabi-Yau\footnote{Type IIB theory admits\cite{Gal,Hol} soliton 
solutions in the NS and RR sector of the theory called p-branes.    
They are extended objects in a theory with p spatial dimensions.
They arise if there is $p+1$ form in the theory coupled to the 
$p+1$ dimensional world volume. 
For those p-branes associated with the RR sector, $p=-1,1,3,5,7$.
Recently, the dynamics of the RR sector p-branes was\cite{Pol} 
associated to objects referred as Dp-branes, extended objects with    
mixed boundary conditions(B.C) refering to as Neumann or Dirichlet at the 
worldsheet boundary, in the type I theory. Here D stands for
Dirichlet B.C.  Neumann B.C
exist in p-dimensions and Dirichlet one's in the remaining.
D-branes are the carriers of the RR charge, predicted from string-string
duality in six dimensions.}
spaces. 
Type IIB in ten dimensions admits extremal black holes  
solutions in the RR sector of the theory. They represent BPS saturated 
p-brane solitons.
Compactification of type IIB on a Calabi-Yau space produces 
$h^{(1,1)} + 1$ supermultiplet moduli with  +1 associated with the 
dilaton and 
$h^{(2,1)}$ vector multiplets and the graviphoton.
In addition, it gives the abelian gauge group $U(1)^{h_{(2,1)} + 1}$.
In general special geometry\footnote{See next section.}
applied to the compactification of type IIB on the 
Calabi-Yau space in four dimensions,
requires that the scalar component Z   
and the prepotential F of the vector multiplets to be given by
the period of the three form $\Omega$ over the canonical
homology cycles ${a_I}, b_I$ as 
\begin{eqnarray}
Z_I = \int_{a_I} \Omega,\;\;\;\frac{\partial F}{\partial Z_I}= 
\int_{b_I}\Omega,\;\;\;I=1,\dots,h^{(2,1)}.
\label{kotera1}
\end{eqnarray}
Here, $\Omega$ is the holomorphic three form describing the 
complex structure of the Calabi-Yau space.
BPS states are $\propto |-\hat{\nu}_e^I Z_I + \hat{\nu}_m^I
F_I|$. The integers $\hat{\nu}_e, \hat{\nu}_m$ are the 
electric and magnetic charges of the threebrane wrapped around
the three surfaces ${a_I}, b_I$.
The appearance of a logarithmic singularity in the 
K\"{a}hler metric at the conifold point $Z{=}0$, involved in 
the compactification of type IIB on the Calabi-Yau space, is then 
identified\cite{stro1} with the extremal three brane black hole 
becoming massless.  In analogy with SW theory, 
the three-brane becomes massless when the associated cycles 
vanish. 
The appearance of the singularity, when the corresponding
3-cycles along the 3-surfaces vanish, is then identified with
the existence\cite{stro1} of a massless black hole solution in the
metric of type IIB for the 3-brane.
This is the analog of   
Seiberg-Witten appearance of the massless monopole singularity.

Among the various equivalences between the different
perturbative string theories, we mention
the S-duality conjecture\cite{SDUAL1} for which more details will be 
given in chapter five.

In this chapter we are interested only in the proposal 
of \cite{kava,fhsv} which
provided evidence for the exact nonperturbative equivalence
of the 
heterotic string compactified on $K_3 \times T_2$, with 
IIA compactified on a Calabi-Yau threefold.
The proposal identifies the moduli spaces of heterotic string and its 
dual IIA as ${\cal M}_V^{heterotic}= 
{\cal M}_V^{IIA}$ and ${\cal M}_H^{heterotic}={\cal M}_H^{IIA}$, where
the subscripts refer to vector multiplets and hypermultiplets 
respectively. 
In this sence the complete prepotentials for the vector
multiplets for the two "different" theories match, including perturbative
and non-perturbative corrections, and ${\cal F}^{het}= {\cal F}^{IIA}$.
Compactification
of the heterotic string on $K_3 \times T_2$ and of type IIA
on a Calabi-Yau threefold produce models
with N=2 supersymmetry in four dimensions.

Lets us assume that we have a heterotic string compactified 
on the $K_3 \times T^2$ manifold for which the dual type IIA 
compactified on a Calabi-Yau threefold exists.
Then the following 
non-renormalization\cite{stro1} theorem holds:

\begin{quote}
\em
Since in $N=2$ heterotic strings the dilaton is part of the vector 
multiplet, in reality 
of the vector-tensor multiplet as we will see in the next section,   
the prepotential of the vector multiplets for the heterotic model
is getting corrected beyond tree level of perturbation theory while 
that the hypermultiplet superpotential is exact and equal to the tree 
level result.
For the heterotic dual realization in type IIA, the prepotential of the 
vector multiplets at tree level is exact, while the hypermultiplet 
superpotential is getting corrected beyond tree level.
Since there is no coupling between vector multiplets and 
supermultiplets\cite{DWLVP} at the perturbative level\footnote{At the
level of effective $N=2$ supergravity, the vector multiplets are 
coordinates on a special K\"ahler manifold ${\cal M}_V$ while the
hypermultiplets parametrize\cite{DWLVP} a quaternionic 
manifold ${\cal M}_H$.}, we can   
extend\cite{stro1} this argument at the non-perturbative level
and conclude the following :   
The exact vector multiplet prepotential for 
a heterotic model, which has a type II dual, can be derived by
the calculation of the tree level vector
multiplet prepotential of the type IIA side. The exact supermultiplet
prepotential
for the heterotic side is equal to its tree level result.
\end{quote}
The heterotic model receives perturbative
and non-perturbative corrections to its prepotential of the vector 
multiplets in the heterotic side. 
In this chapter, we will calculate the one loop correction to the  
perturbative prepotential of the vector multiplets
for the heterotic string compactified on a six dimensional orbifold.
It comes from the solution of a partial differential equation.
The one loop correction to the perturbative prepotential has already
been calculated before in \cite{afgnt} from string amplitudes. 
Our procedure is complementary to \cite{afgnt} since we calculate, 
contrary to \cite{afgnt} where the third derivative of the prepotential 
with respect to the T moduli was calculated, the third derivative of the 
prepotential with respect to the U moduli. 
Furthermore, we {\em establish a general procedure} for calculating
one loop corrections to the one loop prepotential, not only
for heterotic strings compactified on six dimensional orbifolds,
which has important implications for any compactification of the 
heterotic string having(or not) a type II dual. This procedure is an 
alternative way to the calculation of the prepotential which was 
performed in \cite{mouha}. However, the procedure in \cite{mouha} 
seems to us more complicated.

In addition, it was further proposed \cite{klm1} that the  
existence of heterotic-type IIA duals\cite{kava}
can be traced back to the $K3$ fiber structure or the 
elliptic fibration\footnote{These structures were further elaborated
at \cite{fava1,lla,lla1}.}.
In section 4.4, we will discuss the
issue of $K3$ fiber structure. Elliptic fibrations will be discussed
in section 4.5.

A general result\cite{fava1} concerning the geometry
behind the existence of heterotic duals, is that   
the Calabi-Yau manifold in the IIA side can be written instead, as
a fibre bundle with base $P^1$\footnote{
$P^1$ is the complex projective space with homogeneous coordinates
$[x_o, x_1]$.} generic fiber the $K_3$ surface.
The previous result was further elaborated in \cite{lla}. 
Existence of the IIA dual in the Calabi-Yau threefold phase
with the dual heterotic string admitting a weekly coupled phase while
the dual type IIA realization is in the strongly coupled phase, was 
proved that can happen only when\cite{fava1,lla} the generic fibre is 
the $K_3$ surface and the base is $P^1$. 

At this chapter we will continue the 
work of\cite{afgnt,anto3,wkll,afgnt}.
We will calculate the
prepotential of $N=2$ vector multiplets of heterotic string when
the T,U moduli subspace exhibits an $SL(2,Z)_T \times SL(2,Z)_U 
\times Z_2^{T \leftrightarrow U}$
duality group\footnote{However, due to factorization
properties of the $T^2$ subspace of the heterotic Narain lattice,
the same result can be applied to any heterotic string 
compactification on $K_3 \times T^2$, with different 
embeddings of the $K_3$ instantons with instanton number k on the 
gauge group.}. 
The one loop K\"{a}hler metric in the moduli space of 
vector multiplets in $N=2$ six dimensional orbifold\cite{dhvw}
compactifications of the heterotic string follows directly
from this result.
Furthermore, we calculate the
prepotential of $N=2$ vector multiplets of heterotic string for 
the case of $N=2$ sectors
in $(2,2)$ symmetric non-decomposable Coxeter orbifolds. 

For the description 
of general properties of the basic theory of 
$N=2$ theories, we will follow closely in the beginning sections the 
work of \cite{wkll} while 
in the description of calculating the prepotential of vector 
multiplets 
from string amplitudes\footnote{We will draw heavily from these 
works.}
we will follow the work of \cite{afgnt,anto3}.
In section 4.2 we will describe general properties of $N=2$ heterotic 
strings. In addition, we describe properties of the moduli space of
compactification of 
the heterotic string on a $K_3 \times T_2$ manifold.
In section 4.3 we will discuss the special K\"ahler geometry 
describing
$N=2$ locally sypersymmetric theory of the heterotic string 
with emphasis on the couplings
of vector multiplets. 
In sction 4.4 We give descriptions of the low energy theory
for the classical 
and quantum theory for both the heterotic string compactified on
a six dimensional orbifold. We also discuss the
$K_3 \times T_2$ manifold and its dual type II on a Calabi-Yau 
3-fold.
In section $4.5$ we describe our results for compactification of
heterotic
string on a six dimensional internal manifold. Here, we assume that
the action of the lattice twist decomposes the torus in the form
$T^2 \oplus T^4$.
The calculation comes from the use of string amplitudes
of \cite{anto3,afgnt}.  We will calculate
one-loop
corrections to the K\"ahler metric for the moduli of 
the usual vector multiplet T, U moduli fields of the $T^2$ torus
appearing in $N=2$ heterotic 
strings compactified on orbifolds. The calculation
on the quantum moduli space takes into consideration points of 
enhanced gauge symmetry. 

The one-loop correction to the prepotential
for the vector multiplets then follows directly.
In section $4.5$, we will 
describe our results for the case of $N{=}2$ six dimensional 
orbifold 
compactification of the heterotic string, where the underlying 
torus lattice does not decompose as $T^2 \oplus T^4$.
The moduli of the unrotated complex plane has a modular symmetry 
group that is a subgroup of $SL(2,Z)$. In particular, we consider 
this modular symmetry
to be one of those appearing in non-decomposable orbifold 
compactifications.

%*****************************************************************
%*******************************************************************
%********************************************************************
%********************************************************************
%*******************************************************************

\subsection{Properties  of $N=2$ heterotic string and Calabi-Yau 
vacua}

A well known future of heterotic string vacua is the existence of 
their internal sector.
The asymmetric nature of the heterotic string can be made apparent 
by the left-right asymmetry of the world-sheet degrees of
freedom. In the critical dimension which is 10 for the heterotic 
and the type II,
the number of the  critical dimension comes from the
fact that the the central charges of the matter system and the ghost
system of the Virasoro algebra vanishes.
For a generic vacuum of the heterotic string the contribution to the 
central charge from the ghost degrees of freedom is $- 22$ for the
left side 
and $- 9$ for the right side. This has to be balanced from an 
appropriate internal CFT.
 
In general, for classical vacua of the heterotic 
string one
replaces the internal manifold with a $c=(9,9)$, $N=2$ 
superconformally
invariant theory on the world sheet together with four
$N=(0,1)$ free world-sheet superfields
that give finally rise to the four dimensional space-time. 
Furthermore, we are left with a $c=13$ trace anomaly
to the left moving sector which is saturated from free
bosons moving on a maximal torus 
$SO(10) \bigotimes E_8$ Kac-Moody algebra that is responsible 
for part 
of the gauge group. 
The list of $(2,2)$ vacua includes the Calabi-Yau
compactifications \cite{chsw}, orbifolds \cite{dhvw,imnq}, tensor
products of minimal models \cite{ge1} or generalizations \cite{kasu}.
We exclude the $(0,2)$ models\cite{cve} since no corresponding 
type II theory exists.

For abelian $(2,2)$ orbifolds constructed
by twisting a six dimensional torus, the point group rotation
is accompanied by a similar rotation in the gauge degrees of freedom.
The four dimensional gauge group in this case is enlarged beyond
$G= E_6 \otimes E_8$ by a factor that can be a $U(1)^{2},\;SU(2) 
\times U(1)$, if $P=Z_4$ or $Z_6$, or $SU(3)$ if $P=Z_3$.
If we symbolize by $h_{(1,1)}$ the number\footnote{For 
compactifications on Calabi-Yau manifolds, $h_{(1,1)}$ and $h_{(2,1)}$ 
represent the Hodge numbers of the manifold.}
of $(1,1)$ moduli in the
untwisted sector then we have respectively 
\newline 
$h_{(1,1)}= 3,5$ and $9$.
Twisted moduli are not neutral with respect to G and are not moduli 
of the orbifold.
Abelian $(2,2)$ orbifolds can flow to a Calabi-Yau vacuum, by
blowing up the twisted moduli fields, by giving them vacuum expectation
values\cite{HV}.

On the other hand compactification of the heterotic string 
on a six dimensional compact manifold can 
put restrictions on the allowed manifolds, which demend on the 
number of supersymmetries we want to preserve in four dimensions.
The supersymmetry transformations of the fermionic fields in  
ten dimensions give\cite{Calabi1,gsw2}, assuming that the 
supersummetry generator $\eta$ leaves the vacuum invariant,
\begin{equation}
\delta \psi_M = D_M\; \eta = 0
\label{preser1}
\end{equation}
\begin{equation}
\delta \chi^{\alpha}=\Gamma^{MN} 
F^{\alpha}_{MN}\; \eta = 0,\;\;\;\;M,N=1,\dots,10,
\label{preser2}
\end{equation}
where $\psi_M$ is the gravitino, $\eta$ is the supersymmetry
generator, $F_{MN}$ the gauge field strength of the gauge fields
and $D_M$ the covariant derivative. 
Eqn.(\ref{preser1}) represents the fact that the spinor $\eta$ is 
covariantly constant. 
For compactifications of the ten dimensional target space
in a manifold $M_4 \times K$, condition (\ref{preser1}) imply, 
via the integrability condition $[D_M,D_N]=0$, that
$M_4$ space is Minkowskian. Furthermore, 
these conditions are associated with the existence 
of the
compact manifold K to be a complex K\"{a}hler manifold.
This means that it admits a metric of U(N) holonomy. 
In local form, the metric comes from 
the equation 
$g_{ij}={\partial^2 K}/(\partial z^i \partial{\bar z}^j)$, where
K is the K\"{a}hler potential.
Holonomy\footnote{The holonomy group of the complex manifold 
is the group which leaves invariant the representation of the
Majorana-Weyl spinor $\eta$.}group
$SU(n)$ implies that in the four dimensional 
space-time we get $N=1$ supersymmetry and the complex manifold is a
Calabi-Yau n-fold. $SU(2)$ holonomy is associated with the four 
dimensional $K_3$, while $G_2$ holonomy and Spin(7) one with seven 
and eight dimensional compactification manifolds respectively.
Manifolds which satisfy the conformal invariance conditions admiting 
Ricci flat metrics are called Calabi-Yau manifolds\cite{Calabi1}.
From the mathematical point of view, Ricci-flatness is associated 
with the vanishing first Chern class\footnote{For mor   e on these issues 
see \cite{hub}. }of the tangent bundle of the 
manifold.
In the case of a six-dimensional orbifolds\cite{dhvw}, by "blowing up" 
the quotient singularities we recover the corresponding smooth 
Calabi-Yau manifold.

The $N=2$ superconformal algebra - after we expand the fields in modes 
as 
$T(z)=\sum L_n z^{-n-2}$,$T_{F}^{\pm}(z)=\sum G_r^{\pm} z^{-r-3/2}$  
and 
$J(z)=\sum J_n z^{-n-1}$ - takes the form ($G_r^+=(G_{-r}^-)^{\dag}$)
\begin{equation}
[L_{n},L_{m}]=(n-m)L_{n+m} + \frac{c}{12}(n^{3}-n){\delta}_{n+m,0} ,
\label{jki10}
\end{equation}
\begin{equation}
[G_{r}^{\pm},G_{s}^{\pm}]=2L_{r+s}{\pm}(r-s)J_{r+s}+{\frac{c}{3}}
(r^2-{\frac{1}{4}}) {\delta_{(r+s,0)}}\;,[G_r^{\pm},G_s^{\pm}]=0  .
\label{jki11}
\end{equation}
\begin{eqnarray}        
\left[L_n,G_r^\pm\right] = (\frac{n}{2}-r)G_{n+r}^{\pm}& , 
\left[L_n, J_m\right] = - m J_{n+m}&\\
\left[J_n,J_m\right ] = \frac{c}{3} n \delta_{m+n,0} &, 
\{J_n,G_r^\pm\}=\pm G^{\pm}_{n+r}&.\nonumber
\end{eqnarray}
is generated by four generators: one conformal weight(CW) 2 
energy momentum tensor $T$, one abelian current $J(z)$ and two
fermionic conformal
weight(CF) $3/2$ supercurrents $G^{\pm}(z)$ with a abelian 
charge $\pm$1. 
The current algebra  $J(z)$ corresponds to a free boson in
two dimensions 
$J(z) = \left(\frac{c}{3}\right)^{\; 1/2} \partial_{z} \phi$ 
and
$c$ represents the  
trace anomaly. 
If $r \in Z+ 1/2$, then the N=2 algebra describes the NS sector, while
if $r \in Z$ we are in the Ramond sector. Acting on a state $|\phi>$
with the generators $L_o$, $J_0$ and using the relations
$J_o |\phi> = q|\phi>,\;L_o|\phi>= h|\phi>$, where q the $U(1)$ charge 
and
h the conformal weight , we get the constraints $h \geq \frac{|q|}{2}$
for the NS sector and $h \geq \frac{c}{24}$ in the Ramond sector.
 The left moving $U(1)$ charge combines with the $SO(10)$ Kac-Moody 
 algebra
to accomplish the  left-moving gauge group enlargement group to 
the $E_6$. 
$N=1$ space-time supersymmetry requires $N=2$ supersymmetry on the 
worldsheet while an additional constraint comes from the condition 
that the right moving NS primary fields\footnote{
Take for example the bosonic
string. Assuming that the energy-momentum tensor has an expansion in
modes
as $T(Z)=\oint \frac{dz}{2\pi i} z^{n+1}T(Z)$,  we have
$[L_n,{\tilde \phi}(z)] \stackrel{def}{=}z^n [z \partial_z +(n+1)h]
{\tilde \phi}(z)$. The conformal weight h is defined via the operator
product expansion of the primary field with $T(Z)$.
namely
\begin{equation}
T(Z){\tilde \phi}(Z)\stackrel{def}{=}=\frac{{\tilde \phi}(w)}{(z-w)^2}
+ \frac{\partial_w {\tilde \phi}(w)}{(z-w)}.
\label{prima1}
\end{equation}
}
must have integral $U(1)$ 
charge \cite{BD}. The last statement is really the demand that the 
operator product of the gravitino vertex operator 
\begin{equation}
\psi_{\alpha}^{i}(z) = e^{-{\varphi}/2}S^{\alpha} e^{i\sqrt3/2 
\phi(z)} 
\end{equation}
with a physical state to be local.
The ghost field is $\varphi$, $ S $ is the $SO(4)$ space-time spin 
field 
and the exponential comes from the dependence from the internal sector,
the Ramond ground states with conformal weight $3/8$.
If we have N-extended space-time supersymmetries then we have $N$ 
supercharges 
$Q_\alpha^i$ ($i=1,\dots , N$) 
which obey
\begin{eqnarray}
\{ Q_\alpha^i,\bar Q_{\dot\beta j}\} & =2i\delta^i_j
\gamma^{\mu}_{\alpha {\dot \beta}}\;P_{\mu},\;&
\{Q_{\alpha}^i,Q_{\beta}^j \} =2C_{\alpha \beta}\,Z^{ij}.
\label{quadru1}
\end{eqnarray}
where $Z^{ij}$ denotes the central charges. The supersymmetry charges 
are defined in general as
\begin{equation}
Q_\alpha^i=\oint {{\rm d}z\over
2\pi i}V_\alpha^i(z)\,,\qquad \bar Q_{\dot\alpha i}=\oint
{{\rm d}z\over 2\pi i}  \bar V_{\dot\alpha i}(z)\,,
\end{equation}
where  $V_\alpha^i(z)$ and $\bar V_{\dot\alpha i}(z)$ are the vertex 
operators in the $-1/2$ ghost picture.

When the heterotic vacuum has $N=2$ space time supersymmetry the nature 
of
the supersymmetry algebra implies that the right moving algebra splits 
into a sector with $c=6$ and $N=4$ SCFT 
and a free SCFT field theory $c=3$ piece with $N=2$ supersymmetry. 
On the 
other hand on the the left moving side of the heterotic string there 
is no
world-sheet supersymmetry but instead we have a bosonic CFT, with 
a space-time sector consisting from four free bosons and an internal 
sector
with ${\bar c}$ = 22. This is in contrast with the type II models where 
there is $N=2$ world sheet supersymmetry in both sectors.

The massless spectrum of the $N{=}2$(space-time) $d{=}4$ heterotic 
string consists, 
among other particles, of the graviton
$G_{\mu \nu}$, the antisymmetric tensor $B_{\mu \nu}$ and the dilaton
which are all created from vertex operators of the form 
$\propto {\bar \partial X}_{\mu}(\bar z) {\partial X}_{\nu}(z)$ .  
In addition it contains two gravitini and two dilatini which are 
created 
from vertex operators of 
the form $ {\bar \partial X}_{\mu}(\bar z) V_{\alpha}^i(z)$ and   
the gauge boson generators of the group $U(1)^2_R$ with vertex 
operators
$ {\bar \partial X}_{\mu}(\bar z)\partial X^{\pm}(z)$.
In the supergravity multiplet there are included together with the 
graviphoton(the spin one gauge boson of the supermultiplet), the 
graviton and two
gravitini.
However, the dilaton is included in the vector-tensor 
multiplet\cite{wkll}. It contains
the dilaton, the antisymmetric tensor, the two dilatini 
and a $U(1)$ gauge field. The vector-tensor multiplet consists of a 
N=1 vector multiplet and a $N=1$ linear\cite{hiov} multiplet.
This on shell structure must be described as 
well 
from a $N=2$ vector multiplet via a duality transformation on the 
antisymmetric tensor field.
In the dual description, the antisymmetric tensor is converted to the 
axion $\tilde a$ via a supersymmetric duality transformation, and the 
dilaton D
and the axion combine to form the complex scalar $S = e^{\phi^{\prime}} 
+ i {\tilde a}$.
Off-shell the 8 + 8 component structure of the  vector-tensor multiplet
can be realized\cite{sost,cla} in the presence of the central charge.
In this case, after 
we linearize the Lagrangian for the vector-tensor multiplet we  
obtain \cite{wkll} :
\begin{equation}
{\cal L} = -{\textstyle{1\over 2}}(\partial {\phi^{\prime}})^2
-{\bar \lambda}^i {\partial \llap/} \lambda_i +
{\textstyle{1\over 2}}H^2 - {\textstyle{1\over 4}}F^2
+{\textstyle{1\over 2}}{\prime D}^2   \,.    
\label{vtaction}
\end{equation}
Here, $D^{\prime}$ is a real scalar auxiliary field, $\phi^{\prime}$
is the dilaton, F is the self-dual field strength of the 
gauge field, H the antisymmetric two
form antisymmetric tensor and $\lambda$ a doublet of Majorana spinors.
The action (\ref{vtaction}) has the same degrees of freedom as the 
action for a vector multiplet.
 
Other particles that are present at the massless spectrum of the 
heterotic 
string
include $N=2$ vector multiplets with gauge 
bosons $A_{\mu}^{\alpha}$, together with their superpartners, the 
two gauginos
${\lambda}_{i}$ and a complex scalar $C^{\alpha}$.
The vertex operator for a generic vector multiplet are
\begin{equation}
\left(A_{\mu}^a,{\lambda}^a_{i\;\alpha},{C}^a \right) \sim
\left( J^a({\bar z})\; {\partial X}_{\mu} (z),
J^{a}(\bar z)\, V_{i\, \alpha} (z),
J^{a}(\bar z)\; {\partial X}^{\pm} (z)\right)
\end{equation}
and the multiplet itself and the currents $J^{\alpha}$ transform 
in the adjoint representation of the non-abelian gauge group $G$ 
created from the
zero modes of the Kac-Moody currents. In fact we will see later that 
the scalars 
of vector multiplets in the moduli space can be divided to moduli and 
matter. The maximal rank for $G$ is 22.

Since  we will be describing compactifications of heterotic vacuua 
on a 2-torus $T_2$, it is necessary to give some details. 
The moduli of the torus is parametrized from
the relations $T = 2 ( B + i {\sqrt G}) $ and 
$ U = 1/G_{11}(G_{12} + i {\sqrt G})$ where $G_{ij}$ is the metric of 
the 
$T_2$, $\sqrt G$ is its determinant and $B$ the constant antisymmetric 
tensor background field.
At the classical level, the moduli space of orbifold compactification 
of the N=2 heterotic string compactified in a six dimensional
torus corresponds
to the coset space ${\frac{SO(2,2)}{SO(2) \times SO(2)}}|_{T,U}$. 
The same type of moduli appears when we compactify\cite{walt,mouha} the 
heterotic string on the manifold $K_3 \times T_2$.

The subspace of the vector moduli space corresponding to the 
T, U moduli is associated with the lattice 
$\Gamma^{2,2}$ part of the Narain lattice $\Gamma^{22,6}$ based 
on compactifications of even Lorentzian lattices of the heterotic 
string. 
The gauge group, using the standard embedding by equating the spin
connection with the gauge connection, gives at generic points in the 
moduli space of the torus, the gauge group 
$E_8 \times E_7 \times U(1)^4$. The first $U(1)^2$ is associated
with the moduli of the $T^2$ parametrizing the torus, while the other
$U(1)^2$ comes from the dilaton in the vector-tensor multiplet and the 
graviphoton. 
The heterotic prepotential for this model at the semiclassical limit
$S \rightarrow \infty$ has been calculated in \cite{wkll,afgnt,mouha}.
Various tests have been performed for the heterotic prepotential
with several Calabi-Yau IIA duals. In most of the tests, the weak 
coupling limit expansion of the heterotic prepotential
has been matched with the corresponding prepotential of the type
IIA side using\cite{grple} its mirror in type IIB. 
Mirror symmetry supports the existence of
for every Calabi-Yau manifold $\tilde E$, the mirror partner $E$ such 
that $h^{(1,1)}({\tilde E}) = h^{(2,1)}(E)$ and $h^{(1,1)}(E) = h^{(2,1)}
({\tilde E})$.
Orbifold compactifications of the heterotic string\cite{dhvw}, in four
dimensions on a six dimensional torus $T^2 \oplus T^4$,  
produces a $T^2$ subspace.
The classical moduli space, of r vector multiplets in the $T^2$
subspace, is the group 
\begin{equation}
\left. \frac{SU(1,1)}{U(1)}\right|_{\makebox{dilaton}} \left. \times 
\frac{O(2,r)}{O(2) \times O(r)}\right/ O(2,r;Z).
\label{quadru2}
\end{equation}
For our case, the classical duality group comes with $r=2$.
Here, $O(2,2;Z)$ is the target space duality group. The theory enjoys 
the non-trivial global invariance
i.e identifications  under target space duality symmetries
\cite{DUAL,dis,gg} the
$PSL(2, {\bf Z})_{T} \times PSL(2, {\bf Z})_{U}$ dualities acting as
\begin{equation}
T\rightarrow {aT-ib\over icT+d}\,,\qquad
U\rightarrow{a^{\prime}U-ib^{\prime}\over ic^{\prime}U+d^{\prime}}.
\label{tutransf}
\end{equation}
At special points in the moduli space of the torus the associated gauge
group becomes enhanced to $SO(4)$ or $SU(3)$ as we have already 
mentioned in the introduction. 
In the presence of Wilson line moduli, associated with the internal 
torus $T_2$ the modular group acts as
\begin{eqnarray}
T \rightarrow \frac{aT-ib}{ icT+d},&
U \rightarrow U - {ic}& \frac{BC}{ icT + d},\nonumber\\  
B \rightarrow {\frac{B}{icT + d}},&C \longrightarrow &{\frac{C}{icT + d}},
\label{treele}
\end{eqnarray}
with a,b,c,d, $a^{\prime}$, $b^{\prime}$, $c^{\prime}$, $d^{\prime}$ are
integers and  $ad-bc=1$, $a^{\prime} b^{\prime} -c^{\prime} d^{\prime}=1$.
The same transformation law appears also to the transformation law of 
the matter fields for $(2,2)$ compactifications.

Let us discuss now Calabi-Yau manifolds.
In general a Calabi-Yau manifold refer to Ricci-flat K\"{a}hler 
manifolds of vanishing Chern class. The condition of vanishing Chern 
class 
on a compact manifold has as a consequense the existence of a Ricci 
flat K\"{a}hler metric. 
The condition of vanishing Chern class\footnote{ 
By definition the 1st Chern class $c_1(X)$ is defined  
as $c_1 = tr R$, where R is the Ricci tensor. }
is associated with the 
existence of a two form $\rho$ equal to $\rho = R_{i{\bar j}} 
dz^i \wedge d{\bar z}^{\bar j}$, with $R_{i{\bar j}}$ the Ricci tensor.
Vanishing Chern class means, Ricci flat metric, that $R_{i{\bar j}} 
= 0$.  

The massless modes coming from the compactification on a Calabi-Yau 
manifold are associated\cite{gsw2} to the zero modes of the
Laplace operator on the compactification manifold.
The inner product is defined as $\textstyle{<\gamma, \delta> =
\int \gamma \wedge {\ast \delta}}$. For our purpose it is enough to know
that action of the Poincare duality operator $\ast$, in an n-dimensional
manifold, transforms a p-form to an n-p form.   
The number of the linearly independent p-forms associated to the
zero modes in the action is now the
number of linearly independent p-forms that are closed but not exact.
This is defined as the Betti number $b_p$, namely $\textstyle{b_p = 
\sum_{p+q=r} h^{p,q}}$.
The vector space of the closed p-forms modulo the exact forms is the 
cohomology group $\textstyle{H^p(M,R)}$ on the manifold M, with dimension 
equal to the Betti number and  $\textstyle{H^p(M,R)}$.
In general, we can define a $(p,q)$ from the wedge product. So if
we have a p-form A and a q-form B, we define their product to be a $p+q$
form
$\textstyle{(a \wedge B)_{i_1 \dots i_{p+q}}={p!q! \over (p+q)!}(A_{i_1
\dots i_p} B_{i_1 \dots i_q})\pm permutations}$, where the permutations 
are completely antisymmetric in all $p+q$ indices.
Forms with $(p,q)$ indices which are closed but not exact generate the
Dolbeault cohomology groups $H^{(p,q)}(X)$.
On Calabi-Yau manifolds, and in general in K\"{a}hler manifolds,
the Hodge-de Rham Laplacian $\textstyle{\bigtriangleup_{\bar \partial} =
{\bar \partial} {\bar \partial}^{*}+ {\bar \partial}^{*}{\bar \partial}}$
annililates the $(p,q)$ forms. The associated cohomology groups 
$H^{(p,q)}(X)$ decompose as $\textstyle{H^{k}(X)= \oplus_{p+q=k} 
H^{(p,q)}(X)}$. Here, $\textstyle{H^k (X)}$ is the De Rham cohomology
group\cite{griha} which annihilates $(p,0)$ forms with Laplacian
$\textstyle{\bigtriangleup =d d^{*}+ d^{*}d}$. 
The dimensions of $\textstyle{H^{(p,q)}(X)}$ are the Hodge numbers
$\textstyle{h_{(p,q)}}$ and satisfy $\textstyle{h_{(p,q)}=h_{(q,p)}}$
and from Poincare duality $\textstyle{h_{(p,q)}=h_{n-p,n-q}}$.
The Euler number is given by $\textstyle{\chi =\sum_{p,q}
(-1)^{p+q}h_{(p,q)}}$. 
For the Calabi-Yau three folds, $\textstyle{h_{(1,0)} = h_{(2,0)}=0}$
and $\textstyle{h_{(0,0)}=h_{(3,0)}=1}$ and the Euler number
is $\textstyle{\chi = 2(h^{(1,1)} - h^{(2,1)}})$. Remember that 
for a Calabi-Yau threefold the Euler number is
is two times the net number of chiral generations.
For the $K_3$ surface, the Hodge numbers are $\textstyle{h_{0,0}=h_{2,0}=
h_{2,2}=1}$ and $\textstyle{h_{1,1}=20}$, so $\chi(K_3) =24$.
A choice of complex coordinates\cite{can,hub,cande,cgh} in a Calabi-Yau 
space defines a complex
structure. Complex structure deformations are parametrized by the so
called complex structure moduli which are associated with
the variations of the metric $\textstyle{\delta g_{ij}, 
\delta g_{\bar {ij}}}$.
In addition, there is an additional form of moduli, the K\"{a}hler
class moduli associated with mixed indices variations of the 
Ricci flat  K\"{a}hler metric, i.e $\textstyle{\delta g_{i{\bar j}}}$.
Variation of the metric of the Calabi-Yau space in order to
preserve Ricci flatness, associates the quantities
$\textstyle{i dg_{i{\bar j}}dz^i \wedge dz^{\bar j}}$ to harmonic $(1,1)$
forms
and $\textstyle{i \Omega_{ij}^{\bar k}dz^i\wedge dz^j 
\wedge d{\bar z}^{\bar k}}$ to harmonic $(2,1)$ forms. Here, 
$z^i$ are the complex coordinates\cite{hoth} of the Calabi-Yau manifold 
and
$\textstyle{\Omega_{ij\rho}= g_{\rho {\bar k}} \Omega_{ij}^{\bar k}}$ 
is the constant three form. Naturally, harmonicity means that $\textstyle{
i dg_{i{\bar j}}dz^i \wedge dz^{\bar j} = \sum_{i=1}^{h_{1,1}}
\epsilon^1_i \psi_i}$ and $\textstyle{{\psi_i} \in H^{1,1}}$.
In addition, $\textstyle{i \Omega_{ij}^{\bar k}dz^i\wedge dz^j
\wedge d{\bar z}^{\bar k}= \sum_{i=1}^{h_{2,1}} \epsilon^2_i \delta_i}$
and $\textstyle{\epsilon^2_i \in H^{2,1}}$.
The four dimensional fields associated to the parameters $\textstyle{
\psi_i}$ and $\textstyle{ \delta_i}$ are the moduli of the 
low energy effective action. 
In other words, the variations of the metric associated with $H^{1,1}$
cohomology correspond to the K\"ahler class moduli and
variations of the metric associated with $H^{2,1}$ cohomology to 
complex structure moduli.
For compactifications on $K_3$ manifolds, the moduli space of metrics
with $SU(2)$ holonomy associated 
to complex and K\"ahler deformations is 
$\textstyle{{\cal M}\frac{SO(19,3)}{SO(19) \times SO(3) \times
SO(19,3;Z)} \times R^{+}}$, where $R^{+}$ is  
associated\cite{asmo} with the volume of $K_3$.
Adding the moduli coming from deformations of the antisymmetric
tensor we get the moduli space of $K_3$ 
$\textstyle{\frac{SO(20,4) \times SO(20)}{\times SO(4)}/SO(4,20;Z)}$.

The low energy $N=1$ supergravity of type I and heterotic string 
theories
is subject to amonalies coming from hexagon diagrams
which prevent it from describing an anomaly free string theory. 
In this case anomalies are cancelled\cite{LWi,gs1,Calabi1} by the 
addition of
appropriate counterterms which modify the supersymmetry structure. 
Similarly, in six dimensions the total anomaly is associated to 
the eight form 
\begin{equation}
I_8 = {\tilde \theta}_1 tr R^4 + 
 {\tilde \theta}_2 (tr R^2 ) + {\tilde \theta}_3 tr R^2 tr F^2 
+ {\tilde \theta}_4  {{(tr F^2)}^2}
\label{eightf1}
\end{equation}
where
${\tilde \theta}_1$,${\tilde \theta}_2$, ${\tilde \theta}_3$,
 ${\tilde \theta}_4$ are 
numbers depending on the spectrum\cite{gswe,erle} of the theory. 
Cancellation of anomaly requires 
$\textstyle{{\tilde \theta}_1 = n_H - n_V + 29 n_T -273 = 0}$, where
$n_V$, $n_H$, $n_T$ are the numbers of vector multiplets, 
hypermultiplets and antiselfdual tensor multiplets respectively.
Because in six dimensions we have one tensor multiplet,
which incorporates the dilaton, a Weyl spinor and an antiself-dual
antisymmetric tensor, the last constraint becomes $n_H - n_V =244$.
Now Green Schwarz mechanism factorization of anomalies is at work
with $I_8 \propto -\frac{1}{(2 \pi)^3 16} {\cal G}
{\tilde {\cal G}}$, ${\cal G} = tr R^2 - \sum_a \upsilon_a (tr F^2)$ 
and\footnote{Here, R, F are the gravitational and gauge field
strengths. 
The coefficients $\upsilon_a$, ${\tilde \upsilon}_a$ depend on the
particle content and the sum is over the gauge group G factors $G_a$.}
${\tilde {\cal G}} = tr (R \wedge R) - \textstyle{\sum_a 
{\tilde \upsilon}_a } tr (F \wedge F)_a$.
Cancelation of anomalies requires modification of the antisymmetric field 
stregth H as
\begin{equation}
H=dB + \omega^L - \sum_a {\upsilon}_a \omega_a^{YM},\;
\omega_L = tr (\omega R - \frac{1}{3} \omega^3),\;
\omega^{YM} = tr(A F - \frac{1}{3}\omega^3).
\label{modif1}
\end{equation} 
Here, $\omega_L$ , $\omega^{YM}$
are the Yang-Mills and Lorentz Chern-Simons three forms,
A the gauge field, R the Riemann tensor and $\omega$ the spin 
connection. However, because H is globally defined on $K_3$,
$\textstyle{\int_{K_3} dH=0}$.  As a result, 
we get that the constraint 
\begin{equation}
\sum_a n_a = 24,\; n_a = \sum_a \int_{K_3} (tr F^2)_a,= \int_{K_3}tr R^2 =24.
\label{modify2}
\end{equation}
Here, the instanton number $n_a$ becomes equal to the Euler number of 
$K_3$.  
Intially, in ten dimensions the unbroken group is $E_8 \times  E_8 \times 
U(1)^4$,
where the $U(1)$'s are associated with the $T^2$ and the graviton and the 
graviphoton. 
The spectrum can be derived from index theory\cite{kava,gswe}. 
The spectrum of the theory after compactification on $K_3 \times T^2$
can be calculated\cite{gswe} using index theory.
The gauge group G can be broken to a subgroup H, 
by vacuum expectation values of $K_3$ gauge fields in $\cal G$, where 
$H \times {\cal G} \subset G$.
The gauge group G breaks into the subroup H,  
which is the maximal subgroup commuting with the $\cal G$ subroup,
the commutant of $\cal G$.
We perform the decomposition $adj G = \sum_{i} (R_i, M_i)$, where
$R_i$, $M_i$ representations of the gauge groups H and $\cal G$ 
respectively.
Then the number of left-handed spinor multiplets transforming in the
$R_i$ representation of H is given by 
\begin{equation}
N_{R_i}=\int_{K_3} -\frac{1}{2}tr_{R_i}F^2 + \frac{1}{48} dim_{M_i}
tr R^2 = dim_{M_i} -\frac{1}{2} \int_{K_3} c_2(V) index(M_i),
\label{index1}
\end{equation}
where V is the $\cal G$ bundle parametrizing the expectation
values(vev's) of the vacuum gauge fields on $K_3$. By $c_2(V)$ we 
denote the
second Chern class of the gauge bundle V and $dim_i$ the dimension
of the representation i.
In addition, the dimension of the moduli space of gauge bundles is
$4 h_a - dim({\cal G}_a)$, where $h_a$ is the Coxeter number of
${\cal G}_a$ and dim its rank. 
In a general situation we allow for the gauge group G to break to 
the commutant of $\otimes \cal G$, by embedding the gauge connections
of a number of a product of gauge bundles $V_a$ with gauge group 
${\cal G}_a$ into G, 
resulting
in the breaking of G into the commutant of $\otimes_a {\cal G}_a$.
In this way, we identify, for manifolds of $SU(2)$ holonomy, 
the spin connection of $K_3$ with the gauge group $\otimes_a 
{\cal G}_a$, breaking the G symmetry into H. This is the analog of
breaking the gauge group $E_8$, in manifolds of $SU(3)$ holonomy,
by the standard embedding\cite{iss1} of the $SU(3)$ gauge connection 
into the spin connection, to the 
phenomenogically interesting $E_6$ gauge group.
Embedding an $SU(2)$ gauge bundle\footnote{Here, 
$h(SU(2))=24$.}
into one of the $E_8$'s, we get 45 hypemultiplet scalars plus a 
contribution of $20$ from the gravitational multiplet, making a total of 
65 hypermultiplets. In addition, we get a number of $56$'s in $E_7$ 
giving $N_{56} =10$.

%$$$$$$$$$$$$$$$$$$$$$$$$$$$$$$$$$$$$$$$$$$$$$$$$$$$$$$$$$$$$$$$$$$$
%************************************************
%***************************************
%$$$$$$$$$$$$$$$$$$$$$$$
%$$$$$$$$$$$$$$$
%$$$$$$$$$$$$$$$$$$$$$$$$$$$$$$$$$$$$$$$$$$$$$$$$$$$$$$$$$$$$$$$$$$$

\subsection{Special Geometry and Effective Actions}

In this part of the Thesis we will describe properties of the low 
energy
effective actions of $N=2$ effective string theories.
In $N=2$ supersymmetric Yang-Mills theory the action is described
by a holomorphic prepotential $F(X)$, where $X^A$ ($A=1,\ldots,n$)
are the complex scalar components of the corresponding vector
superfields. 
Two different functions
$F(X)$ could correspond to equivalent equations of
motion. 
In general such 
equivalences involve symplectic reparametrizations combined with
duality transformations.

The couplings of the classical vector multiplets with supergravity 
are
determined by a holomorphic function $F(X)$, the prepotential function 
which is a holomorphic function of
$n+1$ complex variables $X^I$ ($I=0,1,\ldots,n$) and 
it is a homogeneous function of degree 
two\cite{DWVP} in the fields  $X^I$ .
The general action for vector multiplets coupled to $N=2$ supergravity
was first obtained with the superconformal tensor calculus.

In $N=2$ supergravity theories, supersymmetry demands an additional 
vector
superfield $X^0$ which account for the accommodation of the 
graviphoton.
It stands for the $\textstyle{I=0}$ component of the vector multiplets and it
belongs to a compensating multiplet.
The graviphoton is the vector component of the compensating 
multiplet and is the spin one gauge boson of the 
supergravity multiplet.
The coordinate space of physical scalar fields
belonging to vector multiplets of an $N=2$
supergravity is described from special K\"ahler 
geometry \cite{DWVP,special},
with the K\"ahler metric
$\textstyle{g_{AB}=\partial_A \partial_{\bar B} K(z,\bar z)}$
resulting from a K\"ahler potential of the form
\begin{equation}
K(z,\bar z)=-\log\Big(i{\bar X}^I(\bar z)\,F_I(X(z) -iX^I(z)\;,
\bar F_I({\bar X}({\bar z})\Big),\;F_I=\frac{\partial F}{\partial X_I} 
,\;{\bar F}_I =\frac{\partial F}{\partial {\bar X}_I}
\label{KP}
\end{equation}
and the Riemann curvature tensor 
satisfying\cite{BEC}
\begin{equation}
{R^A_{BC}}^{D} = 2{\delta}^A_{(B} {\delta}^D_{C)}  - e^{2K}
{\cal W}_{BCE} {\bar {\cal W}}^{EAD},
\end{equation}
where $ W_{aBC} $ is a holomophic 3-index symmetric tensor given by
\begin{equation}
{\cal W}_{ABC} = F_{IJK}(X(z)) \frac{\partial X^{I}(z)}{\partial z^{A})}
\frac{\partial X^J(z)}{\partial z^{B}} \frac{\partial X^{K}(z)}
{\partial z^{C}}.
\label{intesection}
\end{equation}
By choosing inhomogeneous coordinates $z^A$
the so called, {\em special} coordinates, defined by
$\textstyle{ z^A =X^A/X^0,\;A=1,\dots,n}$
or by $\textstyle{X^0(z)=1,\;X^A(z)=z^A}$,
the K\"ahler potential can be written as\cite{SU}
\begin{equation}
K(z,\bar z) = -\log\Big(2({\cal F}+ \bar{\cal F})-
(z^A-\bar z^A)({\cal F}_A-\bar{\cal F}_A)\Big)\,,
\label{Kspecial}
\end{equation}
where $\textstyle{{\cal F}(z)=i(X^0)^{-2}F(X)}$.
Up to a phase, the proportionality factor between
the $X^I$ and the holomorphic sections $X^I(z)$ is given by
$\exp\big({1\over 2} K(z,\bar z)\big)$.
The kinetic energies of the gauge fields are
\begin{equation}
{\cal L}^{\rm gauge}\ =\ -{\textstyle{i\over 8}}
\left( {\cal N}_{IJ}\;F_{\mu\nu}^{+I}F^{+\mu\nu J}\
-\; {\bar{\cal N}}_{IJ}\;F_{\mu\nu}^{-I} F^{-\mu\nu J} \right),
\end{equation}
where 
$F^{\pm I}_{\mu \nu}$ represents the selfdual and anti-selfdual 
 $\textstyle{F^{\pm}_{\mu \nu}=(1/2)(F^I_{\mu \nu} \pm 
{\tilde F}^I_{\mu \nu})}$
field strengths proportional to the symmetric tensor
\begin{equation}
{\cal N}_{IJ}={\bar F}_{IJ}+2i \frac{{\rm Im}(F_{IK}){\rm Im}(F_{JL})\,
X^K X^L}{ {\rm Im}(F_{KL})\,X^K X^L},\;\;F_{IJ}=\frac{\partial^2 F}{
{\partial X_I} {\partial X_J}},\;F_I={\cal N}_{IJ}X^J.
\label{Ndef}
\end{equation}
Here, $\cal N$ is the field-dependent tensor of the gauge involved
in the gauge couplings g, $g^{-2}_{IJ}= {i\over 4}({\cal N}_{IJ}
-\bar {\cal N}_{IJ})$. The generalized $\theta$ parameters
$\theta_{IJ}= 2\pi^2({\cal N}_{IJ}+\bar {\cal N}_{IJ})$.
Subscripts on the F variable denote derivatives and repeated indices,
as usual, are summed.

The equivalence of equations of motion under different functions 
$F(X)$ could describe equivalences under electric-magnetic
dualities of the field strengths, and not local 
gauge transformations to the vector potentials $A^I_{\nu}$.
Because for the non-Abelian case, such a duality is meaningless
since the equations of motion cannot be made invariant under the symplectic
transformations which will be defined in (\ref{FGdual}), 
we will work with abelian gauge fields. Note, that a non-abelian gauge
field have only electric charge. 
In this way, when all the fundamental fields are neutral, one can 
freely
choose any integral basis for the electric and magnetic charges.

Let us define the tensors \cite{DWVP,DWVVP} $G^{\pm}_{\mu\nu I}$ as
\begin{equation}
G^+_{\mu\nu I}={\cal N}_{IJ}F^{+J}_{\mu\nu}\,,\quad G^-_{\mu\nu
I}=\bar{\cal N}_{IJ}F^{-J}_{\mu\nu}.
\label{defG}
\end{equation}
Then the set of Bianchi identities and equations of motion
for the abelian gauge fields is expressed as
\begin{equation}
{\partial}^{\mu} \left(F^{+I}_{\mu \nu} -F^{-I}_{\mu\nu}\right)
=0,\;\;
{\partial}^{\mu} \left(G_{\mu \nu I}^{+} -G^{-}_{\mu \nu I}\right)=0.
\label{Maxwell}
\end{equation}
These are invariant under the symplectic $Sp(2n+2,R)$ 
transformations
\begin{equation}
\displaystyle{
F^{+I}_{\mu \nu} \longrightarrow \tilde F^{+I}_{\mu \nu} =
U^I_J\, F^{+J}_{\mu \nu}+Z^{IJ}\,G^{+}_{\mu \nu J}\;\;
G^{+}_{\mu \nu I} \longrightarrow \tilde G^+_{\mu \nu I}= V_I^J\;
G^{+}_{\mu \nu J}+ W_{IJ}\,F^{+J}_{\mu \nu} },
\label{FGdual}
\end{equation}
where $U$, $V$, $W$ and $Z$ are constant, real,  $(n+1)\times(n+1)$
matrices.
Alternatively, 
\begin{eqnarray}
\left(\begin{array}{c}
F^{+I}_{\mu \nu}\\
G^{+I}_{\mu \nu}
\end{array}\right) \rightarrow
\left(\begin{array}{cc} U & Z \\
[1mm] W & V 
\end{array}\right) 
\left(\begin{array}{c}
F^{+I}_{\mu \nu}\\
G^{+I}_{\mu \nu}
\end{array}\right),
\label{koufme1}
\end{eqnarray}
with
\begin{eqnarray}
{\cal O}\ \stackrel{\rm def}{=}
\left(\begin{array}{cc} U & Z \\W & V\end{array}\right),\;\;\in 
Sp(2n+2, R)
\label{uvzwg}
\end{eqnarray}
and
\begin{equation}
{\cal O}^{-1} = \Omega\, {\cal O}^{\rm T} \,\Omega^{-1} \;
\mbox{and} \; \Omega =\left(\begin{array}{cc}
0&1 \\[1mm]
-1 & 0
\end{array}\right).
\label{spc}
\end{equation}
The matrices $U$, $V$, $W$ and $Z$, satisfy
\begin{equation}
\displaystyle{U^{\rm T} V - W^{\rm T} Z = V^{\rm T} U - Z^{\rm T}W =
{\bf 1},\;\;
U{\rm T}W = W^{\rm T}U\;\; Z^{\rm T}V= V^{\rm T}Z\;}.
\label{spc2}
\end{equation}
The kinetic term of the vector fields does not
preserve its form under general $Sp(2n+2,R)$ transformations and only 
the equations of
motion and Bianchi identities are in fact equivalent.
In the case of abelian gauge fields, one can always choose a
coordinate basis
$X^I$ for which the prepotential $F$ does exist.
On the other hand, in string theory, the 
dilaton dependence of the gauge couplings is explicit only in a
basis where $F$ does not exist.

Target-space duality transformations can always be implemented as
$Sp(2n+2,{R})$ transformations of the period vectors $(X_I, F^I)$ of 
special geometry. For target-space duality transformations 
the Lagrangian is left invariant by the subgroup
that satisfies $W=Z=0$ and $V^T = U^{-1}$.
The presence of  $Z=0$ and 
$V^{\rm T}=U^{-1}$ together with the condition that $W^{\rm T} U$ 
has to be a symmetric matrix provides the semiclassical 
transformations 
\begin{equation}
\displaystyle{{\tilde X}^I = U^I_J X^J,{\tilde F}^{\pm I} =
 U^I_J F^{\pm J},\;
{\tilde F}_I =[U^{-1}]^{J}_I F_J +W_{IJ} X^J,\;
{\tilde{\cal N}}=[U^{-1}]^{\rm T}{\cal N}U^{-1} +WU^{-1}},
\label{Tsymp12}
\end{equation}
which may be implemented as Lagrangian symmetries of the
vector fields $A^I_{\mu}$.
The last term in (\ref{Tsymp12}) amounts to
a constant shift of the theta angles 
at the quantum level.     
Because such shifts are quantized, the
symplectic group must be restricted to $Sp(2n+2,{\bf Z})$.
Such shifts in the $\theta$-angle do occur
whenever the one-loop gauge couplings have logarithmic singularities
at special points in the moduli space where massive modes
become massless. 
It will be confirmed by our results as well later in this chapter.
Constant shifts in the theta angle occur, when we encircle the 
singular line $T=U$ at the quantum moduli space.
As a result such symmetries are associated with the 
semi-classical one-loop monodromies around such singular points.
An other form of duality transformations interchanges the
field-strength tensors $F^I_{\mu \nu}$ and $G_{\mu \nu I}$ and
correspond to electric-magnetic dualities. These
transformations appear as
$U=V=0$ and $W^{\rm T}=- Z^{-1}$, and 
$\tilde{\cal N} = - W \,{\cal N}^{-1}\,W^{\rm T}$,
so that they give rise to an inversion of the gauge couplings and
hence must be non-perturbative.
In the heterotic string theory, such transformations represent the 
$S$-dualities.

The classical rigid field theory is not associated with field
dependence of the physical observables.  However, by introducing a
cut-off at the Planck scale, in the quantum theory, the superheavy 
states are integrated out leaving only the light fields.  
Integration of the heavy fields induce moduli dependence in
the effective theory.
In real
terms, to properly describe the low energy theory of the 
physical vacuum, the field-dependent couplings of the EQFT should 
be written as
complete analytic functions of the moduli fields and the dependence on
all the other fields must be described by a truncated power series.

Dividing the scalars as ${z^{A}} = {X^{A}}/ {X^{0}}$ belonging to
vector multiplets into 
moduli ${\Phi}^{a} = -i z^{a}$ and ``matter'' scalars
${\Upsilon}^{k} = -i z^{k}$,
we expand the prepotential $\cal F$ of the theory as a truncated power
series in the matter scalars as
\begin{equation}
{\cal F}{(\Phi,{\Upsilon})} = h({\Phi}) +\sum_{cd} 
f_{cd}({\Phi}){\Upsilon}^{c} {\Upsilon}^{d}.
\label{zorika1}
\end{equation}
All scalars in the non-Abelian vector multiplets may be considered
as matter \cite{wkll} and not as moduli, since their
vacuum expectation values can induce a non zero mass
for some of the non-Abelian fields.
For such non-Abelian matter, the gauge symmetry of the prepotential
requires for the gauge kinetic function of each non-abelian gauge group
factor $(a)$, $f_{ab}(\Phi)\ = \delta_{ab} f_{(a)} (\Phi)$.
Scalars in vector multiplets neutral under an abelian symmetry must be
considered as moduli, otherwise as matter.
For hypermultiplets in the effective theory charged
under an Abelian gauge symmetry, the scalar superpartner of that
gauge boson should be regarded as matter since its
vacuum expectation value can in principle give masses
to all charged hypermultiplets.
But if all the light particles are neutral with
respect to some Abelian gauge field, then its scalar superpartner is
a moduli.                                                              
So we divide the Abelian vector multiplets into ${\Phi}^{a}$
and ${\Upsilon}^a$ such that all the light of the EQFT
are exactly massless for ${\Upsilon}^a=0$ and arbitrary ${\Phi}^{a}$.
In this limit the heterotic string moduli space factorises in
the product\cite{cefegi} form $\textstyle{{\cal M}_{het}= {\cal M}_{IIA} 
\times {\cal M}_{IIB}}$, where the moduli spaces for the type IIA and 
IIB represent vector multiplets. 
The effective quantum field theory must satisfy several constraints.
In particular, the Wilsonian prepotential of an $N=2$ supersymmetric
theory must be a holomorphic function and expanding, 
\begin{equation}
K (\Phi,{\bar \Phi}, U, {\bar U}) =  K({\Phi},{\bar \Phi}) + 
{\sum}_{ab} Z_{ab}({\Phi},{\bar \Phi}) U^{a}{\bar U}^{b} +{\dots},
\label{part1}
\end{equation}
with 
\begin{equation}  
K(\Phi,{\bar \Phi}) = -\log \left(2(h  + {\bar h})
-{\sum}_{\alpha} ({\Phi}^{\alpha} +{\bar \Phi}^{\alpha})
({\partial h}_{\alpha}+
{\bar \partial}{h}_{\alpha})\right)
\label{part2}
\end{equation}   
and
\begin{equation}
\quad Z_{ab}(\Phi,{\bar \Phi})\ 
=\ 4e^{K(\Phi,{\bar \Phi} )}\,{\rm Re}\, f_{ab}(\Phi).
\label{Zexp}
\end{equation}

The Wilsonian gauge couplings follow from eqn.(\ref{Ndef}).
In addition the vector superpartners of $\Phi$ do not mix 
with the graviphoton
and hence the Wilsonian gauge couplings are simply
$(g^{-2}_{ab})^{\rm W}={\rm Re} f_{ab}(\Phi)$
and for  non-Abelian gauge fields
$(g^{-2}_{(a)})^{\rm W}={\rm Re} f_{(a)}(\Phi)$.
In contrast, the vector superpartners of the moduli
mix with the graviphoton
and as a result the Wilsonian gauge couplings
$(g^{-2}_{ab})^{\rm W}$, $(g^{-2}_{a0})^{\rm W}$ and
$(g^{-2}_{00})^{\rm W}$ exhibit explicit
non-holomorphic function moduli dependence.
The complete result for the Wilsonian prepotential
of $N=2$ theories, gives that it is only renormalized 
only up to one loop order of perturbation theory on analogy with
the rigid case.
Thus,
\begin{equation}
{\cal F}={\cal F}^{(0)}+ {\cal F}^{(1)}+{\cal F}^{(NP)} ,
\label{Floop}
\end{equation}
where ${\cal F}^{(0)}$ represents the tree level prepotential,
${\cal F}^{(1)}$ is the one loop correction while ${\cal F}^{(NP)}$
receives corrections from world-sheet instantons and other
non-perturbative effects.
The perturbative one-loop correction to 
the prepotential
of vector multiplets in decomposable and non-decomposable orbifold
constructions of the heterotic string will be calculated later.
Note, that the one loop correction to the prepotential of the vector
multiplets has been calculated before, indirectly via 
its third derivative, in \cite{wkll,afgnt}.
Analytically,
\begin{equation}
\displaystyle{h(\Phi) 
= h^{(o)}(\Phi) + h^{(1)}(\Phi)\;\;,
f_{ab} 
= f^{(o)}_{ab}(\Phi) + f^{(1)}_{ab}(\Phi)},
\label{fhexp}
\end{equation}
and for the non-Abelian gauge group factors involved in the theory
the Wilsonian gauge couplings read
\begin{equation}
(g^{-2}_{(a)})^{W} = {\rm Re} f^{(o)}_{ab} (\Phi)
+\;\;\; {\rm Re} f^{(1)}_{ab}(\Phi).
\label{gna40}
\end{equation}
Renormalization is up to one loop order, as it
happens in the $N=1$ Wilsonian couplings of effective field 
theories.
In Calabi-Yau manifolds, special geometry is associated with
the description of their moduli spaces. We will give more details 
at the end of the next section.

%******************************************************************
%******************************************************************
%******************************************************************
%******************************************************************

\subsection{ Low energy Effective theory of N=2 Heterotic 
superstrings and related issues}
In this section, we will describe the low energy theory
of $N=2$ symmetric orbifold compactified heterotic superstrings.
In addition, we will describe properties of the effective theory of 
type II supestrings compactified on a Calabi-Yau three fold.
For heterotic strings compactified on a six dimensional orbifold, we
consider the case where the internal torus lattice action correspond
to the topus decomposition $T^4 \oplus T^2$.
The moduli space of the torus is parametrized by the usual moduli
$T$ and $U$. These moduli are part of the vector
multiplet moduli space. 
Properties concerning the moduli space of such 
theories
have already been discussed in section 4.2.  Theories, with
the same structure including e.g $N{=}2$ orbifold compactifications of 
the heterotic strings\cite{dhvw} and $N{=}2$ heterotic string 
compactified\cite{kava} on the $K_3 \times T^2$.
The low-energy theory describing 
any classical $(2,2)$ vacuum includes the gravitational sector, 
containing the
graviton, dilaton the axion and the superpartners, together with
the $E_8 \otimes E_8$ gauge multiplets and a set of chiral superfields
which constitute the ${\bf 27}$, ${\bar {\bf 27}}$ representations
of $E_6$ matter fields.    
In addition, world sheet supersymmetry demands that each $\bf 27$,
${\bar {\bf 27}}$ supermultiplet of matter fields is accompanied by an
$E_6$ singlet moduli superfield, representing the moduli \footnote{They 
are the highest components of 
chiral primary fields of the left moving superalgebra.}. 
These moduli in the case of Calabi-Yau threefolds correspond 
to the deformation parameters of the K\"ahler and complex structure.
Note, that $(2,2)$ symmetric orbifold compactifications of the 
heterotic string flow to their Calabi-Yau counterparts, after blowing up
the twisted moduli scalars\cite{Di1,HV} associated with the fixed points
of the orbifold.
Twisted moduli are not neutral with respect to the gauge group of
the $(2,2)$ theory and will not considered here, as they will not
be involved in our discussions. 
The K\"{a}hler function K characterizing the general heterotic $(2,2)$
compactifications has the following power expansion\cite{dkl1} in the 
matter
fields $K= \Sigma + \dots$, where $\dots$ represent a power expansion 
in terms of
matter fields and $\Sigma$ has a block diagonal structure in $(1,1)$ 
and $(2,1)$ moduli, i.e, $\Sigma = \Sigma^{(1,1)} + \Sigma^{(2,1)}$ 
The neutral moduli of heterotic string compactifications are 
coordinates in
a manifold with real dimension $2(h^{(1,1)} + h^{(2,1)} + 1)$.
The additional complex dimension refers to the dilaton axion system. 
In reality, in all heterotic $(2,2)$ compactifications the moduli 
spaces for the
$(1,1)$ and $(2,1)$ moduli spaces are special K\"{a}hler spaces 
and the K\"ahler potential must be treated using the lanquage
of special geometry.

The axion is subject to the discrete Peccei-Quinn symmetry
to all orders of perturbation theory. Since the axion is connected 
through a duality 
transformation to the antisymmetric tensor field, whose
vertex operator decouples at zero momentum, this means that
every physical amplitude involving $B_{\mu \nu}$ at zero momentum 
is zero.
As a result the effective theory of the heterotic superstring
is independent of the field $B_{\mu \nu}$ at zero momentum and the 
coupling of 
field appear only through its derivative. 
The dilaton and the axion belong to a vector multiplet.
Since the axion couples to the dilaton D via the complex scalar
S, which we will refer next as the dilaton, 
we conclude that any dependence
of holomorphic quantities, e.g the Wilsonian gauge couplings,
will be through the combination $S+{\bar S}$.
However, these arguments\footnote{Related discussion 
related to the expected corrections to the 
holomorphic superpotential will be discussed in chapter five.}   
are not valid non-perturbatively.
The structure of the heterotic vector multiplet moduli space is given by
the coset 
manifold 
based on the symmetric space\cite{special}
\begin{equation}
\frac{SU(1,1)}{U(1)} \otimes \frac{SO(2,n-1)}{SO(2) \times SO(n-1)}.
\end{equation}
The first factor corresponds to the dilaton. The prepotential
for this space reads
\begin{equation}
F(X)= -{X^1\over X^0}\Big[X^2X^3-\sum_{I = 4}^n(X^{I})^2\Big]\,.
\label{ourF}
\end{equation}
while the values of the moduli are identified as 
\begin{equation}
S=-i {X^1\over X^0}\,,\quad T=-i {X^2\over X^0}\,,\quad
U=-i {X^3\over X^0}\,,\quad \phi^i=-i {X^{i+3}\over X^0}\ ,\quad
(i=1, \ldots, P)\ ,
\label{modid}
\end{equation}
with the remaining  $X^I,C^{a} = -i X^{a+P+3}/ X^0, a = p+4,\ldots,n$ 
to correspond \cite{SFetal,Ceresole} to matter scalars.
Fron the values of the moduli previously given, it follows that the
the K\"ahler potential is
\begin{equation}
K= -\log \left( (S+\bar S) [(T+\bar T)(U+\bar U)-\sum_{i}
({\phi}^i+{\bar \phi}^i)^2 -\sum_{a}(C^a+{\bar C}^a)^2 \right)\, 
\label{Ktree}
\end{equation}
while we get in terms from quantities defined previously that
\begin{eqnarray}
h{(o)} =\ - S\left(TU- \sum_i(\phi^i)^2 \right),\;\; 
f^{(o)} = S,\;\;K= -\log(S+ {\bar S})-\log [(T+\bar T)(U+ 
{\bar U}) \nonumber\\
- \sum_{i}(\phi^i+\bar\phi^i)^2],\;
Z = {2\over (T+{\bar T})(U+ {\bar U})-\sum_{i} ({\phi}^i+{\bar 
\phi}^i)^2}.\nonumber\\
\label{treecoupl2}            
\end{eqnarray}
Especially for the non-Abelian factors in
the gauge group $G$ (or more generally any non-moduli vector
 multiplets)
the tree-level gauge coupling is universal\cite{Wit}. 
In the language of special geometry, comparing 
(\ref{treecoupl2},\ref{gna40}) we conclude that
dilaton's vacuum expectation value, $g^{-2}_{(a)} = {\rm Re} S$.

If we examine the various couplings for the vector 
superpartners of the moduli we see that the couplings
involving the coupling of the dilaton with itself and the moduli T,U or
the graviphoton  
are not  
become weak in the large dilaton limit as they should be. 
This is a sign that we are using a wrong symplectic basis. By changing 
to an
other symplectic basis, e.g replacing the $F_{S}^{\mu \nu}$ with
its dual field strength, we find that the couplings are now weakly 
coupled in the large-dilaton limit. In this way, we are using a 
basis and
$(X^{I}, F_{J}) \rightarrow ({\tilde X}^{I}, {\tilde F}_{J})$ where
\begin{equation}
{\tilde X}^I = X^I \; for   I \neq 1,\;{\tilde X}^1 =F_1,\;
{\tilde F}_I = F_I\;{for}\;I \neq 1,\;{\tilde F}_1 =- X^1
\label{newbasis}
\end{equation}
and the components of the symplectic matrix ${\tilde {\cal O}}$ 
are defined as
\begin{eqnarray}
\left(\begin{array}{c}
{\hat{X}}^I \\
{\hat{F}}_J 
\end{array}
\right)= {\tilde {\cal O}}
\left(\begin{array}{c}
X^K\\F_L
\end{array}
\right).
\label{karoump2}
\end{eqnarray}
The elements of ${\tilde {\cal O}}$ are as in (\ref{uvzwg}) and obey
\begin{equation}
U^{I}_{ J} = V^{J}_I = {\delta}^I_J \; 
{\rm for}\;  I,J \neq 1\;,\; Z^{11}=1\;, W_{11} =-1
\label{iok1}
\end{equation}
In this new basis, the prepotential does not exist, since in the 
new basis 
the matrix $S^{I}_{J}$ has zero determinant and the definition of a
prepotential is meaningless \cite{Ceresole}.
In the transformed basis, the K\"ahler potential for the moduli and 
the gauge couplings are found to be\cite{wkll} 
\begin{eqnarray}
{\hat {K}}_{\Phi}  = K_{\Phi} = - \log(S+\bar S) - \log(2(\hat {z}^J
\eta_{JI} {\hat {\bar z}}^I), &
{\hat{\cal N}}_{IJ} = -2i {\bar S} {\eta}_{IJ} + 2i(S + 
\bar S)\nonumber\\
\times  \frac{{\eta}_{IK}\;{\eta}_{JL}\;({\hat z}^{K} 
{\hat{\bar z}}^{L} + 
{\hat{\bar z}}^{K} {\hat{z}}^{L})}
{ {\hat{z}}^{K} {\eta}_{KL} {\hat{\bar z}}^{L}}.
\label{goods}
\end{eqnarray}
In the transformed basis the couplings behave strongly in the small 
dilaton limit. 
In this limit, the target space dualities of 
$N=2$ heterotic string vacua leave the classical lagrangian 
invariant, under transformations when $\hat{W} = \hat{Z} = 0 $ and
$\hat{U}$, $\hat{V} \in SO(2,2+P)$.
In fact, it is clear that the K\"ahler potential is invariant under
 symplectic
transformations which act on the $(X^{I},F_{I})$. 
Moreover, in the absence of the one loop correction to the
prepotential, we can use the $PSL(2, Z)_{T}$ target space
duality symmetry subroup of the full symmetry group 
of toroidal compactifications to study 
the transformation behaviour of the period vectors of special 
geometry.
In sum, in the symplectic basis 
${\hat{X}}^{I},{\hat{X}}_{J}$, we get that the corresponding 
symplectic matrices are given by 
\begin{equation}
\hat U= \pmatrix{d&0&c&0&0 \cr 0&a&0&-b&0 \cr b&0&a&0&0 \cr 
0&-c&0&d&0\cr
0&0&0&0&{\bf 1}_P}\,,
\quad
\hat V= (\hat U{}^T)^{-1} = \pmatrix {a&0&-b&0&0 \cr 0&d&0&c&0
\cr -c&0&d&0&0 \cr 0&b&0&a&0\cr 0&0&0&0&{\bf 1}_P}\,,
\label{dualmat}
\end{equation}
while $\hat W=\hat Z =0$.
Especially under the generator$g_1 : T \rightarrow \frac{T}{T+1} \in 
{\Gamma }^{o}_{3}$, we get that $\hat U$ is defined as follows
\begin{eqnarray}
g_1 :\;{\hat{U}}&=&\left(\begin{array}{cccccc}
1&0&1&0\\
0&1&-1&0\\
0&0&1&0\\
0&-1&0&1
\end{array}\right)\quad
V = ({\hat U}^{T})^{-1} = \left( \begin{array}{cccccc}
1&0&0&0\\
0&1&0&1\\
-1&0&0&0\\
0&0&0&1 
\end{array}\right)\label{matrices}.\\
\label{karoump1}
\end{eqnarray}
In a similar way we can derive the matrices corresponding to the
generators for ${\Gamma }^{o}_{3}$, $g_2 : T \rightarrow T + 1$.
Considerations involving the calculation
of the full quantum duality group ${\Gamma}^{o}(3)$ will not 
included here.

In Calabi-Yau manifolds, special geometry is associated with
the description of their moduli spaces. In type IIB,
the $H^{2,1}$ cohomology describes the deformation of the complex
structure of the Calabi-Yau space $\cal M$. Now the K\"ahler metric for
the $(2,1)$ moduli is defined\footnote{In the rest of the section
the notation for the special coordinates is as follows, $Z^i= -i
X^i/X^o$.}
 from the
Weyl Peterson metric\cite{can,cande,cande1,cgh}
$\sigma_{ij}$, namely
\begin{equation}
G_{ij} = \sigma_{ij}
/ (i(\int_{{\cal M}} \Omega \wedge {\bar \Omega})),
\label{homo1}
\end{equation}
where
\begin{equation}
{\varphi}_{i}=(1/2){\varphi}_{i k \lambda{\bar \rho}}dx^{k}dx^k
dx^{\lambda} dx^{\bar \rho},\;\;
\sigma_{ij} = \int_{{\cal M}} \varphi_i
\wedge {\bar \varphi}_{\bar j}
\label{homo2}
\end{equation}
and ${\varphi}_{ik\lambda{\bar
\rho}}= (\partial g_{{\bar \rho}{\bar \xi}}/ {\partial t^i})
{\Omega}^{\bar xi}_{k \lambda}$.
Here, $t_i=1,\dots,b_{2,1}$ and $G_{ij}=-\partial_i
\partial_j( i \int_{\cal M} \Omega \wedge {\bar \Omega})$.
The three form tensor $\Omega$ is given in terms of the
holology basis $\alpha$, $\beta$ as $\Omega = X^{I}\alpha_I + i F_I
\beta^{I}$.   
The complex structure is described by the periods of the holomorphic
three form $\Omega$ over the canonical homology basis.
%In the symplectic basis $(\alpha^I, \beta_I)$ the
Here, the periods are given by $X^I =\int_{A^I}\Omega$,
$iF_I =\int_{B^I}\Omega$ the integral of the holomorphic three form
over the homology basis. The K\"ahler potential comes from the
moduli metric
\begin{equation}
G_{ij}=-i\partial_i \partial_j \{ i \int \Omega \wedge {\bar
\Omega}\},\;\;
K=-\log(X^I {\bar F}_I + {\bar X}^I F_I).
\label{pol1}
\end{equation}
Now the Riemmann tensor is defined as
\begin{equation}
R_{i{\bar j}k{\bar l}}= G_{i{\bar j}} G_{k{\bar l}} +G_{i{\bar l}}G_{k{\bar
j}}-{\bar {\cal C}}_{ikn}{\bar {\cal C}}_{{\bar j}{\bar l}{\bar n}}
G^{n{\bar n}}e^{2K},  
\label{pol2}
\end{equation}
where the expression of the Yukawa couplings in a general coordinate
system are given by $ {\cal C}= \int \Omega \wedge 
\partial_i \partial_j   \partial_k \Omega $, $\partial_i = 
\partial / \partial z^i$.
The holomorphic function F does not receive quantum
corrections from world-sheet instantons and as a consequence neither the
the K\"ahler potential derived from it. 
Calabi-Yau threefolds can be constructed among other ways as 
a hypersurface or as a complete intersection of hypersurfaces in a
weighted
projective space $P^N(\vec w)$.
Remember, that the complex projective space $CP^N$ is the space defined
by the homogeneous complex coordinates $Z_1,\dots,Z_{N+1}$ which obey
$(Z_1,\dots,Z_{N+1}) \stackrel{\lambda \neq 0}{\equiv}(\lambda^{w_1}
Z_1,\dots,\lambda^{w_{n+1}} Z_{N+1})$ for complex $\lambda$.
The threefold is obtained from the $CP^4$, the quintic with\cite{klemmos} 
the general
equation $\sum_{k_i} \alpha_{k_1 k_2 k_3 k_4 k_5}x_{k_1}x_{k_2}
x_{k_3}x_{k_4}x_{i_5} =  0$ while the $K_3$ can be obtained from
the $\sum_{k_i} \alpha_{k_1 k_2 k_3 k_4} x_{k_1} x_{k_2} x_{k_3}
x_{k_4}=0$, in projective $P^3$ and $P^2$ respectively(rp).
They describe complex manifolds parametrized by 135 and 35
complex coefficients $a_{k_i}$ rp, which after removing an               
overall redundancy they give 101, 19 elements of $H^{(1,1)}$ rp. 
The weighted projective space  $P^N(\vec w)$ is defined by the conditions
on the 
homogeneous coordinates
$(Z_1,\dots,Z_{N+1})\stackrel{\lambda \neq
0}{\equiv}(\lambda^{d_1},\dots,\lambda^{d_{N+1}})$ and 
$P^N(\lambda^{d_1}Z_1,\dots,\lambda^{d_{N+1}}Z_{N+1})\stackrel{def}{=}
\{c^{*}=C^{N+1}/(Z_1=0,\dots,Z_{N+1}=0)\}$. The last condition, exludes 
the origin
of the complex space. The $d_i$ are the weights and the sum of the
weights is the degree of the variety.

%%%%%%%%%%%$$$$$$$$$$$$$$$$$$$$$$$$$$$$$$$$$$$$$$$$$$$$$$$$$$
%$$$$$$$$$$$$$$$$$$$$$$$$$$$$$$$$$$$$$$$$$$$$$$$
%$$$$$$$$$$$$$$$$$$$$$$$$$$$$$$$$$$$$$$$$
%$$$$$$$$$$$$$$$$$$$$$$$$$$$$$$$
%$$$$$$$$$$$$$$$$$$$$$$$$
%$$$$$$$$$$$$$$$$$$$$$$$$$$$$$$$$$$$$$$$$$$$$$$$$$$$$$$$$$$
%$$$$$$$$$$$$$$$$$

\subsection{* One loop correction to the prepotential from string 
amplitudes}

\subsubsection{One loop contribution to the K\"{a}hler metric -
Preliminaries}

The one-loop
K\"{a}hler metric for orbifold compactifications of the heterotic 
string,
where the internal six torus decomposes into $T^2 \oplus T^4$, was 
calculated in \cite{anto3}. In this section, we will use the general 
form of the 
solution for the one loop K\"{a}hler metric appearing in
\cite{anto3,afgnt}
to calculate the 
one loop correction to the prepotential of $N{=}2$ orbifold 
compactifications of the heterotic string. 
While the one loop prepotential has been calculated with the use 
of string amplitudes in \cite{afgnt}), in my Thesis I will provide 
an alternative 
way of calculating the one-loop correction\cite{wkll,afgnt}
to the prepotential
of the vector multiplets of the $N{=}2$ orbifold compactifications
of the heterotic string.
Note that in the following we will change notation, following the 
spirit
of the calculation in \cite{afgnt}, namely all moduli fields, 
including
the dilaton, are rescaled by a factor of i, $ P \rightarrow iP$.

In this section, we will describe the background theory 
of the one-loop contribution to the K\"{a}hler metric.
For this purpose, we will use not the standard 
supergravity\cite{cre}
lagrangian up to two derivatives in the bosonic fields,
described by the superconformal action formula
\begin{equation}
e^{-1}{\cal L}=-\frac{3}{2}[S_o{\bar S}_o e^{-\frac{1}{3}
G(Z,{\bar Z})}]_D +\left([S_o^3]_F +h.c\right)+
\frac{1}{4}\left([f_{ab}(Z)W^aW^b]_f +h.c\right)
\label{superco1}
\end{equation}
with matter decsribed by chiral multiplets 
$Z_i$ 
only. Instead, we will
use the linear multiplet formulation\cite{dfkz,cefergi1,anto3}.
Note that both formulations are equivalent, since the linear multiplet
can always be transformed in to a chiral multiplet by a 
supersymmetric duality transformation.
In eqn.(\ref{superco1}), $S_o$ is the chiral compensator field, and
$G(Z,{\bar Z})=K(Z,{\bar Z}) +\log|w(Z)|^2$ the gauge kinetic
function, where K is the 
K\"{a}hler potential and ${w}$ the holomorphic superpotential. In 
addition,
$W^a$ is the chiral spinor superfield of the Yang-Mills field 
strength
$F^a_{\mu\nu}$, and D, F subscripts refer to the vector density and 
chiral density in superspace.  

In the superconformal formalism\cite{kuue}, the action for the 
linear multiplet is given up to one loop order by
\begin{equation}
{\cal L}=-{(S_o {\bar S}_o)}^{3/2}(\frac{\hat{L}}{2})^{-\frac{1}{2}}
e^{-\frac{G^{(o)}}{2}} + (\frac{\hat{L}}{2})G^{(1)} +{(S_o^3 w)_F}
\label{linear1}
\end{equation}
where now the gauge kinetic function is given by
$G^{(o)}(z,{\bar z}) + lG^{(1)}(z,{\bar z})$.
The vev of $l$ is the four 
dimensional gauge coupling constant $g^2$.

Eqn.(\ref{linear1}) does not have the the gravitational kinetic energy
$\propto R$ term to its canonical form. Instead, the chiral compensator
field is used to properly normalising its coefficient, procedure which
fixes the value of the compensator field.
The advantage of using the linear multiplet instead of the chiral 
multiplet in eqn.(\ref{superco1}) is that it provides
an easy way of calculating\cite{anto3} one loop corrections 
to the K\"{a}hler metric. An easy way to see this comes from the 
following equation\footnote{Coming by expanding 
eqn.(\ref{linear1}).}, which includes the
bosonic kinetic energy terms,
\begin{equation}
{\cal L}_{bosonic}=-\frac{1}{4 l^2}\partial_{\mu}l \partial^{\mu}l
+\frac{1}{4 l^2}h^{\mu}h_{\mu}-G_{i{\bar j}} \partial_{\mu} z^i
\partial^{\mu} {\bar z}^{\bar j} -\frac{i}{2}(G_{ij}\partial z^j
 -G_{i{\bar j}}\partial_{\mu} {\bar z}^{\bar j})h^{\mu}.
\label{linear2}
\end{equation}
The last term in eqn.(\ref{linear2}) reveals that the one loop 
correction to the K\"{a}hler metric $G_{z,{\bar z}}$ will come
by calculating the CP-odd part of the amplitude between the complex 
scalars and the antisymmetric tensor $b^{\mu \nu}$
\begin{equation}
<z(p_1){\bar z}(p_2)b^{\mu \nu}(p_3)>_{odd} = i \epsilon^{\mu \nu
\lambda \rho}p_{1\lambda}p_{2\rho}G^{(1)}_{z{\bar z}}.
\label{correk1}
\end{equation}
Here, G is the
K\"{a}hler metric and
$h^{\mu}=\frac{1}{2} {\epsilon}^{\mu \nu \lambda \rho}
\partial_{\nu} b_{\lambda\rho}$
is the dual field strength of the antisymmetric tensor field
$b_{{\lambda}{ \rho}}$.

The amplitude receives contributions only from $N{=}2$ sectors.
We are not considering contributions to the 
K\"{a}hler metric 
which arise from $N=1$ sectors, since these contributions  
arise only in $N{=}1$ orbifold compactifications of the  
heterotic string.  Here, we are only interested in the geometry
underlying the $N=2$ sectors.

Lets us suppose that the internal six dimensional lattice 
decomposes into $T^2 \oplus T^4$, with the $T^2$  inside
the unrotated plane.
Compactifications of the heterotic string in four dimensions with 
$N=2$ supersymmetry involve a $U(1) \times U(1)$ gauge group
from the untwisted $T^2$ unrotated subspace. This plane is
parametrized\cite{dis} in terms of moduli T, U. For
special points in the moduli space, namely the $T=U$ line
the gauge group becomes enhanced to $SU(2) \times U(1)$.
It can become enhanced to $SO(4)$ or $SU(3)$ along the 
$T=U=i$ or $T=U= e^{{2\pi i}/3}$ lines respectively.
In this subspace of the Narain moduli space, we will 
be interested mostly, to calculate the moduli dependence
of the one loop correction to the prepotential. 
Denote the untwisted moduli from a $N=2$ sector 
by P, where P can be the T or U moduli
parametrizing\cite{dis} the two dimensional unrotated plane.
Then the one loop contribution\cite{anto3} to the 
 K\"{a}hler metric is given by 
\begin{eqnarray}
G^{(1)}_{P\;{\bar P}}=\frac{1}{(P + {\bar P})^2} {\cal I},\;\;\;
{\cal I}= \int_{\cal F} \frac{d^2\tau}{{\tau}^2_2} \partial_{\bar \tau}
(\tau_2 Z) F({\bar \tau}).
\label{metrik1}
\end{eqnarray}
Here, the integral is over the fundamental domain,
and the factor $\frac{1}{(P + {\bar P})^2}$ is the tree level moduli 
metric $G^{(0)}_{P\;{/bar P}}$. Z is the partition
function of the fixed torus
\begin{equation}
Z = \sum_{(P_L,P_R) \in \Gamma_{2,2}}
q^{P_L^2 /2} {\bar q}^{P^2_R /2} ,\;q \equiv e^{2 \pi \tau} \;
\tau = \tau_1 + \tau_2 ,
\label{partik1}                                                                
\end{equation}
and $P_L, P_R$
are the left and right moving momenta associated with this plane.
$F(\tau)$ is a moduli independent meromorphic form\footnote{A 
function f is meromorphic at a point A if 
the function h, $ h(z) \stackrel{def}{=} (z- A) f(z)$ is 
holomorphic (differentiable) at 
the point A. In general, this means that the function h is 
allowed to have poles.}   
of weight $-2$ with a single pole at infinity due to the tachyon
at the bosonic sector.
The function F was fixed in \cite{afgnt} to be 
\begin{equation}
F(\tau)= -(1/\pi)\frac{j(\tau) [j(\tau) -j(i)]}{j_{\tau}(\tau)},\;\;\;
\;\;\;
j_{\tau }\stackrel{def}{=}\frac{\partial j(\tau)}{\partial \tau},
\label{partik2}
\end{equation} 
where j the modular function for the group $SL(2,Z)$.

%$$$$$$$$$$$$$$$$$$$$$$$$$$$$$$$$$$$$$$$$$$$$$$$$$$$$$$$$$$
%$$$$$$$$$$$$$$$$$$$$$$$$$$$$$$$$$$$$$$$$$$$$$$$
%$$$$$$$$$$$$$$$$$$$$$$$$$$$$$$
%$$$$$$$$$$$$$$$$$$$$$
%$$$$$$$$$$$
%$$$$$$$$$$$$$$$$$$$$$$$$$$$$$$$$$$$$$$$$$$$$$$$$$$$$$$$$$$$$

\subsubsection{* Prepotential of vector multiplets/K\"ahler metric }
 
For the calculation of the prepotential of the vector multiplets
we will will follow the approach of \cite{afgnt}.
Recalling the general form of the prepotential eqn.(\ref{part2})
\begin{equation}
K= -\ln (iY),\;\;\;\;\;\;F = STU + f(T,U).
\label{partik3}
\end{equation}
The lagrangian (\ref{linear1}) may be related to the chiral 
multiplet one (\ref{superco1}), by a duality transformation.
We introduce the dilaton S as a Lagrange multiplier into
(\ref{linear1}), e.g $({\cal L} -L (S+{\bar S})/4)_D$. 
Using the equation of motion for S we get
\begin{equation}
({\cal L} -L \partial {\cal L})_D \equiv \frac{-3}{2}S_o {\bar S}_o
e^{-\frac{K}{3}}.
\label{partik21}
\end{equation}
In this form the K\"{a}hler potential has an expansion
as
\begin{equation}
K=-\ln\{(S - {\bar S}) -2G^{(1)} \} + G^{(o)}
\label{partik31}
\end{equation}
Expanding {\ref{partik1}) 
\begin{eqnarray}
K^{(1)}_{P{\bar P}}=\frac{2i}{(S-{\bar S})}G^{(1)}_{P{\bar P}},\;\;
\;\;G^{(1)}_{T\;{\bar T}}=\frac{i}{2(T-{\bar T})^2}
\left(\partial_T -\frac{2}{T-{\bar T}}\right)\left(\partial_U - 
\frac{2}{U - {\bar U}}\right)f + c.c.
\label{partik4}
\end{eqnarray}

Using the equations for the momenta
\begin{equation}
{p_L}=\frac{1}{\sqrt{2ImT ImU}}(m_1+ m_2{\bar U} + n_1 {\bar T}+n_2 
{\bar U}
{\bar T}),\;\;p_R =\frac{1}{\sqrt{2ImT ImU}}(m_1+ m_2{\bar U}+ n_1 T+
n_2 T{\bar U})
\label{partik5}
\end{equation}
we can prove that $ \cal I$ satisfies\cite{anto3} the following
differential equation
\begin{equation}
\{\partial_T \partial_{\bar T}+\frac{2}{(T-{\bar T})^2}\}{\cal I}=
-\frac{4}{(T-{\bar T})^2}\int d^2\tau {\bar F}({\bar \tau})
\partial_{\tau} (\partial_{\bar \tau}^2+\frac{i}{\tau_2} 
\partial_{\bar \tau})
(\tau_2 \sum_{P_L,P_R} q^{P_L^2 /2} {\bar q}^{P^2_R /2}).
\label{partik6}
\end{equation}
The integral representation of eqn.(\ref{partik6}) is a total 
derivative 
with respect to $\tau$ and thus zero.
However, the integral can give non-vanishing contributions at the 
enhanced symmetry points $T{=}U$. 
Solving (\ref{partik6}) away of the 
enhanced symmetry points gives 
\begin{equation}
\{\partial_T \partial_{\bar T}+\frac{2}{(T-{\bar T})^2}\}{\cal I}=
\{\partial_U \partial_{\bar U}+\frac{2}{(U-{\bar U})^2}\}{\cal I}=0.
\label{partik55}
\end{equation}
The singularity structure of (\ref{partik55}) at the enhanced symmetry
 point\footnote{Enhanced symmetry point behaviour at a general point
in the moduli space has been examined
in chapter three. Direct application to the momenta of 
eqn's.(\ref{partik5}),
shows that they correspond to the lattice points $m_2=-n_1=\pm 1$,
$m_1 = n_2 = 0$ and gauge group enhancement from $U(1)  \times U(1)
\rightarrow U(1) \times SU(2)$.}
will be taken into consideration later in its integral
representation. 
The general solution of the (\ref{partik5}) is\cite{afgnt}
\begin{equation}
{\cal I}=\frac{1}{2i}\left(\partial_T - \frac{2}{(T-{\bar T})}\right)
\{(\partial_U  - \frac{2}{U - {\bar U})})f(T,U)+
(\partial_{\bar U} + \frac{2}{U - {\bar U}}){\bar f}(T,{\bar U}\}+c.c.
\label{partik7}
\end{equation}
It can be shown\cite{afgnt} that $\bar f$ is zero. Note that 
f represents the one-loop correction to the prepotential of the 
vector multiplets T, U and determines via eqn.(\ref{partik4}) the one 
loop correction to the K\"{a}hler metric for the T, U moduli.
In \cite{afgnt} it was shown function $f(T,U)$ of (\ref{partik7})
satisfies the differential equation     
\begin{equation}
-i(U -{\bar U}) D_T \partial_T \partial_{\bar U} 
{\cal I}=\partial_T^3 f,
\label{partik8}
\end{equation}
where $D_T = \partial_T + \frac{2}{(T - {\bar T})}$ is the 
covariant derivative. 
Expansion of the l.h.s and integration by parts results in
\begin{equation}
f_{TTT}=4 \pi^2 \frac{U-{\bar U}}{(T -{\bar T})^2}\int d^2 \tau
{\bar F}({\bar \tau})\sum_{P_L,P_R} P_L {\bar P}_R^3
 q^{P_L^2 /2} {\bar q}^{P^2_R /2}.
\label{partik9}
\end{equation}
Examination of the behaviour of the r.h.s of eqn.(\ref{partik9}) under 
separately modular transformations $SL(2,Z)_T$ , $SL(2,Z)_U$, together
with examination of its singularity structure at the enhanced symmetry
point $T{=}U$, uniquely determines the well known solution
of the third derivative of the vector multiplet prepotential.
Remember that we examine the behaviour of the prepotential
including the region of the moduli space where we have gauge symmetry
enhancement to $U(1) \times SU(2)$. 

For
$N{=}2$ heterotic strings compactified on decomposable orbifolds
\begin{equation}
f_{TTT}= -{2i\over\pi}\frac{j_T(T)}{j(T)-j(U)}
\left\{\frac{j(U)}{j(T)}\right\}
\left\{\frac{j_T(T)}{j_U(U)}\right\}
\left\{\frac{j(U)-j(i)}{j(T)-j(i)}\right\}.
\label{partik10}
\end{equation}
In \cite{wkll,afgnt} $f_{TTT}$ was determined by the property
of behaving as a meromorphic modular form of weight 4 in T. 
In addition, $f_{TTT}$ had to vanish at the order 2 fixed point
$U{=}i$ and the order 3 fixed point $U{=}\rho$ of the 
modular group $SL(2,Z)$. Moreover, it had to transform with modular 
weight $-2$ in U under $SL(2,Z)_U$ transformations and
exhibit a singularity at the $T{=}U$ line. 
The prepotential function for the $f_{UUU}$ is obtained by the 
replacement $T \leftrightarrow U$. 

Here, we find the the equation for $f_{UUU}$. In simple form
\begin{equation}
\partial_{U}^3f = f_{UUU}= -i(T-{\bar T})^2 \partial_{\bar T}D_U 
\partial_U {\cal I},
\label{oux1}
\end{equation}
where $D_U= \partial_U + \frac{2}{U -{\bar U}}$, the covariant derivative
with respect to U variable, transforms with modular weight 2 
under $SL(2,Z)_U$ modular transformations, namely
\begin{eqnarray}
U\stackrel{SL(2,Z)_U}{ \rightarrow} \frac{ aU + b}{cU + d},\;\;\;\;D_U \rightarrow (cT+d)^2 D_U.
\label{covartibat}
\end{eqnarray}
We should notice here, that because of the symmetry exchange $T
\rightarrow U$, the result for $f_{UUU}$ comes directly from
(\ref{partik10}), by the replacement $T \rightarrow U$. However, this 
can be confirmed by the solution of (\ref{oux1}).  
In addition, we will find the differential equation for the function f.
The calculation of the prepotential f comes from the identity
\begin{equation}
(f)_{proj}=  2i(T-{\bar T})^2 (U-{\bar U})^2 \partial_{\bar U}\partial_{\bar 
T}{\cal I}.
\label{oux2}
\end{equation}
Explicitly, 
\begin{equation}
(f)_{proj}= 2i(T-{\bar T})^2 (U-{\bar U})^2 \partial_{\bar U}\partial_{\bar
T} \int_{\cal F} \frac{d^2\tau}{{\tau}^2_2} \partial_{\bar \tau}
(\tau_2 Z) F({\bar \tau}).
\label{ouxx2}
\end{equation}
As we can see 
the one loop correction to the holomorphic prepotential comes by 
taking derivatives of ${\cal I}$ with respect to the conjugate 
moduli variables
from which the holomorphic prepotential does not have any dependence.
The holomorphic prepoetential is defined projectively, by taking the
action
of the conjugate moduli derivatives on the holomoprhic part of the 
one loop K\"ahler metric integral $\cal I$.
In this way, we 
{\em always produce} the differential equation for the f function 
from the string amplitude. In addition, the solution of this equation 
calculates the one loop correction to the K\"ahler metric.  
Using now, the modular transformations of the momenta
\begin{eqnarray}
(P_L, {\bar P}_R)\stackrel{SL(2,Z)_T}{\rightarrow} \left({\frac{cT+d}{
c{\bar T}+d}}\right)^{\frac{1}{2}} (P_L, {\bar P}_R),\;\;
(P_L, {\bar P}_R)\stackrel{SL(2,Z)_U}{\rightarrow}
\left({\frac{cU+d}{
c{\bar U}+d}}\right)^{\frac{1}{2}} (P_L, {\bar P}_R),
\label{oux5}
\end{eqnarray}
we can easily see that the one loop prepotential has the correct modular
properties, it transforms with modular weight $-2$ in T and $-2$ in U.
Eqn.(\ref{oux2}) is the differential equation that the one loop
prepotential
satisfies. The solution of this equation determines 
the one loop correction
to the K\"ahler metric and the K\"ahler potential for $N=2$ orbifold 
compactifications of the heterotic string. 
Compactifications of the heterotic string on $K_3 \times T_2$,
appears to have the same moduli dependence on T and U moduli, for 
particular classes of models\cite{afgnt,kava,klm1,afiq1}.
Formally, the same routine procedure, namely {\em taking the 
derivatives with 
respect to the conjugate T and U moduli on} ${\cal I}$, can be applied 
to any heterotic string amplitude between two moduli scalars and  
antisymmetric tensor, in order to isolate from the general solution 
(\ref{oux2}) the term $f(T,U)$. 
The solution for $f_{TTT}$ in 
eqn.(\ref{partik10}) was derived for $N=2$ compactification of the  
heterotic strings in \cite{afgnt} via the modular properties of the one 
loop 
prepotential coming from the study of its integral representation
(\ref{partik9}). Specific application for the model based on the 
orbifold
limit of $K_3$, namely $T^4/Z_2$, was given in \cite{mouha}. 
At the orbifold limit of $K_3$ compactification of the heterotic string 
the Narain lattice was decomposed into the form $\Gamma^{22,6}=
\Gamma^{2,2} \oplus \Gamma^{4,4} \oplus \Gamma^{16,0}$. 
It was modded by a $Z_2$ twist on the $T^4$ part together with a $Z_2$ 
shift $\delta$
on the $\Gamma^{(2,2)}$ lattice. For reasons of level matching 
$\delta^2$ was chosen to be $1/2$. 
By an explicit string loop calculation via the one loop gauge couplings
in \cite{mouha}, from where the one loop prepotential was extracted with
an ansatz, they were able to calculate the third derivative of the
prepotential. It was found to agree with the result 
of \cite{wkll,afgnt} which was calculated for the S-T-U subspace  
of the vector multiplets of the orbifold compactification of the 
heterotic string.

In reality, $F(\tau)$ is the trace of $F (-1)^F q^{L_o - 
\textstyle{\frac{c}{24}}} {\bar q}^{{\bar L}_o -
\textstyle{\frac{c}{24}}}/{\eta}({\bar \tau})$ over the Ramond 
sector boundary conditions of the remaining superconformal blocks.
For the S-T-U model with instanton embedding $(d_1,d_2)=(0,24)$ 
the supersymmetric index was calculated in \cite{mouha}
in the form
\begin{equation}
\frac{1}{\eta^2} {Tr_R} F{(-1)}^F q^{L_o - \frac{c}{24} }
{\bar q}^{{\bar L}_o -\frac{c}{24}}\; =\; -2i \frac{E_4\; E_6}{\Delta},
\label{partik233}
\end{equation}
where F is the right moving fermion number. 
Expanding $\cal I$ we get that
\begin{equation}
{\cal I}=(-i \pi)\int \frac{d^2 \tau}{\tau_2}(p_R^2 -\frac{1}{2\pi \tau_2})
{\bar F}({\bar \tau}).
\label{polik1}
\end{equation}
Specific tests of dual pairs were performed, in the spirit of 
\cite{mouha}, in \cite{dose1,dose2,dose3,dose4,dose5}.
Particular examples of calculating the prepotential for dual pairs
will not be performed here.

We have said that one important aspect of the expected duality is that
the vector moduli space of the heterotic string must concide at the 
non-perturbative level with the tree level exact vector moduli space 
of the type IIA theory. 
For the type IIA superstring compactified on a Calabi-Yau space
X the internal $(2,2)$ moduli space, $N=2$ world-sheet supersymmetry for
the left and the right movers,
is described, at the large complex structure limit of X, by
the K\"ahler\footnote{Let us consider the target space of a complex 
manifold
${\cal M}$ with dimension n. Choose coordinates on ${\cal M}$, $\phi_m$
and ${\bar \phi}_m$. Then ${\cal M}$ 
admits 
a K\"ahler structure if we can define a $(1,1)$ form J with the 
property $J= iG_{l {\bar m}} d\phi_m  \wedge d{\bar \phi}_l$ where for a 
 K\"ahler manifold the metric is $G_{l {\bar m}}= \partial_{\phi^m}
\partial_{{\bar \phi}_l}K=\frac{\partial}{\partial \phi^m}
\frac{\partial}{\partial {\bar \phi}_l} K$, and the  K\"ahler potential is K.}
moduli,  namely  $B+iJ \in H^2(X,C)$,
where $B+iJ = \sum_{i=1}^h{(1,1)} (B+iJ)_a e_a$ with $B_a$, $J_a$ real
numbers and $t_a =(B+iJ)_a$ representing the special coordinates and 
$e_a$ a basis of $H^2(X,C)$.
In the content of the moduli of the Calabi-Yau space of X, the
holomorphic
prepotential at the large radius limit takes the form
\begin{equation}  
F= -\frac{i}{6}\sum_{\alpha,\beta,\gamma}(D_{\alpha} \cdot D_{\alpha}
\cdot D_{\gamma})t_{\alpha} t_{\beta} t_{\gamma}
-\frac{\chi \zeta(3)}{2 (2\pi)^3}\sum_{(d_i)_{i=1,\dots,n}} n_{d_1,\dots,d_n}
{\cal L}i_3 (\Pi_{i=1}^{n} q_i^{d_i}),
\label{koufme1}
\end{equation}
where ${\cal L}i_3(x)= \textstyle{\sum_{j \geq 1}\frac{x^j}{j^3}}$. The
first term in eqn.(\ref{koufme1}) is a product of the the Calabi-Yau 
divisors D, associated to the basis $e_a$, 
and the second term\cite{can} represents world-sheet instanton 
contributions. The $n_{d_1,\dots,d_n}$ are the world sheet instanton 
numbers.
Performing duality tests between a heterotic model and its possible
type IIA dual is then equivalent to comparing the weak coupling limit 
of the prepotential\cite{kava} of the vector multiplets for the 
heterotic string with the large radius limit of (\ref{koufme1}).
After identifing the heterotic dilaton with one of the vector moduli
of the type IIA model in the form $t_s =(B+iJ)_s = 4\pi i S$,
the type IIA prepotential takes the general form\cite{cande}
\begin{equation}
{\cal F}_{IIA} =-\frac{1}{6}C_{ABC} t^A t^B t^C -\frac{\chi \zeta(3)}
{2(2\pi)^3}\sum_{d_1\dots,d_n} n_{d_1,\dots,d_n} {\cal L}_{i_3}e^{2\pi 
i[\sum_k d_k t^k]},
\label{koufeme2}
\end{equation}
where we are working inside the K\"ahler cone\cite{can,hoso2} 
$\sigma\stackrel{def}{=}\{\sum_{\rho} t^{\rho} J_{\rho}|
t^{\rho}> 0 \}$, where $J^{\rho}$ generate the cohomology group 
$H^2(X,R)$ of the Calabi-Yau three fold X.
In a particular symplectic basis eqn.(\ref{koufeme2}) can be 
brought in the form\cite{cardoo1}
\begin{equation}
{\cal F}_{IIA} = -\frac{1}{6} C_{ABC}t^A t^B t^C +\sum_{A=1}^{h_{1,1}}
\frac{c_2 \cdot J_A}{24} t^A + \dots,
\label{koufeme3}
\end{equation}
where $\textstyle{c_2 \cdot J_A = \int_X c_2 \wedge J_A}$ is the 
expansion of the second Chern class of the Calabi-Yau three fold in terms
of the basis $J^{*}_A$ of the cohomology group $H^4(X,R)$. The
cohomology group $H^4(X,R)$
is dual to the $H^2(X,R)$, namely 
$\textstyle{\int_X J^{*}_A \wedge J_B = \delta_{AB}}$.   
In \cite{klm1} it was noticed that the nature of type II-heterotic sting
duality test has to come from the $K_3$ fiber structure over $P^1$
of the type IIA side.                                                                                                                                    
The form of the $K_3$ fibration can be found\cite{klm1,afiq1} 
by taking for example
the CY in $P^4(1,1,2k_2,2k_3,2k_4)$ and then set $x_o =\lambda x_1$. 
After rescaling $x_1 \rightarrow x_1^{1/2}$ we arrive at the equation
for the hypersurface 
\begin{equation}
F(\lambda)Z_1^d +Z_2^{d/{k_2}} + \cdots = 0.
\label{koufeme4}
\end{equation}
The degree $d=1+k_2 + k_3 + k_4$. For generic values of $\lambda$ 
eq.(\ref{koufeme4}) is a $K_3$ surface in weighted $P^3$. So 
$P^4(1,1,2k_2,2k_3,2k_4)$ is a $K_3$ fibration
fibered over the $P^1$ base which is parametrized by $\lambda$.
At the large radius limit
of X, in (\ref{koufeme3}), the heterotic dilaton S
is identified as one of the vector multiplet variables  
as $t_s= 4 \pi i S$.
Confirmation of duality between dual pairs is then  
equivalent to the identification\cite{louisne}
\begin{equation}
\textstyle{{\cal F}_{IIA}={\cal F}_{IIA}(t^s, t^i) +
{\cal F}_{IIA}(t^i)={\cal F}_{het}^o(S, \phi^I) +
{\cal F}_{het}^{(1)}(\phi^I)}.
\label{koufame11}
\end{equation}
Here,
we have expand the prepotential of the type IIA in its large radius
limit, namely large $t_s$. In the heterotic side, we have the 
tree level classical contribution 
as a function of the dilaton S and the vector multiplet moduli 
$\Phi^I$, in addition to the one loop correction as a function 
of only the $\Phi^I$.
The general differential equation for the one loop correction to the
heterotic prepotential ${\cal F}_{het}^{(1)}$ was given before by
eqn.(\ref{oux2}).
Summarizing, the existence of a type II dual to the weak coupling phase 
of the heterotic string is exactly the existence of the 
conditions\cite{lla}
\begin{equation}
D_{sss}=0,\;\;\;\;D_{ssi} =0\; for\;every\;i,\;.   
\label{koufame12}
\end{equation}
Moreover, from eqn.(\ref{treecoupl2}) we see that the tree level 
heterotic prepotential can be expanded \cite{FVP} in the form 
\begin{equation}
{\cal F}^{(o)} =  -S(\eta_{ij} M_i M_j -\delta_{ij}Q^I Q^J),
\;\;\eta_{ij} = diag(1,-1,\dots,-1),
\label{koufame13}
\end{equation}  
from which we can infer that 
\begin{equation}
sign(D_{sij}) = (+,-,\dots,-)=sign(\eta_{ij}).
\label{koufame14}
\end{equation}
However, there is another condition which will be usefull.
It originates from the higher derivative gravitational couplings
of the heterotic vector multiplets and the Weyl multiplet
of conformal $N=2$ supergravity\cite{gavos1}. The relevant couplings
originate from terms in the form $g_n^{-2} R^2 G^{2n-2}$,  where
R is the Riemann tensor, G the field strength of the graviphoton.
The $g_n$ couplings obey $g_n^{-2}=Re {\tilde F}_n(S,M^i) + \dots$.
The same of couplings appear in type II superstring\cite{gavos1,berda}. 
In the heterotic side they take the form
\begin{equation}
{\tilde F}_n = {\tilde F}^{(0)}(S, M_i) + {\tilde F}^{1}(M^i)
+ {\tilde F}^{NP}(e^{-8\pi^2 S}, M^i),\;\;{\tilde F}_1=24 S,\;\;
{\tilde F}^o_{n \geq 1} = const,
\label{koufame141}
\end{equation}
where S is the heterotic dilaton and $M^i$ the rest of the vector
multiplets moduli.
Such terms appear as well in the effective action of type II vacua 
and they have to match with heterotic one's due to duality.
In the large radius limit the higher derivative
couplings satisfy(the lowest order coupling)
${\tilde F}_1 \rightarrow -\frac{4 \pi i}{12}\sum_{a} 
(D_a \cdot c_2) t_a $, 
where $c_2$ is the second Chern class of the three fold X.
Because at the tree level, $g_1^{2}= Re {\tilde F}_1$ we can infer the 
result that
\begin{equation}
D_a \cdot c_2(X) =24.  
\label{koufame15}
\end{equation}
The last condition represents\cite{lla} the mathematical
fact that the Calabi-Yau threefold X admits a fibration $\Phi$ such as
there is a m ap $X \rightarrow W$, with the base being $P^1$ and generic 
fiber the $K_3$ surface.
Furthermore, the area of the base of the fibration gives
the heterotic four dimensional dilaton.

%$$$$$$$$$$$$$$$$$$$$$$$$$$$$$$$$$$$$$$$$$$$$$$$$$$$$$$$$$$$$$

\subsection{* One loop prepotential - perturbative aspects}

Since we have finished our discussion of the general properties of the  
$N=2$ heterotic strings, we will now discuss the calculation of the 
perturbative corrections to the one loop prepotential.

Let us expand at the moment the expression of eqn.(\ref{zorika1})  
around small values of the non moduli scalars $C_a$ as in
(\ref{Ktree}) and (\ref{treecoupl2})
\begin{equation}
F = -S(TU - \sum_i \phi^i \phi^i) + {h}^{(1)}(T,U,\phi^i),\;\;
f_{(a)} = S + {f_{(a)}}^{(1)}(T,U,\phi^i),
\label{hfexp}
\end{equation}
where the functions ${h}^{(1)}$ and $ f_{(a)}^{(1)}$  
enjoy a non-renormalization theorem, namely they receive perturbative
corrections only up to one loop order. Its higher loop corrections
in terms of the $1/{(S+ \bar S)}$ vanish,
due to the surviving of the discrete Peccei-Quinn 
symmetry to all orders of perturbation theory as a quantum symmetry.
For the same reason, $f_{(a)}^{(1)}$ receives corrections, only
up to one loop level. 
The one loop prepotential, if we expand it in the general form
$F(X)= H^{(0)}(X) + H^{(1)}(X)$ with
$H^{(1)}(X) = -i(X^{0})^{2} {\Omega}^{(1)}$
where the superscripts denote tree lavel
and one loop corrections respectively gives us through relations 
related
to the basis $X^{I}, F_{J}$ that the following relations are valid
\begin{equation}
{\hat F_{I}} = -2 i S_{IJ} {\hat X^{J}} + {H^{(1)}}_{I}\;,
H^{(1)}(X) = \frac{1}{2}{\hat F_{I} } {\hat X^{I} } 
\end{equation} 
with 
\begin{equation}
{H^{(1)}}_{I} = {\partial {H^{(1)}}}/{\partial X^{I} }
\label{router1}
\end{equation}
The loop corrections to the  prepotential have to take into account 
the generation
of the discrete shifts in the theta angles due to monodromies around 
semi-classical singularities in the quantum moduli space where 
previously massive states become massless.
In this way, the classical transformation rules of 
(\ref{dualmat}) become modified to 
\begin{equation}
{\hat X}^{I}  \rightarrow {{\hat U}^{I}}_{J} \;
{\hat F_{I} } \rightarrow {{\hat V}_{I}}^{J} +  {\hat W}_{IJ} P^J
\label{oltrans}
\end{equation}
with
\begin{equation}
{\hat V} = {\hat U}^{\rm T\;-1},\;
{\hat W} = {\hat V }{\Lambda},\; {\Lambda}={\Lambda}^{T}.
\end{equation}
and $\hat U \in SO(2,P + 2,Z) $. In the classical theory $\Lambda = 0$
but in the full quantum theory around a singularity, the closed 
monodromy
gives rise to 
${\hat F_{I}} \rightarrow {\hat F_{I}} + {\Lambda}_{IJ} {\hat F_{I}} $
and the transformation rule of the one loop prepotential becomes
\begin{equation}
H^{(1)}{(\tilde X)} = H^{(1)}(X) + 1/2 {\Lambda}_{IJ} {\hat X}^{I}
{\hat X}^{J}.
\end{equation}
As a consequence, 
the one loop prepotential changes by a quadratic polynomial
in T and U when moving around a semi-semiclassical singularity.
In the lanquage of special geometry this reads\cite{wkll}
\begin{equation}
f(T,U) \rightarrow (icT + d)^{-2} ( f(T,U) + {\Pi}(T,U).
\label{second1}
\end{equation}
A special aspect of the theory is related   
to the transformation rule of the dilaton.
At the level of string theory the dilaton vertex has a 
fixed relation to the vector tensor multiplet and it is 
invariant under any symmetries of the string theory. However
when we are discussing the vector multiplet which is dual to the 
vector tensor multiplet the dilaton is no longer invariant
under 
the perturbative symmetries of string theory and
is receiving perturbative corrections. It follows via the relations
(\ref{router1}), (\ref{oltrans}) and the relation 
$X^1 = -{\hat F}_{1} = -i S {\hat X}^{o}$
that the dilaton transforms as
\begin{equation}
S \rightarrow {\tilde S} = S + \frac{i { {\hat V}_{1}}^{J}
({H_{J}}^{(1)} +
\Lambda_{JK}{\hat X}^{K})}{{\hat U}^{o}_{I} {\hat X}^{I}}.
\label{router2}
\end{equation}
But if we insist in keeping the dilaton invariant then we can define
an 'invariant' dilaton as
\begin{equation}
S^{inv} = S + \frac{1}{2(P+4)} \left[ i {\eta}^{IJ} {H_{IJ}}^{(1)} + 
L \right],
\label{dilinv}
\end{equation}
where $L$ obeys $ L \rightarrow  L - i {\eta}^{IJ} {\Lambda_{IJ}}$.
We will find now the transformation properties of the non-moduli gauge 
couplings $f_{(a)}(\Phi)$.
When we are discussing about the physical properties of a low energy 
theory,
we have to distinquish about the momentum dependent physical gauge 
couplings and the Wilsonian gauge couplings. 
The effective gauge couplings account for all the quantum effects both
at high and at low energy. As a result the low energy effects due to 
massless particles give rise 
to non-holomorphic moduli dependence of the effective gauge couplings
and to all orders of perturbation theory it has been found \cite{kalou1}
that
\begin{equation}
{g_{a}}^{-2}(\Phi,{\bar \Phi},p^2) =  Ref_{(a)}(\Phi) +
{\frac{b_{a}}{16 {\pi}^2} }( \log \frac{M_{PL}^{2}}{p^2} +
K_{\Phi}(\Phi, {\bar \Phi})) + constant.
\end{equation}
Finally the Wilsonian couplings $f_{(a)}$ transform as
\begin{equation}
f_(\Phi) \rightarrow\ f_(\Phi)
-\ {b_{(a)} \over 8 \pi^2} \,\log({{\hat U}^0}_{J} X^J/X^0).
\label{fcompp}
\end{equation}

In that case, under target space duality 
\begin{eqnarray}
T \rightarrow {\frac{aT-ib}{icT+d}},\;\;& U \rightarrow U, \nonumber\\
h(T,U) \rightarrow {{h(T,U)+ \Xi(T,U)} \over (icT+d)^2},\;&
f_{a}(S,T,U) \rightarrow f_{a}(S,T,U) - \frac{b_{a}}{8\pi^2}
\log(icT+d)
\label{Ton1h}
\end{eqnarray}
and a similar set of transformations under $PSL(2,Z)_{U}$. The net
result is that ${\partial}_{T}^{3} h^{(1)}(T,U)$ is a singled valued 
function of weight $-2$ under U-duality and $4$ under T-duality. 

We turn now our previous discussion to the case of $N=2$ orbifold
compactifications of six dimensional heterotic string vacua\cite{dhvw}.
The one loop correction to the prepotential of vector multiplets
for the subspace of the Narain lattice corresponding to the 
T, U moduli of the decomposable $T^2$ torus has already been calculated
in \cite{wkll,afgnt}. In explicit form may be derived from 
eqn.(\ref{partik10}).
In this section of the Thesis we will discuss the calculation of 
the prepotential for the case where the moduli subspace of the 
Narain lattice associated with the T, U moduli exhibits a modular 
symmetry group $\Gamma^o(3)_T \times \Gamma^o(3)_U$. 
The same modular symmetry group appears\cite{deko1} in the $N=2$ sector
of the $N=1$ $(2,2)$ symmetric non-decomposable $Z_6$ orbifold defined 
on the lattice $SU(3) \times SO(8)$. In the third complex plane associated
with the square of the complex twist $(2,1,-3)/6$ the mass operator
for the untwisted subspace was given to be
\begin{equation}
m^2= \sum_{m_1, m_2, n_1, n_2\;\in\;Z} \frac{1}{2T_2U_2}
|TU^{\prime}n_2 + T n_1 - U^{\prime}m_1 + 3m_2|^2.
\label{arga1}
\end{equation}

Any $(2,2)$ orbifold will flow to its Calabi-Yau limit  after
giving vacuum expectation values to its twisted "moduli" 
scalars\cite{HV}.
In this limit, the corresponding Calabi-Yau phase exist.
Let us forget the $N=1$ orbifold nature of the apppearance
of this $N=2$ sector. Then its low energy supergravity theory 
is described by the underlying special geometry. 
The question now is if calculating the prepotential using its modular 
properties and the singularity structure, as this was calculated
for decomposable orbifold compactifications of the heterotic 
string\cite{wkll}, there is a type II dual realization.
We believe that it is the case.
In the analysis of the map between type II
and heterotic dual supersymmetric string theories\cite{liya,klm1}  
it was shown that subgroups of the modular group do appear.
In particular in one modulus deformations of $K_3$ fibrations
the modular symmetry groups appearing are all connected
to the $\Gamma_o(N)_{+}$, the subgroup
of the $PSL(2,Z)$, the $\Gamma_o(N)$ group together
with the Atkin-lehner involutions $T\rightarrow \frac{-1}{NT}$.
In certain\cite{klm1} type II models, e.g the surface 
$X_{24}(1,1,2,8,12)^{-480}_{3}$, 
the $K_3$ fiber is a two moduli system $X_{12}(1,1,4,6)$.
In a certain complex structure limit the $K_3$ fiber 
degenerates to a $K_3$
elliptic fibration $X_6(1,2,3)$, it look locally as a torus, over $P^1$  with modular 
groups connected to e.g $\Gamma(3)$ and $\Gamma(2)$.
We expect that the same prepotential, beyond describing the  
geometry of the $N=2$ sector of $Z_6$ in exact analogy to the  
decomposable case, may come form a compactification of the heterotic
string on the $K_3 \times T^2$. An argument that seems to give some support
to our conjecture was given in \cite{fava1}.
%In general, if we compactify a ten dimensional string theory 
%to four dimensions on $T^2 \times X$, with X any four manifold,
%we get a duality group $SL(2,Z)_T \times SL(2,Z)_U$. 
It was noted by Vafa and Witten 
that if we compactify a ten dimensional string theory on $T^2 \times X$,
where X any four manifold, acting with a $Z_2$ shift on the Narain lattice 
we get the modular symmetry group $\Gamma_o(2)_T \times \Gamma_o(2)_U$.
They even describe the Narain lattice that exhibits this symmetry.
In this respect it is obvious that our calculation of the prepotential  
which we present in this Thesis, may come from a shift
in a certain Narain lattice of $T^2$. We suspect that this is a $Z_3$
shift but we were not able to prove it.
From the mass operator (\ref{arga1}) we deduce that at the point $T=U$
in the moduli space of the $T^2$ torus of the untwisted plane, with
$n_=m_1=\pm 1$ and $n_2=m_2=0$, its $U(1) \times U(1)$ symmetry
becomes enhanced to $SU(2) \times U(1)$. Moreover, the third derivative
of the prepotential has to transform, in analogy to the 
$SL(2,Z)$ case, with modular weights -2 under ${\Gamma^o(3)}_U$ and
4 under $\Gamma^o(3)_T$ dualities.
The Hauptmodul function, the analog of $SL(2,Z)$ j-invariant,  for
$\Gamma^o (3)$ is the function $\omega$ described in chapter three.
The function $\omega^o(3)$ has a single zero at zero
and a single pole
at infinity. In addition, its first derivative
has a first order zero at zero, a pole at infinity and
a first order zero at $i{\sqrt{3}}$.
The modular form F of weight k of a given subgroup of the 
modular group $PSL(2,Z)=SL(2,Z)/{Z_2}$ is calculated 
from the formula 
\begin{equation}
\sum_{p \neq 0, \infty }{\nu_p} + \sum_{p=0,\infty} \frac{1}{m}
\times (order\;of\;the\;point) = \frac{\mu k}{12}.
\label{eksiso1}
\end{equation}
Here, $\nu_p$ the order of the function F, the lowest power in the  
Laurent expansion of F at p and m is the
ordre of the subgroup fixing the point.
 The index $\mu$ for $\Gamma^o(3)$
is calculated from the expression\cite{shoe}
\begin{equation}
[\Gamma: \Gamma_o(N)] =N \Pi_{p/N}\;(1+p^{-1})
\label{extresi1}
\end{equation}
equal to four. The sum of the widths at all cusps is equal to the index
of the subgroup of $PSL(2,Z)$.
The width at infinity is defined as the smallest integer
such as the transformation $z \rightarrow ( z + \alpha )$ is in the group,
where $\alpha \in Z$. 
The width at zero is coming by properly transforming the width at infinity
at zero.
For $\Gamma^o(3)$ the width at $\propto$ is 3 and the width at zero is 1.
The holomorphic prepotential can be calculated easily if we examine its
seventh derivative. The seventh derivative has modular weight
12 in T and 4 in U. In addition, it has a sixth order pole at the 
$T=U$ point whose coefficient A has to be fixed in order to produce the
logarithmic singularity of the one loop prepotential.
As it was shown\cite{afgnt,wkll} the one loop 
prepotential as T approaches $U_g = \frac{aU + b}{cU + d}$, where 
g is an $SL(2,Z)$ element\footnote{The same argument works for the 
subroups of the modular group, but now there are 
additional restrictions on the 
parameters of the modular transformations. }
\begin{equation}
f \propto -\frac{i}{\pi} {\{(cU+d)T -(aU + b)\}^2} \ln(T - U_g).
\label{opli11}
\end{equation}
The seventh derivative of the prepotential is calculated to be
\begin{equation}
f_{TTTTTTT} = A \frac{\omega(U)_U^3  \omega(U)^5 (\omega^{\prime}(U))^3}
{(\omega(U) -\omega(\sqrt{3}))^2 (\omega(U) -\omega(T))^6} X(T),
\label{teleios1}
\end{equation}
where $X(T)$ a meromorphic modular form with modular weight 12 in T.
The complete form of the prepotential is
\begin{eqnarray}
f_{TTTTTTT} = A \left(\frac{[\omega(U)_U^3  \omega(U)^5 
(\omega^{\prime}(U))^3}{
{(\omega(U) -\omega(\sqrt{3}))^2 [(\omega(U) -\omega(T))^6}]}\right)
\left(\frac{\omega(T)_T^6}{\omega^2(T)\{(\omega(T)-
\omega (\sqrt{3}))^4 \}}\right).
\label{teleios2}
\end{eqnarray}

The two groups $\Gamma^o(3)$ and $\Gamma_o(3)$ are conjugate to each 
other. If S is the generator 
\begin{eqnarray}
S= \left(\begin{array}{cc}
0&-1\\
1&0
\end{array}\right),\;\;we\;have\;\Gamma^o(3)=S^{-1} \Gamma_o(3) S.
\label{eftasa1}
\end{eqnarray}
So any statement about modular functions on one group is a statement 
about the
other. We have just to replace everywhere $\omega(z)$ by $\omega(3z)$ 
to go from a
modular function from the $\Gamma^o(3)$ to the $\Gamma_o(3)$.
In other words, the results for the heterotic prepotential with
modular symmetry group $\Gamma^o(3)$ may well be described by the
prepotential of the conjugate modular theory.

We have calculated the prepotential of a heterotic string
with a $\Gamma^o(3)_T \times \Gamma^o(3)_T \times Z_2^{T 
\leftrightarrow U}$ classical duality group.
The same dependence on the T, U moduli and its modular symmetry  group
appears in the
$N=2$ sector of the $Z_6$ orbifold defined on the six dimensional      
lattice $SU(3) \times SO(8)$, namely the $Z_6$-IIb.
The effective theory of the T, U moduli $N=2$ sector of 
the $Z_6$-IIb
orbifold, appears in $N=1$ symmetric orbifold compactifications of the
heterotic string. 
Recall now the discussion in (\ref{koufame14}).
In general one expects $sign(d_{sij})=(+,-,\dots,-,0,\dots,0)$, where
the non zero entries correspond to the moduli from the generic $K_3$ 
fibre. The
zero entries correspond to singular fibres, fibers which degenerate
at points in the moduli space to non $K_3$ surfaces like a smooth 
manifold, and correspond to the heterotic side to strong  
coupling singularities\cite{witta}.
Because of the maximum number of $K_3$ moduli 20, the number of 
generic fibers is constrained to be less than 20. 
The perturbative heterotic vacuua correspond to moduli of the type IIA
coming from the generic fibres. 

We believe that the nature of the lattice twist
of non-decomposable orbifolds is such that its form when acting
on the $N=2$ planes may correspond to orbifold limits of $K_3$.
In this phase, the $K_3$ surface can be
written as an orbifold of $T^4$. The fixed points of $T^4$ under the 
orbifold action are the singular limits of $K_3$ because the metric
on the fixed point develops singularities. The singularities 
of $K_3$ follow an ADE classification pattern. In fact, because
at the adiabatic limit\cite{fava1}, 
we can do even the reverse, we can map the type II phase to the heterotic 
one. In the limit where the base of the fibration has a large area, but 
the volume of the $K_3$ fiber is of order one, we can replace the 
$K_3$ fibers with $T^4$ fibers. In this form, the heterotic $K_3 \times T^2$
compactification is replaced by a heterotic string description of the $T^4$
fibers, namely the Narain lattice $\Gamma^{20,4}$.

\newpage
\vspace{5cm}
\begin{center}
{\bf CHAPTER 5}
\end{center}
\newpage
\section{\bf Superpotentials with T and S-duality and Effective 
$\mu$ terms.}

\subsection{Introduction}
Superstring theory, if it is to have any chance to be 
consistent with
real world, has to make definite predictions which will be 
subsequently verified by the experiment or even predict some new 
phenomena.
However in order to accomplish such a role, the theory
has first of all to solve its own 
problems. Beyond any doubt the biggest problem of all 
is the question of $N=1$ space-time supersymmetry breaking.
The breaking, due to the presence of the gravitino in the
effective action, must be spontaneous and not explicit.
This problem is crucial for the theory to make contact with
the low energy physics and to correctly predict the particle
masses.
It is expected, that the breaking will correctly create the 
hierarchy between 
the light particles by predicting exactly the Yukawa couplings of 
the light particles with the Higgs scalars.
As a result, at the level of supergravity theory  the masses of the 
physical
particles, directly connected to the soft term 
generation created
by the supersymmetry breaking, will be predicted.

String theory as the only candidate for a theory which can
consistently incorporate gravity, has still some problems.
It has a huge number of consistent vacua \footnote{We are
speaking of
supersymmetric solutions of string theory since supersymmetric 
vacua
don't suffer from stability problems \cite{giva} and furthermore 
they are known to provide a solution for the hierarchy problem.} 
without a associated mechanism which singles out one of them.
Another problem is related to the determination of physical couplings
and masses of the theory, which becomes
a complicated dynamical problem, since they depend on 
the vacuum expectation values of the dilaton and the moduli.
In the absence
of a perturbative method to exactly fix their values, this problem 
is left to be fixed from the process of supersymmetry breaking.
A third problem is associated with the calculation of
physical mass and couplings of the quark and  lepton superfields
and Higgs doublets after supersymmetry breaking.
In non-supersymmetric theories like those coming from the
standard 
model or extensions 
of it\footnote{grand unified or technicolor theories},
the scalar masses remain unprotected against 
quadratic divergencies, thus creating the gauge hierarchy problem. 
Supersymmetric gauge field theories solve
technically the gauge hierarchy problem with the introduction of  
 terms \cite{giracri} that explicitly break supersymmetry, the so called 
 soft terms.

In gauge theories the Higgs sector in unprotected against large
radiative corrections which can give very large masses to the Higgs 
particles,
due to quadratic divergencies, therefore creating a hierarchy problem.
In supersymmetric gauge theories the hierarchy problem is technically
solved since the theory is free of quadratic divergencies.
However, spontaneous breaking of global supersymmetry with the
introduction of soft breaking terms does not produce very realistic 
models. In locally supersymmetric theories
the soft breaking terms arise naturally in the low energy supergraviy
lagrangian of spontaneously broken supergravities coupled to matter
multiplets\cite{cre,nille,lah}.
Hierarchy remains stable against radiative corrections\cite{fkzw}
only when $m_{3/2} \leq\le 1$ Tev. This means that the Higgs sector
of the theory is protected against large perturbative corrections
as long as the gravitino mass obey this constraint.

However, for the spontaneously broken $N=1$ locally 
supersymmetric effective superstring theory 
\footnote{that is coming from the superstring vacuum
in the  
limit of keeping ${m_{3/2}}^2$ fixed and $k \rightarrow 0$, with k the 
gravitational coupling.} the contributions\cite{rck} to the one loop 
effective potential take the generic form
\begin{equation}
V_{1}= V_{0}+ {\frac{1}{64{\pi}^{2}}}Str{\cal M}^{0} \times
 {\Lambda}^{4}{\log \frac{{\Lambda}^{2}}{{\mu}^{2}} +
{\frac{1}{32 {\pi}^{2}}} Str{\cal M}^{2}} {\Lambda}^{2} +
{\frac{1}{64{\pi}^2}} Str {\cal M}^{4} \log \frac{{\cal M}^2}
{{\Lambda}^{2}} + \dots,
\label{effect}
\end{equation}
where $V_o$ is the classical potential of the theory or order of the 
electroweak scale. 
The general form of
\begin{equation}
Str {\cal M}^{n} \equiv \sum (-)^{2 J_{i}} (2 J_{i} + 1) m_{i}^{n}.
\label{sup}
\end{equation}
It depends on the Higgs masses and represents the sum over powers of the
field dependent mass eigenvalues of the different degrees of freedom.
The divergencies depend on the metric of the chiral and
gauge superfield content of the underling theory, are 
field dependent and are not guarantee to be vanishing.

The effects on the gauge hierarchy problem, after the
spontaneous breaking of local supersymmetry,
receive  contributions related to the the quadratically
divergent corrections to the effective potential
\begin{equation}
Str {\cal M}^2 (z, \bar z) = 2 ( n-1- G^{I} H_{I{\bar J}} G^{\bar J})
m_{3/2}^2
\label{ferraros}
\end{equation} 
with
\begin{equation}
{H_{I{\bar J}}}={\partial}_{I} {\partial}_{\bar J}
{\log det} G_{M{\bar N}}\\ 
- {\partial}_{I} {\partial}_{\bar J} \log det Re f_{ab}.
\end{equation}
So, (\ref{ferraros}) can be different\footnote{   
Here, K is the K\"ahler potential, $J_i$ counts the spin of the i-th
particle, and $f_{ab}$ the function which determines the gauge kinetic 
terms and we have assumed that the scalar manifold of the theory has 
n fields. } from zero \cite{fkzw}. 
However these contributions can be vanishing, if we 
demand
that the moduli fields transform in a scale invariant way i.e 
under the target space duality symmetries
 in the large moduli limit of 
the underling $\sigma$-model.
Of course such a scenario puts constraints \cite{nano1} on the low 
energy content of
the theory based on the need to stabilize the gauge hierarchy 
and not based on physical properties coming from an underling principle.
This is a general problem of all the string 
models\cite{fer1,fer2,fer3,dhvw,imnq} constructed up to know.
Only special classes of models are compatible with the
phenomenological requirements required to single out a particular 
vacuum \cite{iblu,fkzw,ILR}.
Other problems connected to the breaking of supersymmetry, is the 
question of the smallness of the cosmological constant problem 
and as a 
result the question of the selection mechanism which could proliferate 
string vacua. 
Non vanishing contributions to the cosmological constant
may come from the part of the quadratically divergent contributions
related to the gauge hierarchy problem as well as the 
non-perturbative moduli dependent part of the vacuum of the theory.
At present we will not be concerned with the cosmological constant
problem,
but instead we will concentrating our efforts to
the moduli of the dependent part of the potential.

String theory is a theory of only one scale, the string scale.
Physical quantities in string theory are not input free parameters
as in supergravity models. They depend on dynamical fields
whose value depend on the vacuum expectation value of the dilaton
and the moduli fields.

While a lot of work have been done at the level of the 
effective theory in order to solve the problem of
supersymmetry breaking and possibly to predict 
the value of the dilaton, the majority of the scenarios in the 
works involved have failed to properly incorporate its value.
The dilaton value is limited from the LEP \cite{CE} measurements,
giving support to extrapolations of the values of the gauge coupling 
constants, in consistency with the picture \cite{Dim} of grand unification
idea at the scale ${g_{uni}}^{2} \sim {4 \pi}/26$.
In string theory the gauge coupling constants are 'unified', by
connecting the value of the tree level gauge \footnote{see relation (3.2)}
coupling constants to the Newton coupling constant \cite{gi,al3}.
Recently, it was shown\cite{newt} that the strong coupling limit of the 
$E_8 \times
E_8$ heterotic string is given by an orbifold of the eleven 
 dimensional M-theory, known to have
as a low energy limit the eleven-dimensional supergravity.
In this picture 
the unification of the coupling constants with gravitation happens at the
grand unification scale.

The tree level value of the dilaton at the unification scale is 
ReS $\sim$ 2 and is expected that such a value will be
determined from an action which incorporates
S-duality \cite{F,SDUAL1} as well as T-duality 
invariance \cite{cfilq,Mag,LT}.

Several mechanisms have been used to break consistently supersymmetry.
The main flow of research has been concentrated in three main
 directions.
The tree level coordinate dependent compactification \cite{fkp,fkpz,kr}
mechanism - CDC,
 which 
extended the "Scherk-Schwarz" mechanism \cite{SC} in string theory, 
the non-perturbative gaugino condensation \cite{Nil,drsw,Mag,LT}
mechanism and via magnetized tori \cite{bacha1,bacha2}.

In the CDC mechanism, spontaneous supersymmetry breaking
is achieved by coupling the lattice momenta of toroidal
compactifications to the charges of a $U(1)$ current. The latter does
not
commute with the gravitino vertex operator and it therefore breaks
supersymmetry.
The net result of the investigations so far, show
that in this case CDC gives no-scale models with vanishing
potential and zero cosmological constant at the tree level of 
string theory.
The problem with in this approach is that supersymmetry is broken
but the values of the
moduli parameters are not fixed. The hope is, that they
will be fixed from radiative corrections or from non - perturbative 
phenomena. Furthermore contributions to
the cosmological constant can arise at the one loop
level.

In the magnetized tori approach, a magnetic field associated with a 
$U(1)$ gauge group generates mass splittings among the
hypermultiplets, which carry non-zero $U(1)$ charges.
The two previous approaches are distinquished 
from the fact that in the latter case, at tree level the potential
of the theory is different from zero and the gauge group
$SO(32)$ can be broken down to standard model. 
Common future of the previous two mechanisms is the impossibility
of fixing the value of the dilaton.

At this work, we will be mainly concerned with the gaugino condensation
approach. Our primary concern, is the dynamical fixing of the value of 
the dilaton. This problem is a complicated one in string theory and one
solution involves the use of multiple condensates \cite{Kras} to stabilize
its vacuum expectation value. 
This approach, in contrast with the previous one's is not a 'stringy' one
but field theoretical.

In this scenario
the non-perturbative generated superpotential for the 
composite
vector supermultiplets is responsible for the creation of the 
required hierarchy.
Whenever the gauge interactions become strong, the condensate forms
and breaks supersymmetry.
As a result, an effective potential for the moduli is generated after
the 
integration of the gauge degrees of freedom associated with the
gaugino bound states.
Then, with a typical value of the hidden SUSY breaking sector scale 
of order ${\Lambda}$ equal to $10^{13}$ Gev, the hierarchy in the low 
energy gauge sector of the 
model is stable against quantum corrections, if the gravitino mass is 
of order of
$\frac{{\Lambda}^{3}}{M_{P}^2}$ i.e of
one Tev.

Our study, uses the Hauptmodul functions of chapter three.
The existence of the S-duality 
symmetry of string theory was conjectured in \cite{FILQ,F}.
In section $(5.2)$ we will see how different parametrizations of
the non-perturbative effects, which combine T-S duality, 
provide constraints \cite{bines} that severely constraint the form of 
the effective action.  Moreover in section $(5.3)$
we want to propose a possible supersymmetry breaking 
scenario, conjectural, which 
use S-duality \cite{homo,nil1,nil2}, 
to fix the value of the dilaton. 
Furthermore in section $(5.4)$ we will discuss the 
$\mu$ term generation in orbifold compactifications .
We must say here these $\mu$ terms are part of the 
soft supersymmetry breaking terms of the effective low energy
lagrangian of the orbifold compactificxations of the heterotic 
string\cite{dhvw}.
We will see later in section $(5.4)$ the general form of the 
soft supersymmetry breaking terms. They include the "trilinear"
A-terms and the "bilinear" B-terms.
The resulting $\mu$ terms receive contributions from the 
non-perturbative superpotentials of chapter three.
Because the relevant contributions to the $\mu$ terms arise
in the one loop corrected effective action of orbifold
compactifications of the heterotic string,
they enjoy all the invariances of its effective theory
in the linear representation of the dilaton.

%*******************************************************************
%
%*********************************************************
%
%*****************************************************************

\subsection{*Constraints from duality invariance on the superpotential
and the K\"ahler potential for the globally and the locally 
supersymmetric
theory}

The effective low energy theory of string compactifications with 
$N=1$ supergravity
up to 
two space time derivatives, is described from the following functions
of chiral superfields, the gauge invariant K\"{a}hler potential
K, the  superpotential W and the gauge kinetic function f, which is 
associated with the kinetic terms of the fields in the vector multiplets.
The K\"{a}hler potential has to be a real analytic function of
chiral superfields.
The K\"ahler potential and the 
superpotential are connected together via the\footnote{We
consider for simplicity
dependence on one modulus field.}
 relation \cite{CRE}
\begin{equation}
G(z,{\bar z}) = K(z,\bar z) + \left(\log W|(z)|\right)^{2}.
\label{opi}
\end{equation}
Because the spectrum and interactions of the string vacuum are invariant 
under the appropriate T duality group and T-duality has been proven to be
a good
symmetry in any order of perturbative string theory \cite{alos},
the
effective low energy theory of the orbifold compactification
of the heterotic string on a torus has to be
invariant under the $SL(2,Z)$,$\;$T-duality group with
\begin{equation}
T \rightarrow \frac{\alpha T - i b}{i c T + d}\;.
\label{kolosaman}
\end{equation}
When considering a global supersymmetric theory,
the constraints from modular invariance on the K\"ahler potential 
and the superpotential of the theory gives 
that while K has to be invariant up to a K\"ahler 
transformation the superpotential has to be modular invariant. 
Accordingly, we can choose for 
the superpotential \cite{flst} any polynomial 
of the modular function ${\omega}$ where ${\omega}$ 
is one of the j functions for the congruence
subgroups of the $PSL(2,Z)$ which have been listed in chapter three.
In the case of local supersymmetry the constraints are different.
For effective low energy superstring theories with
$N=1$ supersymmetry the action contains the terms
\begin{eqnarray}
e^{-1} {\cal L} = e^{G} [\;3 - G_{t} G^{t {\bar t}} G_{{\bar t}}
\;]+  
e^{G/2}{\bar{\xi}}_{{\mu}R}\;{{\sigma}^{\mu \nu}}{\xi}_{{\nu}R} 
+\dots,
\label{jiok}
\end{eqnarray}
where the first factor in parenthesis is the effective potential of  the
theory and the last term depend on ${\xi}_{\mu}$  the gravitino.

From the term involving the gravitino ${\xi}_{\mu}$,
we can see that $G$ has to be modular invariant.
This can be implemented either as a separately modular invariant  
superpotential and K\"ahler potential or by keeping the whole $G$ 
expression modular invariant. 
Here we will be interested in congruence subgroups of the modular
group
$\Gamma$ and especially those appearing in the non-decomposable
orbifold constructions of the heterotic string.
In fact by using \footnote{we are following the notation 
of \cite{flst,cfilq}.} the expression \cite{shoe}
\begin{eqnarray}
\int^{T} dt \Psi(t, N)&= 4 \pi \log \frac{\eta(N T)}{\eta(T)} + C&
={\pi/6} \log \frac{\Delta(N T)}{\Delta(T)} + C,
\label{kra1a}
\end{eqnarray}
we can identify the latter modular invariant expression 
as part of $\frac{\Delta(N T)}{\Delta(T)}  =\omega(NT)$
the non-perturbative $G$ function for the non-decomposable orbifolds
based on the subgroup ${\Gamma}_{o}(N)$.  The result for the 
$\Gamma^o(N)$ group easily follows by replacing in (\ref{kra1a})
$N \rightarrow 1/N$.
Note, that we have used the relation
\begin{equation}
\Delta(T) = \eta^{24}(T),
\label{kra1akr}
\end{equation}
where $\eta$ is the Dedekind function.
The function $\Psi(T,N)$ is defined as 
\begin{equation}
\Psi(\tau, N) = N G_2(NT) - G_2(T),
\end{equation}
where T is a modulus field appearing in the low energy lagrangian,
and under T-duality transformations is transforming covariantly 
\begin{equation}
\Psi(A T, N) \rightarrow \left(c \tau + d\right )^{2} 
\Psi(T,N).
\end{equation}
The value of the $G_{2}(T)$ function is given by the 
Eisenstein series
\begin{equation}
G_2(T) = \sum_{n_1,n_2\;\in\; Z}
(i n_1 T + n_2)^{-2}=
\frac{\pi^2}{3} -8 {\pi}^{2} \sum_{m, m_{1} \succeq 1} m
q^{m m_{1}},
\end{equation}
\begin{equation}
{\tilde G_2(T)}= G_2(T) - \frac{\pi}{Re T}.
\label{kra1b}
\end{equation}
It transforms inhomogenously under $SL(2,Z)$ transformations
\begin{equation}
G_2 \rightarrow (i c T + d)^{2} G_2 -\;2 \pi i c(i c T + d).
\end{equation}

Take for example now, the non-decomposable orbifold Z6-IIb\cite{deko}.
The tree level K\"ahler potential for the untwisted subsector with
target space duality group 
$\Gamma^{o}(3)_T  $ is $K(T,{\bar T})=-\log(T +{\bar T})$.  
By using the expression 
\begin{equation}
G_{T}(T,{\bar T}) = -(1/(T + {\bar T})) + \frac{\partial \log \cal W}
{\partial T}
\end{equation}
we will be able to fix the leading term in eqn.~({\ref{kra1c}). 
The leading terms of the non-perturbative superpotential $\cal
W$ were calculated in chapter 3. The term which was associated with the 
contribution of particular point in the moduli space of the 
non-decomposable orbifold corresponding to symmetry enhancement was
found
in the leading order as follows
\begin{equation}
{\cal W} = {\omega(T)}{\eta(T)}^{-2}{\eta(U)}^{-2}+ {\dots}
\label{kra1c}
\end{equation} 
By identifying
\begin{equation}
\partial_{T}{\log \cal W} = r \frac{3}{\pi}[\frac{1}{3} G_{2}(\frac{T}{3})
- G_2(T)]+
\frac{3}{2 \pi}G_{2}(\frac{T}{3}) + \frac{1}{2 \pi}G_{2}(U) 
\end{equation}
we can recover back the result of eqn. ({\ref{kra1c}).
The previous function transforms in the proper way under modular
transformations, has modular weight $ -1$
and is the leading term in the expansion of eqn. ({\ref{kra1c}}).

\subsection{* S- and T- dual supersymmetry breaking}

While trying to solve the problems of string theory from the perturbative
framework\footnote{see chapter 4 of the thesis.},
it is the non-perturbative status of string theory which can 
at the moment give some definite answers \cite{seiwi1}.
At the level of the creation of the non-perturbative
superpotential that could give rise to dynamical determination of the 
vacuum expectation values of the dilaton and moduli fields, resulting
in
hierarchical supersymmetry breaking, the gaugino condensation mechanism 
was suggested \cite{drsw,Kras,Nil,Ross,CLMR,Dix} as a mechanism for a 
realistic supersymmetry 
breaking in string theory. The conseptual difficulty
in the above approach is that gaugino condensation by itself is a 
field theoretical phenomenon and does not provide for a consistent
skeleton which would incorporate non-perturbative effects at the
small radius limit in the $\sigma$-model sence.
Furthermore, in the original approach \cite{drsw} a vacuum with vanishing
vacuum energy and broken supersymmetry was only possible if a constant
c, coming from possible non-perturbative effects was present in the
superpotential of the theory. However, this constant is 
quantised\cite{wirohm}.

Later, we will make use of the target space modular invariance
\cite{flst,Am} together with the assumption of the existence of 
of S-duality for $N=1$ vacua, to dynamically study a way
of creating a modular superpotential with the correct
modular invariance properties for the
moduli fields coming from the compactification of our high
energy vacuum.  
At this part of the thesis, we are using the principle 
of S-duality
to examine possible dynamical mechanisms for fixing the value of 
the dilaton.
We will not give   emphasis to mechanisms which are concentrating 
only in the use of T-duality \cite{FILQ,Bin,Mag,Love1,Love2} as it 
was the approach up to now.
Models which are based solely on T-duality are clearly
not
satisfactory and the main drawback of the models existing in the 
literature\cite{homo,nil1,nil2}
is the difficulty to fix the value of the dilaton.

In general, there are two different approaches for the
non-perturbative gaugino condensation.
These are the effective lagrangian approach \cite{Mag,Tay,TGS},
where we can use a
gauge singlet gaugino bilinear superfield as a dynamical degree
of freedom and the effective superpotential approach, which
was used with
superpotentials transforming covariantly under T-duality
\cite{FILQ,Bin}, by replacing the condensate
field by its vacuum expectation value.
In the models of \cite{homo} the value of the dilaton is fixed at 
a realistic value but supersymmetry is unbroken at the minimum,
while at the models of \cite{nil1,nil2} the value of the dilaton
is at a fairly good level but the cosmological constant is negative.
Here we will use the principle of S-dual gaugino 
condensation \cite{nil1}
to describe models based on subgroups of the modular 
group. 
We will not consider the presence of hidden matter \cite{LT,LM,CCC}.
The dynamics of the effective theory of gaugino condensation
is described by the composite superfield $U = {\delta_{ab}} 
W^{a}_{\gamma} {\epsilon}_{\gamma \lambda} W^{b}_{\lambda}$,
which at the lowest order contains the gaugino bilinear as
its scalar component. Let us consider the 
superpotential of \cite{nil1}, which generalize the 
Veneziano-Yankielowicz
superpotential incorporating both $SL(2,Z)_T$ duality and $SL(2,Z)_S$
duality 
\begin{equation}
W = \frac{\Psi^3}{\eta(S)} \times
\left(\frac{1}{2 \pi}{\ln {(j(S)}} + 
3 b \ln (\Psi \eta^{2}(T)/\mu) + c\right),
\label{super}
\end{equation}
where $ {\Psi}^3 = {W}_{a}  {W}^{a}$
the value of the condensate, $\mu$ is the scale of magnitude
at which that the condensate forms. 
In \cite{nil1} the value of c was fixed from the requirement that
for S,T equal to 1 the gaugino condensate gets an expectation value
equal to $\mu$. In fact we will see that we can do more. 
We regognize j as the j-invariant modular function for $SL(2,Z)$.

The effective K\"ahler potential includes the chiral superfield
$\Psi$ which transforms under T-duality with a modular weight $-1$
and which we choose it to be
$K = -\log(S + \bar S)- 3 \ln(T +{\bar T} - {\Psi}{\bar \Psi})$. 
At the weak coupling limit $S \rightarrow \infty$ the S-duality 
superpotential (\ref{super}) must flow to
the global limit of Veneziano-Yankielowicz models, namely 
$W \approx Y^3 S$.
In this case, we have to adjust the modular prefactor in front, to
correctly recover this limit. 

In general, as it was suggested in \cite{CLMR,LT,CCC} working with 
the weak coupling limit of (\ref{super}) is equivalent to working 
with the condensate integrated out of the form of the 
superpotential.
If we integrate out the gaugino bilinear the resulting superpotential
becomes
\begin{equation}
W= \frac{\mu^3 (-c-b)e^{-\frac{c}{b}-1}}{{j(S)}^{\frac{1}{2\pi b}} 
\eta^2(S) \eta^6(T)}=\frac{ \mu^3 {\alpha}_o}{j^{\frac{1}{2\pi b}} 
\eta^2(S) \eta^6(T)},
\label{ion1}
\end{equation}
where $\alpha_o= (-c-b)e^{-\frac{c}{b}-1}$. 
Here, the constant b is equal, assuming $E_8$ gauge group, 
to $b=\beta_o(E_8)/{96 \pi^2}$.

The auxiliary fields which when their vacuum expectation value
is non-vanishing break local supersymmetry are given by
\begin{equation}
h^i = e^{\frac{1}{2} G} G^i = |W| e^{\frac{1}{2} K} \left(
K^i + \frac{W^i}{W}\right),
\label{sda221}  
\end{equation}
where K is the K\"ahler potential and W is the superpotential and
$W^S$ denoted the derivative of the superpotential with respect to
the i variable, either S ot T moduli.
The S-duality invariant superpotential will break supersymmetry
if one of the auxiliary fields, either S or T, gets non-vanishing
vacuum expactation value.
We are mostly interested if the $h^S$ will break supersymmetry.
The scalar potential of the theory is given by
\begin{equation}
V= |h^S|^2 G_{S {\bar S}^{*}}^{-1} + |h^T|^2  G_{T {\bar T}^{*}}^{-1}
-3 e^{G}.
\label{ghyu}
\end{equation}

At the moment there is some control on the $N=1$
non-perturbative aspects of heterotic string theory.
Non-perturbative contributions can appear in $N=1$
heterotic strings in the form\cite{topola} of higher weight interactions 
$\Pi^n W^{2g}$, involving chiral projections of non-holomorphic 
functions of chiral superfields. A typical amplitude at genus g 
involves 2g-2 gauginos and 2 gauge bosons.
In $N=2$ compactifications of the heteroric 
string\cite{kava,fhsv} on $K_3 \times T^2$, the non-perturbative 
contributions to 
the 
prepotential of the heterotic side are calculated\cite{mouha} from the 
exact result of the type IIA dual pair. In this way world sheet instanton
effects on type IIA are mapped on spacetime instanton effects on the 
heterotic side.
In addition, in $N=4$ non-perturbative contributions involve comparison
\cite{grGut,anpio} of
M-theory predictions with the loop dependence of $R^4$ terms 
in the effective action of type IIB or IIA.
Here, we demand that S-duality is a good symmetry of, possibly
of a formulation of string theory in a different form, string theory
when the all non-perturbative corrections are taken into account.
Since we assume that S-duality holds at the $N=1$ heterotic string 
theory, it has to hold at the level of the effective action as well.
This means that the G function of $N=1$
supergravity has to be S-duality invariant.

There are some comments that we want to make at this point.
At the time that S-duality was claimed to be valid as an  
symmetry of the $N=1$ string effective action 
the j-invariants for the subgroups of the modular group
$PSL(2,Z)$, which were clearly 
indentified in this Thesis, were completely unknown to the authors. 
In fact, a relevant 
comment of the authors in \cite{cfilq} confirms this argument.
In order to understand why S-duality could involve subgroups of the
modular group, we must first understand that there is nothing special
about $PSL(2,Z)_S$. 
All the evidence for $PSL(2,Z)_S$ duality involve $N=4$ heterotic
strings.              
So the conjectural $PSL(2,Z)_S$ for $N=1$ is a scenario
of convenience, since it gives us the dynamical mechanism
to fix the value of the dilaton.

In general it is possible to discuss supersymmetry breaking in the 
presence 
of matter fields. However, we believe that the low energy potential
of the theory\footnote{We are speaking about the requested form of 
the final 
lagrangian which can be coming from the M-\cite{var} of 
F-theory\cite{vafakos}.},
which will determine the value of the continuous parameters
of the theory, must not include matter fields in order that the 
spontaneous supersymmetry breaking to be model independent.
We should note that in Seiberg-Witten pure $SU(2)$ theory
the quantum symmetry groups $\Gamma^o(2)$ and $\Gamma_o(2)$ 
appear when the number of the hypermultiplets 
is equal to zero amd two respectively\cite{klemmos}.
So, if we imagine that this quantum symmetry group
is the low energy limit of the duality group of the theory,
then if there is S-duality present in pure $SU(2)$ Yang-Mills
it has to be $\Gamma^o(2)$ or $\Gamma_o(2)$.
This argument provides support to our claim
that the associated high energy S-duality group of the
string model might be $\Gamma^o(2)_S$ or $\Gamma_o(2)_S$.
The $N=2$ supersymmetric Yang-Mills appears at the $\alpha 
\rightarrow 0$ limit of the associated string theory vacuum.

We will now discuss the potential coming from the superpotential
\begin{equation}
W_{I} = \frac{\Psi^3}{\eta(2S)} \times
\left(\frac{1}{2 \pi}{\ln {(\omega (S)}} +
3 b \ln (\Psi \eta^{2}(T)/\mu) \right),
\label{supera}
\end{equation}
where $ {\Psi}^3 = {W}_{a}  {W}^{a}$
the value of the condensate, $\mu$ is the order of magnitude
that the condensate forms .
The prefactor of $\eta^2(2S)$ was used to provide the correct
modular weight of W and not to fix
its large S limit of (\ref{super}) following  \cite{nil1}.
We should notice at this point that the value of $\omega(S)$
represents the value of j-invariant for the congruence
subgroups of the modular group $\Gamma_{o}(2),\Gamma_{o}(3),
\Gamma^{o}(3)$ and $\Gamma^{o}(2)$ which appear in the case of
$(2,2)$ symmetric non-factorizable orbifold models, when no
continuous Wilson lines are involved.

We assumed that the superpotential has $\Gamma_0(2)$
S-duality and the gauge kinetic function f is $\Gamma_o(2)_S$
duality invariant. This means that under strong-weak coupling
duality , $1/g^2_{non-pertur} 
\stackrel{S \leftrightarrow 1/S}{\rightarrow}1/g^2_{non-pertur}$.
This implies \cite{SDUAL1,F} 
S-duality invariance of the effective actions under the 
$\Gamma_{o}(n)_S$ or $\Gamma^{o}(n)_S$ in general.
Take for example $\Gamma_o(2)$ invariance.
This means that $\textstyle{S \rightarrow \frac{S}{S+1}}$.

Integration of the bilinear condensate gives the superpotential
\begin{equation}
W_{I}= \frac{\mu^3 \alpha_0}{\eta^6(T) \eta^2(2S) (\omega(S))^{\alpha_1}},
\label{bdvdb}
\end{equation}
where $\alpha_1 = \frac{1}{2 \pi b}$ and $\alpha_o \equiv -b e^{-1}$.
The K\"ahler potential is
$K= -\log (S+ {\bar S}) - \log( T+ {\bar T} - Y {\bar Y})$.
The potential coming from (\ref{bdvdb}) is
\begin{eqnarray}
V_{I} =\frac{|W_{I}|^2}{S_R T_R^3} 
\{{ S_R^2 \left( \frac{1}{2 \pi}(G_2(2S)) 
- \alpha_12 \pi i[E_2(S) - 2 E_2(2S)]\right)}^2 -3\} + 
\frac{3 \mu^6 \alpha_o^2}{4 \pi^2 S_R T_R^3
\eta^{12}(T) }&\nonumber\\ 
\times T_R^3 G_2^2(T) \frac{1}{ \eta^4(2S)\omega(S)^{2\alpha_1}},
\label{bdvdb1}
\end{eqnarray}
where 
\begin{equation}
G_2(2S) = -4 \pi \frac{\partial_S \eta(2S)}{\eta(2 S)} - \frac{2
\pi}{S_R},\;
G_2(T) = -4 \pi \frac{\partial_T \eta(T)}{\eta(T)} - \frac{\pi}{T_R},
\label{bdvdb2}
\end{equation}
$S_R =( S + {\bar S})$ and 
\begin{equation}
T_R = T+ {\bar T} - 
\frac{\mu^6 \alpha_o^2}{\omega^{2 \alpha_1} \eta^4(2S) \eta^{12}(T)}.
\label{bdvdb3}
\end{equation}
In the decompactification limits $T_R \rightarrow \infty$, 
and its dual limit $T_R \rightarrow 0$, the potential diverges 
$V_I \rightarrow \infty$.
As a result, for gaugino condensation to happen, it is necessary that the
theory is forced to be compactified. 
The potential at the limit $S_R \rightarrow \infty$ goes to zero. It 
becomes a free theory only when $2 \pi b < 6$ holds.
This means at the weak coupling limit the dilaton cannot 
be determined from gaugino condensates.
We should note that the latter condition is more
restrictive that the analogous condition\cite{nil1} for the modular group 
$PSL(2,Z)$, namely $2 \pi b < 12$, where large gauge groups in the 
hidden sector were required to satisfy the constraint.
Because for the $E_8$ gauge group we get that  
$b|_{E_8}=90/{96 \pi^2}\approx 0.09508$ and $b< 0.9554$ we need a large
gauge group to satisfy the constraint.
Stringy constaints on the possible hidden sector gauge groups allowed
to break supersymmetry can come by the use
of higher order subgroups of the $\Gamma_o(n)$ group. Namely, for the 
$\Gamma_o(n)$ group the general form of the constraint 
$b < \frac{12}{n \pi}$ single out at least one $E_8$ group factor, only
for the modular groups 
$\Gamma_o(3)$ and $\Gamma_o(5)$. If we demand that the
form of the allowed S-duality modular symmetry group 
at the weak coupling limit to be constrained only from modular invariance,
then we could use
$\Gamma_o(7)$ or $\Gamma_o(13)$ modular groups as well. 
Of course, nothing prevents us from using, instead for $\Gamma_o(n)$, 
the $\Gamma^o(n)$ subroups of $PSL(2,Z)$ mentioned in chapter three.

The singular points of the potential can be read from the orbifold points.
The latter are extrema of the potential\cite{flst,cfilq}. 
In complete analogy, we do expect the point $S= \frac{1+i}{2}$, the fixed point 
of the modular group $\Gamma_o(2)$, to be an extremum of the 
potential. 
The auxiliary field $F_S =\textstyle{exp(\frac{1}{2}G)G_S}$ at the orbifold
point vanishes, since the function $G_2(2S)$ vanishes at the same 
point\footnote{It was note in \cite{Love1} that it can be shown
numerically that the latter holds.} 
Alternatively, we could calculate the first derivative with respect to
the T-variable.

We did not include in the fixing of the modular weight of the 
superpotential the prefactor $\eta^2(S)$.
Alternatively, if we want the prefactor in front of W to
have the correct modular weight and the
weakly coupled limit as in Veneziano-Yankielowicz models, we may have
\begin{equation}
W_{II} = \frac{{\Psi}^3}{\eta^2(S) {\omega(S)}^{\frac{1}{12}}}
\left(\frac{1}{2 \pi} {\ln {(\omega (S)}} +
3 b \ln (\frac{\Psi \eta^{2}(T)}{\mu}) \right).
\label{super2222}
\end{equation}
In this case, integrating out the condensate we get
\begin{equation}
W_{II} =  \frac{\mu^3 \alpha_o}{\eta^2(2S) {\omega(S)}^{\frac{12+2\pi b}
{24 \pi b}} \eta^6(T)}
\label{super2223}
\end{equation}
with scalar potential
\begin{eqnarray}
V_{II} =\frac{|W_{II}|^2}{S_R T_R^3} \{ \left( \frac{1}{2\pi}(G_2(2S))
+ \alpha_2 2 \pi [E_2(S) - 2 E_2(2S)]\right)^2 -3 \} +
\frac{3 \mu^6 \alpha_o^2}{4 \pi^2 S_R T_R^3 \eta^{12}(T) 
\eta^4(2S)}&\nonumber\\
\times T_R^3 G_2^2(T) \frac{1}{{\omega(S)}^{\frac{12+2\pi b}{24\pi b}}},
\nonumber\\
\label{afasa1}
\end{eqnarray}
where $\alpha_2 \equiv \frac{12 + 2 \pi b}
{24 \pi b}$, and $T_R = T + {\bar T} - \frac{\mu^6 \alpha_o^2}{
\omega^{2\alpha_2} \eta^4(2S) \eta^{12}(T)}$.

Note that the following identities hold for the Hauptmodul
of $\textstyle{\Gamma_o(2), \frac{\Delta(S)}{\Delta(2S)}}$.
\begin{equation}
\frac{\partial_S \Delta(S)}{\Delta(S)}=(2 i \pi)E_2(S),\;
\frac{\partial_S \Delta(2S)}{\Delta(2S)}=2(2 i \pi)E_2(2S).
\label{sda1}
\end{equation}
and
\begin{equation}
E_2(S)=1 - 24 \sum_n \frac{n e^{2i \pi z}}{(1 - e^{2i \pi z})}=
1-24 \sum_{i=1}^{\infty} \sigma_1(n) q^n,\;
E_S(S)=\frac{d }{dS}\log(\eta(S))         ,
\label{sda2}
\end{equation}
\begin{equation}
\partial_S \omega(S) = 2i \pi \left(E_2(S) - 2 E_2(2S)\right),\;
E_2(T)= 1- 24q -72 q^2 -96q^3-168q^4 + \dots.
\label{sda22}
\end{equation}
Here, $\sigma_{p-1}(n)$ is the divisor\cite{kobl}  $\sigma_{p-1}(n) = 
\sum_{d/n} d^{p-1}$.  
Using a numerical routine, the question of weather
supersymmetry breaking can be solved completely. Since the expressions
for the potentials are known, we can determine whether
or not the auxiliary fields connected with the modulus S or T
breaks local supersymmetry.
Numerical minimization of the potentials $V_I$, $V_{II}$ leads to same 
value $T= T_1 + iT_2 = 1.03 + i 0.54$ and $S= S_1+i S_2 =0.505 + i 0.50$.
In fact, the only difference between the two potentials 
is the different value of the $\alpha_1$ coefficient.
We observe that the minimun of the potential along the S-direction
is near the fixed point of the modular group $\Gamma_o(2)$ group.
The auxiliary S-field at the minimum breaks supersymmetry along the 
S-direction. 
S-duality invariant superpotentials can be studied alternatively from
the superpotentials of (\ref{supera}), (\ref{super2223})  
by replacing $T \rightarrow 2T$.

\subsection{* Effective $\mu$ term in orbifold compactifications}

\subsubsection{Generalities}

The hierarchy problem is solved technically in the case of $N=1$
globally
supersymmetric lagrangians with the addition of soft
breaking terms, namely soft scalar masses and trilinear and bilinear
scalar terms and soft gaugino masses.
In general spontaneously broken locally\cite{cre} supersymmetric quantum
field theories, soft terms arise
naturally from the expansion of the supergravity scalar potential
\begin{equation}
V = e^{G}[ G_{\alpha} (G^{-1})^{\alpha}_{\beta} G^{\beta} - 3].
\label{pot}
\end{equation}
Supersymmetry is spontaneously broken by the vacuum expectation values
of the hidden fields which are gauge singlets under the "observable"
gauge group. The hidden fields interact only gravitationally with
the
observable sector fields and their decoupling from the effective action 
produces the soft terms. 
The real gauge invariant K\"ahler function G is given as usual
\begin{equation}
G(z_{\alpha}, z^{*}_{\alpha})= K(z_{\alpha}, z^{*}_{\alpha})+
\log|W(z_{\alpha})|^{2},
\label{pot1}
\end{equation}
where $z_{\alpha}$ represent all scalar fields of the theory,
including observable  and hidden one's.
We assume for the K\"ahler potential and the superpotential 
has the general form
\begin{eqnarray}
K = K_{o}(h_{i},h^{*}_{i})+ K_{ij}{\phi}_{i}{\phi}^{*}_{j} +
(Z_{ij}{\phi}_{i}{\phi}_{j}+ h.c)+&\dots&\nonumber\\
W = W_{o}(h_{l}) + {\mu}_{ij} {\phi}_{i}{\phi}_{j} + Y_{ijk} 
{\phi}_{i}{\phi}_{j}{\phi}_{k} + &\dots,&
\label{pot3}
\end{eqnarray}
where the fields $h_{i}$ and ${\phi}_{i}$ correspond to the hidden
and observable sector scalar fields respectively. The ellipsis 
correspond to terms
of higher order in the fields ${\phi}_{i},\ {\phi}^{*}_{i}$. The 
terms ${\mu}_{ij},Y_{ijk},K_{ij}$ and $Z_{ij}$ depend on the
hidden sector scalar fields $h_{i}, h^{*}_{i}$.

Soft terms involve mass terms for the gauginos $\lambda_i$ and the
scalars $\phi_i$, the A term with couplings to trilinear superpotential
terms and the B term with couplings to bilinear superpotential terms.
The general form of the effective Lagrangian for the soft terms
 derived
from the expansion of the potential (\ref{pot}) is given by
\begin{equation}
L_{soft} = \frac{1}{2}\sum_{\alpha} M_{\alpha} {\tilde \lambda}
{\tilde \lambda} - \sum_{i} m_{i}^{2} |{\tilde \phi}|^{2} -
(A_{ijk} {\tilde Y}_{ijk} {\tilde \phi}_{i} {\tilde \phi}_{j}
 {\tilde \phi}_{k} + B_{ij} {\tilde \mu}_{ij} {\tilde \phi}_{i}
{\tilde \phi}_{j} + h.c),
\label{pot4}
\end{equation}
where \footnote{The tilde are canonically normalized quantities
appearing when passing  to the low energy lagrangian,
${\lambda}$ is the gaugino field.}
\begin{eqnarray}
{\tilde \phi}_{i} = K_{i}^{\frac{1}{2}} \phi_{i},\;
{\tilde \lambda}_{\alpha}={(Re f_{\alpha})}^{\frac{1}{2}}\lambda_{\alpha},\;
{\tilde Y}_{ijk}= Y_{ijk}\frac{W^{*}_{o}}{|W_{o}|}
e^{\frac{K_{o}}{2}}{(K_{i} K_{j} K_{k})}^{-\frac{1}{2}}.
\label{pot5}
\end{eqnarray}

Let us assume that our low energy theory is that of the minimal
supersymmetric standard model.
In that case the expansion of the K\"ahler potential and the
superpotential reads
\begin{equation}
K = K_{o}(h_{l}, h_{l^{*}}) + \sum K_{i}\phi_{i} \phi_{i}^{*} +
(Z H_1 H_2 + h.c),
\label{lou1}
\end{equation}
\begin{equation}
W = W_{o}(h_{l})  + 
\sum(\lambda_{e}^{ab} L^{a} E_{c}^{b} H_1 + \lambda_D^{ab}Q^{a} 
D_{c}^{b}H_1 + \lambda_U^{ab}Q^a U^b_c H_2 + \mu H_1 H_2).
\label{lou2}
\end{equation}
The summation is over all generations of fermions\footnote{
Here, $Q^a:=(3,2,1/6)$ is the left handed quarks, $U^a_c:= ({\bar
3},1,-2/3)$ the left handed antiquarks or right handed quarks,
$D^a_c:=({\bar 3},1,1/3)$ the left handed antiquarks, $L^a:=(1,2,-1/2)$
the left handed leptons and $E^a_c:=(1,1,1)$ the right handed leptons.
The $\lambda_{e}^{ab},
\lambda_D^{ab}, \lambda_U^{ab}$ are the Yukawa coupling matrices. 
The masses of the quarks and the leptons will be generated by vacuum
expectations values of the Higgs multiplets $H_1:=(1,2,-1/2)$, 
$H_2:= (1,2,+1/2)$,
in the effective low energy theory. The number in the parenthesis
represent the quantum number with respect the $SU(3)\times SU(2) 
\times U(1)_Y$, while the last entry is the weak hypercharge.}.
In eqn.(\ref{lou2}) we observe that there is a mixing term
between the two Higgs fields.
The appearance of the mass mixing term for the two Higgs fields of 
the standard model, which is necessary for the correct electroweak 
radiative breaking of the electroweak symmetry, must not happen 
through the mixing,
$W_{tree} = \mu H_{1} H_{2}$ at the superpotential $W_o$ of
the theory.
If it happens this means that the low energy parameter $\mu$, of the 
electroweak scale, is identified with a parameter of order 
of the Planck scale something unacceptable. In this case,
the $\mu$-term introduces the hierarchy problem.
On the other hand the value of the $\mu$ term cannot be zero at the
electroweak scalar potential.
If $\mu$ is zero,
the lagrangian poccess the Peccei-Quinn symmetry\cite{peqi} which  
after spontaneous symmetry breaking leads to the unwanted axion\cite{wei}. 
Take for example the potential of the supersymmetric standard model
along the neutral direction
after electroweak symmetry breaking.  Then
\begin{equation}
V(H_1,H_2)= \frac{1}{8} (g^2 + g^{\prime\;2})(|H_1|^2 - |H_2|^2)^2 +
\mu_1 |H_1|^2 + \mu_2 |H_2|^2 - \mu_3(H_1 H_2 + h.c),
\label{uat1}
\end{equation}
where 
\begin{equation}
\mu_{1,2}^2= m_{3/2}^2 + V_o +{\tilde \mu}^2,\;\;\mu_3^2= -B m_{3/2}{\tilde
\mu},\;\;{\tilde \mu}=e^{\frac{1}{2}K_o} \mu \frac{W_o^{*}}{|W_o|},
\label{uat2}
\end{equation}
$V_o$ is the cosmological constant.
Here, we have assumed that $g_3 = g_2=g_1= \sqrt{5/3} g^{\prime}$
at the unification scale and ${\tilde \mu}$ is the Higgsino mass.
From the renormalization group equations we derive that if $\mu$ is zero
then it remains zero in all energy scales.

If this is happen then such an appearance can have disastrous results
since the minimum of the potential is at $H_1=0$. In this case, 
the d-type quarks and e-type leptons stay massless, which does not
happen in reality. 
The last problem, related to the appearance of $\mu$, taken together 
with its other problem where its mass can 
 be of order $M_{Planck}$, something unphysical
for a Higgs potential of the order of the electroweak scale,
constitutes the well known $\mu$ problem and several
scenaria have appeared in previous years, providing a solution.
Clearly the presence of such a term in the superpotential of the theory,
is essential in order to avoid the breaking of the 
Peccei-Quinn \cite{peqi}
symmetry and the appearance of the unwanted \cite{wei} axion and to give 
masses to the d-type quarks and e-type leptons which otherwise will
remain massless.

Here, we explore the origin of $\mu$ terms in orbifold compactifications
of the heterotic string.
We discuss particular solutions to the $\mu$ problem related
to the generation of the mixing terms
between Higgs fields and neutral scalars in the K\"ahler potential.
We will examine the contribution of the $\mu$ terms
to the effective low energy lagrangian of N=1 orbifold   
compactifications of the heterotic string.
Alternative mechanisms for the generation of the $\mu$ term
make use of gaugino condensation \cite{Am}, to induce an effective
$\mu$ term \cite{camuno} or the presence of mixing terms in the
Ka\"hler potential \cite{gicaumas}, which induce after supersymmetry
breaking
an effective $\mu$ term given from the last two terms in
eqn.(\ref{term}) of order ${\cal O}(m_{3/2})$. The similarity of
the gaugino condensation with our appoach will be shown later.
Another solution, applicable to supergravity models, makes use of 
non-renormalizable terms (fourth or higher
order) in the superpotential. They have the form
$M_{Pl}^{1-n}A^{n}H_1 H_2$ and generate a contribution \cite{ckni} to
the $\mu$ term of order
${\tilde \mu} \sim {\cal O}(M_{Pl}^{1-n} M_{hidden}^n)$ after the hidden
fields A acquire a vacuum expectation value.

Here, we explore the origin of $\mu$ terms in orbifold compactifications
of the heterotic string.
We discuss particular solutions to the $\mu$ problem related 
to the generation of the mixing terms 
between Higgs fields and neutral scalars in the K\"ahler potential.
We will examine the contribution of the $\mu$ terms
to the effective low energy lagrangian of N=1 orbifold
compactifications of the heterotic string.
Lets us explain the origin of such mixing terms in 
superstring theory\cite{ANTO}. We assume that our effective theory of 
the massless modes after compactification is that of the heterotic 
string preserving $N=1$ supersymmetry.
The superpotential of the effective theory involves the moduli
$M_{i}$ and the observable fields $\Pi^{I}$ and has the 
general form \cite{kloui}
\begin{equation}
W = W^{tr} + W^{induced},  
\end{equation}
where
\begin{eqnarray}
W^{tr}(M_{i},{\Pi}^{I})=\frac{1}{3}{\tilde Y }_{IJL}{\Pi}^{I}
{\Pi}^{J}{\Pi}^{K}
+ ..,\;and\;&W^{in}= {\hat W}(M) + \frac{1}{2}{\tilde \mu}_{IJ}(M){\Pi}^{I}
{\Pi}^{J} + ..
\end{eqnarray}
with $W^{tr}$ the usual classical superpotential and 
$W^{induced}$ the superpotential describing our theory 
at energies below the  
the condensation scale.
The K\"ahler potential, after expanding it in powers of the matter fields 
$\Pi^{I}$ and $\bar \Pi^{I}$, takes the generic form
\begin{equation}
K = {\kappa}^{-2}{\hat K({M},{\bar M})} + Z_{{\bar I}J} {\Pi}^{\bar I}
{\Pi}^{J}
+(\frac{1}{2}H_{IJ}({\Pi},{\bar \Pi}){\Pi}^{I}{\Pi}^{J} + 
c.c)+higher\;order\;terms\;in\; \Pi,
\label{loukou}
\end{equation}
where ${\kappa}^{-2}=8 \pi/ M_{Pl}^2$.
The quantity $Z_{IJ}$ appearing in the previous equation,
represents the normalization matrix of the 
observable superfields and is renormalized to all orders of 
perturbation theory.
The corrections to the $\mu$ term that we are interested will appear
below the scale of
supersymmetry breaking.  
The calculation of the effective Lagrangian \footnote{neglecting the 
effects of electroweak symmetry breaking} 
for the moduli fields
\begin{eqnarray}
V^{eff}({\Pi},{\bar \Pi}) = {\kappa}^{-2} K_{ij} F^{i} {\bar F}^{\bar j}
- 3 {\kappa}^{2} e^{\hat K} |{\hat W}|^{2}\;,&F^{\bar j} = 
{\kappa}^{2} e^{{\hat K}/2} {\hat K}^{{\bar j}i} ({\partial_{i}}
{\hat W} + {\hat W} {\partial_{i}} {\hat K}),
\label{koukouva}
\end{eqnarray}
where $F^{j}$ the auxiliary component of the individual modulus,
gives after substituting the moduli fields with their
vacuum\footnote{at the flat limit 
$M_{Pl} \rightarrow \infty$ while keeping $M_{3/2}$ fixed.}
expectation values,
the following expressions for the masses of the observable matter
fermions and Yukawa couplings
\begin{equation}
{\mu}_{IJ} = {\tilde \mu}_{IJ} + m_{3/2} H_{IJ}
-{\bar F}^{\bar j}{\bar \partial}_{\bar j} H_{IJ},
\label{term}
\end{equation}
\begin{equation}
Y_{IJK} =  e^{{\hat K}/2} {\tilde Y}_{IJK}.
\label{termo}
\end{equation}
The previous expressions induce the effective superpotential  
\begin{equation}
W^{eff} = \frac{1}{2} {\mu}_{IJ} {\Pi}^{I}{\Pi}^{J} +
\frac{1}{3} Y_{IJK}{\Pi}^{I}{\Pi}^{J}{\Pi}^{K}.
\label{termm}
\end{equation}

After supersymmetry breaking 
the effective scalar potential for the observable 
superfields of the theory becomes \cite{sowe,bafs} equal to 
\begin{eqnarray}
V^{eff} = \sum {g_{a}}^{2} /4
\left( {\bar \Pi}^{\bar I} Z_{{\bar I}J}T_{a} {\Pi}^{J} \right)^{2}
+ \partial_{I} W^{eff}Z^{I{\bar J}} {\bar \partial}_{J} 
{\bar W}^{eff} +&\nonumber\\
m^2_{I{\bar J}} {\Pi}^{I} {\bar \Pi}^{J} +\; 
(\frac{1}{3} A_{IJK} {\Pi}^{I} {\Pi}^{J} {\Pi}^{K}
+ \frac{1}{2}B_{IJ} {\Pi}^{I} {\Pi}^{J} + h.c),
\end{eqnarray} 
with the first line to represent the scalar potential of the unbroken
rigid supersymmetry and the second line to represent the so called
soft breaking terms 
\begin{eqnarray}
m^2_{I{\bar J}} = m^2_{3/2} Z_{I{\bar J}}\ -\ F^i {\bar F}^{\bar j}
R_{i{\bar j}I{\bar J}}\;,&A_{IJL} = F^i \; D_i \;Y_{IJL},\;\nonumber\\
&B_{IJ}  = F^i D_i \mu_{IJ}\ -\ m_{3/2} \mu_{IJ},&
\label{louki}
\end{eqnarray}
and 
\begin{eqnarray}
R_{i{\bar j}I{\bar J}}= \partial_i {\bar \partial}_{\bar j} Z_{I{\bar J}}
    -\ \Gamma_{iI}^N Z_{N{\bar L}} \bar{\Gamma}^{\bar{L}}_{{\bar j}\bar
J}\;,\;&\Gamma_{iI}^N =\ Z^{N{\bar J}} \partial_i Z_{{\bar J}I},\nonumber\\
D_{i} Y_{IJL}=\ \partial_i  Y_{IJL}\ +\ \frac{1}{2}{\hat K}_i Y_{IJL}\
    -\ {{\Gamma}^{N}}_{i(I}  Y_{JL)N},&\nonumber\\
{D}_i {\mu}_{IJ}=\ \partial_i \mu_{IJ}\ +\ \frac{1}{2} {\hat K}_i  \mu_{IJ}\
    -\ {{\Gamma}^{N}}_{i(I} {\mu}_{J)N},
\end{eqnarray}
responsible for the soft breaking of supersymmetry.

We are interested in the $\mu$-term generation in  $(2,2)$
orbifold compactifications of the heterotic string. 
Let us  fix the notation \cite{dkl1} first.
We are labeling the ${\bf 27}$,$\;{\bf \bar 27}$ with latters from the
beginning (middle) of the Greek alphabet while moduli are associated with
latin characters.
The gauge group is $E_{6} \times E_{8}$, the matter fields are
transforming under the ${\bf 27}$,$\;{\bf (\bar 27)}$ representations
of the $E_{6} $, ${\bf 27}$'s are related to the $(1,1)$ moduli while
${\bf (\bar 27)}$'s are related to the $(2,1)$ moduli in the usual 
one to one correspondence.
The K\"ahler potential is given by
\begin{equation}
K = G + A^{\alpha}A^{\bar \alpha} Z_{\alpha{\bar \alpha}}^{(1,1)} + 
B^{\nu} B^{\bar \nu} 
Z_{\nu{\bar \nu}}^{(2,1)} + (A^{\alpha} B^{\nu} H_{\alpha \nu} + c.c) +
\dots
\label{papa}
\end{equation}
with the A and B corresponding to the  ${\bf 27}$'s and ${\bf {\bar
27}}$'s respectively. 
The additional contribution in the $\mu$ term  
${\tilde \mu}_{IJ}$ which appears in
eqn.~(\ref{term}) is generated from the presence of
higher weight interactions \cite{ANTO}, which are not appearing
in the standard description of the low energy
superconformal supergravity of the $(2,2)$ heterotic string
compactifications. In the superconformal tensor calculus \cite{kuue}, 
parts of the
action are constructed as the F-component parts of chiral superfields
with weight $(3,3)$. The previous notation, is 
understood to represent the general characterization 
of multiplets in the superconformal calculus, with the components of the
weight to represent the conformal and chiral weights of the dilatations
and the chiral $U(1)$ transformations of the respectively.
In this way, the 
lagrangian density for the superpotential is obtained from a term
$\left({\cal \theta}^{3} W  \right)_{F} $. The ${\cal \theta}$ is the 
compensator field with weights $(1,1)$.
The interactions are created by including in the action
chiral projections\footnote{The analog
of ${\bar D}^2$ of rigid supesymmetry} $\Pi$ acting on complex vector
superfields V of weight $(2,0)$. In general, we demand F terms 
in the action to have weights $(3,3)$.
The superpotential W is a function of fields with weights $(0,0)$ so
the lagrangian density is obtained from the F-component of 
${\theta}^{3} W$. As matter as it concerns the $\mu$ term generation in   
$(2,2)$ compactifications of the heterotic string
corresponding to the presence of the $\tilde \mu$ term,
in eqn.~({\ref{term}), the higher weight interactions responsible for 
this task are generated from terms in the form 
\begin{eqnarray}
\left( {\cal\theta}^{-3}P_{1} P_{2} \right)_{F},\;&
P_{n} \equiv \left({\cal\theta}{\bar {\cal \theta}}e^{-K/3} f^{n}\right),
\;&n=1,2.
\label{arrage}
\end{eqnarray}
Here, the subscript F denotes the F-component and $f$ are complex 
functions with weights $(0,0)$. The presence of mixing terms 
$H_{\alpha \nu}$ for the $27$ , $\bf {\bar 27}$ in the K\"ahler potential
(\ref{papa}) generates the contributions of the last two terms in
(\ref{term}). 
The presence of higher weight interactions gives the contribution
\begin{equation}
{\tilde \mu}= - h^{\bar n} {\cal W}_{ABs}G^{s{\bar s}} f_{\bar s}^{(1)} 
f_{\bar n}^{(2)},
\label{arrangene}
\end{equation}
where $W_{ABs}$ the Yukawa couplings between the scalars s and the Higgs 
moduli A, B and $G^{s{\bar s}}$ the inverse K\"ahler metric for 
the s fields. In addition, $ h^{n}$ is the auxiliary field of the 
n-th modulus field. 
We have assumed an expansion of the superpotential in the form
\begin{equation}
{\cal W}\;=\;{\cal W}_{o}\; +\; {\cal W}_{AB}\;A B.
\label{dgggin}
\end{equation}
The superpotential of the theory in the form (\ref{dgggin}) comes 
from non-perturbative effects since
terms in this form don't arise in perturbation theory due to 
non-renormalization theorems\cite{HV,dsww}.
Furthermore, because supersymmetry cannot be broken by 
any continous parameter\cite{BD1}, the origin of such terms
may not come from a spontaneous breaking version of supersymmetry but 
neccesarrily its origin must be non-perturbative.

Contribution (\ref{arrangene}) vanishes if the low energy particle 
content is that of the minimal supersymmetric standard model.
In this case, the fields s either are superheavy as 
with no Yakawa couplings with the Higgs scalars.
If the superpotential of the theory includes the mixing term 
${\cal W}_{AB}$ between the
two Higges then the $\mu$ term receives an addiitional contribution in
the form ${\tilde \mu} =e^{G/2} {\cal W}_{AB}$. In the following, we
assume
that the Higgs fields A, B are coming from the same untwisted orbifold
complex plane.

Let us assume that the low energy content of our theory is that 
of the minimal supersymmetric standard model.
We want to examine possible $\mu$ term contributions coming from
orbifold\cite{dhvw} compactifications of the heterotic string.
Lets us examine for simplicity the non-decomposable orbifold $Z_6-IIb$.
After
taking into account the result for the expression (\ref{first}) for the 
non-perturbative superpotential, the additional contribution 
$\bar \mu$ to the $\mu$ term becomes 
\begin{eqnarray}
{\tilde \mu}e^{-3S/2b}=
[\;( \eta^{-2}(T)(\frac{1}{3}){\eta^{-2}}
(\frac{U'}{3}))(\partial_T \log {\eta^{2}
(T))}(\partial_U \log \frac{1}{3}&\times&\nonumber\\
\eta^{2}(\frac{U'}{3}))\;]{\tilde W}\;+\; [\;({\eta^{-2}(U^{\prime})}
({\eta^{-2}(\frac{T}{3})}){\frac{1}{3}})(
\;((\partial_T \log {\eta^{2}(T))}&\times&\nonumber\\
(\partial_U \log \frac{1}{3} \eta^{2}(\frac{U'}{3})))\;]{\tilde W}
\;+\;{\cal O}((BC)^2),&&\nonumber\\
\label{lopolo}
\end{eqnarray}
while as matter as it concerns the observable fermion masses,
Higgino masses are given by
\begin{equation}
m = m_{3/2} + (T + {\bar T})h_{T} + (U + {\bar U})h_{U} +
(T +{\bar T})(U + {\bar U}){\tilde \mu}.
\end{equation}
In the previous expression, we have used the tree level
expressions for the wave function normalization factors, i.e
$Z_{A{\bar A}} = Z_{B{\bar B}} = ((T+{\bar T})(U + {\bar U}))^{-1}$.
The gravitino mass, which is associated with the scale of the 
spontaneous breaking of the local supersymmetry,
is given by $m_{3/2}= e^{G/2}{\cal W}$.

The presence of higher weight interactions modifies the
special geometry of $(2,2)$ compactifications and incorporates 
now the matter fields A, B associated with the 27, ${\bar 27}$'s.
In particular, the Riemann tensor $R_{\alpha {\bar \beta} {\nu} 
{\bar \mu}}$ is modified as
\begin{equation}
R_{\alpha {\bar \beta} {\nu}{\bar \mu}}= G_{\alpha {\bar \beta}}
G_{{\nu}{\bar \mu}} -W_{\alpha \nu} G^{s {\bar s}}(
e^G {\bar W}_{{\bar \beta}{\bar \mu} {\bar s}} - T_{{\bar \beta} {\bar
\mu}{\bar s}}),
\label{rieman1}
\end{equation}
where T is given by 
\begin{equation}
T_{{\bar \beta} {\bar\mu}{\bar s}}=
( \bigtriangledown f_{{\bar \mu}}^{(1} ) f_{{\bar s}}^{2)},\;
\bigtriangledown_{[{\bar k}} T_{ {\bar j}] {\bar i} {\bar s}}
=\bigtriangledown_{[{\bar k}}\left( e^G {\bar W}_{ {\bar j}]{\bar i}
{\bar s}}\right),\;T_{ {\bar j} {\bar i} {\bar s}}=e^G {\bar W}_{ {\bar
j} {\bar i} {\bar s}}.
\label{rieman2}
\end{equation}

The proposed non-perturbative superpotentials are consistent
with the use of the corrected one-loop effective action
which uses the linear representation of the dilaton.
The expansion of the superpotentials into the form
$W\;=\;W_{o}\; +\; W_{AB}\;A B$, is consistent with the invariance
of the one-loop corrected effective action
under tree level ${\Gamma}^{o}(3)_T$ transformations (\ref{treele}),
which leave
the tree level K\"ahler potential invariant, only if
$ W \rightarrow (i c T + d)^{-1} W$ and
\begin{eqnarray}
{\cal W}_{o} \rightarrow (i c T + d)^{-1}{\cal W}_{o},&{\cal W} _{AB} \rightarrow
(i c T + d)^{-1} W_{AB} + i\;c\;W_{o}.
\label{oups}
\end{eqnarray}

In the discussion so far we have tacitly identify the expression in 
eqn.(\ref{kolloii}), with the non-perturbative generated superpotential
in $(2,2)$ orbifold compactifications. This follows \cite{fklz}
from the viable identification of the expression of the 
topological free energy in  $N=1$ orbifold
compactifications with the determinant of the square mass matrix.i.e
\begin{equation}
F = \log\; \left( det(e^{K} K^{-2}_{i{\bar j}})|det W_{i{\bar j}}|^{2}
\right),
\end{equation}
where we are adopting the notations of eqn.(\ref{opi}).
Especially in a gaugino condensation approach, the gaugino condensate is
$< \lambda {\bar \lambda} > \propto W(T)$.

We must say that the grouping of terms in the form presented 
in (\ref{kolloii}) is a matter of convinience. 
Specifically, grouping together the first with the third term and the 
second with the forth term we get the result (\ref{kolloii})
and the $\mu$ term (\ref{lopolo}).
On the other hand, 
regrouping the third term in (\ref{kolloii}), we 
get\footnote{Remember, that we have changed the notation from
$U^{\prime}$
to U.} 
\begin{eqnarray}
\sum_{n^T m=0;\; q = 0} \log {\cal M} = \log ( \eta^{-2}(T)
\eta^{-2}(\frac{U}{3}) ( \frac{1}{3} ) \eta^{-2}(U)
\eta^{-2}(\frac{T}{3})\frac{1}{3})
( 1 - 4 BC  &\nonumber\\
\times \partial_T \{\log \eta(T))\log  \eta(\frac{T}{3})\}
\partial_U \{\log \eta(\frac{U}{3}) \log \eta(U) \})).
&\nonumber\\
\label{sxolio1}
\end{eqnarray}
In this case, the ${\tilde \mu}$ term contribution is 
\begin{equation}
{\tilde \mu} = \frac{-4e^{G/2} \partial_T \{\log \eta(T))\log \eta
(\frac{T}{3})\} \partial_U \{\log \eta(\frac{U}{3}) \log
\eta(U) \}}{9\eta^{2}(T) \eta^{2}
(\frac{U}{3})\eta^{-2}(U) \eta^{-2}(\frac{T}{3})}.
\label{sxolio3}
\end{equation}
The last expression appears to give the same moduli dependence, in its
numerator, up to numerical factors,
as the ansatz used for the $\tilde \mu$ term contribution to the $\mu$ term
in \cite{Love2}.
 However, the tree level contribution to the 
non-perturbative superpotential coming from (\ref{sxolio3})
does not have 
modular weight $-1$, since in this case 
\begin{equation}
{\cal W}_0 = \eta^{-2}(T)
\eta^{-2}(\frac{U}{3}) ( \frac{1}{3} ) \eta^{-2}(U)
\eta^{-2}(\frac{T}{3})\frac{1}{3}.
\label{sxolio4}
\end{equation}
In \cite{Love2} the square root of the denominator of the expression
(\ref{sxolio3}) was used as an ansatz for the $\mu$ term.
However, here we can see that the term which could produce the 
same numerator dependence arise with the wrong modular weight,
in its denumerator.
Our results saw that the ansatz used in \cite{Love2} does not arise,
from the calculation of the topological free energy of $(2,2)$
compactifications, up to ${\cal O}(AB)$ terms.

Our previous work on candidate non-perturbative 
superpotentials can be further generalized to other classes of 
non-decomposable Coxeter orbifolds.  For instance,
in the case of the $Z_{(4)}$ orbifold \cite{deko} with Coxeter twist
defined on the lattice $SU(4)^{2}$ and exhibiting 
${\Gamma}^{o}(2)_{T} \times {\Gamma}_{U}$ modular symmetry group,
the non-perturbative superpotential is 
\begin{equation}
{\cal W}_{non-pert} = (\frac{1}{2} \eta^{-2}(\frac{T}{2}) \eta^{-2}(U)
(1 - A B (\partial_T \log \frac{1}{2}\eta^{2}(\frac{T}{2}))
(\partial_U \log\eta^{2}(U))\;).
\label{tlikosi1}
\end{equation} 
The corresponding $\tilde \mu$ term is
\begin{equation}
{\tilde \mu}= e^{G/2}\frac{1}{2}\eta^{-2}(\frac{T}{2})\eta^{-2}(U)
 \partial_T\log( \frac{1}{2}\eta(\frac{T}{2}))
(\partial_U \log\eta^{2}(U)).
\label{tlikosi2}
\end{equation}
The complete list of non-perturbative superpotentials for
non-decomposable orbifolds will appear in a preprint version of the 
Thesis results, related to generalized solutions of the $\mu$ problem.

In recent popular phenomenological studies \cite{munoz1} of soft 
breaking terms in string theories\footnote{With effective low energy theory 
spectrum that gives rise to the particle spectrum of the minimal 
supersymmetric standard model.} study of soft supersymmetry breaking terms 
in the case of of $\mu$ term from Ka\"hler mixing reveals that the
effective parameter space of the theory is non-universal in the general
case, while use of tree level physical quantities in dilaton dominated 
scenaria \footnote{This scenario guarantee the smallness of 
flavour-changing neutral currents.} constraints effectively the  
parameter space in terms of two independent parameters.
The presence of K\"ahler mixing is nesessary, if we want
to avoid the appearance of a large $\mu$-term which makes the  
Higgs heavy.

The form of the $\mu$ term that we have proposed can be used 
to test observable CP violation effects in on-decomposable
orbifold compactifications of the heterotic string
in the spirit suggested in \cite{dugan,Love2}.
We should notice that we have calculated the non-perturbative 
superpotential with the correct properties in the one loop corrected
effective action in the linear representation of the dilaton for
exactly the orbifold $Z_6- IIb$ used there.

In conclusion, in ths chapter we have examined 
ansatz superpotentials invariant under a strong-weak coupling duality
based on subroups of the modular group $PSL(2,Z)$.
The values of the dilaton coming from minimization, appears to
have the same problem with superpotentials invariant under $SL(2,Z)_S$
appeared before in the literature\cite{homo,nil1,nil2,FILQ}.
The exact determination of the vacuum expectation value for the 
dilaton remains an unsolved problem. Its final solution may come 
when we will be able to perform the sum over all possible non-perturbative 
effects. 
In addition, we examined contributions to the $\mu$ terms in $(2,2)$
orbifold
compactifications coming form the presence of non-perturbative
contributions to the superpotential of $N=1$ non-decomposable orbifolds.

\newpage
\vspace{5cm}
\begin{center}
{\bf CHAPTER 6}
\end{center}
\newpage

\section{\bf Conclusions and Future Directions}

In string theory, the threshold corrections  are always 
dependent on some untwisted moduli of vector multiplets, which have the 
interpretation as parametrizing the size and shape of the underlying 
torus. This dependence comes from the 
integration of the heavy modes involved in the compactification
process.
In this Thesis, we calculated this dependence in a number
of quantities of physical interest. 

In chapter three, we used modular orbits in target space free energies,
in $N=1$ $(2,2)$ symmetric non-decomposable orbifold
compactifications \cite{dhvw,imnq,erkl}of the
heterotic string, 
to calculate the moduli dependence in 
non-perturbative superpotentials in $(2,2)$ symmetric orbifolds, 
threshold corrections to gauge 
couplings in $(2,2)$ symmetric orbifolds and threshold corrections
to gravitational couplings in $(0,2)$ $N=1$ orbifolds.
We discuss the regions of moduli space, where additional
massives stated become massless, signaling gauge symmetry
enhancement.
The same method, 
using modular orbits, has been appeared before in \cite{lust} in a
different content, where
the calculation involved decomposition of the internal lattice
in the form $T^4 \oplus T^2$.
In addition, we calculated the moduli dependence of threshold corrections
in a class of generalized Coxeter $(2,2)$ symmetric $N=1$orbifolds.
Similar
calculations have been appeared before in \cite{deko}. Our calculation 
completes the calculation of threshold corrections
in non-decomposable orbifolds, from the classification list 
of $(2,2)$ 
symmetric $N=1$ orbifolds in \cite{erkl}.
The NPS's calculated
are of major importance in the phenomenology of superstring
derived models. They may be used, for future research, in supersymmetry
breaking to determine the values of the moduli involved. Note, that
determining the exact values of the moduli
is of particular importance since it eliminates the 
moduli dependence in threshold corrections to gauge couplings.
Once this has been done, renormalization group equations 
can be used to determine whether or not the undelying string model
has any relation with the real world at energies of the 
order of the electroweak scale.
Furthermore, the identical invariance properties of the
superpotential with the invariances of the one-loop corrected
effective action (\cite{ANTO}) in the linear formulation, indicates 
that the topological nature of the superpotential is well inside the 
perturbative regime of the low energy supergravity.
In addition, it appeals very interesting to apply the methodology of
chapter three, to the calculation of the 
NPS in specific models, with dual pairs, 
coming from heterotic strings\cite{kava} compactified on 
$K_3 \times T_2$. 
Furthermore, it appears to us quite interesting to 
calculate NPS's using the method suggested in\cite{ewi}.
NPS's were calculated\cite{ewi} 
using M-theory compactified on Calabi-Yau four-folds, which gives  
$N=1$ supersymmetry in four dimensions. Using F-theory (in twelve 
dimensions)

In chapter four, we have discussed the one loop correction 
to the one loop prepotential of the vector multiplets for $N=2$
heterotic strings compactified on six dimensional orbifolds. 
The importance of our result comes from the fact, that
in $N=2$ supergravity theories the Wilsonian gauge couplings 
and the K\"ahler potential are determined from the 
holomorphic prepotential.
We have established a general procedure for calculating the 
one loop
corrections to the prepotential of the vector multiplets for $N=2$
heterotic strings compactified on six dimensional orbifolds and/or
for any compactification of the heterotic string on 
the $K_3 \times T^2$.
The difference now is that the index in the Ramond sector, of the
internal system with $(0,4)$ superconformal symmetry, 
counts the embedding of the instantons on the gauge 
bundle\cite{mouha,dose1}of $K_3$.
Our solution provides for an alternative solution to the one
appearing in \cite{mouha} where the one loop prepotential was 
calculated indirectly, with an ansatz, from its relation with the 
Green-Schwarz term. 
It should be noticed that the interesting relation between 
$K_3 \times T_2$ 
heterotic string and the dual type $II$, can be explored further.
The most important result to our opinion at this chapter, equation 
\ref{oux2} can be applied to various dual pairs\cite{kava,anpa}
and at present an ongoing investigation is well under way.

In \cite{afgnt} only the differential equation for the third derivative 
of the prepotential with respect to the T moduli was given. 
The result for the U moduli was derived by use of the mirror symmetry,
in the solution of the equation for the third derivative, 
$T \leftrightarrow U$.
In chapter four we have 
provided an alternative differential equation, from \cite{afgnt} 
for calculating the 
third derivative of the prepotential with respect 
to the U moduli. 
ACertainly, it will give the same result.
Its integral representation and the analysis of its 
properties will be given elsewhere. 

In addition, in chapter four, we calculated the heterotic 
prepotential of the $N=2$ heterotic string compactified on 
in the $(2,2)$ symmetric\cite{erkl,deko} non-decomposable orbifold 
$Z_6$ with torus lattice $SU(3) \times SO(8)$.
This model has the modular symmetry $\Gamma^o(3)_T \times 
\Gamma^o(3)$.  Let me call it A model.
The calculation was based on the modular symmetries and the 
singularity structure of the prepotential following \cite{wkll}.
Alternatively, even though there is no heterotic in $K_3 \times T^2$ 
model, known, exhibiting the same modular symmetry, it is not out of 
the question that it will not be found.
Various modular symmetry groups have appeared\cite{klm1} in the 
literature with their Seiberg-Witten theory known, but with their 
heterotic string limit unknown. 
For example take Seiberg-Witten theory. For pure $SU(2)$ theory
with the number of hypermultiplets equal to $N_f = 0$, the quantum 
duality group\cite{seiwi1,klemmos} leaving the dyon spectrum invariant 
is $\Gamma^o(2)$. In the case that 
the theory has $N_f=1$ hypermultiplets the quantum symmetry group is
$\Gamma_o(2)$. 
However, there is no
corresponding string theory model known where these group appear.
However, the Hauptmodul for $\Gamma^o(2)$ and $\Gamma_o(2)$ was 
presented\footnote{We have
calculated the heterotic string one loop correction 
to the prepotential of vector multiplets corresponding to SWT
$N_f=0,1$. The results will appear elsewhere.}
in chapter three of the Thesis.
In our case the situation, is exactly the opposite with the 
Seiberg theory model. We know the modular group of the $N=2$ 
sector of $Z_6$ or the $K_3 \times T^2$ heterotic model, to which the 
duality group corresponds, but we don't know the  
the Seiberg-Witten theory (SWT) with the same 
quantum symmmetry group. 
However, it would be interesting to understand the way that
the $K_3 \times T^2$ models could be classified, so that the 
exact string theory analog could possible found. 

Summarizing, in view of the result of the \cite{lla}, and assuming
that a Calabi-Yau dual model exists,  calculating the 
perturbative one loop prepotential at its weakly phase is equivalent
to the existence of a IIA dual on its large radius limit, or the
large complex structure limit for IIB,  defined on a bundle 
with base $P^1$ and generic fibre the $K_3$ surface.
In general terms, there is not concrete evidence that 
the duality between the heterotic string and the type II
holds everywhere in the moduli space or in specific regions
for a number of models. 
On $N=2$ heterotic strings the gauge group can certainly be
non-abelian and is bounded by its central 
charge to be less than twenty four, where the contribution of twenty 
two units comes from its internal left moving sector. The two units 
left come from the superpartner of the dilaton and the graviphoton.
On the type II side the gauge group is abelian, and non-abelian
gauge symmetry enhancement can happen at specific points in the
moduli space. There is no bound on the gauge group due to
the central charge.  The last property makes it difficult
to imagine a way such that the maximum admisible rank on
type II models match the dual heterotic ones.

In chapter five, we began by examining the way that the modular 
functions of chapter three can help us to the building of a 
theoretical model which incorporates S-duality in its structure.
The model had to obey a number of constraints, involving 
the correct modular transformations and the correct weak coupling 
limit. In fact, this model is supposed to be S-duality invariant,
for example, under the $\Gamma_o (2)_S$ congruence subroup of the 
modular group. 
This group,  appear in Seiberg-Witten theory\cite{klemmos}
for pure $SU(2)$ Yang-Mills with $N_f$ equal to zero .
Our purpose in the first part of chapter five was only 
zto determine whether or not is it possible for the dilaton
to break supersymmetry, or even fix its value, but it was
concetrated as well on the number of consistency requirements
required to build the particular superpotential.
Furthermore, we saw how the modular functions presented in
chapter three can be used to construct superpotentials
able to possible make a  prediction
for the value of the dilaton.  
Note, that predicting the dilaton value is of particular
phenomenological importance since 
its value determines the value of the string unification scale.
Because, experimental data predict that the values of the 
gauge couplings in the standard model 
seem to be unified at an energy $M_{gut}=10^{16}$ GeV, and
$\alpha_{gut}= g^2_{gut}/4\pi \approx 1/26$ is an open question 
or not whether a realistic superstring model can be build
which simultaneously can break local supersymmetry and fix
the value of the dilaton S at the value predicted by the LEP data,
namely $Re S \approx 2$.

In addition, we examined how the
calculation of the topological free energy in chapter three
can affect calculations involving $\mu$ terms 
coming from contributions of the higher weight F-terms in the 
effective theory of orbifold compactifications\cite{dhvw}.
We provide two different examples of calculating the $\mu$ terms.
However, very easily the number of examples may be increased
to cover the whole list of classification of $N=1$ Coxeter twists in
symmetric orbifold compactifications.
Remember, that $N=1$ orbifolds are the more phenomenologically 
interesting since give chiral models in four dimensions.
In addition, another way of exploring the consequences of the 
$\mu$ term contributions may be in determining the effects
on CP violation\cite{Love2,dugan} in specific models coming from 
non-decomposable orbifolds. In particular, the $Z_6 - IIB$ orbifold
which has been discussed already in the literature\cite{Love2}.

\newpage
\vspace{5cm}
\begin{center}
{\bf CHAPTER 7}
\end{center}
\newpage
{\bf  Appendix A}

The homogeneous modular group \(\Gamma' \equiv SL(2,Z)\) is defined as
the group of two by two
matrices whose entries are all integers and the determinant is one.
It is called the "full modular group and we symbolize it by $\Gamma'$.
If the above action is accompanied with the
quotient \({\Gamma} \equiv PSL(2,Z) \equiv \Gamma' /{\{ \pm 1 \} }\)
then this is called the 'inhomogeneous modular group' and we
symbolize it by $\Gamma$.
The fundamental domain of $\Gamma$ is defined \cite{la}
as the set of points
which are related through linear transformations
 $\tau \rightarrow \frac{a\tau+b}{c\tau+d}$.
If we denote $\tau = \tau_{1} + i\tau_{2}$ then the fundamental domain
of
$\Gamma$ is defined  through the relation
${\cal F} = \{ \tau \in C |\tau_{2} > 0,|\tau_{1}| \leq \frac{1}{2},
|\tau| \geq 1 \}$.
One of the congruence subgroup of the
modular group $\Gamma$ is the group $\Gamma_0(n)$.

The group $\Gamma_0(n)$ can be
represented by the following set of matrices
acting on $\tau$ as $\tau \rightarrow \frac{a\tau+b}{c\tau+d}$:
\begin{equation}
\Gamma_0(n)=\{ \left( \begin{array}{lr}
 a&b\\ c&d \end{array} \right),\; ad-bc=1,\;
\left( c=0\; mod\; n) \right\}
\label{groupe}
\end{equation}

However, we are interested  on the group $\Gamma_0(2)$.
It is generated by the elements
$T$ and $ST^2S$ of $\Gamma$. It's fundamental domain is different from
the group $\Gamma$ and is represented from the coset decomposition
$\tilde {\cal F}=\{1,S,ST\}{\cal F}$. In addition the group has cusps
at the
set of points $\{\infty, 0\}$.

We will give now some details about the integration of the integral
that we used so far.
The integration of eqn.(\ref{trigo}) is over a $\Gamma_0(2)$ subgroup of
the modular group $\Gamma$ since (\ref{trigo})
is invariant under a
$\Gamma_0(2)$ transformation $\tau \rightarrow \frac{a\tau+b}{c\tau+d}$
(with $ad-bc=1, c=0 \bmod 2$),
Under a $\Gamma_0(2)$ transformation  (\ref{trigo}) remains invariant
if at the same time we redefine our integers $n_1,n_2,l_1$ and $l_2$
in the following way:

\begin{equation}
\left(\begin{array}{cc}
n'_1& n'_2 \\
l'_1& l'_2
\end{array}\right)=\left(\begin{array}{cc}
a& c/2\\
2b&d
\end{array} \right)  
\left(\begin{array}{cc}
 n_1& n_2 \\
l_1& l_2 \end{array}\right)
\label{trans}
\end{equation}
Clearly, $c \equiv 0\;mod\;2$.
The integral can be calculated based on the method of decomposition
into
modular orbits.

There are three sets of  inequivalent orbits  under the $\Gamma_0(2)$

$1.$ The degenerate orbit of zero matrices, where after integration
over $\tilde {\cal F}=\{1,S,ST\}{\cal F}$
gives as a total contribution $ I_{0} = \pi T_2 /4 $.

$2.$  The orbit of matrices with non-zero determinants.
The following representatives give a contribution to the integral:

\begin{equation}
\left(\begin{array}{cc}
 k& j \\
 0& p \end{array}\right)\ ,\ \left(\begin{array}{cc}
 0& -p \\ k& j
\end{array}
\right)\ ,\ \left(\begin{array}{cc} 0& -p \\ k& j+p \end{array}
\right)\ ,\ \ 0\leq j < k
, \ \ p \neq 0\ .
\label{i2m}
\end{equation}

$3.$ The orbits of matrices with zero determinant

\begin{equation}
\left(\begin{array}{cc} 0& 0 \\ j& p
\end{array} \right)\ ,\ \left(\begin{array}{cc}
j& p \\
0& 0
\end{array}\right)\ ,\ j,p \in Z\ ,\ (j,p) \neq (0,0)\ .
\label{i3m}
\end{equation}

 The total contribution from the modular orbits gives,

\begin{displaymath}
\begin{array}{ll}
I_3=&\displaystyle{-4 {\Re} \ln \eta(U)-\ln
\left(\frac{T_2}{4}U_2\right) +
\left(\gamma_E-1-\ln \frac{8\pi}{3\sqrt{3}}\right)}\\
&\displaystyle{-\frac{1}{2} \times 4 {\rm Re} \ln \eta(U)-\frac{1}{2}
\times
\ln \left(T_2U_2\right)+\frac{1}{2} \times (\gamma_E-1-\ln \frac{8\pi}
{3\sqrt{3}})}
\end{array}
\end{displaymath}

The first matrix in eqn.$A.1$ has to be
integrated over the half--band
$\{\tau \in C \; \tau_2>0\; ,\;|\tau_1|<\;h\}$
as explained in ref. \cite{dkl2}. In contrast
the second matrix has to be integrated over a
half--band with the double width in $\tau_1$.


\newpage





\begin{thebibliography}{700} 
\bibitem{hone}T. Honeya, Prog. Theor. Phys. 51 (1974) 1907.
\bibitem{sce}J. Scherk and J. H  Schwarz, Nucl. Phys. B81 (1974) 118.
\bibitem{yau}S. T. Yau, Proc. Matl. Acad. Sci. 74 (1977) 1798.
\bibitem{gsw1}M. B. Green, J. H. Schwarz and E. Witten, Superstring theory,\\ 
Vol 1, (1987) Cambridge University Press.
\bibitem{gsw2}M. B. Green, J. H. Schwarz and E. Witten, Superstring theory,\\  
Vol 2, (1987) Cambridge University Press.
\bibitem{lustheis}D. Lust and S. Theisen, Lectures in String Theory,
Springer-Verlag, (1989). 
\bibitem{naaa1} K. S. Narain, M. H. Sarmadi and E. Witten, Nucl. Phys. B279\\
(1987) 369.
\bibitem{fer1}H. Kawai, D. C. Lewellen and S. H. H. Tye, Phys. Rev. \\
Lett. 57 (1986) 1832; (E)58 (1987) 429; Nucl. Phys. B288 (1987) 1.
\bibitem{fer2}I. Antoniadis, C. Bachas and C. Kounnas, Nucl. Phys. B289\\ 
(1987) 87.
\bibitem{fer3}W. Lerche, D. Lust and A. N. Schellekens, 
Nucl. Phys. B287 (1987) 477.
\bibitem{fer4}D. Bailin, A. Love and S. Thomas, Mod. Phys. Lett.A3,\\
(1988) 167.
\bibitem{dihar}L. Dixon and J. Harvey, Nucl. Phys. B274 (1986) 93.
\bibitem{ggmv}L. Alvarez-Gaume, P. Ginsparg, G. Moore and C. Vafa,\\
Phys. Lett. B171 (1986) 155.
\bibitem{marti}E. Martinec, Phys. Lett. B171 (1986) 189.
\bibitem{dsww}M. Dine, N. Seiberg, X.-G.Wen and E. Witten, Nucl. Phys.\\
B278 (1986) 769;  Nucl. Phys.B289 (1987) 319. 
\bibitem{dise12}M. Dine and N. Seiberg, Phys. Rev. Lett. 21 (1986) 2625
\bibitem{La}see: P. Langacker, Grand Unification and the standard \\
model, hep-th/9411247.
\bibitem{weinbe}S. Weinberg, Rev. Mod. Phys. 61 (1989) 1.
\bibitem{Gr}D. Gross, J. Harvey, E. Martinec and R. Rohm, Nucl. Phys B256\\
(1985) 253; B267 (86) 75.
\bibitem{SIG}C. C. Gallan, D. Friedan, E. J. Martinec amd M. Perry,\\
Nucl. Phys. B262 (1985) 593; Nucl. Phys. B278 (1986)78;
\bibitem{Gr1}D. J. Gross and J. H. Sloan, Nucl. Phys B291 (1987) 41.
\bibitem{Di1}L. Dixon, in Proc. of the 1987 ICTP Summer Workshop In High\\
Energy Physics and Cosmology, Trieste, Italy, ed. G. Furlan, J. C. Pati, \\
D. W. Sciama, E. Sezgin and Q. Shafi.
\bibitem{magha}J. Maharana and J. H. Schwarz, Nucl. Phys. B330 (1990) 3
\bibitem{dimeni}E. Witten, Phys. Lett. B149 (1984) 351.
\bibitem{Duffos}M. Duff, Clas. Quant. Grav. 5 (1988) 189.
\bibitem{sha}A. Shapere, S. Trivedi and F. Wilczek, Mod. Phys. Lett. A6\\
(1991) 2677.
\bibitem{osbo}H. Osborn, Phys. Lett. B 83 (1979) 321.
\bibitem{Seiber}N. Seiberg, Nucl. Phys. B435 (1995) 129, hep-th/9411149.
\bibitem{ibann}L. E. Ibanez, "The search for a standard model $SU(3)\\
\times SU(2) \times U(1)$ supersting, An introduction to orbifold\\
 constructions", XVII International GIFT Seminar on Theor. Physics ,\\
Escorial, Spain, World Scientific (1987)
\bibitem{vafas}C. Vafa, Nucl. Phys. B273 (1986) 592.
\bibitem{senda1}I. Senda and A. Sugamoto, Nucl. Phys B302 (1988) 291.
\bibitem{ibaniqe1}L. E. Ibanez, H. P. Nilles and F. Quevedo, Phys. Lett.\\
 B192 (1987) 332.
\bibitem{losathoma}A. Love, W. A. Sabra and S. Thomas, Nucl. Phys. \\
B427 (1994) 181.
\bibitem{fonter}A. Font, L. E. Ibanez and F. Quevedo, Phys. Lett. B217\\
(1989) 217. 
\bibitem{Disei}M. Dine and N. Seiberg, Nucl. Phys. B293 (1987) 253.
\bibitem{dis}R. Dijkgraaf, E. Verlinde and H. Verlinde, Commun.\\
Math. Phys. 115 (1988) 669;{\em On Moduli Spaces of Conformal Field\\
Theories with $c \geq 1$}, Proceedings Copenhagen Conference,{\em\\
Perspectives in String Theory},ed.by P. Di Vecchia and J. L.\\
Petersen, World Scientific, Singapore, 1988. 
\bibitem{Kik}K. Kikkawa and M. Yamasaki, Phys. Let. 149B (1984) 357;\\
N. K. Sakai and I. Senda, Progr. Theor. Phys. 75 (1986) 692. 
\bibitem{nair}V. P. Nair, A. Shapere, A. Strominger and F. Wilczek,\\
Nucl. Phys. B287 (1987) 402; B. Sathiapalan, Phys. Rev. Lett. 58 (1987)\\
1597; J. J. Atick and E. Witten, Nucl. Phys. 310 (1988) 291; R. \\
Brandeberger and C. Vafa, Nucl. Phys. B316 (1989) 391; A. Giveon,\\
E. Rabinovici and G. Veneziano, Nucl. Phys. B322 (1989) 167; M. Dine,\\
P. Huet and N. Seiberg, Nucl. Phys. B322 (1989) 301; J. Molera and\\
B. Ovrut, Phys. Rev. 40 (1989) 1146; M.Duff, Nucl. Phys. B335 (1990) 610.
\bibitem{ler}W. Lerche, D. Lust and N. P. Warner, Phys. Lett. B231 (1989)
 417.
\bibitem{witco}E. Witten, Strong coupling and the Cosmological constant,\\
 hep-th/9506101.
\bibitem{bebest}K. Becker, M. Becker ans A. Strominger, Phys. rev. D 51\\
(1995) 6603.
\bibitem{cmo}M. Cvetic, J. Molera and B. Ovrut, Phys. Lett. B248 (1990) 83.
\bibitem{alva}E. Alvarez and M. A. R. Osorio, Phys.Rev40 (1989) 1150.
\bibitem{GPR}A. Giveon, M. Porrati and E. Rabinovici, Phys.Reports 244\\
(1994)77, hep-th/9401139. 
\bibitem{can}P. Candelas, X. de la Ossa, P. Green and L. Parkes,\\
Phys. Lett. B258 (1991) 118; Nucl. Phys. B359 (1991) 21.
\bibitem{rck}S. Coleman and E. Weinberg, Phys. Rev. D7 (1973) 1888\\
S. Weinberg, Phys. Rev. D7 (1973) 2887, J. Iliopoulos, C. Itzykson and\\
A. Martin, Rev. Mod. Phys. 47 (1975) 165.  
\bibitem{kp} C. Kounnas and M. Porrati, Nucl. Phys. B310 (1988).
\bibitem{fkpz}S. Ferrara, C. Kounnas, M. Porrati and F. Zwirner,\\
Nucl. Phys. B318 (1989). 
\bibitem{fkp}S. Ferrara, C. Kounnas and M. Porrati,Phys.Let.206B (88) 25. 
\bibitem{kr}C. Kounnas and B. Rostand, Nucl. Phys. B341:641-665, 1990. 
\bibitem{vafos}C. Vafa, Lectures on Strings and Dualities,hep-th/9702201
\bibitem{dkl1}L. Dixon,V. Kaplunovsky and J. Louis, Nucl. Phys. B329 (1990) 27. 
\bibitem{kap}V. Kaplunovsky, Nucl. Phys. B307 (1988) 145 and Erratum,\\
Nucl. Phys. B382 (92) 436-438, hepth-9205070.
\bibitem{anto1}I. Antoniadis, K. S. Narain and T. Taylor,\\
Phys. Lett. B267 (1991) 37.
\bibitem{anto2}I. Antoniadis, E. Gava and K. S. Narain,\\
Phys. Lett. B283 (1992) 209; Nucl. Phys. B383 (1992) 93.
\bibitem{anto3}I. Antoniadis, E. Gava and K. S. Narain and \\
T. R. Taylor, Nucl. Phys. B407 (1993) 706.
\bibitem{ber}M. Bershadsky, S. Cecotti, H. Ooguri and C. Vafa, Nucl. Phys. B405
\\
(1993) 279 and Comm. Math. Phys. 165 (1994) 311\\
S. Hosono, A. Klemm, S. Theisen and S. T. Yau, Nucl. Phys. B433 (1995) 501.
\bibitem{kj}V. Kaplunovsky and J. Louis, Nucl. Phys. B444 (1995) 501.
\bibitem{deko}P. Mayr and S. Stieberger, Nucl. Phys. B407 (1993) 725;\\
D. Bailin, A. Love, W. A. Sabra and S. Thomas,\\ 
Mod. Phys. Letters. A9 (1994) 67; A10 (1995) 337 .
\bibitem{deko1}P. Mayr and S. Stieberger, Nucl. Phys. B407 (1993) 725;
\bibitem{bai1}D. Bailin and A. Love, Phys. Lett. B292.(1992) 315.
\bibitem{do}L. Dolan and J. T. Liu, Nucl. Phys. B387 (1992) 86-96.
\bibitem{mns}P. Mayr, H. P. Nilles and S. Stieberger,\\
Phys. Lett. B317 (1993) 53.
\bibitem{ms1}P. Mayr and S. Stieberger, Phys. Lett. B355 (1995) 107. 
\bibitem{ms2}P. Mayr and S. Stieberger, Phys. Lett. B355 (1995),\\
hep-th/9504129. 
\bibitem{erkl}J. Erler and A. Klemm, Commun. Math. Phys. 153 (1993) 579. 
\bibitem{oova}H. Ooguri and C. Vafa, Nucl. Phys. B361 (1991) 469. 
\bibitem{fklz}S. Ferrara, C.Kounnas, D.L\"{u}st and F. Zwirner,\\ 
Nucl. Phys. B365.(1991)431.
\bibitem{gi} P. Ginsparg, Phys. Lett. B197 (1987) 39. 
\bibitem{gins} P. Ginsparg, Applied Conformal Field theory;\\
Lectures given at Les Houches Summer School in Theoretical \\
Physics, Les Houches, France, Jun 28 - Aug 5, 1988:1-168.
\bibitem{dkl2}L. Dixon, V. Kaplunovsky and J.Louis,\\
Nucl. Phys. B{355} (1991) 649.
\bibitem{al1}G. Aldazabal, A. Font, L. E. Ibanez, A. Uranga \\
Nucl. Phys. B452 (1995) 3.  
\bibitem{al2}A. Font, L. E.Ibanez, F. Quevedo, Nucl. Phys. B345:\\
389-430, 1990.
\bibitem{al3}A. Font, L. E. Ibanez, H. P. Nilles and F.Quevedo\\
Nucl. Phys. B331 (1990) 421-474. 
\bibitem{ep1}K. R. Dienes and A. E. Faraggi, hep-th/9505018;\\
 hep-th/9505046. 
\bibitem{ep2}U. Amaldi, W. de Boer, P. Frampton, H. Furstenau and J.
Liu;\\
 Phys. Lett.B281 (1992) 374; I. Antoniadis, J. Ellis,\\
 S. Kelley and D.V.Nanopoulos, Phys. Lett. B272 (1991) 31;\\
A. E. Faraggi, Phys. Lett.302 (1993) 202; M. K. Gailard and R. Xiu, \\
Phys. Lett. B296 (1992) 71.
\bibitem{iba93}L.I.Ibanez, Phys. Lett. B318 (1993) 73.
\bibitem{anto}I. Antoniadis, J. Ellis, S. Hagelin and D. V. Nanopoulos\\
Phys.lett; B194 (1987) 231; B205 (1988) 459; B208 (1988) 209. 
\bibitem{laca}I. Antoniadis, J.Ellis, R. Lacaze and D. V. Nanopoulos,\\ 
Phys. Lett. B268: (1991) 188.
\bibitem{dhvw}L. Dixon, J. Harvey, C. Vafa and E. Witten,\\
Nucl. Phys. B261 (1985) 678; Nucl. Phys. B274 (1986) 285. 
\bibitem{imnq}L.E.Ib\'a\~nez, J.Mas, H.P.Nilles and F.Quevedo,\\
 Nucl. Phys. B301 (1988) 157.
\bibitem{na}K.S.Narain, Phys. Lett. B169B (1986) 41; K. S. Narain,\\ 
M. H. Sarmadi and E. Witten, Nucl. Phys. B279 (1987) 369.
\bibitem{imnq}L.E.Ib\'a\~nez, J. Mas, H. P.Nilles and\\  
F.Quevedo; Nucl. Phys. B301 (1988) 157.
\bibitem{dfkz}J. P. Derendinger, S. Ferrara, C. Kounnas and F. Zwigner\\
Nucl. Phys. B372 (1992) 145-188. 
\bibitem{kalou}V. Kaplunovsky and J. Louis, Nucl. Phys. B444 (1995) 191. 
\bibitem{kalou1}V. Kaplunovsky and J. Louis, Nucl. Phys. B422 (1994) 54\\
hep-th/9402005.
\bibitem{sv}M. A. Shifman and A. I. Vainstein, Phys. Lett. B359 (1991) 571
\bibitem{sv1}M. A.Shifman and A. I. Vainstein, Nucl. Phys. B277 (1986) 456
\bibitem{marku}D. Markushevish, M. Olshanetsky and A. Perelemov,\\
Com.Math.Phys.111 (1987) 247.
\bibitem{giio}A. Giveon and M. Porrati, Nucl. Phys. B355 (1991) 422.
\bibitem{kdva}L. Dixon, V. Kaplunovsky and C. Vafa;\\
Nucl. Phys. B294 (1987) 87.
\bibitem{iblu}L. I. Ibanez and D. Lust, Nucl. Phys. B382 (1992) 305.
\bibitem{ni}For a recent discussion see H. P. Nilles, Phenomenological\\ 
aspects of supersymmetry, Lecture at conf."Gauge Theories, Applied\\ 
supersymmetry and Quantum Gravity, Leuven, Belgium, July'95,\\
hep-th:/9511313.
\bibitem{bim}A. Brignole, L. I. Ibanez and C. Munoz,\\
Nucl. Phys. B422: (1995) 125.
\bibitem{kloui}V. Kaplunovsky and J. Louis, Phys. Lett. B306 (1993) 269.
\bibitem{nisti}H. P. Nilles and S. Stieberger, hep-th|9510009.
\bibitem{ni1}H. P. Nilles, "Dynamical gauge coupling constants",\\
Lectures given at the 195 Trieste Summer School, hep-ph/9601241.
\bibitem{difa}K. Dienes and A. Faraggi, preprint IASSNS-HEP-95-24,\\
hep-th/9510009.
\bibitem{sowe}S. K. Soni and H. A. Weldon, Phys. Lett. B126 (1983) 215.
\bibitem{kats}Y. Katsuki, Y. Kawamura, T. Kobayashi, N. Ohtsubo, Y. Ono\\
and K. Tanioka; Nucl. Phys. B341 (1990) 611.
\bibitem{spa1}M. Spalinski,Phys. Lett. B275 (1992) 47.
\bibitem{spa2}M. Spalinski; Nucl. Phys. B377 (1992) 339.
\bibitem{ersp}J. Erler and M. Spalinski, Int.J.Mod.Phys. A9\\
(1994)4407. 
\bibitem{che}M. Chemtob, hep-th/9506178.
\bibitem{bafs}V. Kaplunovsky and J. Louis,  Phys. Lett. B306 (1993) 269.
\bibitem{hand}Handbook of Mathematical functions with Formulas\\
, Graphs and Mathematical Tables, National Bureau os Standards,\\
Applied Mathematical Series, 55 Abramowitz,M and Stegun, IA ,eds.
\bibitem{gigi1}P. Ginsparg, Phys. Rev. D35 (1987) 648.
\bibitem{erja}J. Erler, D. Jungnichel and J.Lauer, Phys. Rev. D 45
(1992) 3651.
\bibitem{apostol}T. M. Apostol, Modular functions and Dirichlet series,\\
 Springer 1976.
\bibitem{kobl}N. Koblitz, 'Introductio to Elliptic curves and modular\\
forms', Springer Verlag, (1983).
\bibitem{moh1}T. Mohaupt, Int. J. Mod. Phys. A9 (1994) 4637.
\bibitem{illt}L. I. Ibanez, D. Lust, W. Lerche and S. Theisen.\\
Nucl. Phys. B352(1991)435.
\bibitem{gipo}A. Giveon and M. Porrati, Phys. Lett. B246 (0019) 54\\
Nucl. Phys. B355 (1991) 422
\bibitem{abak}I. Antoniadis, C. Bachas and C. Kounnas, Phys. Lett.\
B200 (1988)297.
\bibitem{algaa}T. W. Appelquist and J. Carrazone, Phys. Rev.D11 (1975)\\
 2856
\bibitem{ejlm}J. Erler, D. Jungnichel, J. Lauer and J.Mas;
\bibitem{blst1}D. Bailin, A. Love, W. Sabra and S. Thomas;\\
Phys. Lett. B320 (1994) 21.
\bibitem{blst2}D. Bailin, A. Love, W. Sabra and S. Thomas,\\
Mod.Phys. Lett.(1994) 1229.
\bibitem{clm1}G. L. Cardoso, D. Lust and T. Mohaupt, Nucl. Phys. B432 
(1994) 68.
\bibitem{lust}G. L. Cardoso, D. Lust and T. Mohaupt, Nucl. Phys. B450\\
(1995) 115, hep-th 9412209.
\bibitem{wein1}S. Weinberg, Phys. Lett. B 91 (1980) 51.
\bibitem{fegini}S. Ferrara, L. Girardello and H. P. Nilles, Phys. Lett.\\
B125 (1983) 457.
\bibitem{kikou} E. Kiritsis and C. Kounnas, Nucl. Phys. B442 (1995) 472.
\bibitem{zwie1}B. Zwiebach, Phys. Lett. B156 (1985) 315.
\bibitem{grsl}D. Gross and J. H. Sloan, Nucl. Phys. B291 (1987) 41.
\bibitem{thei1}S. Theisen, Nucl. Phys. B263 (1986) 687.
\bibitem{tsey1}A. A. Tseytlin,  Phys. Lett. B176 (1986) 92.
\bibitem{duff1}S. M. Christensen and M. J. Duff, Phys. Lett. B76 (1978) 571.
\bibitem{ri}P. M. Petropoulos and J. Rizos, hep-th 9601037.
\bibitem{mouha}J. Harvey and G. Moore, hep-th 9510182.
\bibitem{le}D. C. Lewellen, Nucl. Phys. B337 (1990) 61.
\bibitem{fklz}S. Ferrara, C. Kounnas, D. Lust and F. Zwigner,\\
Nucl. Phys. B365 (1991) 431.
\bibitem{dere1} J. P. Derendinger, S. Ferrara, C. Kounnas and F.\\
Zwirner, CERN-TH.6004/91 (1991).
\bibitem{dere2}J.P. Derendinger, 3rd Hellenic School on Elementary \\
Particle Physics, Proceedings 1990,World Scientific.
\bibitem{dere3}S. J. Gates, M. T. Crisaru, M. Rocek and W. Siegel,\\
 Superspace, Benjamin/Cummings, Reading, 1983.
\bibitem{L1}J. Louis, talk, 2nd Int.Symp.Boston,PASCOS proceedings,\\ 
P.Nath ed., World Scientific; SLAC-PUB-5527 (1991).
\bibitem{ovr1}G. Lopes Cardoso and B. Ovrut, Nucl. Phys. B369 (1992) 351\\
B392 (1993) 315.
\bibitem{cfilq}M. Cvetic, A. Font, D. Lust, L. E. Ibanez and F. Quevedo,
Nucl. Phys. B361 (1991) 194.
\bibitem{gu}Rc.Guuning, Lectures on modular forms (Princeton,NJ,1962)
\bibitem{la}S. Lang, Introduction to modular forms\\
(Springer, Berlin, 1976).
\bibitem{stiebe1}S. Stieberger, Ph. D thesis, TUM-HEP-220/95.
% ********************************************************
%          REFERENCES CHAPTER   4
%
%
% ********************************************************
\bibitem{seiwi1}N. Seiberg and E. Witten; Nucl. Phys. B426 (1994) 19;\\
Erratum B430 (1994) 485; B431 (1994) 484.
\bibitem{club}A. Klemm, W. Lerche, S. Theisen and S. Yankielowicz;\\
Phys. Lett. B344 (1995) 169,  hep-th, 9412158;P. Argyres and A. Faraggi,\\
Phys. Rev. Lett. 73 (1995) 3931, hep-th 9411057.
\bibitem{Calabi1}P.Candelas, G. Horowitz, A. Strominger and \\
E. Witten, Nucl. Phys. B258 (1985) 46.
\bibitem{hub}T. Hubsch, Calabi-Yau manifolds, World Scientific 1991.
\bibitem{louisne}J. Louis and K. Forger, Holomorphic couplings
in string theory, hep-th/9611184.
\bibitem{trad}D. Finnell and P. Pouliot, Nucl. Phys. B453 (1995) 225,\\
hep-th/9593115; K. Ito and N. Sasakura, Phys. Lett. B382 (1996) 95,\\
hep-th/9602073; N. Dorey, V. A. Khoze and M. Mattis, hep-th/ 9606199,\\
 hep-th/9603136, hep-th/9607202.
\bibitem{kava}S. Kahru and C. Vafa,hep-th 9505105, 
\bibitem{fhsv}S. Ferrara, J. A. Harvey, A. Strominger, C. Vafa,\\
hep-th/9505162.
\bibitem{stro1}A. Strominger, 'Massless Black Holes and Conifolds\\ 
in String Theory, hepth/9504090.
\bibitem{ftheo}C. Vafa, 'Evidence for F-theory', Nucl. Phys. B469 (1996) 
403,hep-th/9510169.
\bibitem{cv1}C. Vafa, A stingy test of the fate of the conifold,\\
hep-th 9505023.
\bibitem{stro2}B. Greene, D. Morrison and A. Strominger, `Black Hole\\
Condensation and the Unification of String Vacua' ,(hep-th/9504145).
\bibitem{cande}P. Candelas, P. Green, and T. Hubsch, Rolling Among\\
Calabi-Yau vacua, Nucl. Phys. B330(1990) 49. 
\bibitem{cande1}T. M. Chiang, B. R. Green, M. Gross, and Y. Kanter,\\ 
Blach Hole condensation and the Web of Calabi-Yau manifolds,\\
Nucl. Phys.(Proc. Suppl.) B46 (1996) 248,hep-th/9511204.
\bibitem{Gal}C. Callan, J.Harvey and A. Strominger, Nucl. Phys. B367\\
(1991) 60.
\bibitem{Hol}G. Horowitz and A. Strominger, Nucl. Phys. B360 (1991) 197
\bibitem{Pol}J. Polchinski, Phys. Rev. Lett.75 (1995) 4724.
\bibitem{stroo}A. Strominger, Open P-branes, Phys. Lett. B383 (1996) 44.
\bibitem{asmo}P. Aspinwall and D. Morrison, U-duality and Integral\\
structures, Phys. Lett. B355 (1995) 141.
\bibitem{aspo1}P. Aspinwall, Enhanced gauge symmetries and $ K_3$\\
 surfaces, Phys. Lett. B357 (1995) 329
\bibitem{cva2}C. Vafa and E. Witten, "A strong coupling test of string
duality",\\
  Nucl. Phys. B431 (1994) 3, hep-th/9408074.
\bibitem{cva1}L. Girardello, A. Giveon, M. Porrati and A. Zaffaroni,\\
Phys. Lett. B234 (1994) 331.
\bibitem{cva3}E. Witten,"On S-duality in abelian gauge
theory',hep-th/9505186.
\bibitem{cva4}C. Vafa and E. Witten, "A one loop test of string duality",\\
Nucl. Phys. B447 (1995) 261, hep-th/9505053.
\bibitem{sen1}A. Sen, "String-String Duality Conjecture in Six \\
Dimensions and Charged Solitonic Strings" , hep-th/9504027.
\bibitem{huto}C. Hull and P. Townsend, "Unity of Superstring\\ 
Dualities",Nucl. Phys. B438 (1995) 109, (hep-th/9410167).
\bibitem{var}E. Witten, String Theory Dynamics in Various\\
Dimensions,Nucl. Phys. B443 (1995) 85, (hep-th/9503124).
\bibitem{vafakos}C. Vafa, 'Evidence for F-theory', Nucl.\\
 Phys. B469 (1996) 403, hep-th/9602022.
\bibitem{hastro}J. Harvey and A. Strominger, hep-th/9504047.
\bibitem{klt} for a clear exposition of these issues see:\\
A. Klemm, W. Lerche and S. Theisen: Non perturbative effective\\ 
actions of $N=2$ Supersymmetric Gauge theories,hep-th-9505150.
\bibitem{wkll}B. de  Wit, V. Kaplunovsky, J. Louis and D. Lust,\\
Nucl. Phys. B451 (1995) 53-95.
\bibitem{afgnt}I. Antoniadis, S. Ferrara, E. Gava, S. Narain,\\
T. R. Taylor, Nucl. Phys. B447 (1995) 35-62.
\bibitem{sost}M. Sohnius, K. Stelle and P. West, Phys. Lett. B92 (1980) 123.
\bibitem{hiov}A. Hindawi, B. Ovrut and D. Waldram, Phys. Lett. B392\\
(1997) 85, hep-th-9609016.
\bibitem{cla}P. Claus, B. de Wit, M. Faux, B. Kleijn, R. Siebelink\\
and P. Termonia,  Phys. Lett. B373 (1996) 81, hep-th/9512143.
\bibitem{walt}M. Walton, Phys. Rev. D37 (1987) 377.
\bibitem{griha}P. Griffiths and J. Harris, Principles of Algebraic \\
Geometry, John Wiley and Sons, New York (1978).
\bibitem{hoth}S. Hosono, A. Klemm and S. Theisen, Lectures on Mirror 
Symmetry,\\
 hep-th/9403096, from the proceedings of the 3rd Baltic
Student Seminar, Helsinki 1993.
\bibitem{cgh}P. Candelas, P. Green and T. Hubsch, Phys. Rev. Lett. 62\\
(1989) 1956.
\bibitem{cefegi}S. Cecotti, S. Ferrara and L. Girardello, Inter. J. Mod.\\
Phys. A, (1989) 2475.
\bibitem{afiq1}G. Altazabal, A. Font, L. E. Ibanez and F. Quevedo,\\
hep-th/9510093, Nucl. Phys. B461 (1996) 85.
\bibitem{cardoo1}K. Behrndt, G. L Cardoso, B. de Wit, R. Kallosh, D.\\
Lust, T. Mohaupt, Nucl.Phys. B488 (1987) 236, hep-th/9610105.
\bibitem{klm1}A. Klemm, W. Lerche and P. Mayr, Phys. Lett. B357 (1995)\\ 
313.
\bibitem{hoso1}S. Hosono, A. Klemm, S. Theisen and S.-T.Yau, Nucl.\\
Phys. B433 (1995) 501.
\bibitem{hoso2} S. Hosono, A. Klemm, S. Theisen and S.-T.Yau, Comm.\\
Math. Phys. 167 (1995) 301. 
\bibitem{hoson3}P. Candelas, X. de la Ossa, A. Font, S. Katz and D.R.\\
 Morrison,Nucl. Phys. B416, (1994) 481.  
\bibitem{fava1}C. Vafa and E. Witten, "Dual sting pairs with $N=1$\\
and $N=2$ supersymmetry",hepth/9507050.
\bibitem{lla}P.Aspinwall and J.Louis, Phys. Lett. B369 (1996) 233,\\
P.Aspinwall, Phys. Lett. B371,(1996) 231; hep-th/9511171, Enhanced gauge\\
symmetries and Calabi-Yau three folds.
\bibitem{lla1}D. Morrison and C. Vafa, Compactifications of F-theory\\
in Calabi-Yau three folds I,II, hep-th/9603161, hep-th/9602114.
\bibitem{grple}B. Greene and R. Plesser, Nucl. Phys. B338 (1990) 15.
\bibitem{gswe}M. Green, J.H. Schwarz and P.C. West, Nucl. Phys. B254,\\
(1985) 327.
\bibitem{gs1}M. Green and J.H. Schwarz, Phys. Lett. 149B (1984) 117.
\bibitem{LWi}L. Alvarez-Gaum\'{e} and E. Witten, Nucl. Phys. B234 (1984) 269.
\bibitem{erle}J. Erler, J. Math. Phys. 35 (1994) 1819.
\bibitem{klemmos}A. Klemm, "On the Geometry behing $N=2$ Supersymmetric\\
Effective actions in four dimensions", hep-th/9705131.
\bibitem{gavos1}I. Antoniadis, E. Gava, K.S. Narain and T.R. Taylor,\\
N=2 type II-Heterotic duality and Higher derivative F-terms,\\
Nucl. Phys. B455, (1995) 109.
\bibitem{berda}M. Bershadsky, S. Cecotti, H. Ooguri and C. Vafa,\\ 
Nucl. Phys. B405 (1993) 279 and Comm. Math. Phys. 165 (1994) 311.
\bibitem{iss1}E. Witten, Nucl. Phys. B286 (1986) 79.
\bibitem{dose1}G. L. Cardoso, G. Curio and D. Lust, Perturbative \\
Couplings and Modular Forms in N=2 String Models with a Wilson Line,\\
 hep-th/9608054.
\bibitem{dose2}V. Kaplunovshy, J. Louis and S. Theisen, Phys. Lett. B357\\ 
(1995) 71.
\bibitem{dose3}T. Kawai, Phys. Lett. B371 (1996) 191, hep-th/9512046.
\bibitem{dose4}T. Kawai, hepth-9607078.
\bibitem{dose5}G. L. Cardoso, G. Curio, D. Lust and T. Mohaupt,\\
hep-th/9603108.
\bibitem{witta}E. Witten, Small instantons in String Theory, hep-th/9511030.
\bibitem{chsw}P. Candelas, G. Horowitz, A. Strominger and E. Witten,\\
Nucl. Phys. B258 (1985) 46.
\bibitem{liya}B. H. Lian and S. T. Yau, hep-th/9411234.
\bibitem{ge1}D. Gepner, Phys. Lett. B199 (1987) 370; Nucl. Phys. B296\\ 
(1988) 757.
\bibitem{anpa}M. Henningson and G. Moore, "Threshold Corrections in 
$K_3 \times T^2$\\ Heterotic String Compactifications", hep-th/9608145.
\bibitem{anpa}I. Antoniadis and H. Partouche, "Exact Monodromy group\\
of $N=2$ Heterotic Superstring", hep-th/9509009.
\bibitem{kasu}Y. Kazama and H. Suzuki, Phys. Lett. B216 (89) 112;\\
Nucl. Phys. B321 (1989) 232.
\bibitem{cve}J. Distler and B. Greene, Nucl. Phys. B304 (1988) 1;
M.Cvetic,\\
 Phys. Rev. Lett. B (1985) 1795, Phys. Rev. Lett. B59 (1987) 2829.
\bibitem{BD1}T. Banks and L. Dixon, Nucl. Phys. B307 (1988) 93.
\bibitem{BD}T. Banks and L. Dixon, Nucl. Phys. B307 (1988) 93;
T. Banks, L. Dixon,\\ D. Friedan and E. Martinec, Nucl. Phys. B299 (1988) 613.
\bibitem{DUAL} For a recent review see for example,\\
A. Giveon, M. Porrati and E. Rabinovici, Phys. Rep.244 (1994) 77.
\bibitem{gg}A. Giveon, N. Maklin and E. Rabinovici,\\
Phys. Lett. B238 (1990) 57.
\bibitem{cefergi1}S. Cecotti, S. Ferrara and M. Villasante, Int. J. Mod.\\
Phys. (1987) 1839.
\bibitem{kuue}T. Kugo and S. Uehara,  Nucl. Phys. B222 (1983) 125.
\bibitem{kokos2}C.Kokorelis, In preparation.
\bibitem{caluo}G. L. Cardoso, D. Lust and B. A. Ovrut, Nucl. Phys. B436\\
(1995) 65-99.
\bibitem{blsth}D. Bailin, A. Love , W. Sabra and S. Thomas,
Mod.Phys.Lett.A10 (1995)337.
\bibitem{NRT}M. Grisaru and W. Siegel, Nucl. Phys. B201 (1982) 192;\\
P. Howe, K. Stelle and P. West, Phys. Lett. 124B (1983) 55.
\bibitem{NS}N. Seiberg, Phys. Lett. B206 (1988)75.
\bibitem{SeibWit}N. Seiberg and E. Witten, Nucl. Phys. B426\\ 
(1994) 19, Nucl. Phys. B431 (1994) 484, (hep-th/9407087).
\bibitem{linear}B. de Wit and J. W. van Holten, Nucl. Phys. B155 (1979) 530.
\bibitem{DWVP}B. de Wit and A. Van Proeyen, Nucl. Phys. B245 (1984) 89\\
B.de Wit, P.G. Lawyers, R. Phelippe, Su S.- Q. and A. Van Proeyen,\\
Phys. Lett.134B (1984) 37.
\bibitem{Ceresole} A. Ceresole, R. D'Auria and  S. Ferrara,\\
Phys. Lett. B339 (1994) 71; A. Ceresole, R. D'Auria, S. Ferrara and A.\\
Van Proeyen, CERN preprints CERN-TH-7510-94 (hep-th/9412200),\\
CERN-TH-7547-94 (hep-th/9502072).
\bibitem{SV}M.A. Shifman and A.I. Vainshtein, Nucl. Phys. B277 (1986) 456;\\
Nucl. Phys. B359 (1991) 571.
\bibitem{HOLAN}S. Ferrara, C. Kounnas and F. Zwirner, Nucl. Phys. B372\\
(1992) 145; G. Lopes Cardoso and B. A. Ovrut, Nucl. Phys. B392 (1993) 305.
\bibitem{FVP} S. Ferrara and A. Van Proeyen, Class. Quantum Grav. 6\\
(1989) L243.
\bibitem{LLT} J. Lauer, D. L\"ust and S. Theisen, Nucl. Phys. B309 (1988) 771.
\bibitem{Seiberg} N. Seiberg, Nucl. Phys. B303 (1988) 286.
\bibitem{NAR} K. S. Narain, Phys. Lett.  B169 (1986) 41;\\
K.S. Narain, M. Sarmadi and E. Witten, Nucl. Phys. B279 (1987) 369.
\bibitem{HIGGS} M. Dine, P. Huet and N. Seiberg,Nucl.Phys\\
B322 (1989) 301;\\
L. Ibanez, W. Lerche, D. L\"ust and S. Theisen, Nucl. Phys.\\ 
{B352} (1991) 435.
\bibitem{CLM2}G. Lopes Cardoso, D.L\"ust and T. Mohaupt, Nucl. Phys. B432 \\
(1994) 68.
\bibitem{ANTO}I. Antoniadis,E. Gava, K. S. Narain and T. R. Taylor\\
,Nucl. Phys. B432 (1994) 187.
\bibitem{DWLVP} B. de Wit,P. Lauwers and A. Van Proeyen,\\
  Nucl. Phys. B255 (1985) 569.
\bibitem{special}S. Ferrara and A. Strominger in {\em Strings '89},\\
eds.R. Arnowitt, R. Bryan, M. J. Duff, D. V. Nanopoulos and\\
C. N. Pope (World Scientific, 1989), p.~245; A. Strominger, Comm.\\
Math. Phys. 133 (1990) 163; L.J. Dixon, V.S. Kaplunovsky and J. Louis,\\
Nucl. Phys. B329 (1990) 27;P. Candelas and X.C. de la Ossa, Nucl.~Phys.\\
B355 (1991)455;L. Castellani, R.D' Auria and S. Ferrara, Phys. Lett.\\
B241  (1990) 57; Cl. Q. Grav. 7 (1990) 1767; R. D' Auria, S. Ferrara\\
and P. Fr\'e, Nucl. Phys. B359 (1991) 705.
\bibitem{BEC} E. Cremmer, C. Kounnas, A. Van Proeyen, J.P.
Derendinger,\\
 S. Ferrara, B. de Wit and L. Girardello, Nucl. Phys.\\
B250 (1985) 385.
\bibitem{SU} B. de Wit, P.G. Lauwers, R. Philippe, Su,S.-Q.\\
and A. Van Proeyen, Phys.~Lett. 134B (1984) 37.
\bibitem{DWVVP}B. de Wit and A. Van Proeyen, Commun.Math.Phys.\\
149 (1992) 307, Phys. Lett. B293 (1992) 94;\\
B. de Wit, F. Vanderseypen and A. Van Proeyen,\\
Nucl. Phys. B400 (1993) 463.
\bibitem{FLT}S. Ferrara, D. L\"ust and S. Theisen,\\
Phys. Lett. B242 (1990) 39.
\bibitem{Shevitz} D. Shevitz, Nucl. Phys. B338 (1990) 283.\\
S. Ferrara, C. Kounnas, D. L\"{u}st and F. Zwirner,\\ 
Nucl. Phys. B365.(1991)431
\bibitem{DY} M. Dine and Y. Shirman,  Phys.Rev. D 50 (1994) 5389.
\bibitem{nilles} H. P. Nilles, Phys. Lett. 180B (1986) 240.
\bibitem{DKL} L. Dixon, V. Kaplunovsky and J. Louis, Nucl. Phys.\\
B355 (1991) 649.
\bibitem{AGN}I. Antoniadis, E. Gava and  K. S. Narain,\\
Phys.~Lett. 283B (1992) 209, Nucl. Phys. B383 (1992) 93.
\bibitem{SFetal}S. Ferrara, L. Girardello, C. Kounnas\\
and M. Porrati,  Phys. Lett. B192 (1987) 368;\\
S. Ferrara and M. Porrati, Phys. Lett. B216 (1989) 289;\\
P. Fr\'e and P. Soriani, Nucl. Phys. B371 (1992) 659.
\bibitem{SEN}  A. Sen, Int. J. Mod. Phys. A9 (1994) 3707.
\bibitem{mool}C. Montogen and K. Olive, Phys. Lett. B72 (1977) 117.
\bibitem{SDUAL1}A. Font, L. Ibanez, D. L\"ust and F. Quevedo,\\
  Phys. Lett. B249 (1990) 35; S. Rey, Phys. Rev. D43 (1991) 256.
\bibitem{SDUAL2}
  A. Sen, Phys. Lett. B303 (1993) 22, Phys. Lett. B329 (1994) 217;\\
J. Schwarz and A. Sen, Phys. Lett. B312 (1993) 105, Nucl. Phys.\\
B411 (1994) 35; J.~Gaunlett and J.~Harvey, hep-th/9402032.
\bibitem{DWVHVP}B. de Wit,J. W. van Holten and A. Van Proeyen,\\
  Nucl. Phys. B184 (1981) 27.
\bibitem{DWPVP}B. de Wit,R. Philippe and A. Van Proeyen,Nucl.\\
  Phys. B219 (1983) 143.
%   **************************************
 %   CHAPTER[ 5  REFERENCES
%
%    **************************************
\bibitem{cre}E. Cremmer,S. Ferrara,L. Girardello and A. Van Proyen,
Nucl. Phys. B212 (1983) 413.
\bibitem{nille}H. P. Nilles, Phys. Rep. 110 (1984) 1.
\bibitem{lah}A. B. Lahanas and D. V. Nanopoulos, Phys. Rep.145 (1987) 1.
\bibitem{flst}S. Ferrara, D. Lust, A. Shapere and S. Theisen;\\
Phys. Lett. B225 (1989) 363.
\bibitem{rohm}R. Rohm, Nucl. Phys. B237 (1984) 253.
\bibitem{giva}P. Ginsparg and C. Vafa, Nucl. Phys. B289 (1987) 414.
\bibitem{giracri}L. Girardello and M. T. Crisaru, Nucl.Phys. B 194\\
(1982) 65.
\bibitem{newt}E. Witten, Nucl. Phys. B471 (1996) 135.
\bibitem{wirohm}E. Witten and R. Rohm, Ann. Phys. 170 (1986) 454.
\bibitem{homo}J. H. Horne and G. Moore, Nucl. Phys. B432 (1994) 109
\bibitem{nil1}Z. Lalak, A. Niemeyer, H. P. Nilles, Phys. Lett. B349 (1995) 99
\bibitem{nil2}Z. Lalak, A. Niemeyer, H. P. Nilles, Nucl. Phys. B453 (1995)100
\bibitem{shoe}B. Shoeneberg, Elliptic Modular Functions. (Berlin 1974)
\bibitem{CE}J. Ellis, S. Kelley and D. V. Nanopoulos, Phys. Lett. B260 \\
(1991) 131; U. Amaldi, W. de Boer and H. Furstenau, Phys. Lett. B260 (1991)\\
447; P. Langacker and M. Luo, Phys. Rev. D44 (1991) 817; R. G. Roberts and \\
G.G.Ross, RAL-92-005 (1992).
\bibitem{fkzw}S. Ferrara, C. Kounnas and F. Zwigner,\\
Nucl. Phys. B429 (1994) 589-625, ERRATUM-ibid. B433 (1995)255.
\bibitem{nano1}J. Lopez and D. V. Nanopoulos, CERN-TH.7519.94,\\
(hep-ph/9412332).
\bibitem{alos}E. Alvarez and M. Osorio, Phys. Rev. D40, (1989) 1150.
\bibitem{Dim}S. Dimopoulos, S. Raby and F. Wilczek, Phys. Rev. D24\\
(1981)1681;L. E. Ibanez and G. G. Ross, Phys. Lett. B105 (1981)\\
439; S. Dimopoulos and H. Georgi, Nucl. Phys. B193 (1981) 375;\\
M. Einhorn and D. R. T. Jones, Nucl. Phys. B196 (1982) 475.
\bibitem{Dine}M. Dine and N. Seiberg, Phys. Rev. Lett. 57 (1986) 2625.
\bibitem{Nil}H. P. Nilles, Phys. Lett. 115B (1982) 193; S.\\
Ferrara, L. Girardello and H. P. Nilles, Phys. Lett.125B (1983)\\
457; J. P. Derendinger, L. E. Ibanez and H. P. Nilles, Phys. Lett. 155B\\ 
(1985) 65; C.Kounnas and M. Porrati, Phys. Lett. B191 (1987) 91.
\bibitem{drsw}M. Dine, R. Rohm, N. Seiberg and E. Witten,\\
Phys. Lett. 156B (1985) 55.
\bibitem{SC}N. Scherk and J. Schwarz, Nucl. Phys. B153 (1979) 161.
\bibitem{bacha1}C. Bachas, "A way to break Supersymmetry"
hep-th:9503030,\\
 CPTH-R349-0395.
\bibitem{bacha2}C. Bachas, Talk presented at the meeting on Topics\\
in the theory of fundamental interactions, Maynooth, Ireland\\
(1995), hep-th/9509067. 
\bibitem{topola}I. Antoniadis, E. Gava, K.S. Narain and T.R. Taylor,\\
'Topological amplitudes in N=1 heterotic superstring theory', Nucl. Phys.\\
 B476 (1996) 133.
\bibitem{grGut}M. Green and M. Gutperle, Nucl. Phys. B498 (1997) 195, hep-th/9701093.
\bibitem{anpio}I. Antoniadis, B. Pioline and T.R. Taylor, 'Calculable 
$e^{1/{\lambda}}$ effects',\\
 hep-th/ 9707222. 
\bibitem{Dix} L. Dixon, talk presented  at the A.P.S.D.P.F.\\
Meeting at Houston (1990); V. Kaplunovsky, talk presented at the\\
"Strings 90" workshop at College Station (1990); L.Dixon,\\
V.Kaplunovsky, J.Louis and M. Peskin, unpublished.
\bibitem{CLMR}J. A. Casas, Z. Lalak, C. Munoz and G. G. Ross,\\
Nucl. Phys. B347 (1990) 243
\bibitem{FILQ}A. Font, L. Ibanez, D. Lust and F. Quevedo\\,
Phys. Lett. B245 (1990) 401; M. Cvetic, A. Font, L. Ibanez,\\
D.Lust and F. Quevedo, Nucl. Phys. B361 (1991) 194.
\bibitem{F}A. Font, L. Ibanez, D. Lust and F. Quevedo, Phys. Lett. B249\\ 
(1990) 35, S. Rey, Phys. Rev. D43 (1991) 256.
\bibitem{F3}S. Kalara and D. V. Nanopoulos, Phys. Lett. B267 (1990) 35\\
A. Sen, Int. Jour. Mod. Phys. A9 (1994) 3707, (hep-th/940202).\\
J. Schwarz and A. Sen, Phys. Lett. B312 (1993) 105, (hep-th/9305185).
\bibitem{bines}P. Binetry and M. Gaillard, LBL-37198, (hep-th/9506207).
\bibitem{CRE}E. Cremmer, S. Ferrara, L. Girardello and A. Van Proyen\\
Nucl. Phys. B212 (1983) 413.
\bibitem{Mag}S. Ferrara, N. Magnoli, T. R. Taylor and G. Veneziano\\
Phys. Lett. B245 (1990) 409.
\bibitem{Bin} H. P. Nilles and M. Olechowsky, Phys.Lett B248 (1990) 268\\ 
P.Binetruy and M. K. Gaillard, Phys. Lett. B253 (1991) 119.
\bibitem{LT}D. Lust and T. R. Taylor, Phys. Lett. B253 (1991) 35
\bibitem{CCC}B. de Carlos, J. A. Casas and C. Munoz,Phys. Lett. B263\\
(1991) 248.
\bibitem{Lo}J. Louis, SLAC-PUB-5645 (1991).
\bibitem{LM}D. Lust and C. Munoz, CERN--TH.6358/91 (1991).
\bibitem{FLST}S. Ferrara, D. Lust, A. Shapere and S. Theisen,Phys. Lett.\\
 B225 (1989)363;S. Ferrara, D. Lust and S. Theisen, Phys. Lett. B233 (1989)\\
 147.
\bibitem{SV} H. P. Nilles, Phys. Lett. B180 (1986) 240;\\
M. A. Shifman and A. I. Vainshtein, Nucl. Phys. B359 (1991) 571;\\
I. Antoniadis, K. S. Narain and T. R. Taylor, Phys.Lett. B267 (1991) 37;\\
J. A. Casas and C. Munoz, Phys. Lett. B271 (1991) 85.
\bibitem{BQ}I. Antoniadis, J. Ellis, A. B. Lahanas and D. V.\\
Nanopoulos, Phys. Lett. B241 (1990) 24; S. Kalara, J. Lopez and\\
D. V. Nanopoulos, CTP--TAMU--69/91 (1991); S. Kelley, J. L. Lopez and\\
 D. V.Nanopoulos, CTP-TAMU-105/91 (1991).
\bibitem{Kras}N. V. Krasnikov, Phys. Lett. B193 (1987) 37
\bibitem{Tay}T. R. Taylor, Phys. Lett. B252 (1990) 59
\bibitem{Crem}E. Cremmer, S. Ferrara, L. Girardello and A.\\
Van Proeyen, Nucl. Phys. B212 (1983) 413.
\bibitem{HV}S. Hamidi and C. Vafa, Nucl. Phys. B279 (1987) 465;\\
L. Dixon, D. Friedan, E. Martinec and S. Shenker, Nucl. Phys. B282\\
(1987) 13;T. T. Burwick, R. K. Kaiser and H. F. M\"{u}ller,\\
Nucl. Phys. B355 (1991) 689; T. Kobayashi and N. Ohtsubo,\\
DPKU--9103.
\bibitem{Love1}D. Bailin, A. Love, W. A. Sabra, and S. Thomas,\\
'Anisotropic solutions for orbifold moduli from dualzity\\
invariant gaugino condensates', Mod. Phys. Lett. A9 (1994) 2543,\\ 
hep-th/9405031.
\bibitem{Love2}B. Acharya, D. Bailin, A. Love, W. A. Sabra, and\\ 
S. Thomas, Spontaneous breaking of CP by Orbifold Moduli, \\
Phys. Lett. B357 (1995) 387, hep-th/9506143.
\bibitem{dugan}M. Dugan, B. Grinstein and L. Hall, Nucl. Phys. B255\\ 
(1985) 413.
\bibitem{ewi}E. Witten, "Non-Perturbative Superpotentials in String 
Theory", hep-th/9604030.
\bibitem{pot}C. Munoz, "Soft supersymmetry breaking terms and the $\mu$ 
problem", FTUAM 95/20, Based on talks at Boston Workshop. 
\bibitem{MK}S. Mahapatra, Phys. Lett. B223 (1989)47; Y. Katsuki,\\
Y.Kawamura, T. Kobayashi, N. Ohtsubo, Y. Ono and K. Tanioka,\\
Nucl. Phys. B341 (1990) 611.
\bibitem{INQ}L. E. Ibanez, H. P. Nilles and F. Quevedo,Phys.\\
Lett. B187 (1987) 25;Y. Katsuki, Y. Kawamura, T. Kobayashi\\
N. Ohtsubo, Y. Ono and K. Tanioka, Prog. Theor. Phys. 82 (1989) 171,\\
 DPKU--8904 (1989).
\bibitem{ILR}L. E. Ib\'anez, D. L\"ust and G. G. Ross, Phys. Lett.\\
B272 (1991) 251.
\bibitem{KLN}S. Kalara, J. Lopez and D. V. Nanopoulos, Phys. Lett.\\
B269 (1991) 84.
\bibitem{AEHN}I. Antoniadis, J. Ellis, J. S. Hagelin and D.V.\\
Nanopoulos, Phys. Lett. B205 (1988) 459, B213 (1988) 56.
\bibitem{CMMM}J. A. Casas, A. de la Macorra, M. Mondragon and C.Munoz,\\
 Phys. Lett. B247 (1990) 50.
\bibitem{kaka}Y. Katsuki, Y. Kawamura, T. Kobayashi, Y. Ono\\
and K. Tanioka, Phys. Lett. B218 (1989) 169.
\bibitem{TGS}T. R. Taylor, G. Veneziano and S. Yankielowicz\\
Nucl. Phys. B218 (1983) 493.
\bibitem{Aff}I. Affleck, M. Dine and N. Seiberg, Nucl. Phys.\\
B241 (1984) 493.
\bibitem{Am}D. Amati, K. Konishi, Y. Meurice, G. C. Rossi and\\
G. Veneziano, Phys. Rep. 162 (1988) 169.
\bibitem{Wit}E. Witten, Phys. Lett. B155 (1985) 151; S. Ferrara,\\
C. Kounnas and M. Porrati, Phys. Lett. B181 (1986) 263; M. Cveti\u{c},\\
J. Louis and B. Ovrut, Phys. Lett. B206 (1988) 227.
\bibitem{Ross} G. G. Ross, Phys. Lett. B211 (1988) 315.
\bibitem{peqi}R. D. Peccei and H. Quinn, Phys. Rev. D 16,(1977) 1791.
\bibitem{wei}S. Weinberg, Phys. Rev. Lett. 40 (1978) 223; F. Wilczeck,\\
  Phys. Rev. Lett. 40 (1978) 229; W. A. Bardeen and S.-H.H. Tye,\\
Phys. Lett. 74B (1978) 229.
\bibitem{camuno}J. A. Casas and C. Munoz, Phys. Lett B306 (1993) 288.
\bibitem{ckni}J. E. Kim and H. P. Nilles, Symmetry Principles toward\\
solutions of the $\mu$ problem, Mod. Phys. Lett. A9 (1994) 3575. 
\bibitem{gicaumas}G. F. Giudice and A. Masiero, Phys. lett. B206, (1986)\\
480.
\bibitem{munoz1}A. Brignole, L. E. Ibanez and C. Munoz, Nucl. Phys.\\
B422. (1994) 125.
\end{thebibliography}
\end{document}